\documentclass[review,onefignum,onetabnum]{siamonline190516}

\usepackage{lipsum}
\usepackage{amsfonts}
\usepackage{graphicx}
\usepackage{epstopdf}
\usepackage{algorithmic}
\usepackage{subfig}
\ifpdf
  \DeclareGraphicsExtensions{.eps,.pdf,.png,.jpg}
\else
  \DeclareGraphicsExtensions{.eps}
\fi
\usepackage{comment}
\nolinenumbers

\def\xv{{\mathbf x}}
\def\yv{{\mathbf y}}
\def\zv{{\mathbf z}}
\def\vv{{\mathbf v}}
\def\wv{{\mathbf w}}
\def\fv{{\mathbf f}}
\def\gv{{\mathbf g}}
\def\hv{{\mathbf h}}

\def\nv{{\mathbf n}}
\def\mv{{\mathbf m}}

\def\tv{{\mathbf t}}
\def\Jm{{\mathbf J}}
\def\xiv{{\mathbf \xi}}

\def\g{\hat{f}}
\def\xiv{{\mathbf \xi}}
\def\R{{\mathbb R}}

\def\S{{\mathcal S}}
\def\I{{\mathcal I}}

\def\RM{{\mathcal R}}
\def\v1{{\mathbf 1}}

\usepackage{enumitem}
\setlist[enumerate]{leftmargin=.5in}
\setlist[itemize]{leftmargin=.5in}

\newsiamremark{remark}{Remark}
\newsiamremark{hypothesis}{Hypothesis}
\crefname{hypothesis}{Hypothesis}{Hypotheses}
\newsiamthm{claim}{Claim}

\headers{Using Decoupled Features for Photo-realistic Style Transfer}{T. D. Canham, A. Martín, M. Bertalmío, and J. Portilla}

\title{Using Decoupled Features for Photo-realistic Style Transfer\thanks{Submitted to the editors July 29th, 2022. \funding{This work has received funding from the Spanish Government grants PID2020-118071GB-I00, PID2020-113596GB-I00 and PID2021-127373NB-I00, and from the European Union’s Horizon 2020 research and innovation program under grant agreement number 952027 (project AdMiRe).}}}

\author{Trevor D. Canham\thanks{Department of Electrical Engineering and Computer Science, York University, Canada 
  (\email{tcanham@yorku.ca}).}
\and Adrián Martín\thanks{Corsmed AD, Stockholm, Sweden (\email{adrianmartinfdez@gmail.com}).}
\and Marcelo Bertalmío\thanks{Instituto de Optica, CSIC, Spain 
(\email{marcelo.bertalmio@csic.es}, \email{javier.portilla@csic.es}).}
\and Javier Portilla\footnotemark[4]}

\usepackage{amsopn}

\ifpdf
\hypersetup{
  pdftitle={An Example Article},
  pdfauthor={D. Doe, P. T. Frank, and J. E. Smith}
}
\fi

\begin{document}

\maketitle

\begin{abstract}
In this work we propose a photo-realistic style transfer method for image and video
that 
is based on vision science principles and on a recent mathematical formulation for the deterministic decoupling of sample statistics.
The novel aspects of our approach include matching decoupled moments of higher order than in common style transfer approaches, 
and matching a descriptor of the power spectrum so as to characterize and transfer diffusion effects between source and target, which is something that has not been considered before in the literature.
The results are of high visual quality, without spatio-temporal artifacts, and validation tests in the form of 
observer preference experiments
show that our method compares very well with the state-of-the-art.
The computational complexity of the algorithm is low, and we propose a numerical implementation that is amenable for real-time video application. 
Finally, another contribution of our work is to point out that current deep learning approaches for photo-realistic style transfer
don't really achieve photo-realistic quality outside of limited examples, 
because the results too often show unacceptable visual artifacts.
\end{abstract}



\section{Introduction}

The ``style'' of an image is challenging to define. This is in part due to the fact that any given feature of an image could be intentionally or unintentionally relevant to its perceived style. This means that the lighting, scene, subject, optics, camera, post processing, and further subdivisions of these categories could all have an essential impact on how an image is appreciated by a viewer. This is further complicated by the unique cognitive and subconscious context of prior experience which impacts how an individual will evaluate imagery.
However, if one begins to isolate artists in the production chain and analyze their procedures, the concept of style can be more clearly defined by the mechanisms and goals employed to alter images. 
For example, cinema colorists, whose job it is to beautify and perfect images in postproduction after they've been shot, often use simple manual adjustments to replicate the color and tone reproduction qualities of reference footage or legacy motion picture film stocks.

Considering this, one can see how the transfer of color rendering characteristics from a target image to a source is an integral part of current workflows. In the field of image processing, this process is called Style Transfer (ST), an umbrella term under which hundreds of methods for its automatic computation have been proposed.
Still today, the vast majority of colorists accomplish this task manually, using few and simple parameters, demonstrating the failure of the field to meet their needs.
This can be explained by the following challenges one faces in the problem of automatic ST: (1) in general, source and target images depict very different scenes, which makes the mapping difficult; (2) the result of the ST procedure must be photo-realistic, which imposes important restrictions on the mapping; (3) there is no `objective' or ground truth result with which to validate or optimize the methods intended to solve this problem, so validation must be done with cumbersome observer tests, leaving no clear way to optimize the methods. As a consequence, the state-of-the-art (SotA) solutions still suffer from important shortcomings that limit their applicability: mostly visual artifacts, and computational complexity in some cases.

The original works in this area, from Kotera et al. \cite{kotera98} and later Reinhard et al. \cite{reinhard01}, aimed to accomplish ST simply by imposing the low level statistics of mean and standard deviation from the target image onto the source. In doing so, the authors could achieve a new version of the source image which reflected some of the target properties like color balance, luminance level, and contrast using simple algebraic operations.
This was a promising tool, because its effects were predictable and robust (with respect to the variety of usable source/target combinations), allowing artists to learn how to use it without becoming frustrated by frequent artifacts. However, in most cases the targeted features failed to describe the style characteristics users were expecting to transfer, like high level scene details (e.g. changing seasons, like the drought effect achieved in the movie ``O Brother, Where Art Thou?'', which was the first film with a fully digital color correction).

Moving forward, Morovic and Sun \cite{morovic98}, and later Pitié \cite{pitie05}  proposed the use of Optimal Transport (OT) methods, which aimed to derive a color mapping between two distributions. In some special cases, the methods were able to produce strong ``alternate reality'' effects, allowing the approach to gain popularity. However, the complexity of the resulting maps often produced artifacts like heavy banding and gamut clipping, prohibiting the approach from achieving the robust performance of the preceding statistical transfer strategy. This marked the beginning of a trend in the field where sensational results were prioritized over usability in a general context, resulting in a departure from the needs of the cinema and photographic communities.

This trend progressed further with the introduction of Neural Style Transfer (NST) by Gatys et al. \cite{gatys13}, whose method employed Convolutional Neural Networks (CNNs) to extract millions of low level features from a target image and impose them onto the corresponding features of the source.
The group found this approach to work especially well in replicating modern painting styles, and the results generated a wave of popularity which dwarfed that of OT and continues to this day, with scores of papers published yearly attempting to advance on the application. While this development has opened up a novel and fruitful dialogue between art and technology fields, the fundamental problems of these methods (and those of deep learning in general \cite{colbrook22}), which make them difficult to use by artists in a photo-realistic context, also persist: a stark loss of performance outside of their training set, and a complete lack of \textcolor{black}{predictability and} corrective recourse due to their black box nature.

In contrast, we present a style transfer methodology which obtains a small set of interpretable feature descriptors, each of which is linked to fundamental parameters of visual response. Following the example set by Reinhard et al. \cite{reinhard01}, we extend the statistical transfer approach to include higher order moments such as skewness and kurtosis, which has not been attempted previously because they are highly correlated. However, by making use of a methodology from \cite{Arxiv2022}, we find transforms along an orthogonal pathway on their respective feature manifolds, which aim to objectively match these target features in the source while making minimal change to the image, avoiding artifacts. Our hypothesis is that these features can well encapsulate the style of target images in a simple and predictable way, since they occupy the most significant positions in the parameter hierarchy of visual adaptation (corresponding to brightness, contrast, shadow/highlight balance, and black/white limits). In doing so, we 
present a robust tool with which artists can build familiarity and can be integrated into their workflows. 
\textcolor{black}{To facilitate this we provide a standalone GUI software which allows for modular control of transferred features, control of image regions for analysis via cropping, and 3D LUT generation such that color and tone transforms can be extracted and imported to any professional editing platform.} 

In what follows we will give an in-depth overview of the competing approaches discussed above and their respective SotA methods, as well as the proposed method and its motivations. Then, we will validate our method quantitatively, via a psychophysical experiment, and qualitatively, via tests on a range of applications. In these tests, we demonstrate that the method produces high quality results with minimal artifacts, compares very well with the SotA, and is of low computational complexity. In addition, we show how the feature decorrelation methodology of \cite{Arxiv2022} can be extended to extract and transfer the relative amount of spatial frequency details, which we employ to simulate optical diffusion filters.

\color{blue}
\color{black}

\section{Related Work}

In the most general sense, ST methods work by isolating feature pairs between source and reference images, and matching them. An early foundational example is the method of Reinhard et al. \cite{reinhard01}, which first transforms images into a decorrelated color representation
(defining luminance ``L'' and chroma $\alpha,\beta$ groups), then extracts the mean and standard deviation values from each channel in the source and target and finally matches them via the following equation, where $C$ corresponds to the color channel, $t$ and $s$ correspond to target and source image, $(i,j)$ signifies the current pixel location, and $\sigma$ and $\mu$ correspond to the standard deviation and mean, respectively. 
\begin{equation*}
    C'_{(i,j)} = {{\sigma^C_t} / {\sigma^C_s}} \times (C_{s(i,j)} - {\mu^C_s}) + {\mu^C_t}
\end{equation*}
Aside from the proposed method, a number of approaches have recently been posited which are effectively extensions of the statistical transfer concept of Reinhard et al. However, instead of extending the statistical transfer mechanism, they simply add pre-processing stages which isolate image regions into groups to be passed through the process described in the above equation. For example, in the method of Wu \cite{wu20}, the image is segmented into salient and non-salient features and the mean and standard deviation are matched between the independent regions. Another instance is the method of Xie \cite{xie20}, where the image is segmented into color object regions by way of K-means clustering and the statistics are matched between the corresponding clusters of the target and source.

Instead of isolating and matching a small number of significant features, NST addresses the problem in a comparatively brute force approach, where millions of trivial features are extracted and the corresponding ones are matched between target and source. NST is by far the most popular avenue of study in the field, with the proliferation of works in recent years exceeding that of any other by at least an order of magnitude. However, if one takes a closer look at these works it can be observed that they are for the most part incremental, simply experimenting with different flavors of network architecture, training strategies, feature derivation, or loss functions. Given the lack of a performance metric in the field, the degree of progress which has been made since the work of \cite{gatys13} is difficult to evaluate.

A recent review from Jing \cite{jing20} also demonstrates that 95\% of these works do not attempt to produce photo-realistic results, further narrowing down the number of relevant methods for professional imaging applications. The review also points out the common drawbacks of these methods which are summarized as follows. \textcolor{black}{First, the methods must be trained, requiring significant storage and computational resources and resulting in a high economic and environmental cost, with some methods recommending the use of 50+ graphics processing units \cite{an20}. Due to this dependence, the photo-realistic results of many existing methods cannot readily be replicated because their code bases, photo-realistic models or datasets are not publicly available. 
Second, they are overfitted to their training sets, such that inputs which deviate from the set will lead to unacceptable results considering professional quality standards. Third, after being trained, many methods require the input images to be segmented and semantically labeled which can be difficult to accomplish \cite{li_2019,yoo_2019}.} Finally, methods in this category are black boxes with no human readable parameters, and thus cannot be tuned in any meaningful way to refine results or correct artifacts.
\textcolor{black}{For representative results of the artifacts that can occur, we refer the reader to the experimental section (Figure \ref{NST}).}

\textcolor{black}{After encountering these caveats}, we identified four significant methods which offer photo-realistic results and could be tested in this context. Among them is the method of Luan et al. \cite{luan17}, which made modifications to the nature of the feature transfer process used by Gatys et al. \cite{gatys13} to maintain integrity and avoid ``painterly'' artifacts. Later, Li et al. \cite{li18} offered an advancement to this work, where the stylization process was followed by a smoothing step to ensure spatial consistency. Another notable recent work was that of Mechrez et al. \cite{mechrez17} which sought to refine the output of other neural style transfer techniques by reinforcing the boundaries of the original image. Finally, we recognize the work of Huang et al. \cite{huang18} as a significant contribution to photo-realistic NST. Despite the fact that they tailored their application for image translation, a slightly different topic which aims to transfer the general content of an input image to a source (not only ``style''), their work demonstrated the capability to make photo-realistic transfers of the variety explored in this paper.

The works of the OT category are similar to NST in the sense that they make use of a loss function to minimize differences between source and target images, but differ in that they create a direct general mapping function to transfer between color distributions instead of matching features.
Morovic and Sun \cite{morovic03} presented the original OT method for ST, which created a complex (one-to-many) mapping between color clusters in the source and target images.
The method presented by Pitié et al. \cite{pitie05} expands this concept to a form which better resembles today's SotA methods, where the task is broken down into successive simultaneous 1D distribution transfer problems, and the simplification to a limited number of data clusters is no longer necessary.

A thorough, recent survey of the topic is also provided by Pitié \cite{pitie20}, where the SotA OT methods, as well as those of other approaches, are tested and compared.
Included among them was the method of Ferradans et al. \cite{ferradans14}, which is based on a regularized OT framework that relaxes the constraints on size consistency between source and target datasets and also adds a penalty to preserve the image structure in the transform.
The method of Rabin et al. \cite{rabin14} advances on this relaxed and regularized discrete OT concept by including a pre-processing step for modeling the spatial distribution of colors within the image domain, and a process for tuning the relaxation parameters automatically.
Another notable recent method is that of Blondel et al. \cite{blondel17}, whose regularization of the OT process allows for smooth approximations between sparse color groups.


Like the methods of optimal transport, the method of Grogan et al. \cite{grogan17} also works on the derivation of a color mapping between two distributions, but instead of using an OT framework it uses the L2 divergence as a cost function to find an optimal Thin Plane Splines transform to perform the color transfer. We also recognize the similar recent work of Li et al. \cite{li19} which segments the image based on region saliency and color clusters, and creates detail preserving dynamic lookup tables to form a mapping to transfer between image segments of the target and source.

Other methods exist which are highly relevant to our application, but do not fall under the categories listed above such as the ``Color Match'' function of Adobe Photoshop, which is the only popular industrial solution for photo-realistic ST that the authors are aware of \cite{photoshop}. We also highlight the work of Zabaleta and Bertalmío \cite{zabaleta21} which transfers the visual appearance of a style target onto a source image in three separate stages. First the luminance properties are transferred via a constrained histogram equalization process, then the color properties are transferred through principle component analysis, and finally a transfer of local contrast properties is accomplished. While the method was originally tailored for cinema applications using unedited camera raw images as sources, it can easily be adapted for transfer between arbitrary source and target images.

\section{Proposed method}

\subsection{Motivation}

The efficient representation principle \cite{Attneave1954,Barlow1961} is
an ecological approach for vision science that has proven to be extremely successful across mammalian, amphibian and insect species and that states that the organization of the visual system in general and neural responses in particular are tailored to the statistics of the images that the individual typically encounters, so that visual information can be encoded in the most efficient way, with minimum redundancy, optimizing the limited biological resources.
Atick \cite{Atick1992} makes the point that there are two different types of redundancy or inefficiency in an information system like the visual system:

\begin{enumerate}
\item
  {\it  If some neural response levels are used more frequently than others.} For this type of redundancy,
 in order for a cell's response to be optimal, the stimulus-response relation should match the input signal distribution, which is achieved by processes of adaptation. Natural images have statistical structure in different statistical moments, and the first two moments,
the mean and the variance,
are particularly important, especially for retinal processing
\cite{Rieke2009}.
Adaptation to the mean, or light adaptation, is fully accomplished in the retina and has the effect of dividing luminance by its average, so that the retina has to encode a signal that has a much smaller range. Adaptation to the variance, or contrast adaptation, begins at the retina, is strengthened at the thalamus and the visual cortex, and has the effect of dividing the response by the local contrast, defined as standard deviation \cite{Mante2005}.
And contrary to what was classically assumed, there is also adaptation to higher order statistics, taking place in the visual cortex \cite{Rieke2009}.
\item
{\it  If neural responses at different locations are not independent from one another.} For this type of redundancy the optimal code is the one that performs decorrelation. There is evidence in the retina, the thalamus and the visual cortex that receptive fields act as optimal ``whitening filters'', decorrelating the signal by flattening the power spectrum.
\end{enumerate}

From these vision science principles, we conclude that if two images have similar statistical moments, then the responses they produce in the visual system will also be similar, and this constitutes the basis of our approach for photo-realistic style transfer: we want to match the statistical moments of the source image to those of the target image, so as to match their overall appearance.

  Furthermore, in a professional scenario, be it for photography, film, or TV, a relevant element of the style of the picture is given by the Point Spread Function (PSF) properties of the lens used in the camera, which has a direct impact on the power spectrum of the resulting image\footnote{A lens with a broader PSF produces images that are more diffuse, and therefore have a power spectrum that is more contained, than in the case of a lens with a sharper PSF.}. For this reason, we've developed an instance of our proposed photo-realistic style transfer method that can also consider the power spectrum.

The novel elements of our approach, in the context of style transfer, are two:

\begin{enumerate}
\item
  While in the color transfer literature the methods that perform moment matching only consider the first two moments \cite{pitie20}, we will consider up to the 4th moment, in a fully decoupled version.
\item
Imitating the effect on the power spectrum produced by the lens, or by a diffusing filter; this is a topic that has not been considered before for style transfer, to the best of our knowledge.
\end{enumerate}

Both issues, matching decoupled moments up to 4th order and matching a descriptor of the power spectrum, are quite challenging problems when we require, as it is our case, results with high visual quality and no artifacts. For this reason, in order to solve them we resort to the recently proposed mathematical formulation of feature decoupling  \cite{Portilla:ICIP:2018,Martinez:ICASSP:2020,MMSP2020,Arxiv2022}, as we will discuss in the following sections.

\begin{figure}[H]
    

    \subfloat{{\includegraphics[width=0.99\textwidth]{selected/isoMom/isoMom_bounded/sources1_3.png.jpeg} }} \\

    \vspace{-8mm}
    \subfloat{{\includegraphics[width=0.825\textwidth]{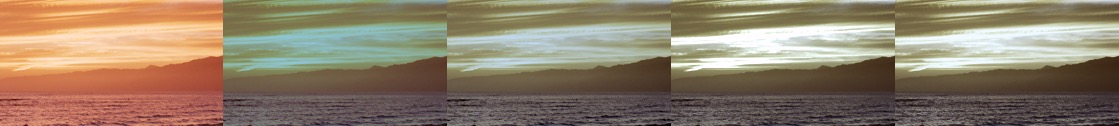} }} \\

    \vspace{-5mm}
    \subfloat{{\includegraphics[width=0.99\textwidth]{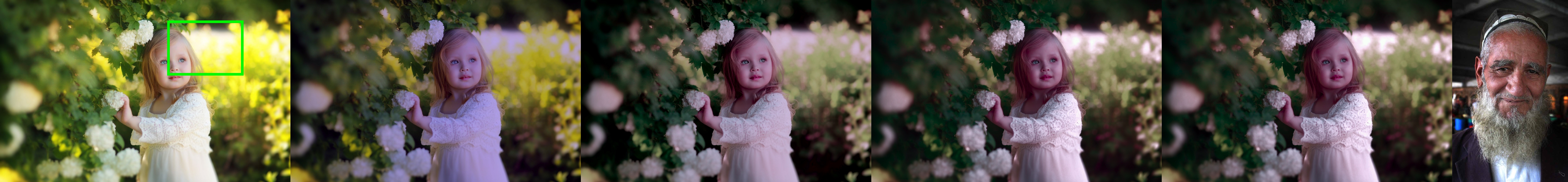} }} \\

    \vspace{-8mm}
    \subfloat{{\includegraphics[width=0.92\textwidth]{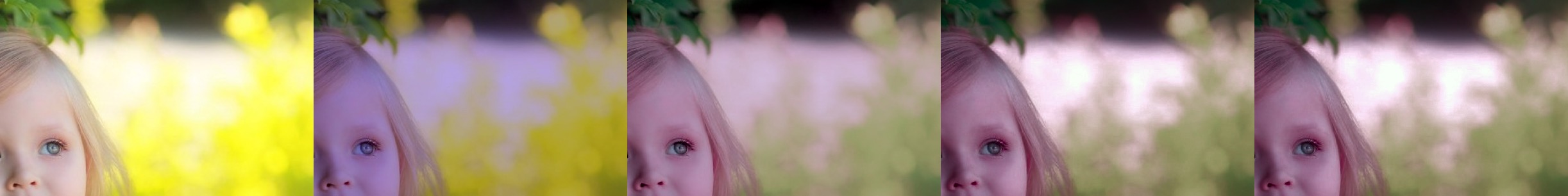} }} \\

    \vspace{-5mm}
    \subfloat{{\includegraphics[width=0.99\textwidth]{selected/isoMom/isoMom_bounded/sources6_8.png.jpeg} }} \\

    \vspace{-8mm}
    \subfloat{{\includegraphics[width=0.825\textwidth]{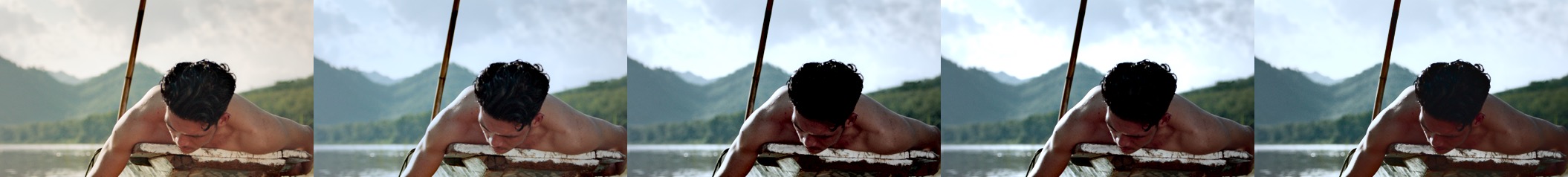} }} \\

    \vspace{-5mm}
    \subfloat{{\includegraphics[width=0.99\textwidth]{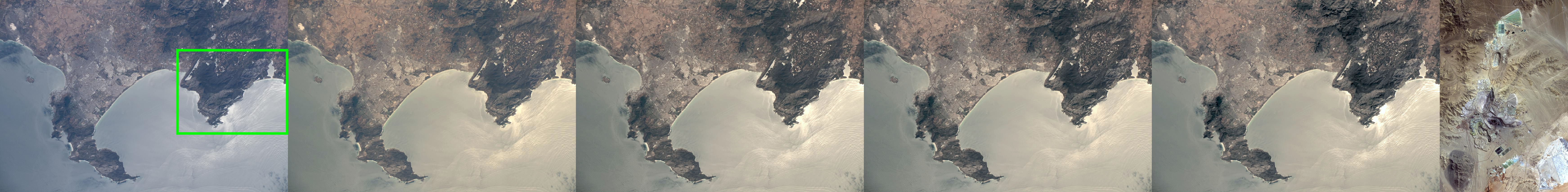} }} \\

    \vspace{-8mm}
    \subfloat{{\includegraphics[width=0.91\textwidth]{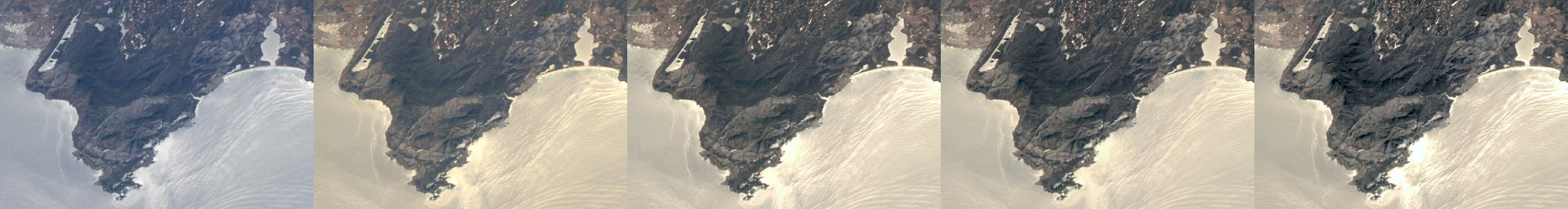} }}
    
    \caption{In comparing the results between each statistical moment applied, it can be seen that the transfer is progressively refined by each step, bringing the source closer to the target image, \textcolor{black}{emphasized in the zoomed regions of every second row}. From left to right: original source image; source mean matched to target on far right; source mean and variance matched; source mean, variance and skewness matched; source mean, variance, skewness and {\em ortho-kurtosis} matched; target image. First and third example source and target provided by ARRI \cite{ARRI}, second example source and target from \cite{LFWTech}, last  example source and target made available by NASA under CC-BY 2.0. Second row source and target from \cite{LFWTech}. Fifth and seventh row source and target provided by ARRI \cite{ARRI}.}
    \label{fig:isoMom}

\end{figure}

\textcolor{black}{In Figure \ref{fig:isoMom} we demonstrate the effect of transferring each statistical feature on image color appearance.} The target image is shown on the right side of each row and the source image is shown on the far left. Then, the statistical moments of mean, standard deviation, skewness, and {\em ortho-kurtosis} (a decoupled version of the kurtosis) are each applied. The hierarchical nature of the statistical moments can be observed here, with the first transfer, the mean, having the greatest impact of all and each subsequent transfer having a progressively subtler effect. Still, a significant impact can be observed with the addition of standard deviation, as the general contrast properties of the image become closer to the target. The skewness parameter has an impact on the balance of the image towards the shadows and the highlights, at times resulting in the clipping or crushing of detail. But the ortho-kurtosis parameter, when matched to the target, acts as a refining step, separating the concentrations of the absolute darkest and brightest pixels in the image from the highlights and shadows. In this way, the decoupled adjustment of these features is useful for exercising more precise control over different areas of the tone scale range, an increasingly important trait in modern color correction as this range is gradually being extended by emerging displays and formats.

\subsection{Overall description of the method}
\label{subsec:overview}
\subsubsection{Transferring the style given by the distribution of brightness, contrast and color values} 

We begin with the case of style transfer for still images, ignoring the power spectrum for the moment and just considering brightness, color, and contrast as the elements of style.

The inputs are two photographs, the source image S and the target image T, and the goal is to obtain a modified version of S that matches the style of T. 
We assume without loss of generality that both S and T are color images in a standard display-ready format like sRGB, which implies a non-linear relationship between the image values and the radiance values captured by the camera sensor that, with a simple Lambertian image formation model, can be approximated thusly:
\begin{equation*}
  Z_{i,j}=f(L \cdot R_{i,j}),
\end{equation*}
where $Z_{i,j}$ is the image value at location $(i,j)$; $f$ is a nonlinear function that approximates brightness perception,
which we assume takes the standard form of a power law of exponent $1/2.2$ (the so-called ``gamma correction'' transform), and 
that the camera applies to the radiance value in order to perform perceptual linearization to optimize the encoding \cite{Bertalmio2019}; $L$ is the scene illuminant; and $R_{i,j}$ is the scene reflectance at the point.

\begin{figure} 
  \centering
  \includegraphics[width=1\textwidth]{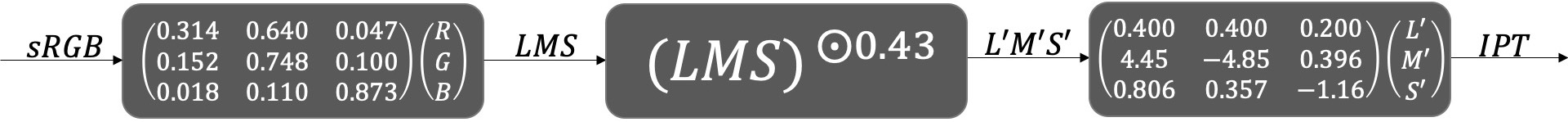}
  \caption{\textcolor{black}{Process for converting radiance (display linear) images with sRGB primaries to IPT space, where we use the symbol ``$\odot$'' for representing an element-wise scalar operation for the vector coefficients (here a power).}}
  \label{fig:iptConversion}
\end{figure}

Our method has the following stages:
\begin{enumerate}
\item
  We undo the gamma correction transformation in S and T by applying the inverse function $f^{-1}$ which yields image radiance values, then we apply the Gray World method \cite{Buchsbaum1980} \textcolor{black}{by taking the channel-wise mean value} to estimate the illuminants $L_S$ and $L_T$ for each radiance image, then we scale the radiance image for S by the scale factor $L_T/L_S$: in this way, we approximate the replacement in S of its original illuminant by that of T.
\item
  Convert the radiance images to the perceptual color space IPT, which has a very good prediction of constant perceived hue \cite{Ebner1998}. IPT is an opponent color space with a luminance channel and two chroma channels. \textcolor{black}{The steps for converting radiance images to IPT space are shown in Figure \ref{fig:iptConversion}.}
\item
  Match the first four decoupled moments of the luminance channel of S to those of T.
\item
  Match the first two moments of the chroma channels of S to those of T.
\item
  Convert S from IPT back to RGB.
\item
  Apply the gamma correction function $f$: this yields the final result $S'$.
\end{enumerate}

When matching moments, it must be noted that standardized moments up to order three (namely, the mean, variance, and skewness) are decoupled, meaning that their gradients are orthogonal everywhere, so we can easily change one by moving along its gradient 
without affecting others.
However, that no longer holds for higher-order standardized moments, in particular resulting in the non-orthogonality or coupling of the gradients of skewness and kurtosis:
not any combination of skewness and kurtosis is algebraically possible, and we will affect one moment when fixing the other ones (see, e.g.,~\cite{Sharma2013,Blest2003,Rosco2015}). 

Therefore, step 3  of our method requires us to perform {\it feature decoupling}, and we do it by using a recently proposed methodology \cite{Arxiv2022}.
Given a set of (possibly coupled) features, this approach produces a set of similar features but with the property of having mutually orthogonal gradients: exactly orthogonal in the whole domain, in favorable cases (like in the case of extending the three first standardized moments with the {\em ortho-kurtosis}), or exactly orthogonal in a high dimensional subset and approximately outside it; these new features, in addition, are identical to the original ones when evaluated within some high dimensional manifolds.
Because, in general, decoupled features sets {\em are not equivalent to their original (coupled) counterparts}, the motivation for using such feature decoupling techniques is not just increasing the efficiency of the feature transfer process, but, more importantly, increasing the descriptive power of the features we transfer. 
In Section \ref{sec:methods} we review the main aspects of the feature decoupling methodology and its 
application to feature transfer, leaving a more detailed technical description for the Appendix.

\color{black}
To compare the behavior of decoupled moments vs. coupled moments 
we first synthesized $100$ discrete signals using a low-pass log-normal model for generating their marginal PDFs which are smooth and positive in the valid normalized range. 
By giving the role of source S and target T to every possible pair of different signals (9,900 in total), we were able to compare the statistics of the 
adjusted signals, 
to the statistics of T and S using decoupled vs. coupled moments\footnote{For transferring the coupled features we directly imposed the target feature values as if we were normalizing and they were the reference values of the normalization, applying Algorithm~ Alg.~\ref{alg:nested_normalization_narrow_path}.}. We used the Kullback-Leibler divergence (KLD) from the densities of the results 
(using coupled vs. decoupled features) to 
the original densities of source and target. This allowed us to measure how likely it is that the adjusted signals were generated from the source and from the target distributions, in each case. 
Note that the divergence to the source is also relevant because we want it as small as possible for a given divergence to the target. 
We have observed first that using the orthokurtosis the probability of expanding the range was larger than that of shrinking it, exactly the opposite of what happens when using the kurtosis. We also observed that the probability of decreasing the KLD to the source when changing the transfer from using kurtosis to orthokurtosis was 55\%, while the KLD to the target did not change on average (50\% chance).  

\begin{figure}
    \centering
    \begin{tabular}{cc}
         \includegraphics[width=0.3\textwidth]{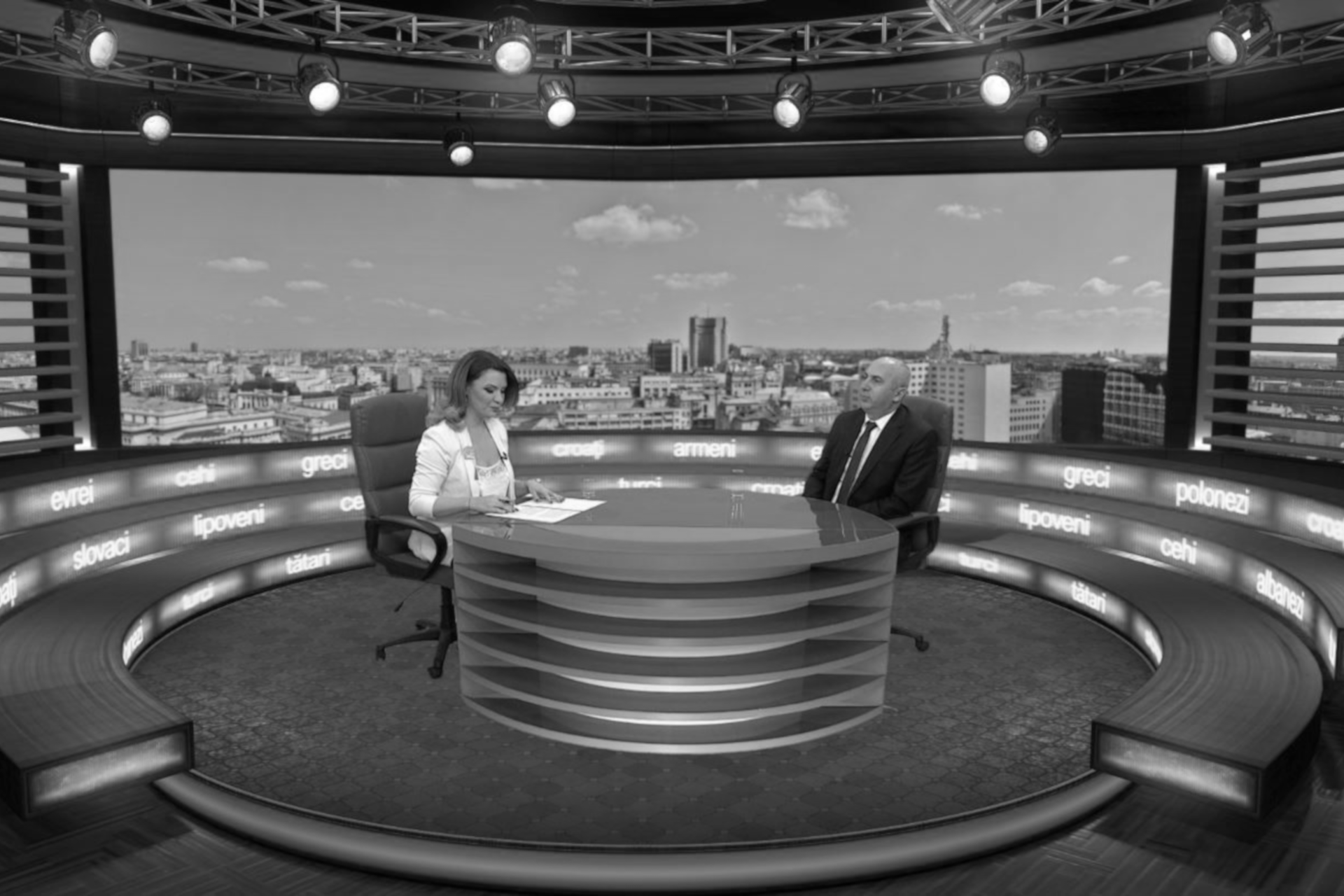} &
         \includegraphics[width=0.3\textwidth]{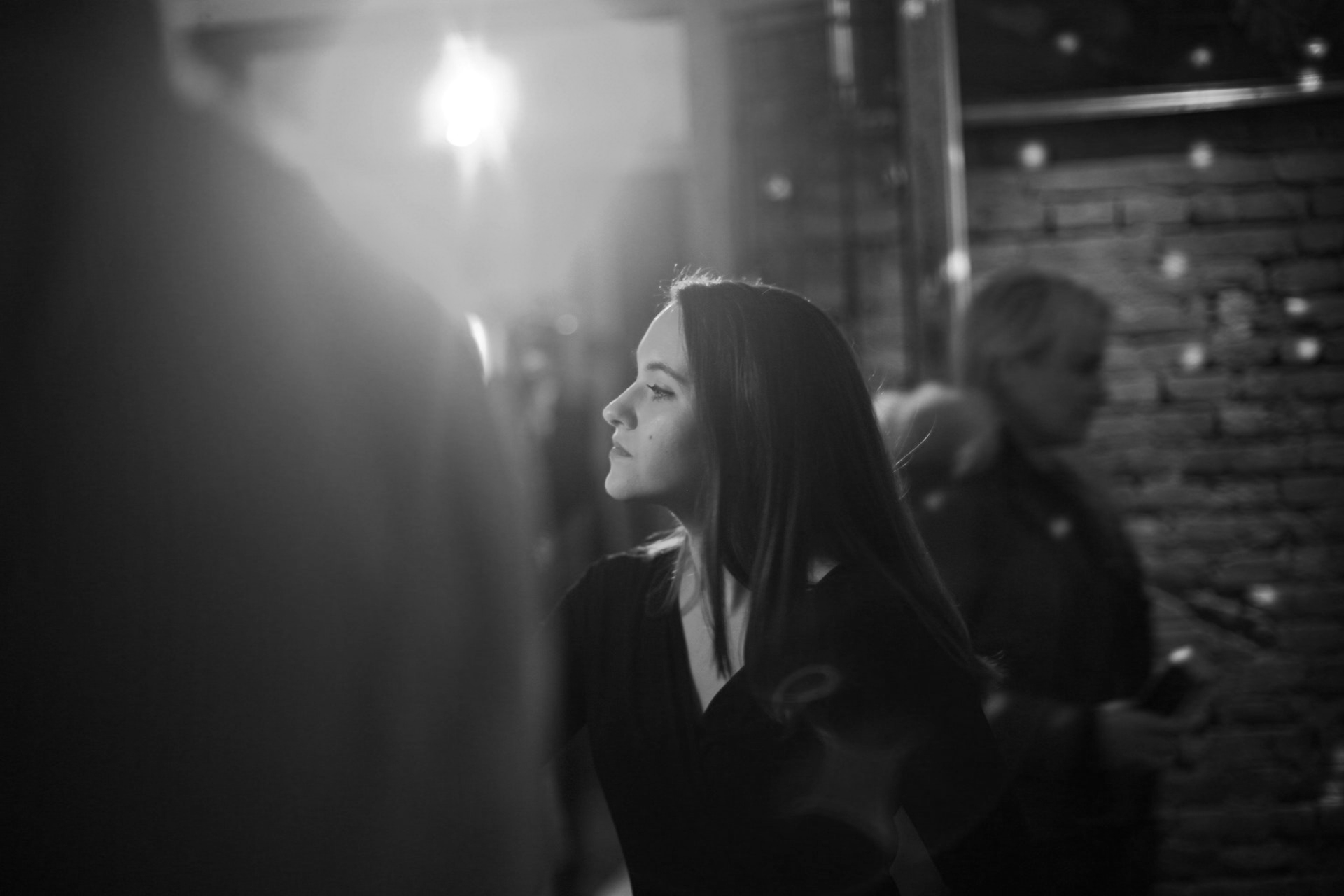}\\
         (a) & (b) \\
    \end{tabular}
    \begin{tabular}{ccc}
        \hspace{-3mm}
         \includegraphics[width=0.25\textwidth]{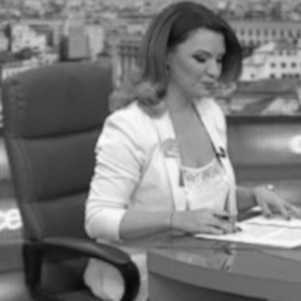} &
         \includegraphics[width=0.25\textwidth]{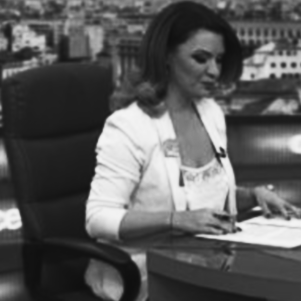} &
         \includegraphics[width=0.25\textwidth]{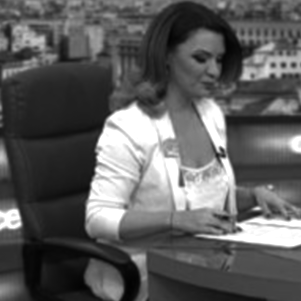} \\
         \vspace{2mm}
         (c) & (d) & (e) \\
         \includegraphics[width=0.27\textwidth]{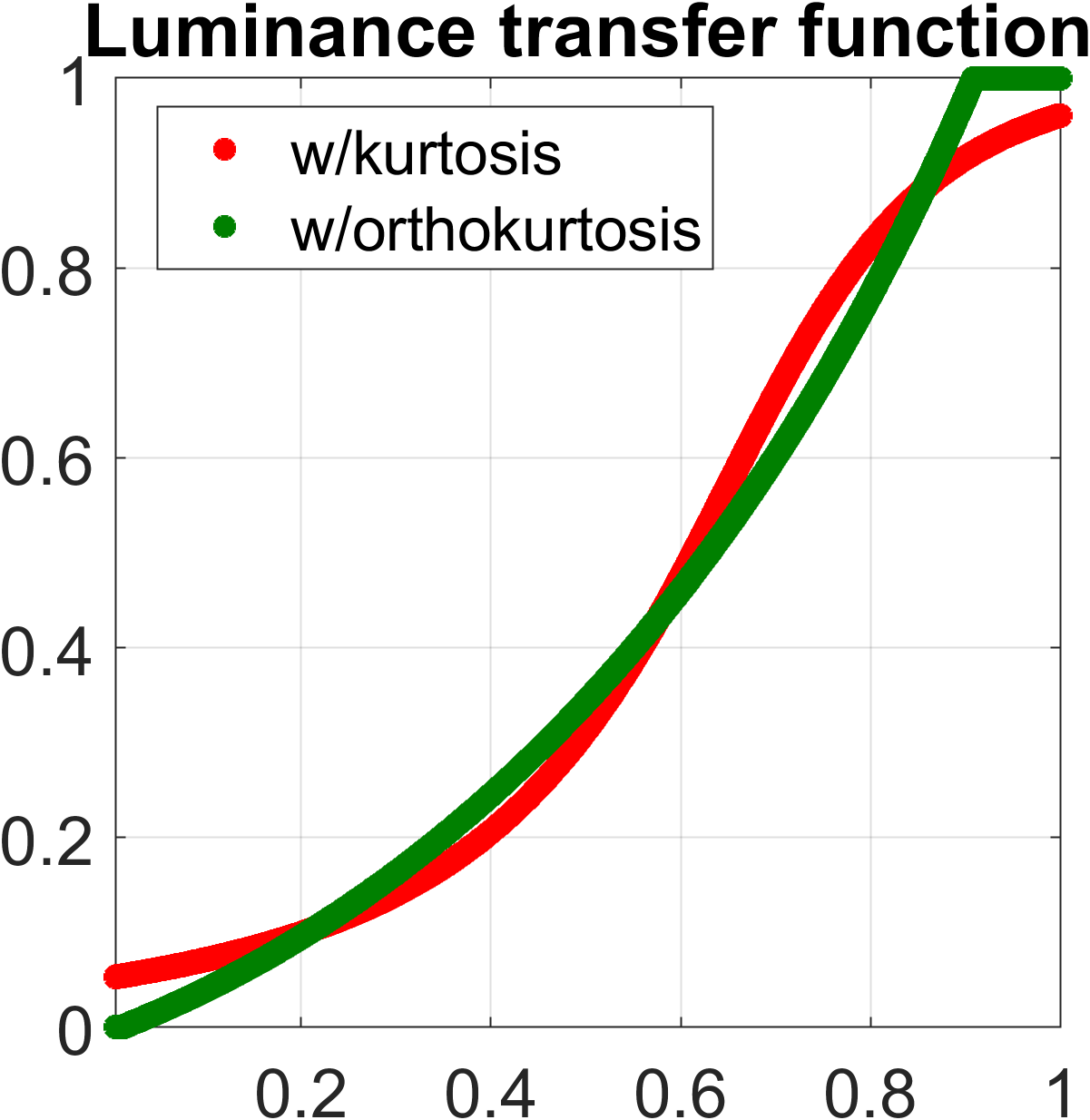} &         \includegraphics[width=0.32\textwidth]{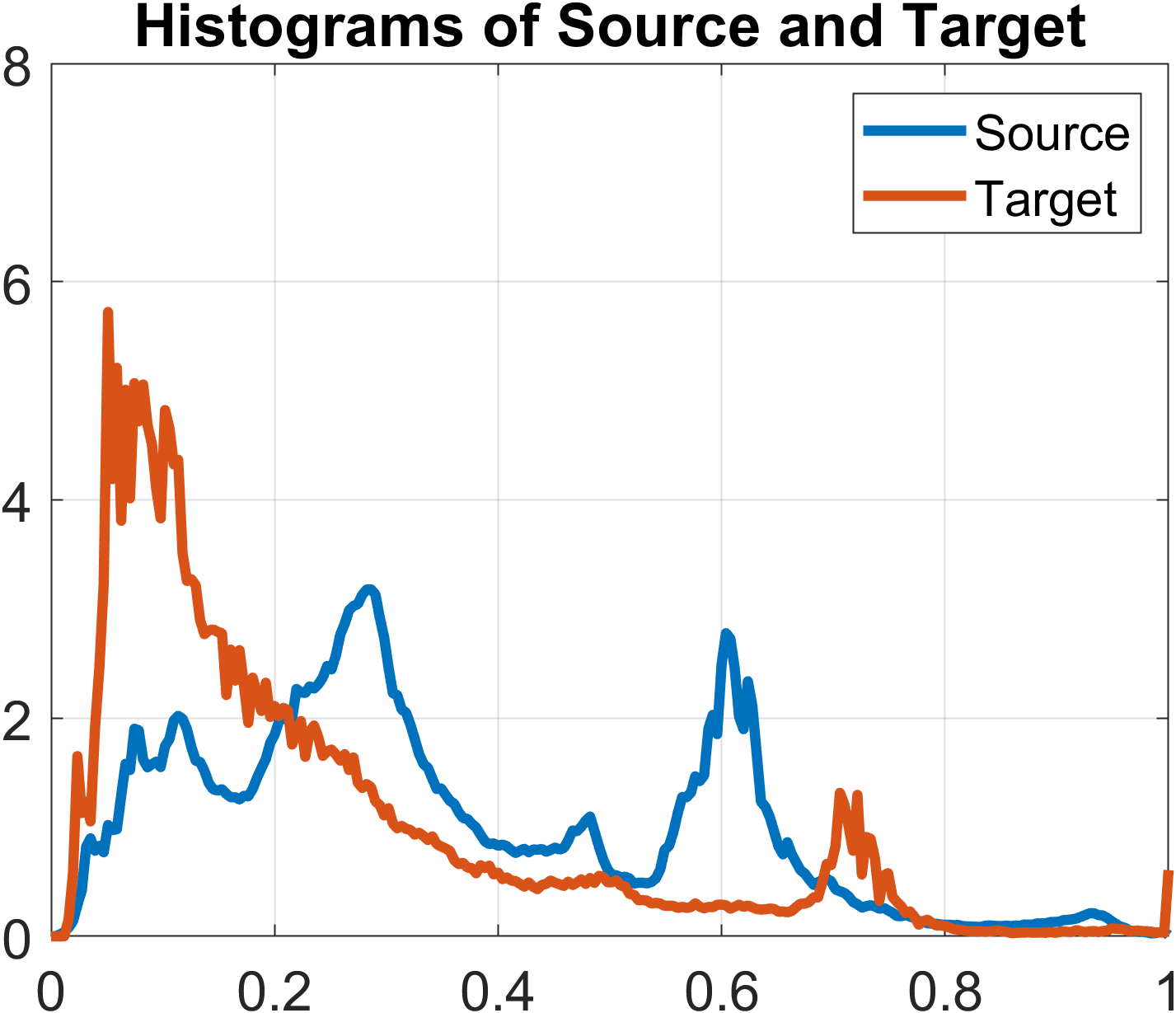} &
         \includegraphics[width=0.32\textwidth]{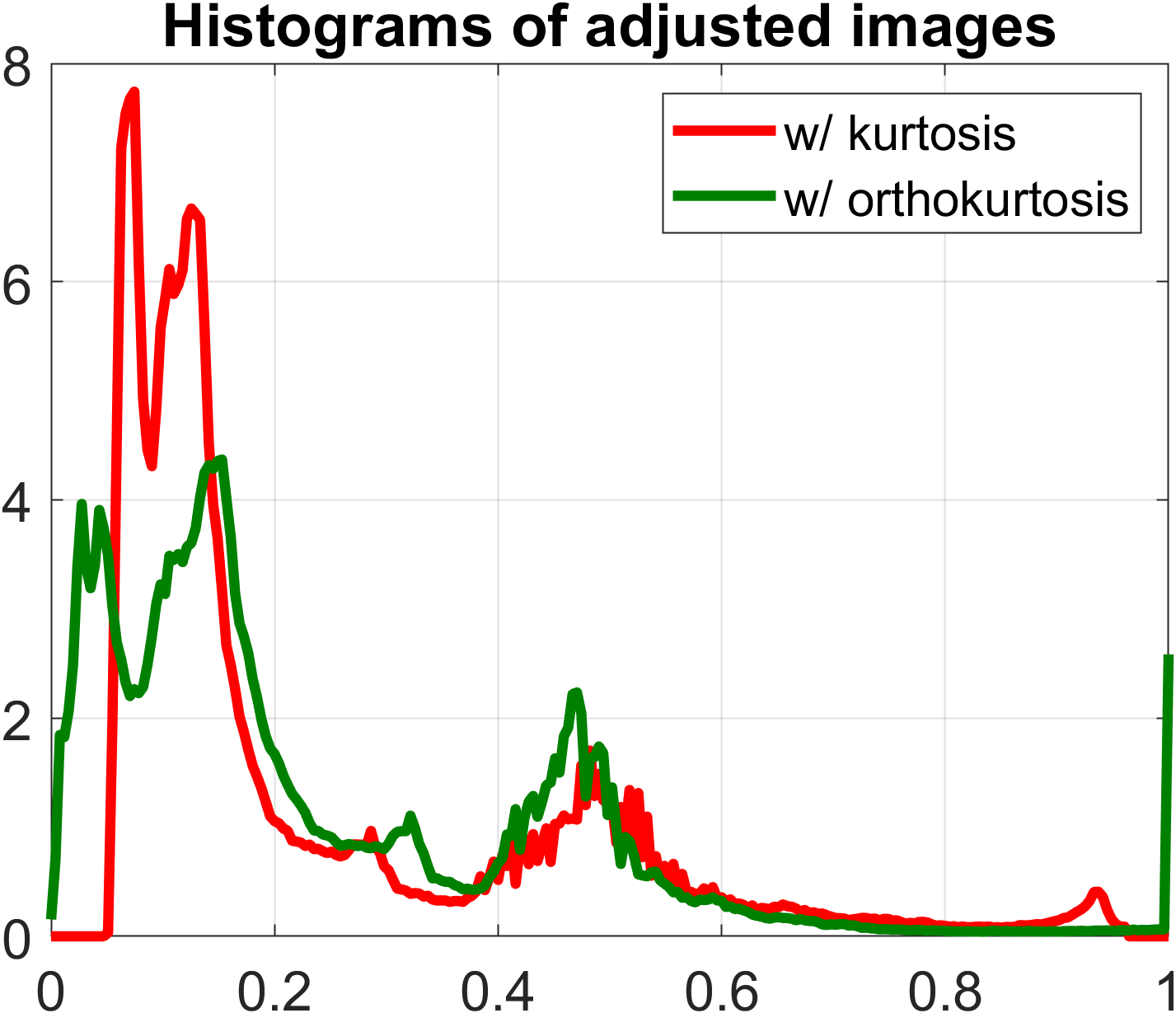} \\
         (f) & (g) & (h)
    \end{tabular}
\caption{\textcolor{black}{Using kurtosis vs. ortho-kurtosis for luminance style transfer. (a) source; (b) target; (c) crop of {\it (a)}; (d) crop of the transfer result
using coupled features; (e) idem using decoupled features; (f) point-wise transfer functions; (g) luminance histograms of panel {\it (a)} (source) and {\it (b)} (target); (h) luminance histograms of the whole images corresponding to panels {\it (d)} (coupled features result, kurtosis) and {\it (e)} (decoupled features result, ortho-kurtosis).}} 
 \vspace{-4mm}
\label{fig:studio_detail_transfer}
\end{figure}

Regarding visual impact, an informal inspection of the ST results of $20$ possible combinations of pairs of different source and target using the images of Figure~\ref{sources}, results in that the majority 
($80\%$) gave very similar visual results using kurtosis or ortho-kurtosis, whereas the rest were favorable to the ortho-kurtosis. 
Concentrating on the latter, we have observed how the higher average entropy loss produced when adjusting the kurtosis in these cases gives rise to 
local contrast fluctuations that 
may both exacerbate and decrease local contrast, to the point of causing unnatural (``posterized'') appearance or visual information loss, respectively.  
In Figure~\ref{fig:studio_detail_transfer} we show a representative example of how the coupled moments' adjustment, with its reduced output range and low slope regions of the red curve transfer function, reduces the visual discriminability of image details (e.g., horizontal lines on the seat and along the table's edge), flattening the volumes' appearance in low and high-luminance areas. On the other hand, its higher gain for mid-gray input levels 
provokes unnatural wide luminance excursions on the skin (e.g., in the face). In this case, and for the whole source and target, the more gentle ortho-kurtosis adjustment allows reducing the KLD to the source 
(from $0.77$ to $0.68$ bits), while having a negligible impact on 
the KLD to the target
(from $0.29$ to $0.30$ bits).  

In summary, adjusting fully decoupled moments up to the fourth order, instead of standardized moments, has a positive effect. 
The achieved reduction of entropy loss w.r.t. the source helps to preserve a more natural appearance of the result, while not affecting negatively the matching of the target's statistics. 
Finally, it must be noted that even if both sets of features provided the same ST performance, using the ortho-kurtosis would still be preferable, because its adjustment is fully reversible and independent from the previous moments. These properties, as explained in the Applications and the Conclusions sections, contribute to the meaningfulness and usability of the proposed method.

\textcolor{black}{While it is possible to extend the approach to decouple moments higher than the fourth order, the addition of more moments makes it more likely for artifacts to occur\footnote{\textcolor{black}{In the extreme case, transferring infinite moments would be akin to histogram matching, which is known to often produce heavy banding artifacts.}}, and, in fact, we have observed in some preliminary experiments that
transferring five moments caused banding.
For this reason, this approach is limited to the first four moments.}

\subsubsection{Transferring the style induced by the optics}
\label{subsec:optics}
\color{black}
In a practical application, we would want our style transfer method to capture optical effects (produced by the PSF of the lens, or by the action of an optical diffusion filter), which, as we explained, have a direct impact on the power spectrum of the image.



If we try to transfer the power spectral features directly from the target to the source in the most general case where the two images are unrelated, \textcolor{black}{artifacts are likely to occur.} 
Therefore, in order to isolate the style element associated with the optical elements we need source and target images 
where the differences in power spectrum are not due to the image content but just to the optics,
e.g. the same scene photographed with and without a diffusion filter. 
From these images we obtain a transfer function that we can then apply to any other source image, which can now be completely unrelated to the ones used to derive the function.

In order to match the shape of the power spectrum of source and target we do not want to do a direct matching of the amplitude of each individual frequency, as that would not be robust and would likely give rise to spectral artifacts. 
A common choice for an economical and robust descriptor of the power spectrum 
is to use a set of overlapping band-pass filters (e.g. see \cite{HB95,PN96,PS00}). Given that there is coupling in the variance measurements at the output of these filters, the estimation of the transfer function also takes advantage of our {\it feature decoupling} method.

As a practical example, shown later on in Figure \ref{diffusion}, we can take two photos of the same static scene (e.g. a tree branch with flowers), one photo S with a lens with a ``regular'', sharp PSF, and another photo T with a shallower PSF lens. From these images we compute a linear kernel K (whose Fourier transform is the transfer function we're determining), which then can be applied to any other source image (e.g. a portrait), yielding a more diffuse result that emulates the effect of the lens used in T.

Our method now has the following stages:
\begin{enumerate}
\item
We're given images S \textcolor{black}{(naturally sharp)} and T \textcolor{black}{(diffused/blurry)} corresponding to the same scene but taken with different optics, and an unrelated source image $S_2$.
\item
Undo the gamma correction transformation in S, T, and $S_2$ by applying the inverse function $f^{-1}$ which yields image radiance values, 
\textcolor{black}{and compute $L_S$ and $L_T$, the luminance of S and T, from the corresponding radiance RGB coefficients 
($L = 0.2126 \times R + 0.7152 \times G + 0.0722 \times B$).}
\item \textcolor{black}{Match the decoupled variances of the four spectral channels (see Figure~\ref{fig:radial_profile_filters}) of $L_S$ to those of $L_T$, obtaining the resulting luminance image ${L_S}'$, a filtered version of $L_S$.} 
\item \textcolor{black}{Compute the spectral transfer function $\Hat{K}(\xiv) = \Hat{L}_{S'}(\xiv)/(\Hat{L}_S(\xiv)+\epsilon)$ (with $\epsilon=10^{-8}$), and obtain its inverse Fourier transform, the linear kernel K modeling the optical PSF. (We used here $\Hat{h}={\cal F}\{h\}$ to denote the Discrete Fourier transform of an array $h$.) }
\item
  Convolve the R, G, and B \textcolor{black}{radiance} channels of $S_2$ 
  with kernel K.
\item
  Apply the gamma correction function $f$; this yields the final result ${S_2}'$.
\end{enumerate}
\textcolor{black}{Note that it is straightforward to combine the two presented types of style transfer, just by adding the spectral matching steps of the luminance right after the inverse gamma correction, in the color-and-luminance adjustment pipeline of the previous subsection.}

\color{black}

\section{Deterministic feature decoupling: analysis and transfer}
\label{sec:methods}

In~\cite{Arxiv2022} one of the authors of this paper and other collaborators presented a new mathematical and algorithmic framework for decoupling sets of given real differentiable functions ({\em features}) applied to $N$-dimensional vectors (representing, e.g., single-channel discretized images having $N$ pixels). Given a set of features ${\cal S} = \{f_k(\xv):\R^N\rightarrow\R, k = 1\dots M\}$ the goal was to find a similar set $\hat{\cal S} = \{\hat f_k(\xv):\R^N\rightarrow\R, k = 1\dots M\}$ with the property of having mutually orthogonal gradients: exactly orthogonal in the whole domain, in favorable cases, or exactly orthogonal in a high dimensional subset, and approximately outside it. The new features, in addition, were enforced to be identical to the original ones within some high-dimensional manifolds. 
In~\cite{Arxiv2022} (which includes, extends and refines previous contributions published in conference papers \cite{Portilla:ICIP:2018, Martinez:ICASSP:2020, MMSP2020}), besides providing an in-depth analysis of the mathematical aspects of the problem, the authors demonstrate, through some data analysis examples (statistical regression and texture classification), that moments' decoupling, either in the original domain or at the output of convolutional filters, greatly improves performance for those tasks compared to using non-decoupled features. 


\subsection{Decoupling via normalization}
The decoupling method is built upon the {\em normalization} concept: if we find a transformation $\hat\xv:\R^N\rightarrow \R^N$ that is invariant to a perturbation along the local gradient span ${\cal T}$ of a set of features, then any differentiable function $g$ of the so-transformed sample $\hat\xv(\cdot)$, $g(\hat\xv(\cdot))$, will be decoupled from each of the features of ${\cal T}$. We choose for $\hat\xv(\cdot)$ a normalization obtained by following the local gradients of the features until reaching a set of reference values $\vv^{ref}$ for those features (e.g. $\vv^{ref}$ would be $[0,1]^T$ for imposing zero mean and unit mean square value). We term ``reference manifold''  $\RM$ the set of vectors having the reference values for their given features.
For the case of having only two features, normalization consists in ``sliding'' along the curves integrating the gradient of the first feature, until reaching the reference (normalizing) value for that feature. In higher dimensions, all possible trajectories reaching the reference manifold provide the same solution (under some conditions discussed in Appendix B).   
The Nested Normalization (NeN) algorithm consists in normalizing the observation each time, imposing one more reference value in the features' values, as cumulative constraints.
The nested structure implies $\RM_{k+1}\subset\RM_k$, where $\RM_k$ denotes the {\em reference manifold} considering the features up to the $k$-th one. This means that the signal is progressively normalized, enforcing that each time $k$ is incremented, a new reference value for the features \textcolor{black}{is obtained} without releasing the reference values of the previous features. The new features are defined as the original ones evaluated on the normalized samples. Thus, the sequential process of obtaining the new features runs parallel to that of the nested normalization process.
This is explained in 
detail in Appendix B (general decoupling and transfer algorithms). 

\subsection{De-normalization and feature transfer} 
\label{subsec:de-normalization}
Although mainly focused on analysis, Ref. \cite{Arxiv2022} also showed how the proposed decoupling method could be used for feature transfer too. In fact, a set of representative results had been presented 
in~\cite{MMSP2020}. 

\begin{figure} 
  \centering
  \includegraphics[width=1.0\textwidth]{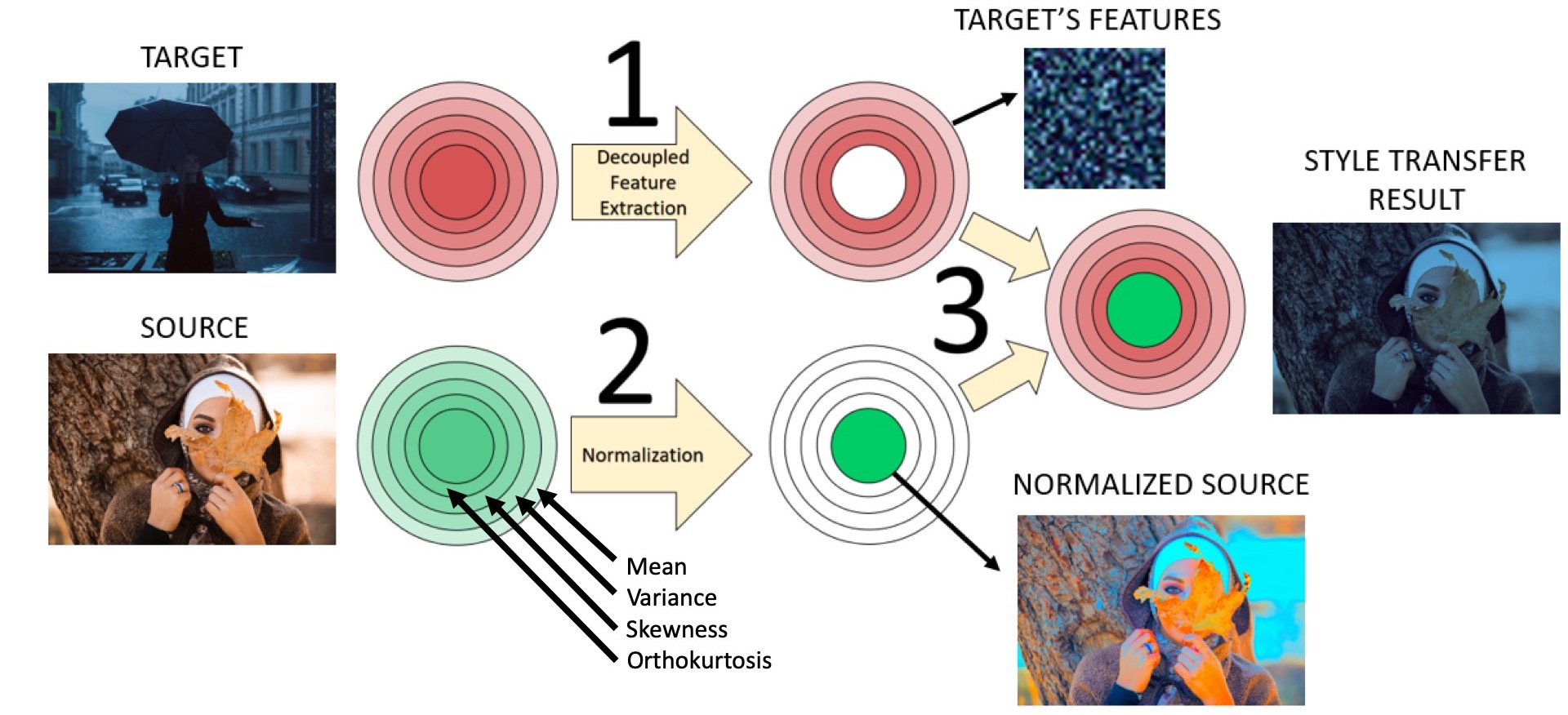}
  \caption{Three-step feature transfer using de-normalization. 
  \textcolor{black}{
  The extracted features (concentric layers, here labeled as the luminance's decoupled moments) 
  capture the style of the image, while the normalized image (central circle) has no 
  style information. 
  Here the target's features 
  are 
  visualized by means of a random image with the same features 
  (``style without scene''). Conversely, the normalized source has 
  pre-fixed values for those features (``scene without style'').
  These two components are fused together 
  into the final result. }} 
  \label{fig:denormalization_scheme}
  \vspace{-5mm}
\end{figure}

The first property to note, at this point, is the reversibility of the normalization operation, \textcolor{black}{resulting from applying a differential equation along a trajectory free from singularities. Provided we keep, along with the normalized sample, the original features' values,
%
%
%
we can undo every step of the normalization loop in reverse order, in what we have called {\em de-normalization}, and recover the original sample from its normalized version: }
\begin{equation}
    \check\xv_{k}(\hat\xv_{k}(\xv;\vv_k^{ref});\hat\fv_{k+1}(\xv)) = \xv, \,\,\, k = 1,\dots, M-1
    \label{eq:denormalization_orig}
\end{equation}
(note the inverted circumflex accent is used to denote de-normalization).
In particular, we would choose $k=M-1$ to impose back all the decoupled features in vector $\hat\fv(\xv)$ to the normalized sample $\hat\xv_{M-1}$. Furthermore, nothing prevents us from changing the decoupled features' values to other ones, 
e.g., to those obtained by analyzing another sample $\yv$, for style transfer purposes:
\begin{eqnarray}
    \zv(\xv;\hat\fv(\yv)) & = &  \check\xv_{M-1}(\hat\xv_{M-1}(\xv;\vv_k^{ref}),\hat\fv(\yv))
        \label{eq:denormalization_transfer_1}
        \\
    \hat\fv(\zv(\xv;\hat\fv(\yv))) & = & \hat\fv(\yv).
    \label{eq:denormalization_transfer} \\
    \hat\xv_{M-1}(\zv(\xv;\hat\fv(\yv))) & = & \hat\xv_{M-1}(\xv;\vv_k^{ref})
    \label{eq:denormalization_transfer_2}
\end{eqnarray}
Equation~\eqref{eq:denormalization_transfer_1} describes a three-step procedure for what we termed {\em controlled feature adjustment} in \cite{MMSP2020}, which allows transferring decoupled features from a (target) sample $\yv$ 
to another (source) sample $\xv$, 
as described in Algorithm~\ref{alg:CFA}.
The result of the transfer, $\zv$, has the features $\hat\fv(\yv)$ of the target $\yv$ (Eq.~\eqref{eq:denormalization_transfer}) and the same normalization $\hat\xv_{M-1}$ as the source $\xv$ (Eq.~\eqref{eq:denormalization_transfer_2}).
\textcolor{black}{
This is illustrated in the scheme of Figure~\ref{fig:denormalization_scheme}. Step 1 is the feature extraction from the target. Step 2 is the normalization of the source. Step 3 is the construction of a new image from the output of the two previous steps, 
through de-normalization.}
%
\begin{algorithm}
\begin{algorithmic}[1]
\REQUIRE
$\xv_{src}, \xv_{tgt} \in  \R^N$ 
,\,
$(\S,\vv^{ref}) = \{f_j, v_j^{ref}, \,\,j = 1,\dots, M \}$
\STATE  Compute {\bf target}'s decoupled features: $\hat\fv(\xv_{tgt})$
\COMMENT{Algorithms~\ref{alg:nested_normalization_broad_path} or~\ref{alg:nested_normalization_narrow_path}}
\STATE Normalize {\bf source} up to the $M-1$ level: $\hat\xv_{M-1}(\xv_{src};\vv^{ref})$
\COMMENT{Algorithms~\ref{alg:nested_normalization_broad_path} or~\ref{alg:nested_normalization_narrow_path}}
\STATE Transfer $\hat\fv(\xv_{tgt})$ on  $\hat\xv_{M-1}(\xv_{src};\vv^{ref})$: $\check\xv_{M-1}(\hat\xv_{M-1}(\xv_{src};\vv^{ref});\hat\fv(\xv_{tgt}))$
\COMMENT{Algorithm~\ref{alg:NeN_reversed}}
\RETURN $\zv(\xv_{src};\hat\fv(\xv_{tgt})) = \check\xv_{M-1}(\hat\xv_{M-1}(\xv_{src};\vv^{ref});\hat\fv(\xv_{tgt}))$
\end{algorithmic}
\caption{Three-step NeN-based feature transfer}
\label{alg:CFA}
\end{algorithm}

Steps 1 and 2 are done using any of the analysis/normalization algorithms (included in Appendix B, section~\ref{subsec:gral_algs_for_analysis}), as these algorithms return both 
the decoupled features of the input and its normalization up to the $(M-1)$-th level.  
Step 3 in~Algorithm~\ref{alg:CFA} is the de-normalization procedure (as explained in subsection A.3, Alg.~\ref{alg:NeN_reversed}). 
\textcolor{black}{Note that this transfer method 
is consistent with the classical way we transfer the mean and variance of one image to another: (i) we measure the target's mean and variance, in that order; (ii) we normalize the source's mean (to 0) and variance (to 1), in that order; (iii) we de-normalize the variance and the mean of the normalized source, in that order, imposing the target values measured in (i). However, the classical procedure tells us neither {\em why} we 
apply subtraction to adjust the mean and 
scaling for adjusting the variance 
nor how to proceed with higher order moments or with other features. 
The framework proposed in~\cite{Arxiv2022} 
provides consistent answers to these questions, for the case of sequentially ordered global differentiable features. 
In the next two subsections, we explain how the decoupled features are computed and transferred from the target to the source images for the specific features used here for style transfer. }




\color{black}

\subsection{Exact moments' decoupling analysis and transfer up to the fourth order} 
Marginal sample moments of an observation $\xv\in\R^N$ are usually defined by:
\begin{eqnarray}
\mu_j(\xv) & = & \frac{1}{N}\sum_{n=1}^{N}{x_n^j}, \,\,\,j\in{\mathbb N} \nonumber \\
& = & \frac{1}{N}{\mathbf 1}^T \xv^{\odot j} \nonumber,
\end{eqnarray}
We have chosen for the reference values the expected values of these sample moments for a zero-mean univariate Gaussian density, and $j = 1,\dots,4$, resulting in $\vv^{ref} = [0,1,0,3]^T$.
\subsubsection{Analysis} 
\label{subsec:moments_analysis}
Following the Algorithm~\ref{alg:nested_normalization_broad_path}, 
we compute the gradients of the first two moments and write the corresponding ODEs for adjusting them to their reference values. By doing that we find that normalizing the first two moments becomes the standard subtraction of the mean and dividing the result by the standard deviation, respectively. By terming $\hat\mu_j$ the $j$-th decoupled moment and applying Step 7 of the algorithm we have:
\[
\hat\mu_k(\xv) = \mu_k(\hat\xv_{k-1}(\xv))
\]
and by applying this and the initialization from the algorithm 
we readily obtain the first three decoupled moments, which are nothing but the standard sample mean, (biased) variance and skewness:
\begin{eqnarray*}
\hat\mu_1(\xv) & = & \mu_1(\xv) \\
\nabla \hat\mu_1(\xv) & = & \frac{1}{N}{\mathbf 1} \\
\hat\xv_1(\xv) & = & \xv - \mu_1(\xv) \\
\left\{\mu_1(\hat\xv_1) = 0 \right\} & \Leftrightarrow & \hat\xv_1(\xv) \in \RM_1\\
\hat\mu_2(\xv) & = & \mu_2(\hat\xv_1) \\
\nabla \hat\mu_2(\xv) & = & \frac{2}{N}\hat\xv_1 
\end{eqnarray*}
\begin{eqnarray*}
\hat\xv_2(\xv) & = & \frac{\hat\xv_1(\xv)}{\sqrt{\mu_2(\hat\xv_1(\xv))}}\\ 
\left\{\mu_1(\hat\xv_2) = 0, \mu_2(\hat\xv_2) = 1 \right\} & \Leftrightarrow & \hat\xv_2(\xv) \in \RM_2\subset\RM_1\\
\hat\mu_3(\xv) & = & \mu_3\left(\hat\xv_2\right).\\
\end{eqnarray*}
To obtain the third normalization $\hat\xv_3(\xv)$ we aim to calculate the gradient of the skewness, in order to follow it until achieving the reference value for this feature (0, equivalent to a null third order central moment). After some calculations, this gradient yields:
\[
\nabla \hat\mu_3(\xv) = \frac{3}{N}(\hat\xv_2^{\odot 2} - 3 \hat\mu_3 \hat\xv_2 - 1).
\]
Unlike in previous cases, this gradient can not be analytically integrated.
However, we see that, in this case, the gradients of the decoupled features up to third order span at each point the same linear subspace as the gradients of the original moments. Therefore, they both define a single invariance submanifold, and we just need to find a path following the original gradients that imposes the three reference values for the moments (zero mean, unit variance, and zero skewness). Any order of integration is valid, as, according to the theory ~\cite{Arxiv2022}, the intersection of the invariance submanifold with the reference manifold (i.e., the normalization, Eq.~\eqref{eq:normalization}) exists and is unique within each connected domain. We have devised the following normalizing path for computing $\hat\xv_3(\xv)$ in four sequential steps:
\begin{enumerate}
    \item Follow $\nabla \mu_1$ until subtracting the mean: $\mu_1 \rightarrow 0$
    \item Follow $\nabla \mu_3$ until cancelling the third-order central moment: : $\tilde\mu_3 \rightarrow 0$, $\mu_1 \rightarrow ?$
    \item  Follow $\nabla \mu_1$ until subtracting the mean: $\mu_1 \rightarrow 0$, $\mu_3 \rightarrow 0$
    \item  Follow $\nabla \mu_2$ until imposing unit variance: $\mu_1 \rightarrow 0$, $\mu_2 \rightarrow 1$, $\mu_3 \rightarrow 0$
\end{enumerate}
Step 1 is just a convenient pre-processing for Step 2, which yields a simple Riccatti equation $x'(t)\propto x^2(t)$, whose solution is $x(t) = x(0)/(1 - t\,x(0))$, with an integration bound that must be numerically computed (see Eq.~\eqref{eq:integration_bound}). Note that (i) Step 3 does not affect the adjustment of the zero third-order central moment achieved in Step 2 because the latter is invariant to adding a constant, and (ii) Step 4 does not change the adjustment of Steps 2 and 3, because it is a simple scaling factor on a sample having zero first- and third-order moments. Therefore, the normalization achieves the three desired reference values for the moments.
More formally, the performed normalization and the corresponding definition of the next decoupled moment, $\hat\mu_4$ is:
\begin{eqnarray}
    \hat\xv_3(\xv) & = & \hat\xv_2\left(\frac{\hat\xv_1}{1 - t_0 \hat\xv_1}\right) 
    \label{eq:integration_ricatti} \\
    \left\{\mu_1(\hat\xv_3) = 0, \mu_2(\hat\xv_3) = 1, \mu_3(\hat\xv_3) = 0 \right\} & \Leftrightarrow & \hat\xv_3(\xv) \in \RM_3\subset\RM_2\subset\RM_1
     \nonumber \\
    t_0 & = & \arg_t \left\{\tilde\mu_3\left(\frac{\hat\xv_1}{1 - t \hat\xv_1}\right) = 0\right\}  
    \label{eq:integration_bound}\\
    \hat\mu_4(\xv) = \mu_4(\hat\xv_3)& = & \mu_4\left(\hat\xv_2\left(\frac{\hat\xv_1}{1 - t_0 \hat\xv_1}\right)\right),
    \label{eq:orthokurtosis}
\end{eqnarray}
where $\tilde\mu_3(\xv) = \mu_3(\xv - \mu_1(\xv))$.
Note that the so-defined {\em ortho-kurtosis} (Eq.~\eqref{eq:orthokurtosis}) is not just decoupled from the mean and the variance (as the skewness and the kurtosis are), but it is also decoupled from the skewness.

\subsubsection{Transfer}
For de-normalization purposes, we operate from inner to outer normalization layers, 
as explained in Alg.~\ref{alg:NeN_reversed}. 
For adjusting the fourth-order decoupled moment (the ortho-kurtosis) one must check first if such an adjustment is well-defined, 
testing the Frobenius condition on the gradients of the decoupled features~\cite{Arxiv2022}. 
Now the four gradient fields no longer jointly fulfill that condition~\cite{Arxiv2022}. This implies that now there is no invariance submanifold associated with the first four decoupled moments 
and, as a consequence, there is no unique adjustment of the four desired decoupled moments  by integrating linear combinations of their local gradients. Although this prevents us 
to obtain an analytical solution, we can still use Algorithm~\ref{alg:NeN_reversed} for the de-normalization, and numerically integrate the gradients by following the narrow path trajectory, as a particular solution for the adjustment. So, we start our ODE excursion in $\hat\xv_3(\xv_{src})$ and integrate numerically from there the gradient of the ortho-kurtosis. Within $\RM_{3}$ that gradient can be computed by projecting the gradient of the fourth order moment onto the orthogonal span of the previous gradients~\cite{Arxiv2022}. We continue the integration until its target value is reached in $\hat\zv_3(\xv_{src},\hat\mu_4(\xv_{tgt}))$, with $\mu_4(\hat\zv_3) = \hat\mu_4(\xv_{tgt})$. Note that $\hat\zv_3$ still belongs to $\RM_3$. 
Then, for adjusting the lower order features we follow the same four-step technique used for computing $\hat\xv_3$, but adapted to adjust the skewness to its target value, by simply substituting the desired value in place of the reference value, and $\hat\mu_3$ (skewness) in place of $\tilde\mu_3$ (third-order central moment), in Eq.~\eqref{eq:integration_bound}. Then, the mean and standard deviation are trivially adjusted, as usual, yielding:
\begin{eqnarray}
    \zv(\xv_{src};\hat{\bf\mu}(\xv_{tgt})) & = & \sqrt{\hat\mu_2(\xv_{tgt})}\hat\xv_2\left(\frac{\hat\zv_3(\xv_{src};\hat\mu_4(\xv_{tgt}))}{1 - t_s(\xv_{src},\xv_{tgt}) \hat\zv_3(\xv_{src};\hat\mu_4(\xv_{tgt}))}\right) + \hat\mu_1(\xv_{tgt})
     \label{eq:denormalization_ricatti}\\
    t_s(\xv_{src},\xv_{tgt}) & = & \arg_t \left\{\hat\mu_3\left(\frac{\hat\zv_3(\xv_{src},\hat\mu_4(\xv_{tgt}))}{1 - t \hat\zv_3(\xv_{src};\hat\mu_4(\xv_{tgt}))}\right) = \hat\mu_3(\xv_{tgt}) \right\}
    \label{eq:denormalization_integration_bound_2}\\
    \hat\mu_j(\zv(\xv_{src};\hat\mu(\xv_{tgt}))) & = & \hat\mu_j(\xv_{tgt}), \,\,\, j = 1,\dots,4.
   \nonumber 
\end{eqnarray}
In our case, for style transfer purposes, we have observed a very regular and robust behavior of all the adjustments. 

As a summary of this section, we enumerate the basic results:
\begin{enumerate}
    \item The computation of the first three decoupled moments, by using Alg.~\ref{alg:nested_normalization_broad_path}, results in the standardized first three moments:  mean, variance and skewness.
    \item The normalization (Eqs.~\eqref{eq:integration_ricatti} and~\eqref{eq:integration_bound}) and adjustment (Eqs.~\eqref{eq:denormalization_ricatti} and~\eqref{eq:denormalization_integration_bound_2}) of the first three decoupled moments is unique, and very efficient by using Algs.~\ref{alg:nested_normalization_broad_path} and~\ref{alg:NeN_reversed} in analytic form. 
    \item From normalizing the previous decoupled moments and measuring the fourth-order moment emerges the new (fourth-order) fully decoupled moment, the {\em ortho-kurtosis} (Eq.~\eqref{eq:orthokurtosis}), which is exactly decoupled everywhere to the previous moments.
    \item The adjustment of the ortho-kurtosis (and, eventually, of higher order {\em approximately decoupled} moments) is not unique. We use the narrow-path algorithm (Alg.~\ref{alg:nested_normalization_narrow_path}) to numerically obtain a particular solution that works well in practice. 
\end{enumerate}
Finally, a relevant aspect to note is that the luminance transformation of Eq.\eqref{eq:denormalization_ricatti} corresponds to applying a non-linear point-wise function to the luminance of each pixel: $z_n = \phi(x_n;\xv)$. If we ignore the dependency of $\phi$ with the global input vector $\xv$, we can build a look-up table (LUT) and {\em freeze} that point-wise transformation to perform a very fast style transfer to similar input samples (e.g., frames in a video sequence). 
The requirement is that these other input samples $\xv'$ have global features similar to $\xv$, i.e., 
 $\fv(\xv')\approx\fv(\xv)$.

\subsection{Approximately decoupling analysis and transfer of the MSV at the output of a set of filters} 
\label{subsec:approx_decoupl_VF}
\begin{figure}[h!]
    \centering
    \includegraphics[width=0.6\textwidth]{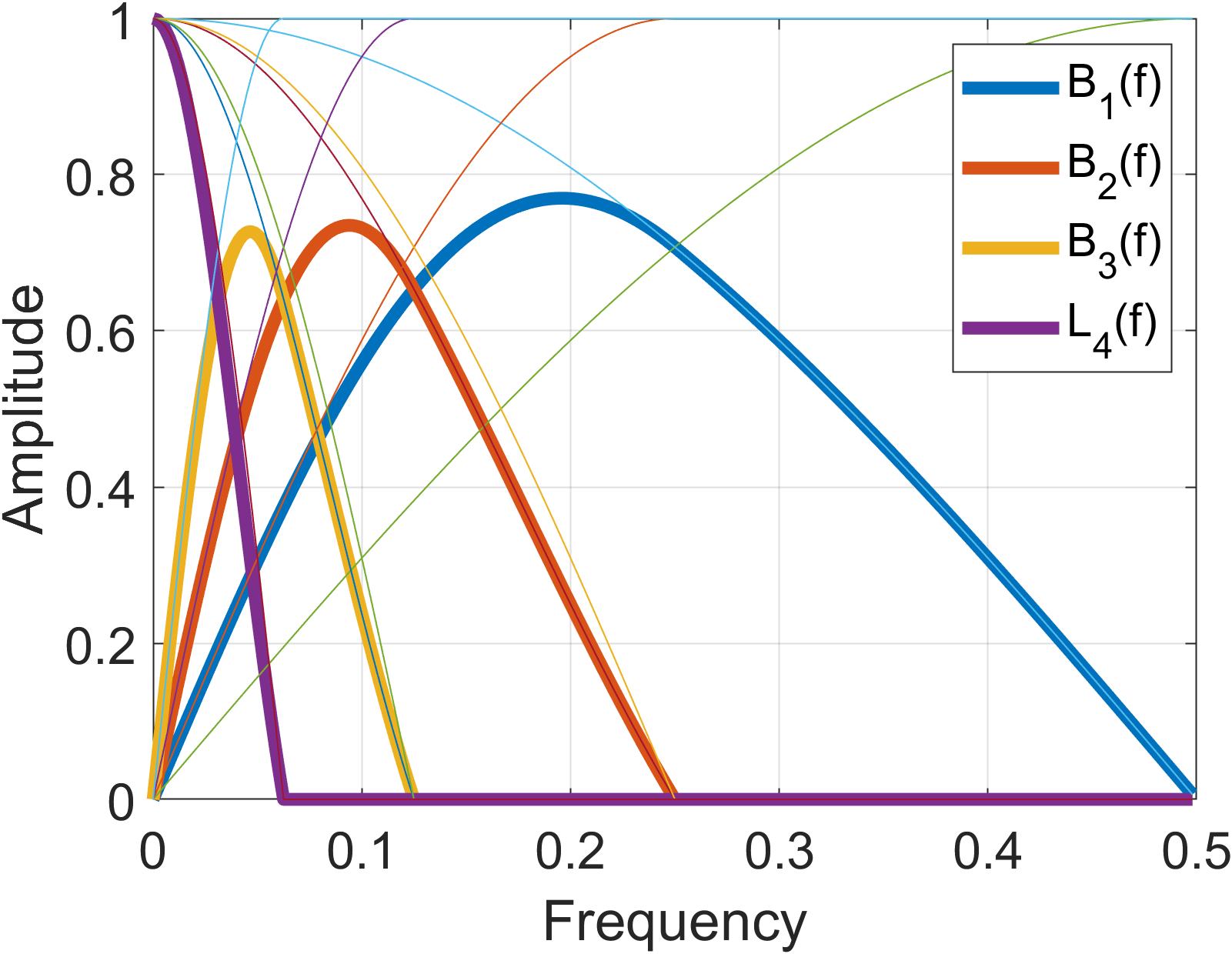}
    \caption{
    \textcolor{black}{
    In thick lines, the spectral profiles of the three band-pass plus one low-pass frequency bands whose mean square values (MSV), together with the total MSV, describe the spectral behavior of the image. They characterize a convolution kernel K imitating an optical PSF when having a sharp image and its diffused version, captured on the same real scene. Then that PSF effect can be transferred to other images. Thin lines correspond to the auxiliary filters used in the recursive construction (see Eqs.~\eqref{eq:filter_recursive_design}).}} 
    
    \label{fig:radial_profile_filters}
\end{figure}

We propose decoupling the mean square value at the output of a set of overlapping isotropic filters defining a Parseval frame, 
as a parsimonious and robust descriptor of the image power spectrum. 
%
We added to these features, the image's mean and MSV.
The set of $M=6$ original features consists of:
\begin{enumerate}
    \item $f_1(\xv) = 1/N\sum_{j=1}^{N}{x_j} = \mu_1(\xv)$, the luminance mean.
    \item $f_2(\xv) = 1/N\sum_{j=1}^{N}{x_j^2} = \mu_2(\xv)$, the luminance mean square value (MSV).
    \item $f_{2+k}(\xv) = 1/N\sum_{j=1}^{N}{(\xv\ast h_{B_k})_j^2} = \mu_2(\xv\ast h_{B_k}), k = 1,2,3$;  MSV at the output of three isotropic band-pass filter at different scales, $B_1$, $B_2$ and $B_3$.
    \item $f_6(\xv) = 1/N\sum_{j=1}^{N}{(\xv\ast h_{L_4})_j^2} = \mu_2(\xv\ast h_{L_4})$:  MSV at the output of a low-pass filter, $L_4$.
\end{enumerate}
Because of constituting a Parseval frame, this representation accomplishes:
\[
MSV_{TOTAL} = MSV_{H_{00}} + \sum_{j=1}^3{MSV_{B_j}} + MSV_{L_4}, 
\]
where $MSV_{H_{00}}$ represents the MSV at the output of a high-pass filter $H_{00}$ complementary of the others. 
We have designed a wavelet-like recursive dyadic scheme for the filters: we first apply a high-pass $H_{00}(f)$ and a low-pass filter $L_{00}(f)$,  where 
$f$ represents the radial frequency. To the result of the low-pass filtering we apply again scaled versions of the low-pass filter $L_{01}(f)$ and high-pass filter $H_{01}(f)$, and so on. 
These pairs of real filters fulfill in the frequency domain $H_{0k}^2(f) + L_{0k}^2(f) = 1$, at every scale $k$.
We initialize the recursion by choosing $H_{00}(f) = \sin(\pi f)$ for $f\leq 1/2$ and 1 otherwise, and apply, for $k$ from 1 to 3:
\begin{eqnarray}\label{eq:filter_recursive_design}
H_{0k}(f) & = & H_{00}(2^k f),\,\,\, f\leq 2^{-(k+1)} \nonumber\\
H_{0k}(f) & = & 1, \,\,\,\,\,\,\,\,\,\,\,\,\,\,\,\,\,\,\,\,\,\,\,\,f> 2^{-(k+1)} \nonumber \\
L_{0k}(f) & = & \sqrt{1 - H_{0k}^2(f)} \nonumber \\
L_1(f) & = & L_{00}(f) \nonumber \\
B_k(f) & = &  L_{k}(f)H_{0k}(f)\nonumber \\
L_{k+1}(f) & = & L_{k}(f)L_{0k}(f)
\end{eqnarray}
%
Figure~\ref{fig:radial_profile_filters} shows the radial profile of the main filters \textcolor{black}{($B_1,B_2,B_3$ and $L_4$)} in thick lines and the auxiliary filters in thin lines.
%
The spectral overlapping of these filters causes the MSV at their outputs to be coupled. 
%
%
We have chosen for the reference values the expected MSV at the output of these filters for a zero mean and variance one white noise input.  
We computed these reference values by numerically integrating the square modulus responses of the filters in the 2D Fourier domain, obtaining: 
$\vv^{ref} \approx 
[0,1,0.1829114,0.0383796,0.0092141,0.0030390]^T$.
As a consequence, normalizing these features will have a whitening effect on the image spectrum.
%

\subsubsection{Analysis}
The first two features (mean and MSV in the pixel domain) just play the role of extracting and normalizing the sample mean and variance of the image.
Therefore, we can write the first three decoupled features as:
\begin{eqnarray}
\g_1(\xv) & = & \mu_1(\xv) \nonumber \\
\g_2(\xv) & = & \mu_2(\hat\xv_1) \nonumber \\
\g_3(\xv) & = & \mu_2(\hat\xv_2\ast h_{B_1}), \nonumber 
\end{eqnarray}
where here the normalized vectors $\hat\xv_1,\hat\xv_2$ 
have the same meaning 
as in Subsection \ref{subsec:moments_analysis}.
From here on, by following the relaxed version of the NeN broad-path Algorithm~\ref{alg:nested_normalization_broad_path}, we need to follow a local linear combination of the original features' gradients to impose the new reference values without releasing the previous ones (the zero mean and the unit variance, initially).
Let us study those gradients and the set of possible solutions they give rise.
%
We have:
\[
\nabla f_j(\xv) = \frac{2}{N} \xv * \tilde{\hv}_j,
\]
where $\tilde{\hv}_j(\nv) = \hv_j(\nv)*\hv_j(-\nv)$.
To simplify this expression and its subsequent integration, we express it in the Fourier domain, by doing the DFT of $\xv$ and $\hv$: \begin{equation}
\label{eq:gradient_variance}
G_j(X(\xiv)) = {\cal F}\{\nabla f_j(\xv)\} = \frac{2}{N} |H_j(\xiv)|^2 X(\xiv),
\end{equation}
where $\xiv$ represents the (2-D vector) discrete frequency, 
and upper case letters represent the Fourier transforms of their original lower-case counterparts (except for $G$, which corresponds to the Fourier transform of the gradients). It is easy to test that any set of gradients of this form fulfills the Frobenius condition~\cite{Arxiv2022}.

We will express the normalization solution w.r.t. all gradients up to $k$ index as a concatenation of 1-D integrating trajectories, one along each gradient $\nabla f_j$, each for a time $t_j$. The ODE equation for a single gradient is: 
\[
\frac{dY_{(j)}(\xiv,t)}{dt} = \frac{2}{N}|H_j(\xiv)|^2 Y_{(j)}(\xiv,t),
\]
whose solution is, in the Fourier domain:
\begin{equation}
    \label{eq:fix_variance}
    Y_{(j)}(\xiv,t) = Y_{(j-1)}(\xiv,t_{(j-1)}) \exp\Big(\frac{2}{N}|H_j(\xiv)|^2t\Big).
\end{equation}
By concatenating these solutions until the $k$-th term, starting from $Y_0(\xiv,t_0) = \hat X_1(\xiv)$ (the original sample with its mean subtracted: $\hat X_1(\xiv\neq{\bf 0}) = X(\xiv)$, $\hat X_1({\bf 0}) = 0$), it yields:
\begin{equation}
    \label{eq:fix_variance_}
    Y_k(\xiv,\tv) = \hat X_1(\xiv) \exp\Big(\frac{2}{N}\sum_{j=0}^{k-2}{|H_j(\xiv)|^2t_{j+2}}\Big),
\end{equation}
where $\tv = [t_2\dots t_k]$.
%
%
%
For our bank of filters we have $H_i(\xiv) = B_{i-1}(\xiv), i = 1\dots 3$, and $H_4(\xiv) = L_4(\xiv).$~\footnote{Now expressed 
as a function of the 2-D frequency vector $\xiv$ instead of on the radial frequency $f$.} 
Note, first, that if the mean is set to zero none of the subsequent transformations of Eq.~\eqref{eq:fix_variance} will affect this condition (because $Y_k({\bf 0},\tv) = \hat X_1({\bf 0}) = 0$, for all $\tv$). Second, the total MSV is a particular case of MSV at the output of a neutral (delta) filter $H_0(\xiv)=1$. The case of $H_0$ is special because there is analytic solution to the problem of adjusting the sample variance by an ODE excursion: 
the MSV is normalized by dividing the transformed sample by its rms value. Thus, Eq.~\eqref{eq:fix_variance} for $k>2$ now becomes:
\begin{equation}
    \label{eq:fix_variance_2}
    Y_k(\xiv,\tv) = \frac{\hat X_1(\xiv) \exp\Big(\frac{2}{N}\sum_{j=1}^{k-2}{|H_j(\xiv)|^2t_{j+2}}\Big)}
    {\sqrt{\sum_{\xiv}{|\hat X_1(\xiv)|^2 \exp\Big(\frac{4}{N}\sum_{j=1}^{k-2}{|H_j(\xiv)|^2t_{j+2}}\Big)}}},
\end{equation}
where now $\tv = [t_3\dots t_k]$, which fulfills  $Y_k({\bf 0},\tv)=0$ ($v_1^{ref}$) and $\sum_{\xiv}{|Y_k(\xiv,\tv)|^2} = 1$  ($v_2^{ref}$), independently of $\tv$.
Now, to obtain 
$\hat\xv_k(\xv;\vv_k^{ref})$ for $k>2$, we solve for
\begin{equation}
\label{eq:ref_condition_VF}
    \tv_k^{ref} =\displaystyle \underset{\tv}\arg
    \Big\{\sum_{\xiv}{|H_j(\xiv)|^2 \,|Y_k(\xiv,t_{j+2})|^2} = v_{j+2}^{ref}\Big\}_{j = 1}^{k-2}.
\end{equation}
We efficiently solve for $\tv_k^{ref}$ in 
Eq.~\eqref{eq:ref_condition_VF} by numerically minimizing the sum of the squared error terms.
Then, $\hat{X}_k(\xiv;\vv_k^{ref}) = Y_k(\xiv,\tv_k^{ref})$, and $\hat\xv_k(\xv;\vv_k^{ref}) = {\cal F}^{-1}\{\hat{X}_k(\xiv;\vv_k^{ref})\}$.
Finally, the new decoupled feature is obtained, as always, by $\g_{k+1}(\xv) = f_{k+1}(\hat\xv_k(\xv))$.

Although not shown here, in this case the gradients of the resulting features $\g_j$ do not fulfill the Frobenius condition beyond $j = K=2$. This implies (see
~\cite{Arxiv2022}) that only the $K+1=3$ first output features 
will have exactly mutually orthogonal gradients everywhere. The rest of the features will be strictly orthogonal only to $\g_1$ and $\g_2$. Nevertheless, empirical tests using another bank of filters and a set of photographic images indicate that the remaining mutual coupling for the rest of the features is expected to be low~\cite{Arxiv2022}. 

\subsubsection{Transfer}

For transferring the decoupled features from one image to another, we apply the three-step Algorithm~\ref{alg:CFA}.
The method for the de-normalization step is analogous to the one of normalization, as described in Section~\ref{subsec:de-normalization}. It is performed by solving the $k-2$ Eqs.~\eqref{eq:fix_variance_2}, for $k= M$ to $k = 3$ (i.e., now in reverse order), leaving the normalized values $v_j^{ref},j=3\dots k-1$ unchanged, but now substituting 
$v_k^{ref}$ by $v_k^{des} =\hat f_k(\xv_{tgt})$, the desired value of the decoupled feature measured in the target. 

It is worth noting that in this case both normalization and de-normalization are filtering operations with their corresponding equivalent filter $\hat H_{eq},\check{ H}_{eq}$  (see Eq.~\eqref{eq:fix_variance_2}). 
By applying first the normalization and then the de-normalization (see Eq.~\eqref{eq:denormalization_transfer_1}), and adjusting the sample mean and variance, we finally obtain:
\begin{eqnarray}
Z_{f0}(X_{src}(\xiv);\hat\fv(\xv_{tgt})) & = & \hat X_1^{src}(\xiv)\hat H_{eq}(\xiv;\{h_j\}_{j=1}^{M-2},\vv^{ref})\check{ H}_{eq}(\xiv;\{h_j\}_{j=1}^{M-2},\hat\fv(\xv_{tgt})) \nonumber \\
\zv_f(\xv_{src};\hat\fv(\xv_{tgt})) & = & \sqrt{\hat f_2(\xv_{tgt})}{\cal F}^{-1}\{Z_{f0}(X_{src}(\xiv);\hat \fv(\xv_{tgt}))\} + \hat f_1(\xv_{tgt}) \nonumber
\end{eqnarray}
When the target corresponds to the same scene as the source but with some optical diffusion applied, the computed kernel  $h_{eq} = \hat h_{eq}\ast\check{h}_{eq}$ will be an approximation of the optical PSF. 


\section{Experiments}
\label{sec:experiments}
\color{black}
The evaluation of ST is as challenging as the definition of the problem itself and for the same reasons. The problem relies on a separation between an image's ``style'' and its ``essence'', such that one image's ``style'' can be applied over the ``essence'' of another.
So, the evaluation of a method's effectiveness first requires the definition of an application relevant or arbitrary threshold which separates the two. For each ST method, this threshold is defined by the features which are extracted and matched, and each user carries their own expectation of what should be transferred from the reference and what source details should stay the same. The user and method thresholds rarely align perfectly, making the judgment of ST quality particularly subjective compared to other image processing problems. For this reason, we believe the most effective way to evaluate these methods is by characterizing their functionality through qualitative evaluations in challenging scenarios. In this section, we will make such comparisons between the proposed method and all of those discussed in the related work section, and present a psychophysical experiment comparing the most relevant approaches. To augment this evaluation, larger test sets corresponding to each example shown here are included in the supplementary material: {\href{https://drive.google.com/drive/folders/15nKBljq0CCA9pm6i1fcISFSUvuYoBhac?usp=share_link}{LINK}}.

\subsection{Method usability}

\begin{figure*}[t]
\centering
\begin{tabular}{cccc}

    \includegraphics[width=0.29\textwidth]{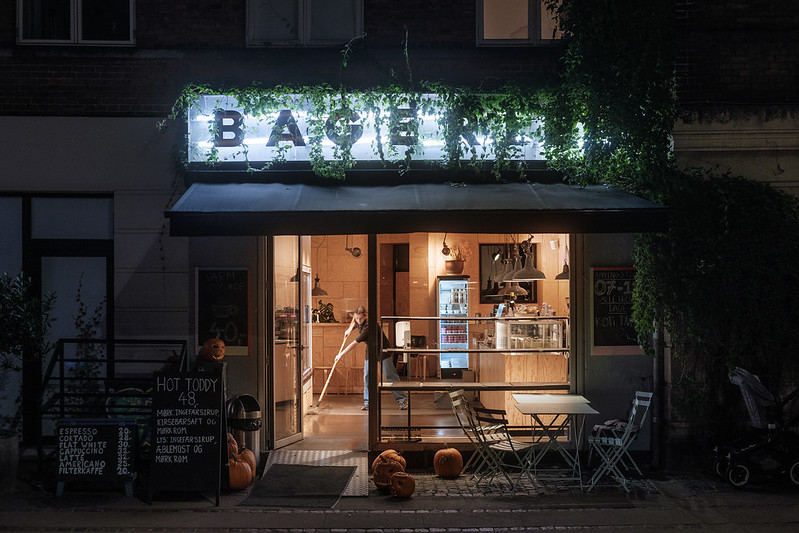}& 
    \includegraphics[width=0.29\textwidth]{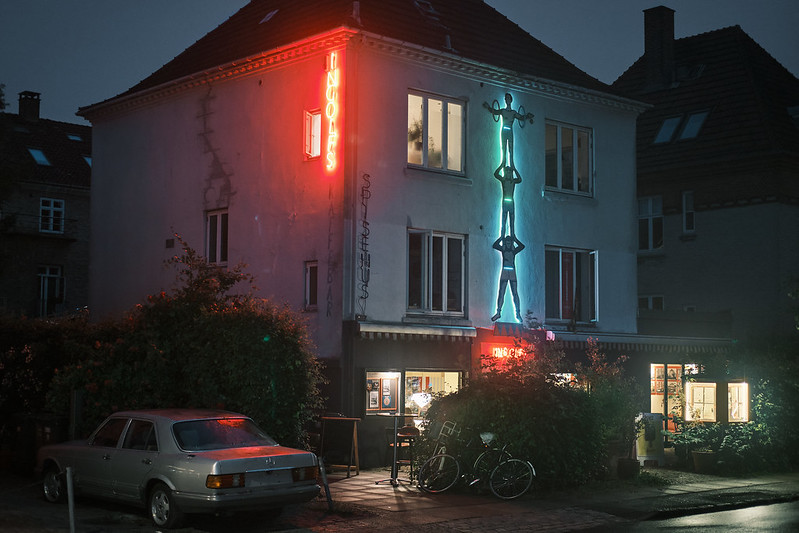}& 
    \includegraphics[width=0.29\textwidth]{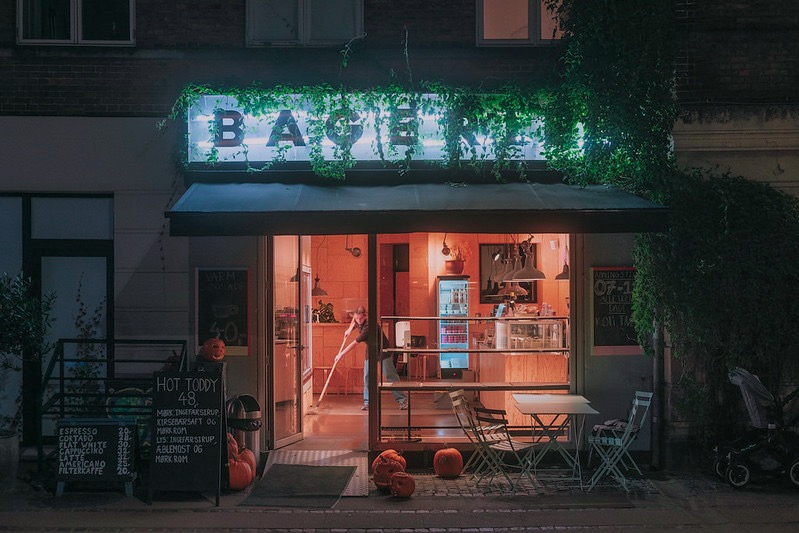} \\
    
    \includegraphics[width=0.29\textwidth]{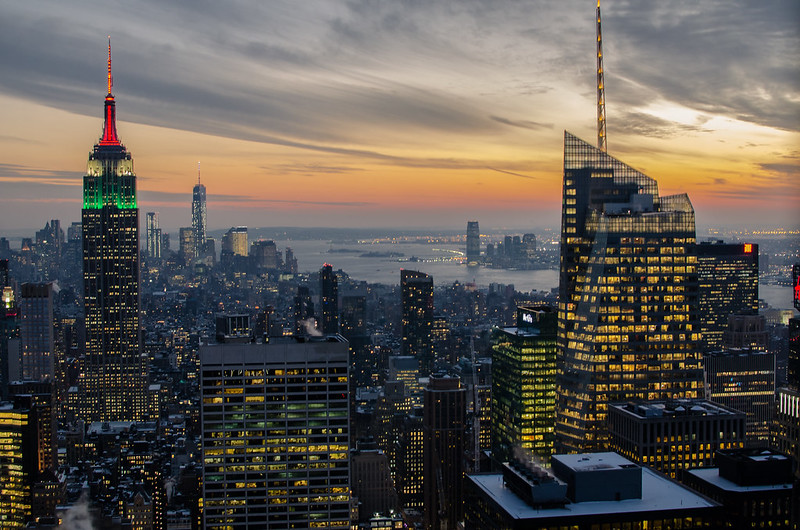}&     
    \includegraphics[width=0.29\textwidth]{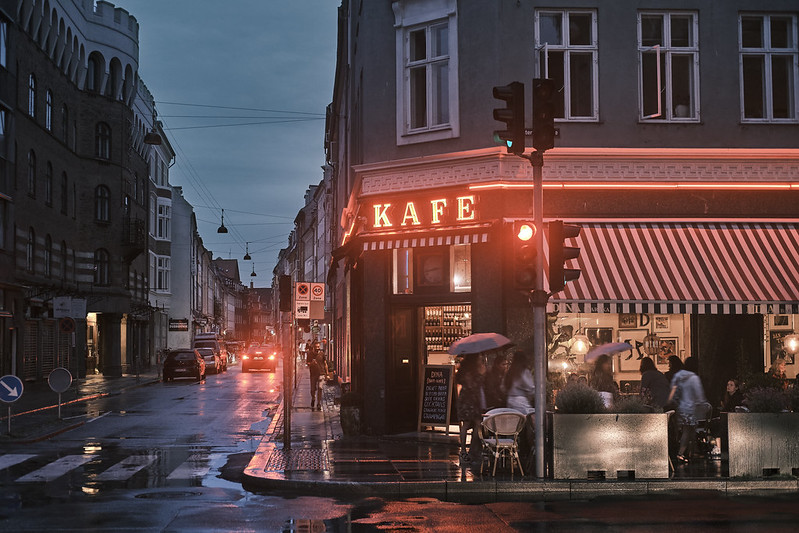}& 
    \includegraphics[width=0.29\textwidth]{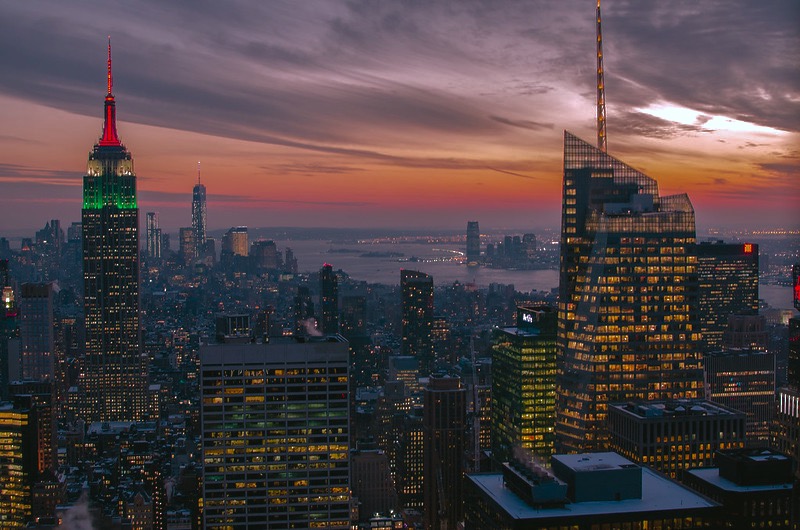}  \\    
    
    \includegraphics[width=0.29\textwidth]{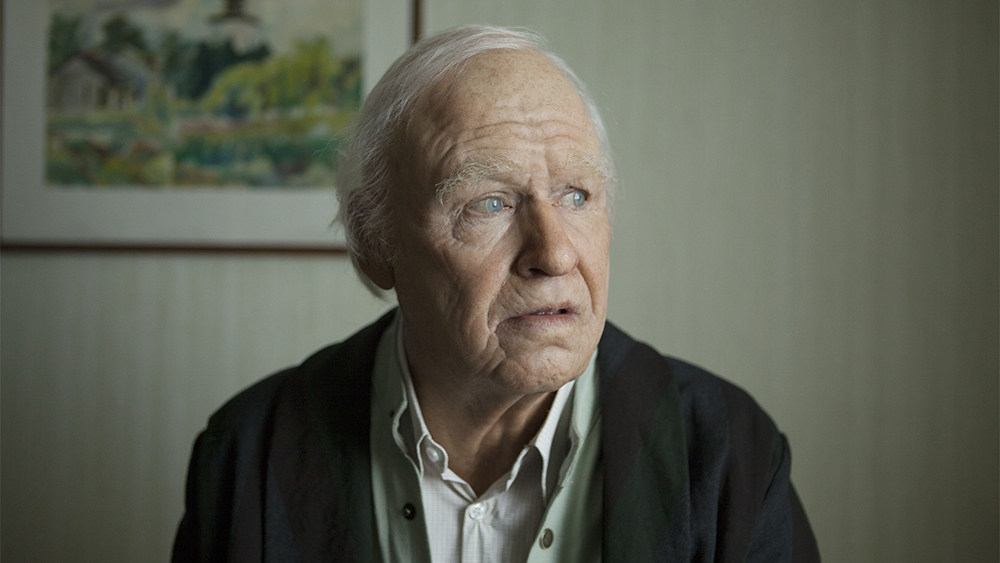}& 
    \includegraphics[width=0.29\textwidth]{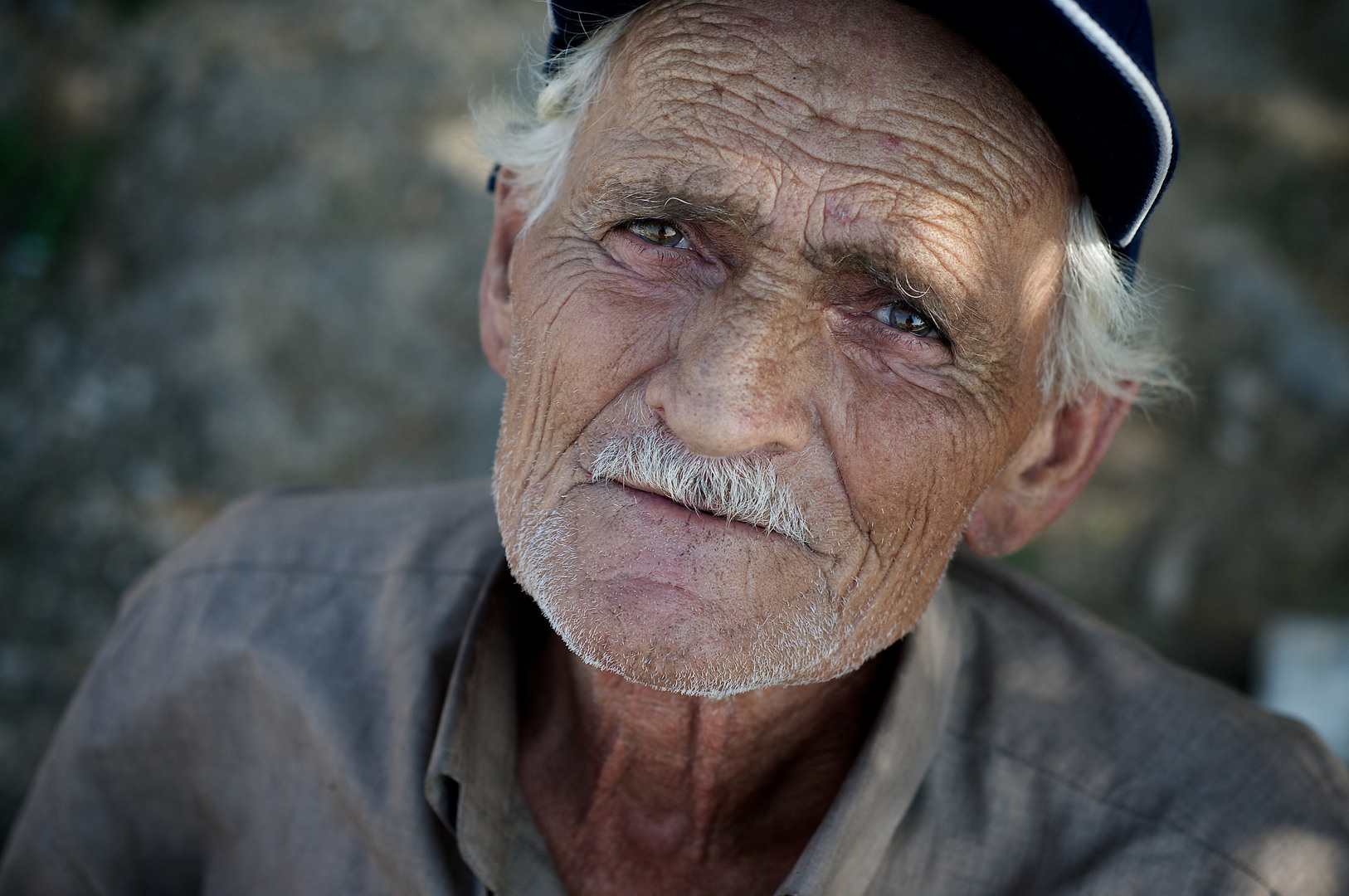}& 
    \includegraphics[width=0.29\textwidth]{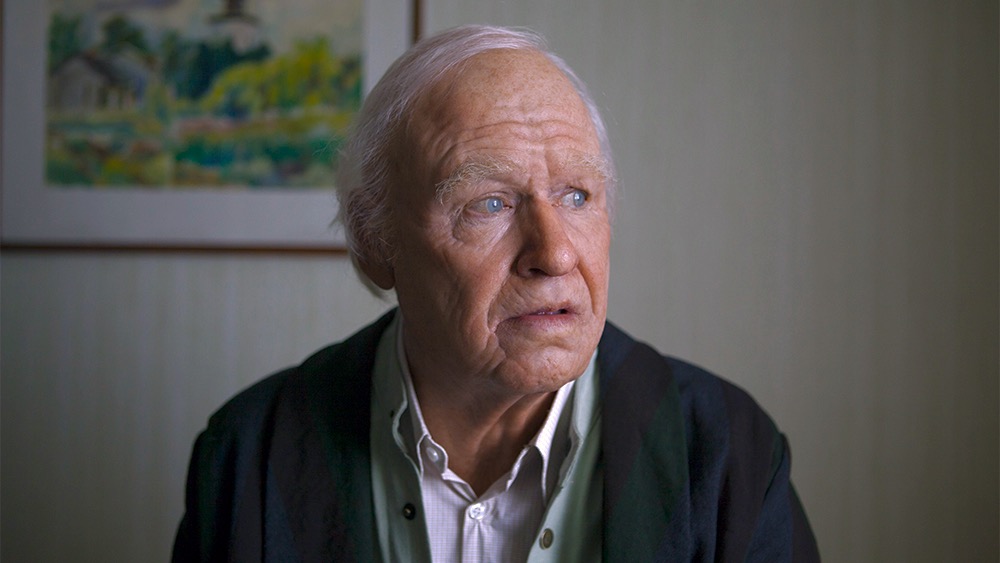} \\
    
    \includegraphics[width=0.29\textwidth]{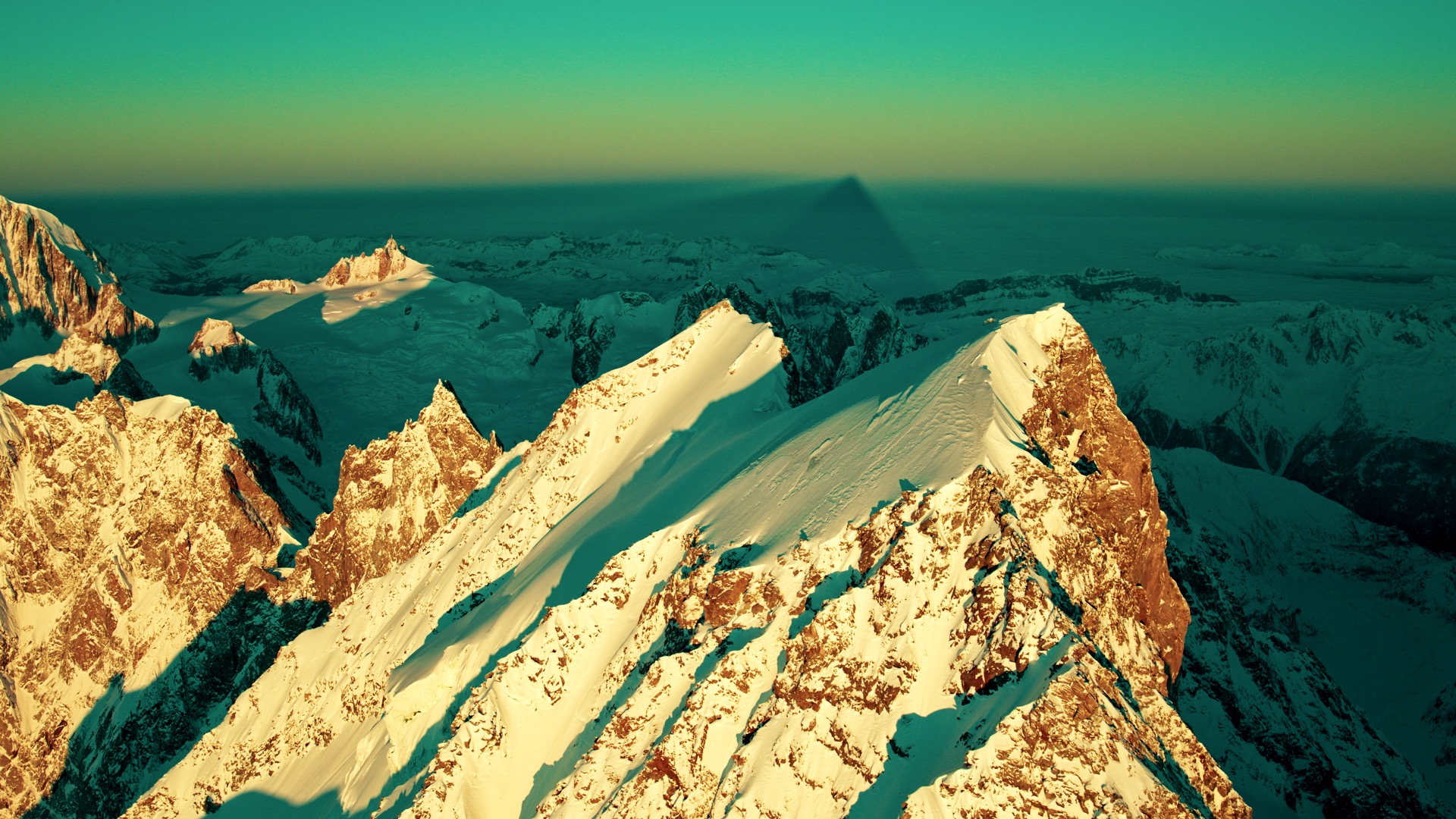}& 
    \includegraphics[width=0.29\textwidth]{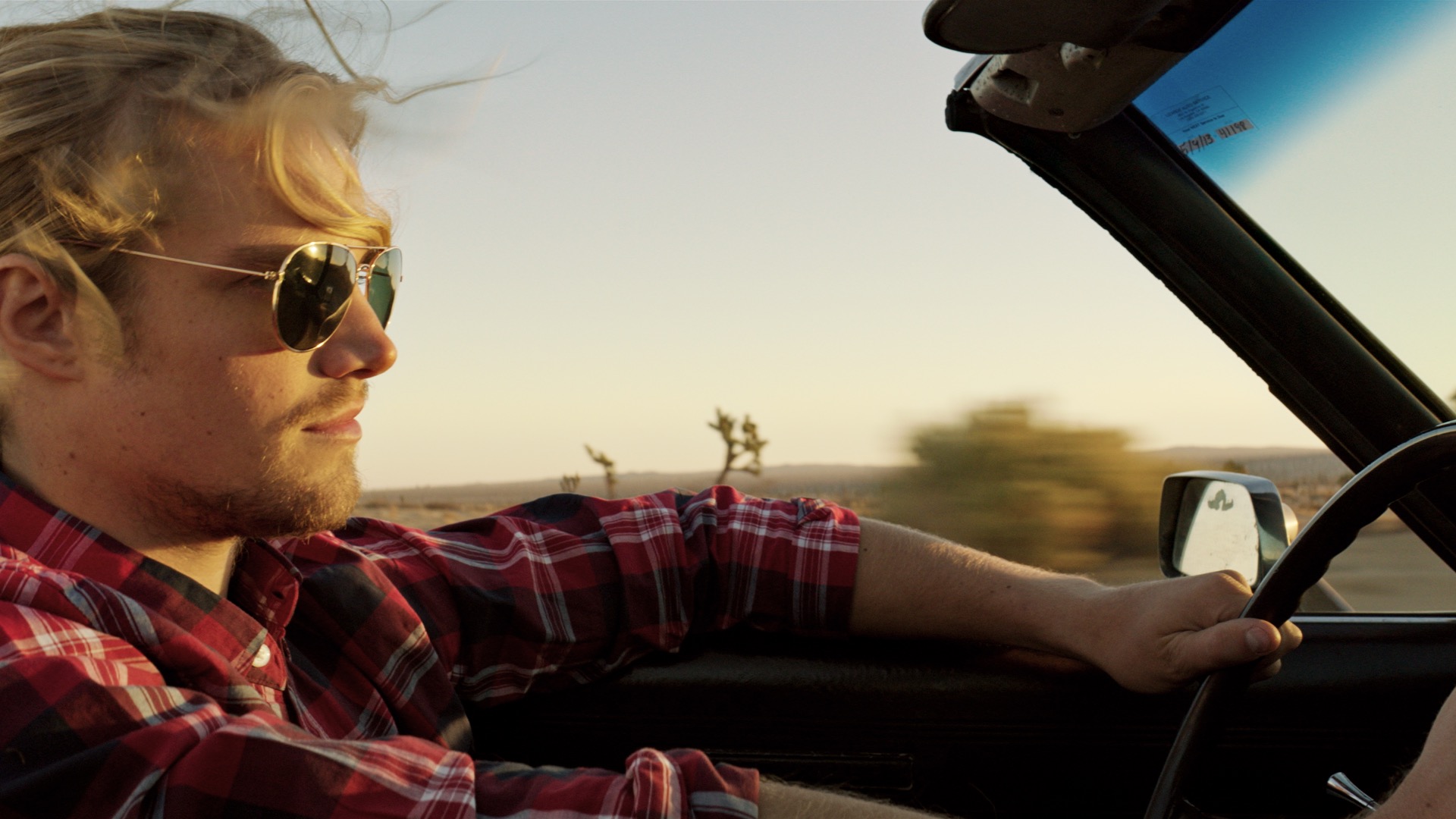}&      
    \includegraphics[width=0.29\textwidth]{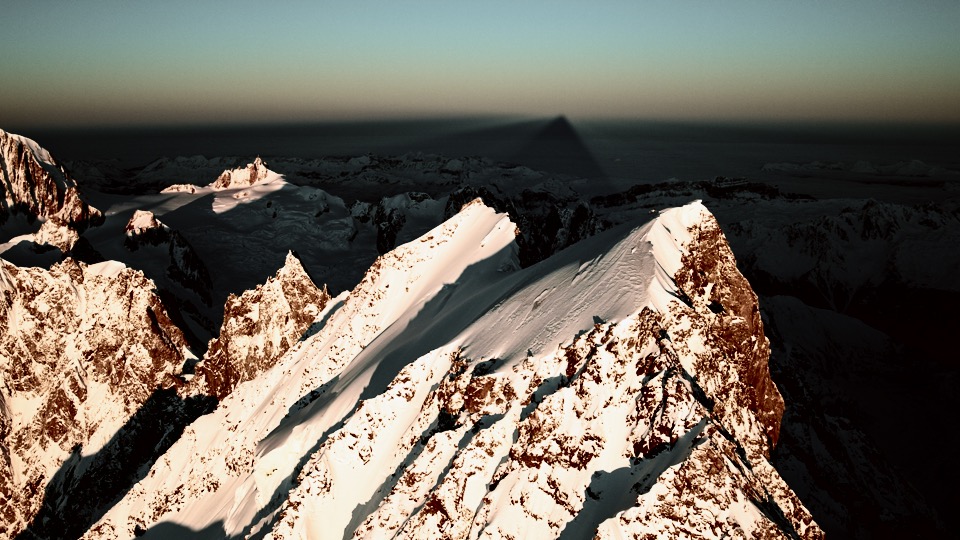} \\

  \end{tabular}
  \caption{From left to right, varied source images, target images, and the results of the default configuration of our method. First row source and target images, and the second row target image uploaded to Flickr by Kristoffer Trolle (CC-BY 2.0). Second row source images uploaded to Flickr by Mk Feeny (CC-BY 2.0). Third row images sourced from \cite{LFWTech}. Bottom row images provided by ARRI \cite{ARRI}.}
  \label{fullMethod}
\end{figure*}

To begin, we offer some remarks on the results and usability of our method. In its \textcolor{black}{default configuration outlined in section \ref{subsec:overview}}, the proposed method is applied to images in an IPT color representation, where it transfers the mean and variance values from target to source for all three channels, and skew and ortho-kurtosis values only on the I channel. This particular configuration was determined to be an optimal balance between \textcolor{black}{avoiding artifacts and} achieving a strong visual similarity to the target. In Figure \ref{fullMethod}, we demonstrate the performance of this automatic configuration on a series of images. It is clear to see from these examples that the results take on a pleasing and visually similar appearance to their targets while preserving important details of the original sources.

\begin{figure*}[t]
\centering
\begin{tabular}{cccc}

    \includegraphics[width=0.22\textwidth]{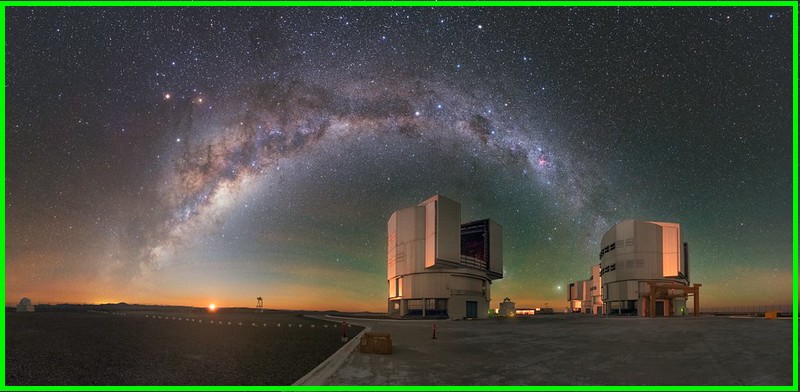}& 
    \includegraphics[width=0.22\textwidth]{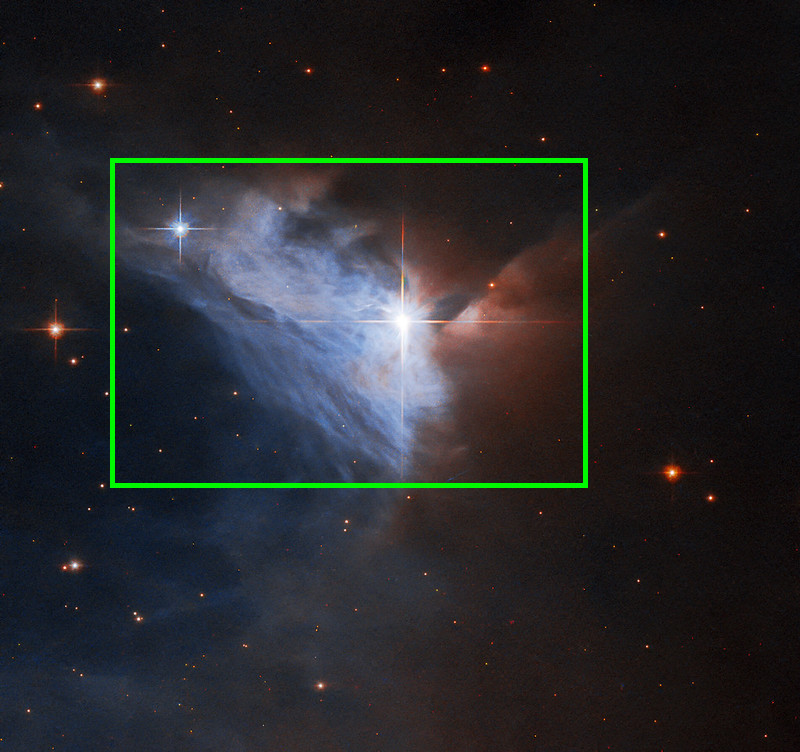}& 
    \includegraphics[width=0.22\textwidth]{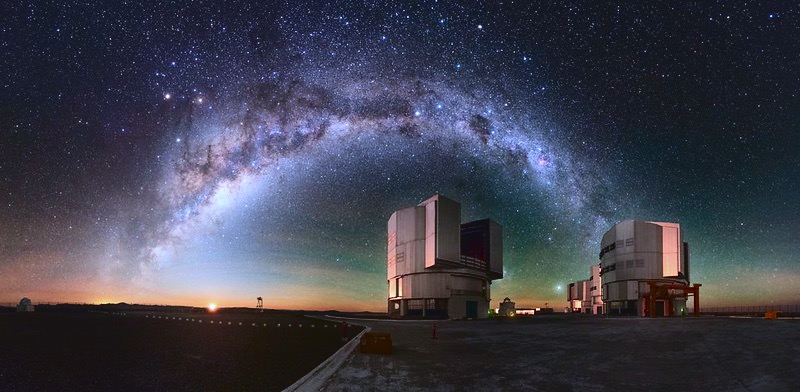}& 
    \includegraphics[width=0.22\textwidth]{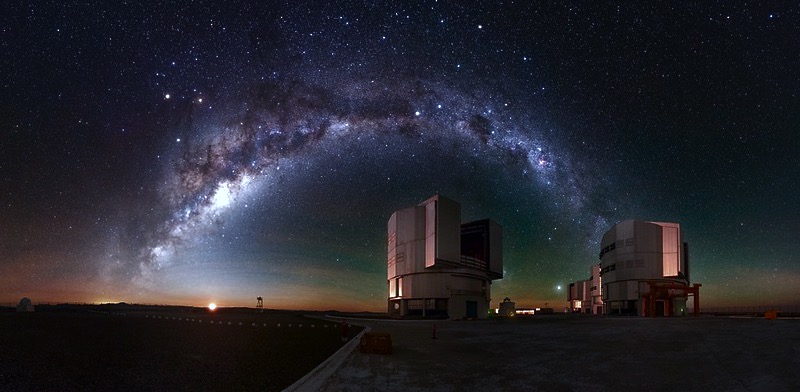} \\ 

    \includegraphics[width=0.22\textwidth]{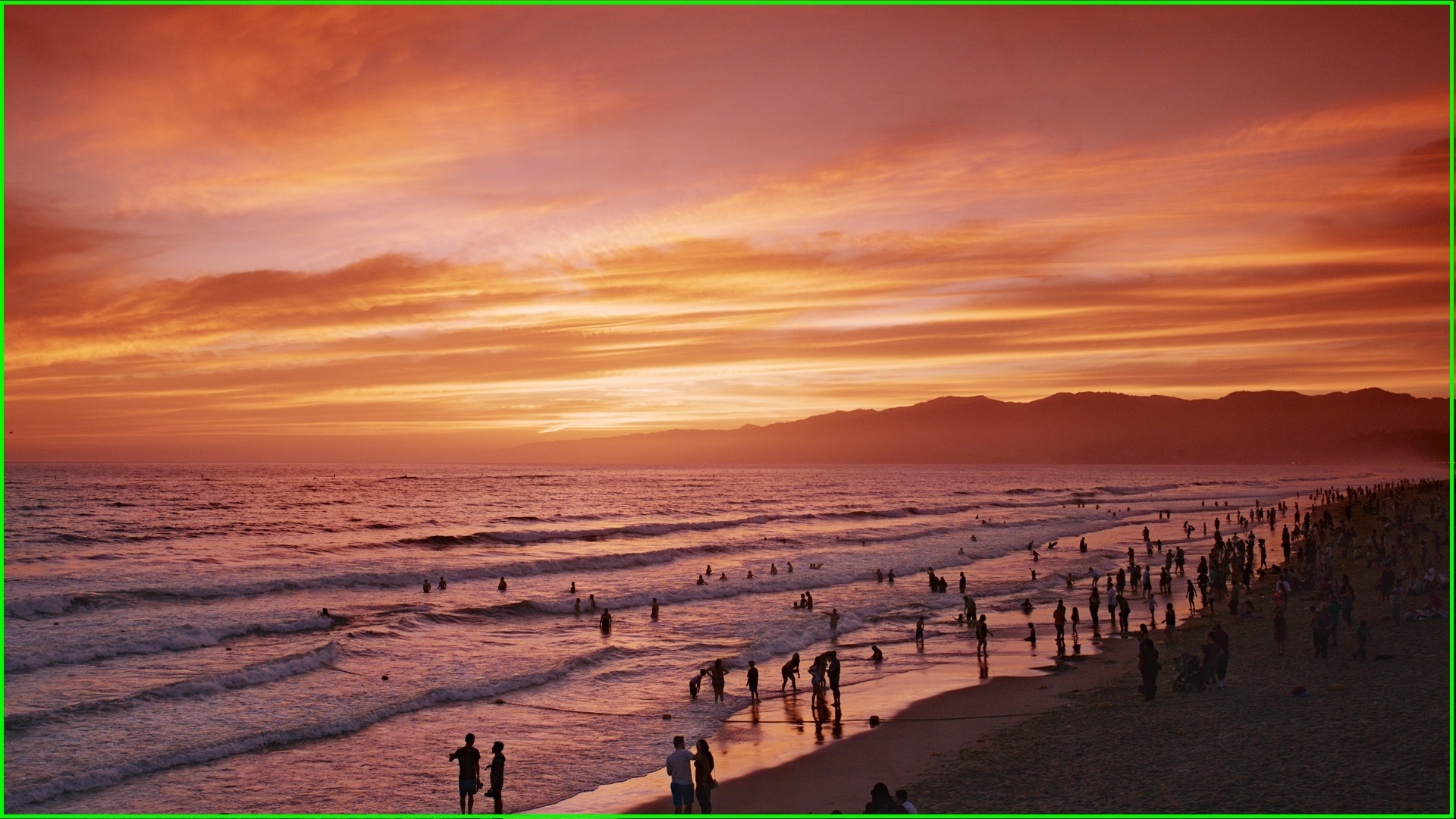}&
    \includegraphics[width=0.22\textwidth]{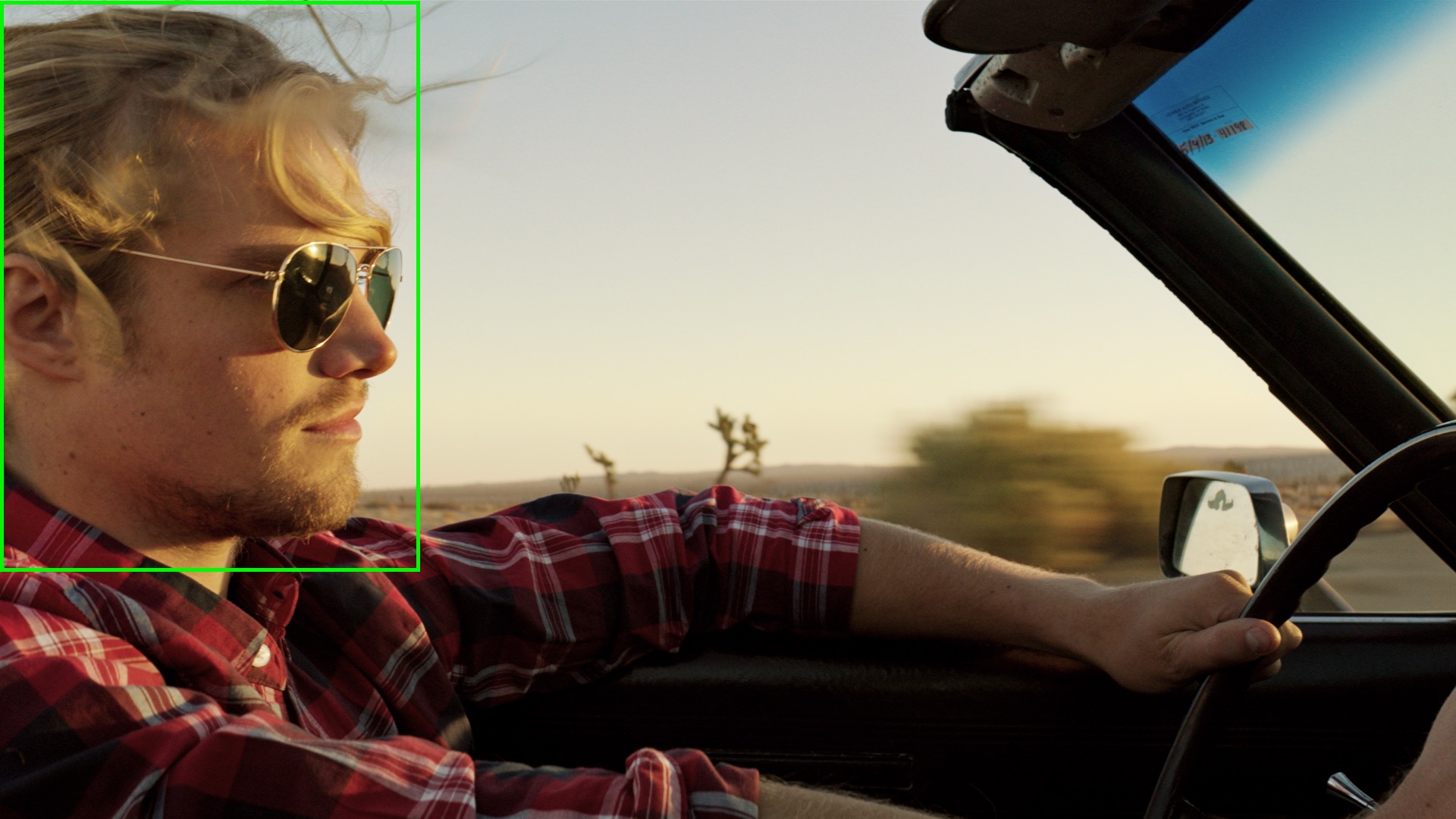}&  
     \includegraphics[width=0.22\textwidth]{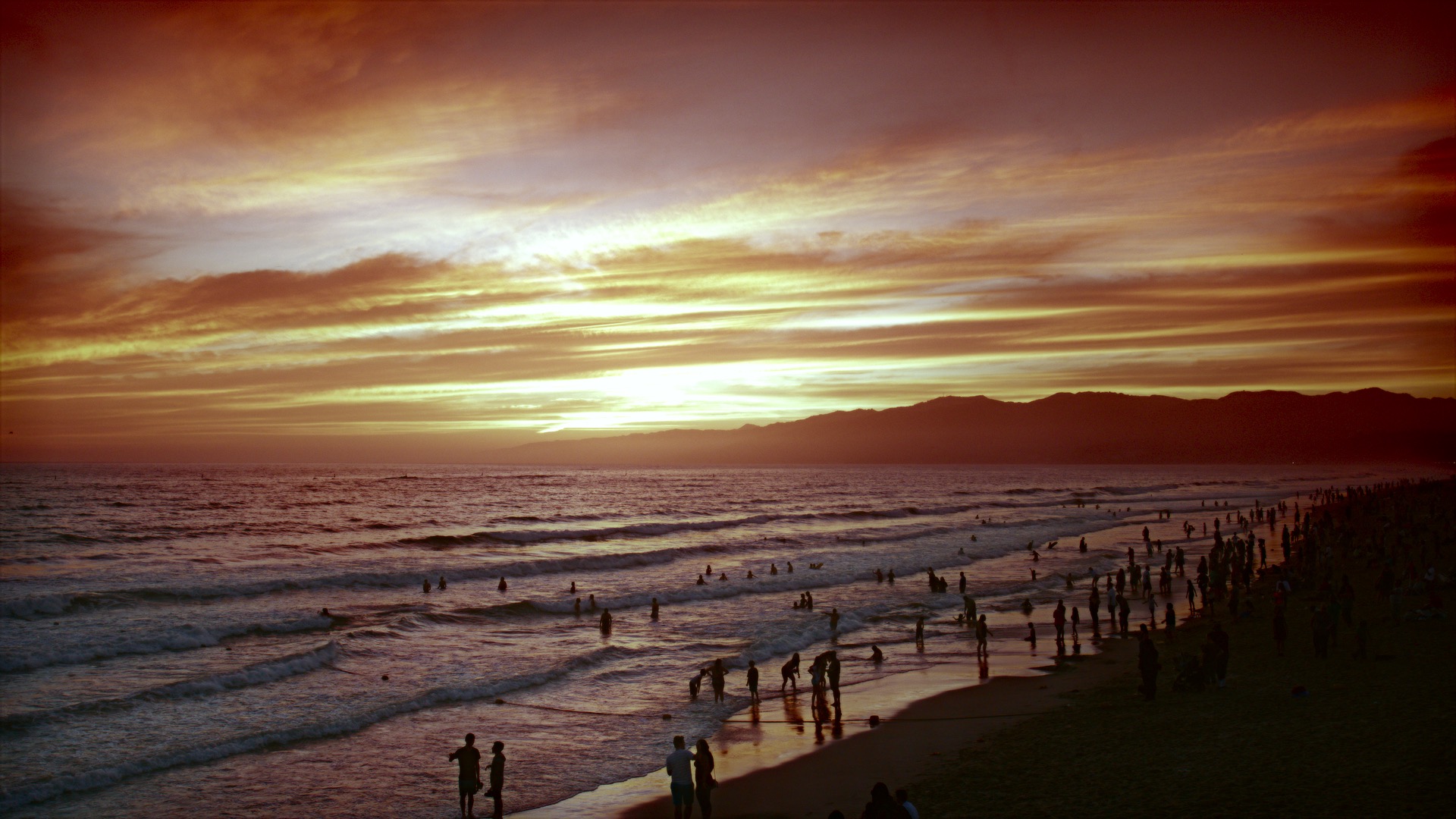}& 
     \includegraphics[width=0.22\textwidth]{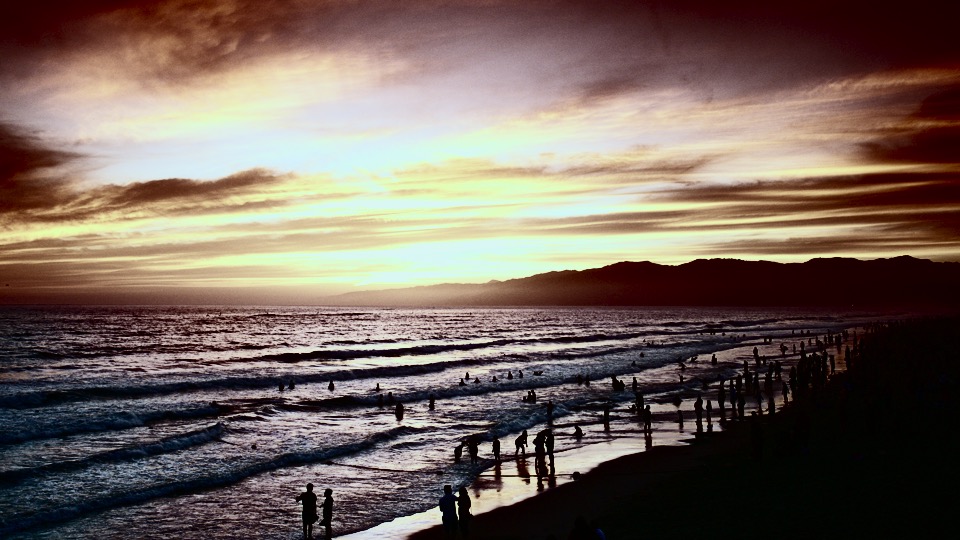} \\
     
    \includegraphics[width=0.22\textwidth]{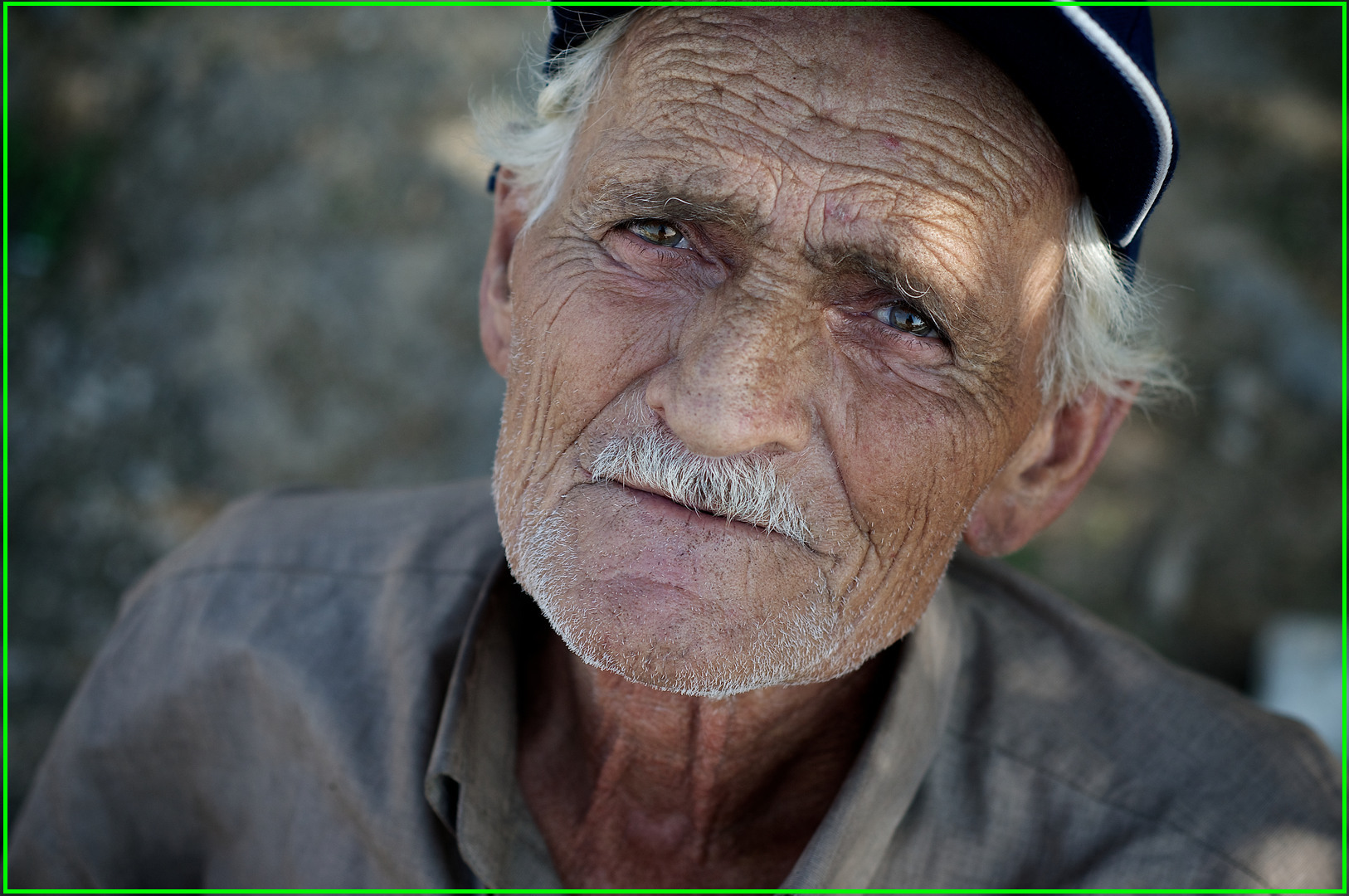}&
     \includegraphics[width=0.22\textwidth]{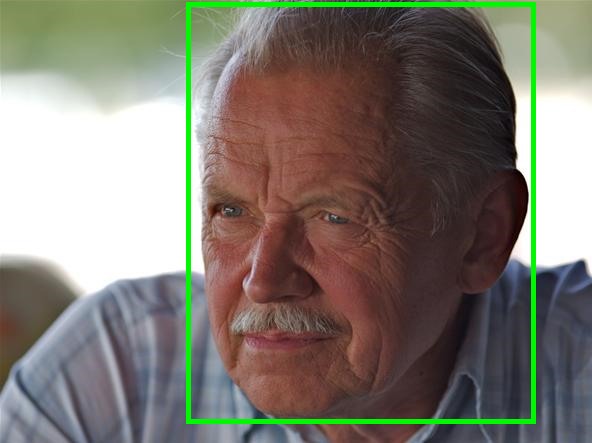}&     
    \includegraphics[width=0.22\textwidth]{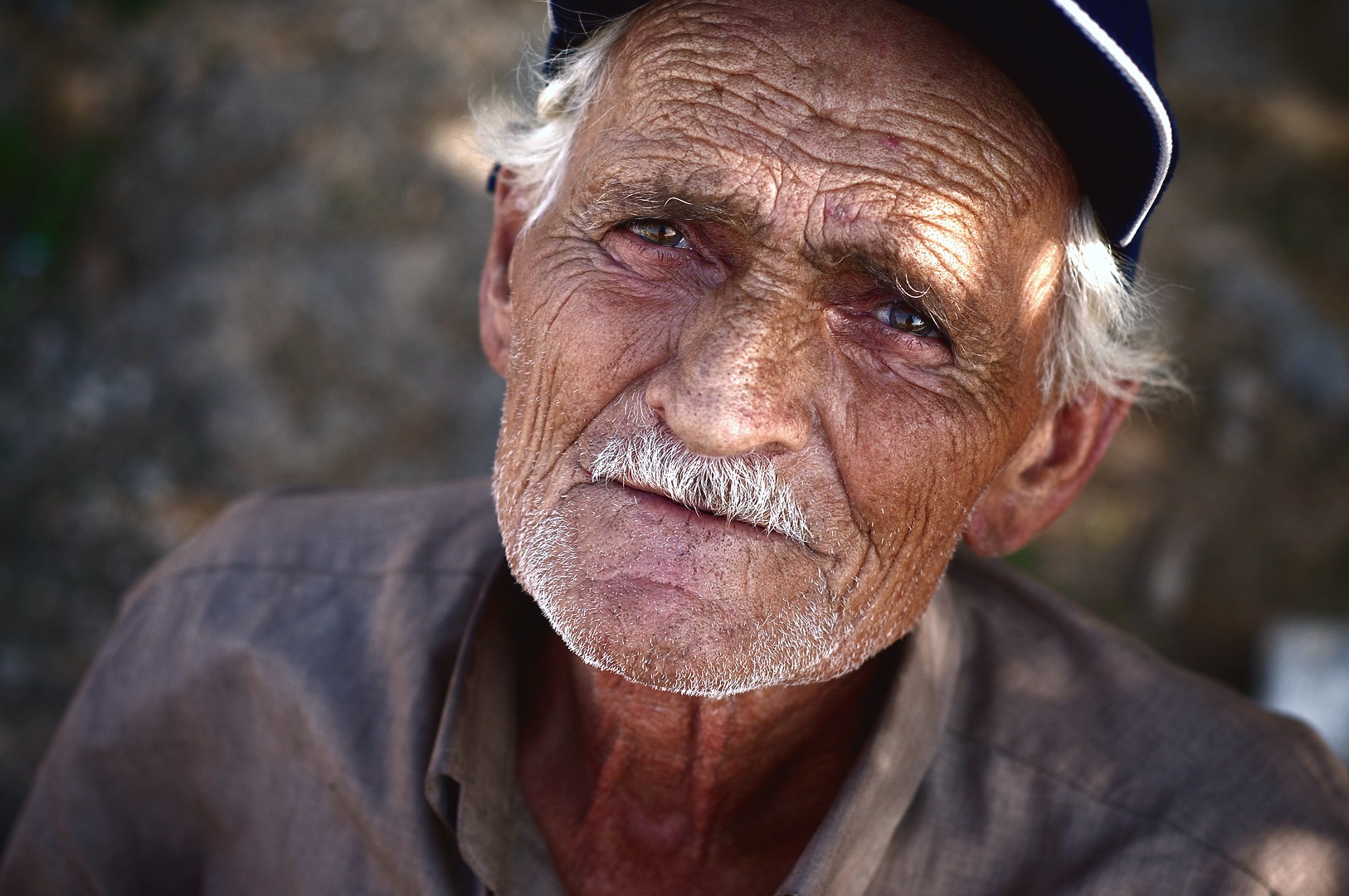}& 
    \includegraphics[width=0.22\textwidth]{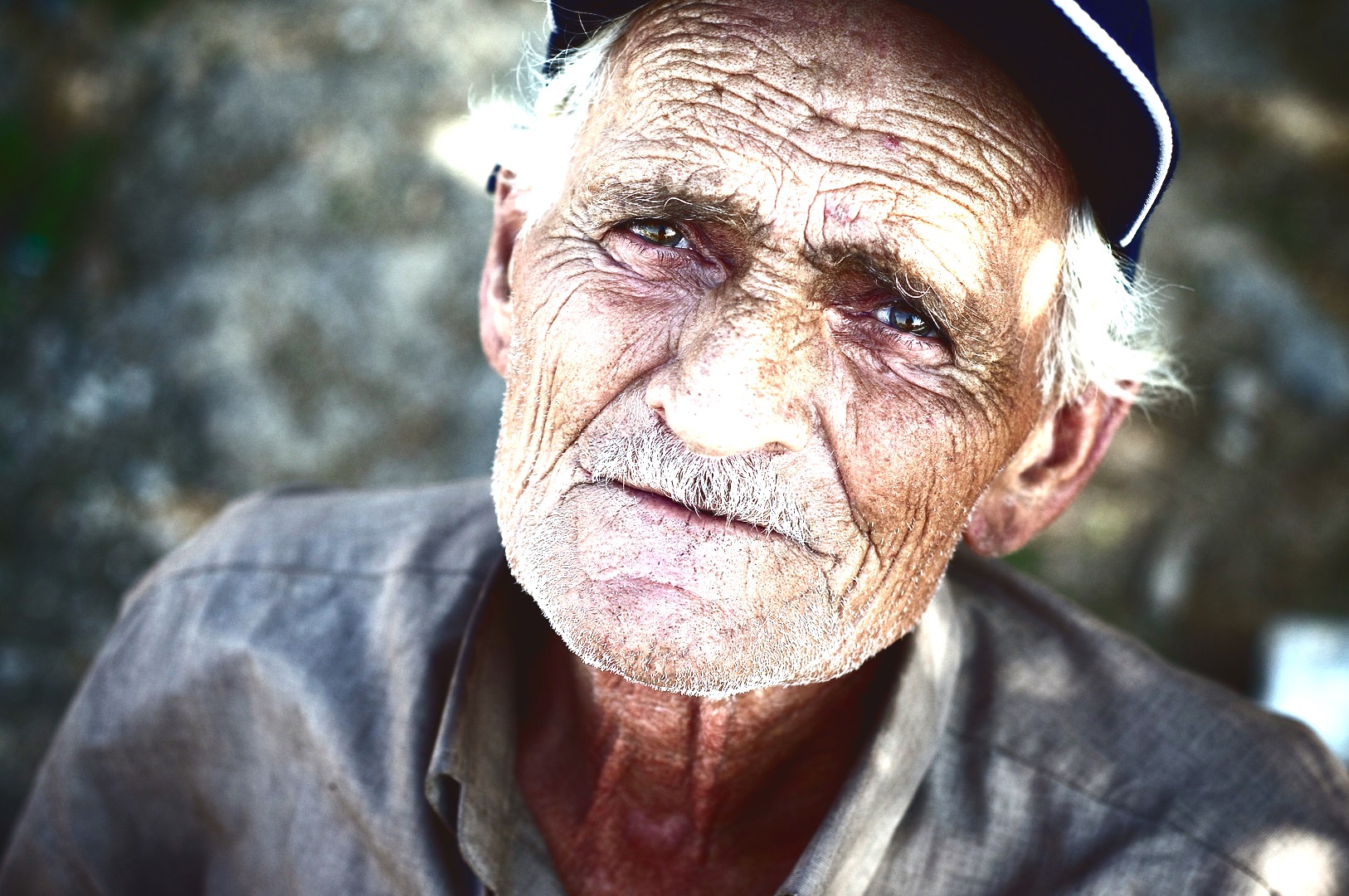} \\
    
    \includegraphics[width=0.22\textwidth]{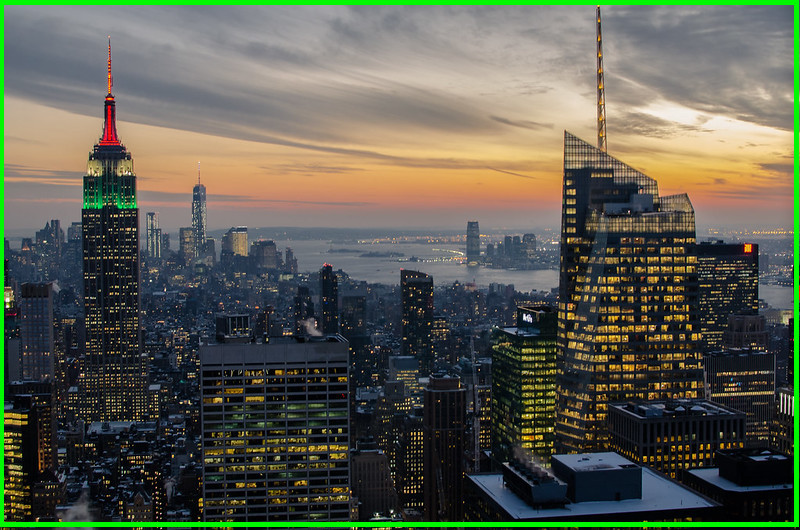}&
    \includegraphics[width=0.22\textwidth]{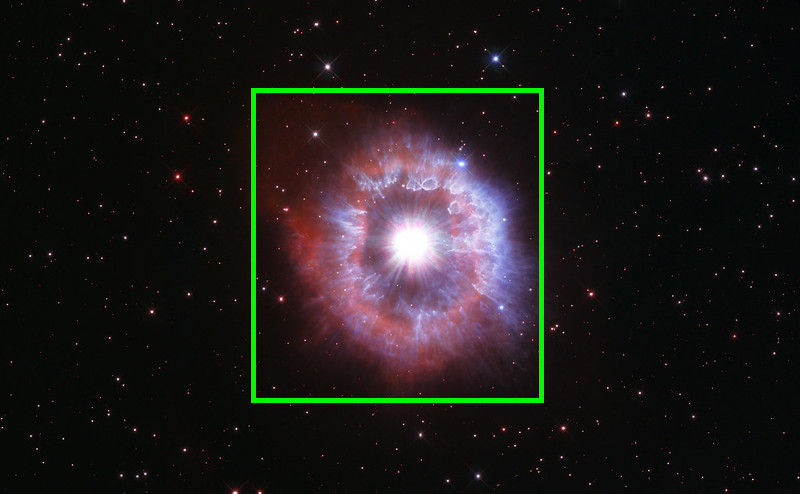}&  
    \includegraphics[width=0.22\textwidth]{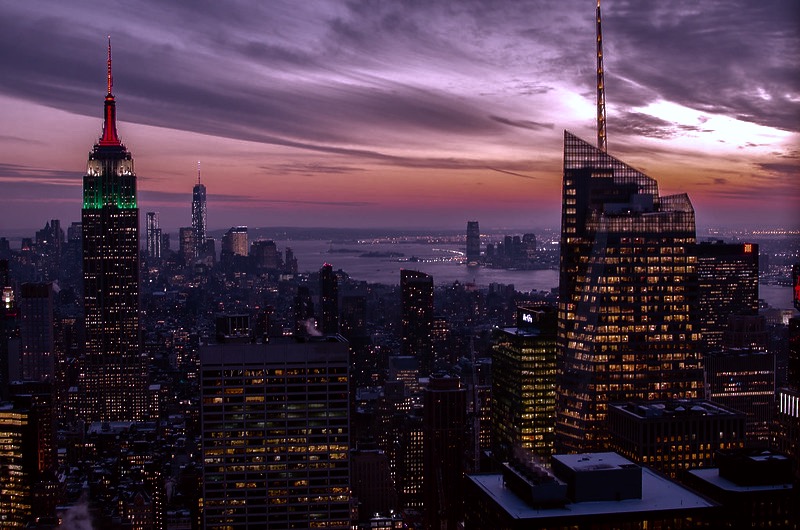}&
    \includegraphics[width=0.22\textwidth]{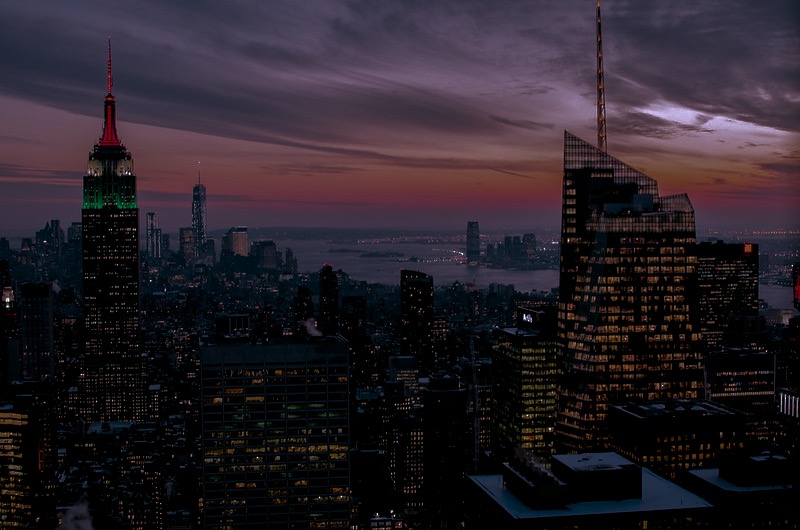} \\

  \end{tabular}
  \caption{From left to right, source image with cropped region bounded in green, target image with cropped region bounded in green, default method transfer result with cropping, default method transfer result without cropping. Images sourcing left to right, top to bottom: European Southern Observatory (ESO), European Space Agency (ESA), ARRI Camera Group (complete second row), Labelled Faces in the Wild ( complete third row) \cite{LFWTech}, MK Feeny, ESA. Licensed via CC-BY 2.0.}
  \label{crop}
\end{figure*}

While the previous examples show the performance of the method in its default state, applied to arbitrary color graded and compressed images without any pre-processing, we suggest several levers which users can employ to find an optimal transform for their purpose. One simple mechanism is the inclusion of more or fewer features on various color channels. In this way, the transfer can be made to reflect the target features more accurately, or better maintain the characteristics of the source. Another effective way to refine results is through manual cropping of the source and target images. Oftentimes when selecting candidates for style transfer we focus only on the appearance of the relevant subject of the image while ignoring the influence of the surrounding area. So by simply including a manual cropping step at the start of the process, users can specifically select what image areas they would like to be included. In Figure \ref{crop}, we demonstrate various cases where limiting the analysis to a region of the image can be beneficial to the results. One can see in these examples that cropping out large bright or dark regions has an effect on the result such that no details are clipped or crushed.

\subsection{Qualitative Comparison}

\begin{figure*}[h!]
\centering
\begin{tabular}{ccccc}

    \includegraphics[width=0.17\textwidth]{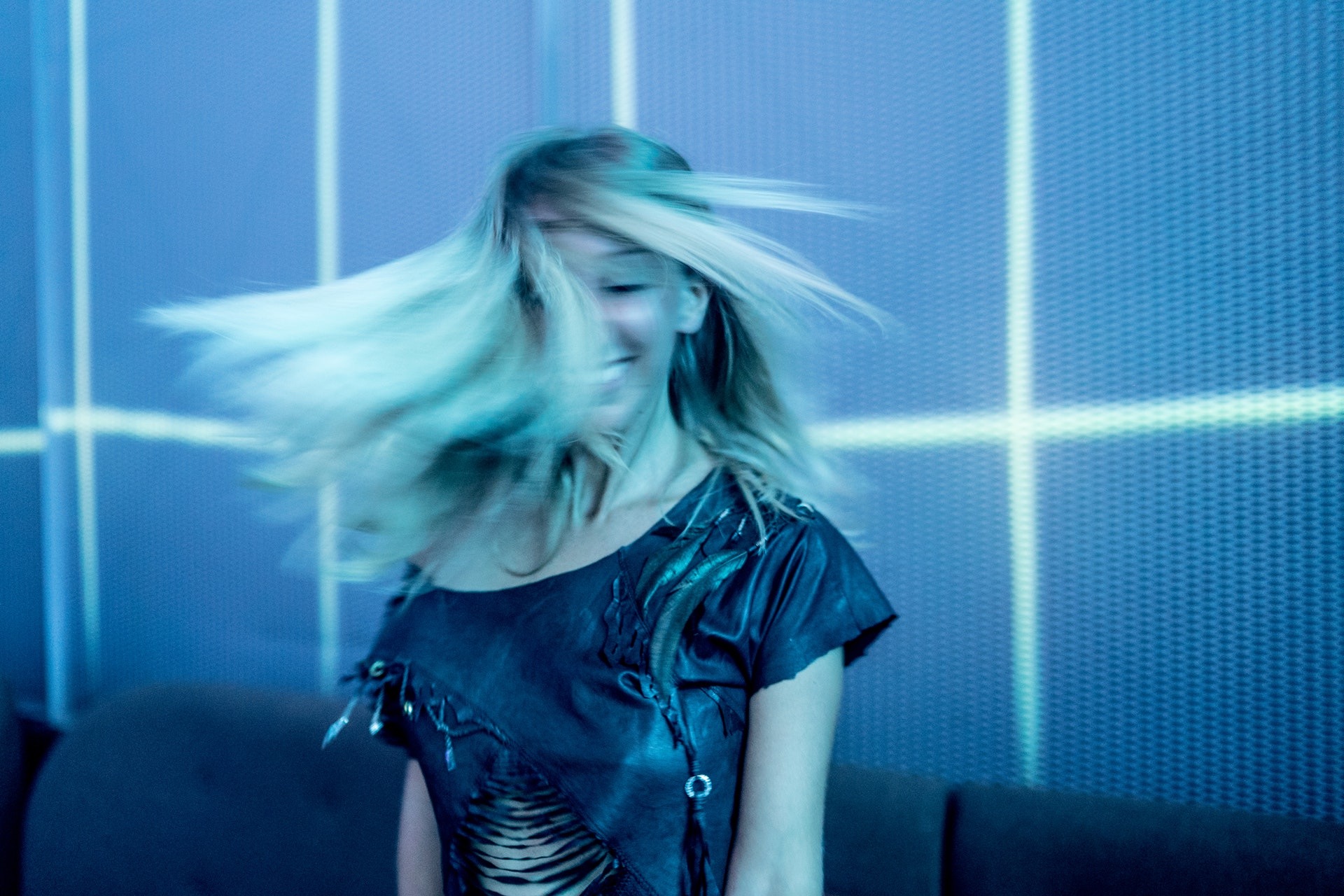}&
    \includegraphics[width=0.17\textwidth]{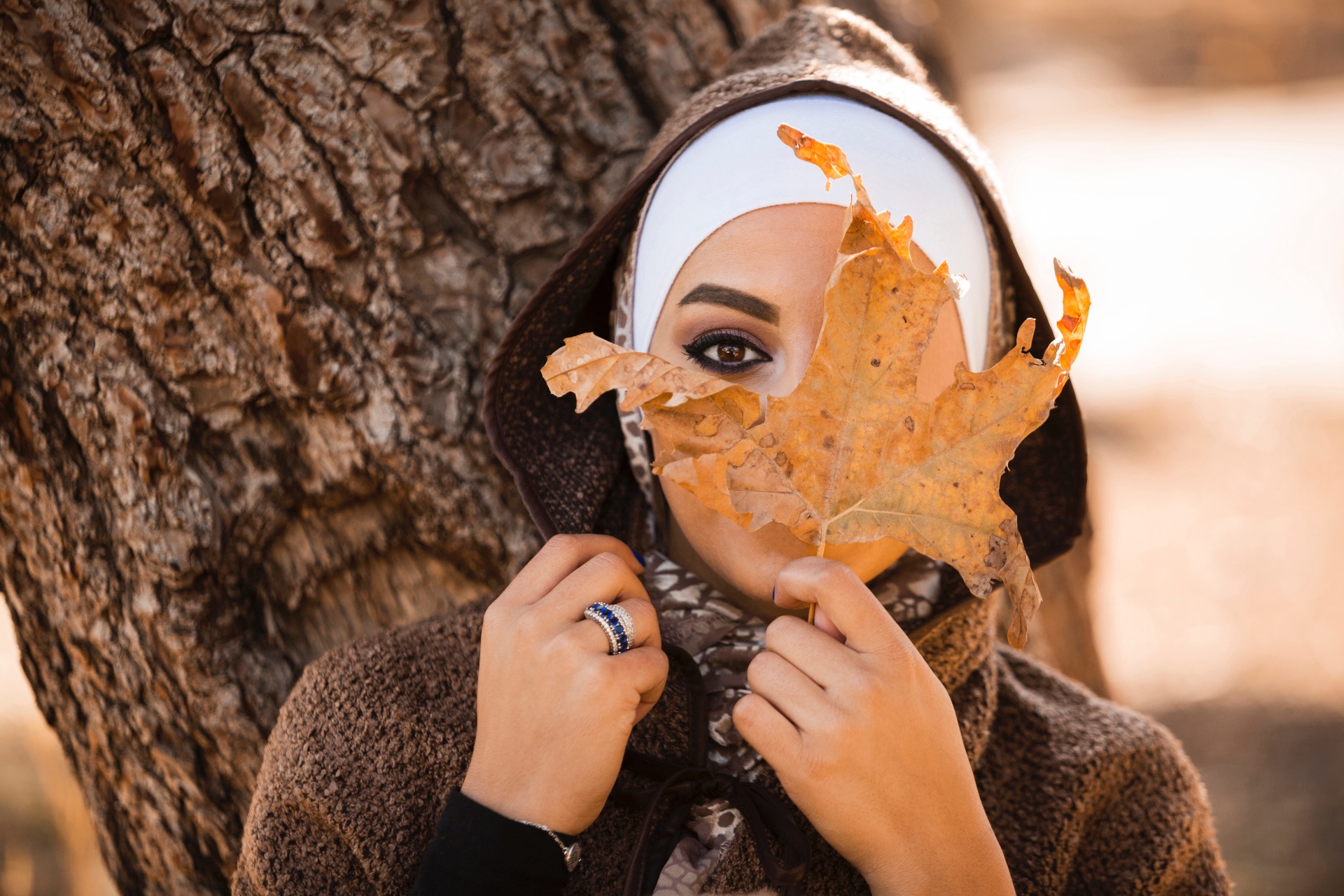}&  
    \includegraphics[width=0.17\textwidth]{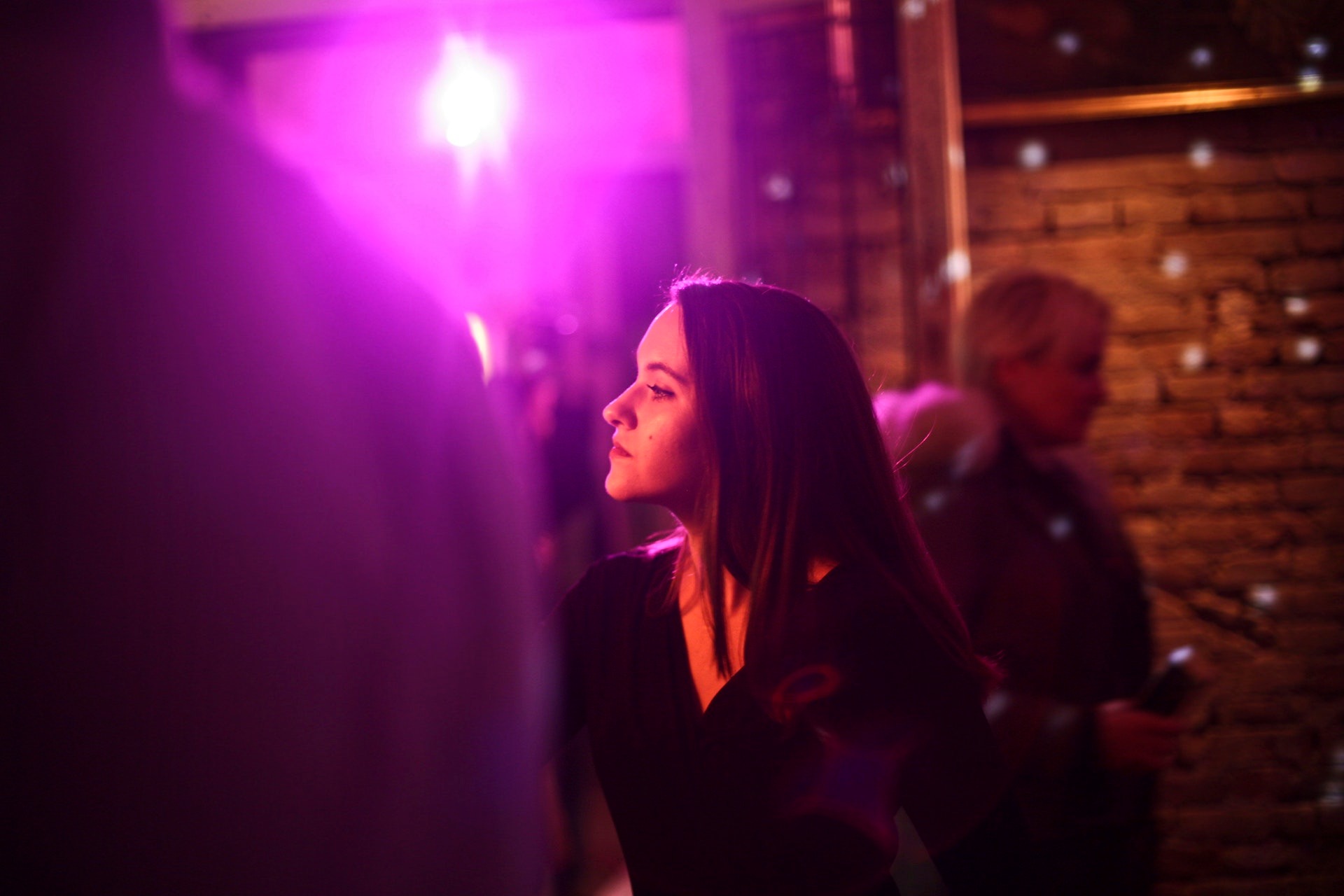}& 
    \includegraphics[width=0.17\textwidth]{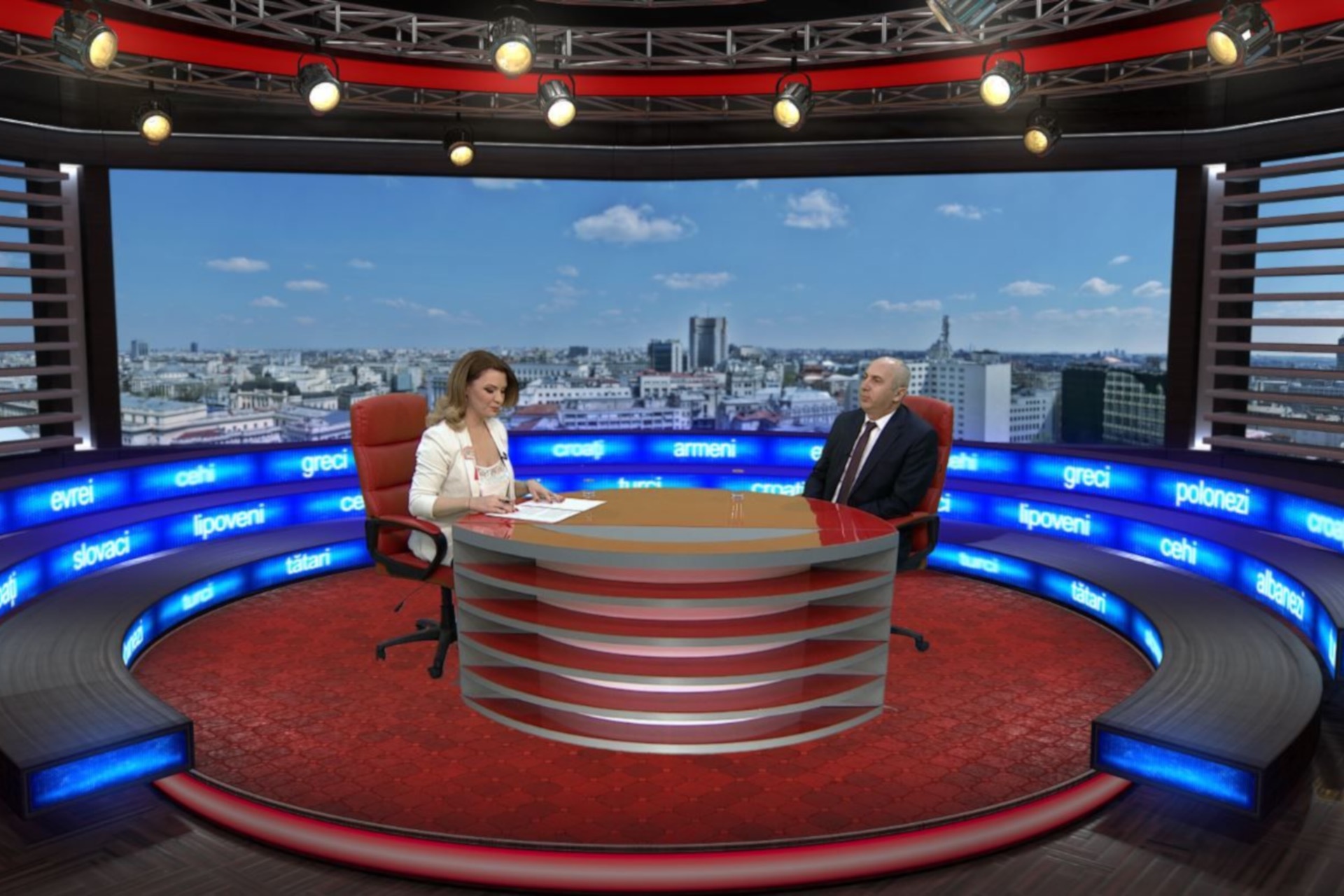}& 
    \includegraphics[width=0.17\textwidth]{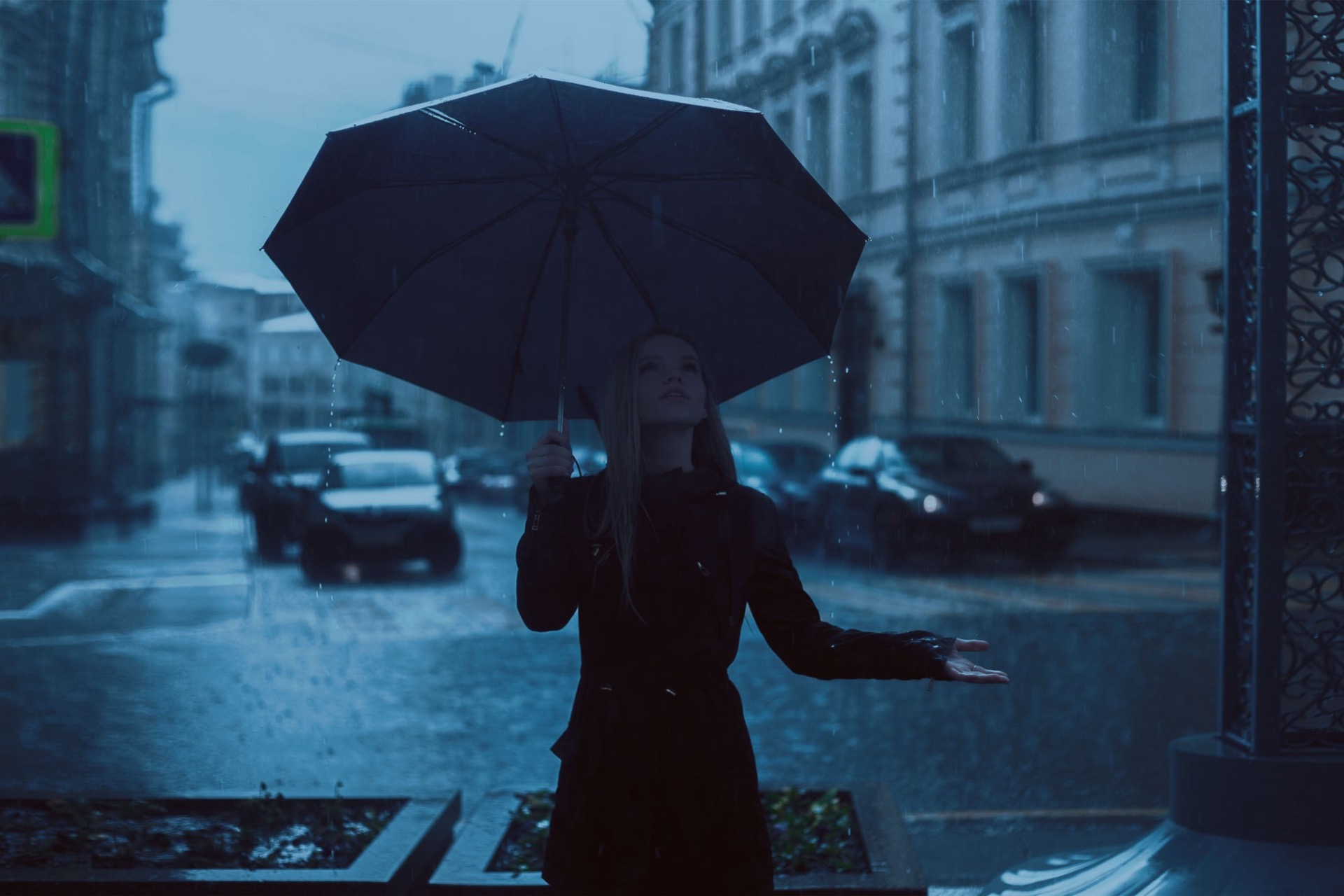} \\

  \end{tabular}
  \caption{Experimental image test set, featuring highly chromatic and stylized scenes. Fourth image made available by TVR Romania, the remainder are made available by Pexels.}
  \label{sources}
\end{figure*}

In the following examples, we qualitatively compare the results of our method to those of the SotA related works. With the exception of \cite{wu20} and \cite{xie20}, which we implemented, we employed demo software provided by the authors of the respective works in all other cases and made use of them in the direct sense that an artist might employ them, simply by replacing their example images with our own. \textcolor{black}{The results of the proposed method are all computed in its default state outlined in section \ref{subsec:overview}.} The original source images are shown in Figure \ref{sources}.

\begin{figure*}[t]
\centering
\begin{tabular}{cccc}

    \includegraphics[width=0.22\textwidth]{selected/relatedWork/sources-references/blue_1920_1280.png.jpeg}&
    \includegraphics[width=0.22\textwidth]{selected/relatedWork/sources-references/orange_1920_1280.png.jpeg}&  
     \includegraphics[width=0.22\textwidth]{selected/relatedWork/sources-references/studio_1920_1280.png.jpeg}& 
     \includegraphics[width=0.22\textwidth]{selected/relatedWork/sources-references/umbrella_1920_1280.png.jpeg} \\

    \includegraphics[width=0.22\textwidth]{selected/relatedWork/sources-references/orange_1920_1280.png.jpeg}&
    \includegraphics[width=0.22\textwidth]{selected/relatedWork/sources-references/studio_1920_1280.png.jpeg}&  
     \includegraphics[width=0.22\textwidth]{selected/relatedWork/sources-references/blue_1920_1280.png.jpeg}& 
     \includegraphics[width=0.22\textwidth]{selected/relatedWork/sources-references/orange_1920_1280.png.jpeg} \\  
     
    \includegraphics[width=0.22\textwidth]{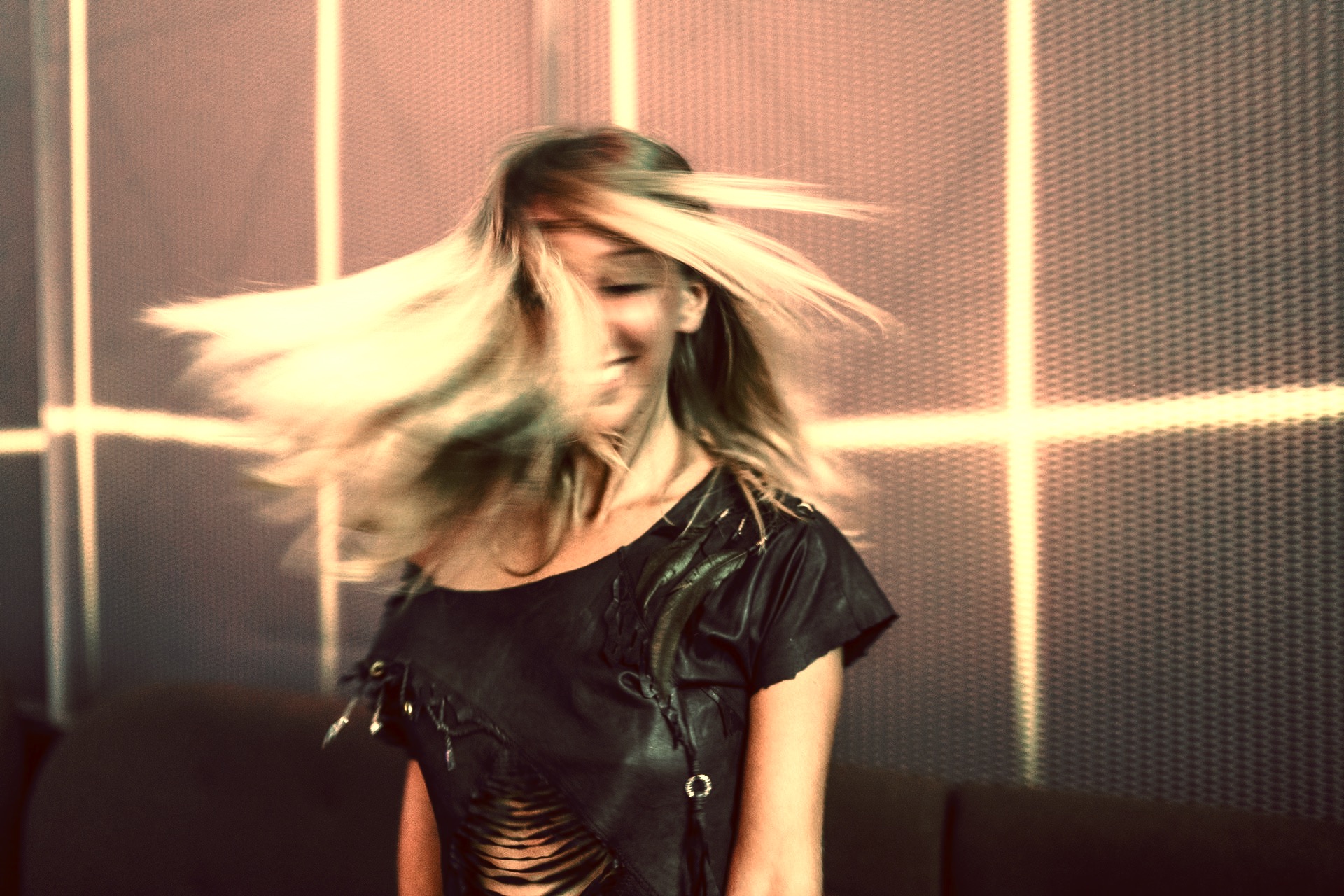}&
    \includegraphics[width=0.22\textwidth]{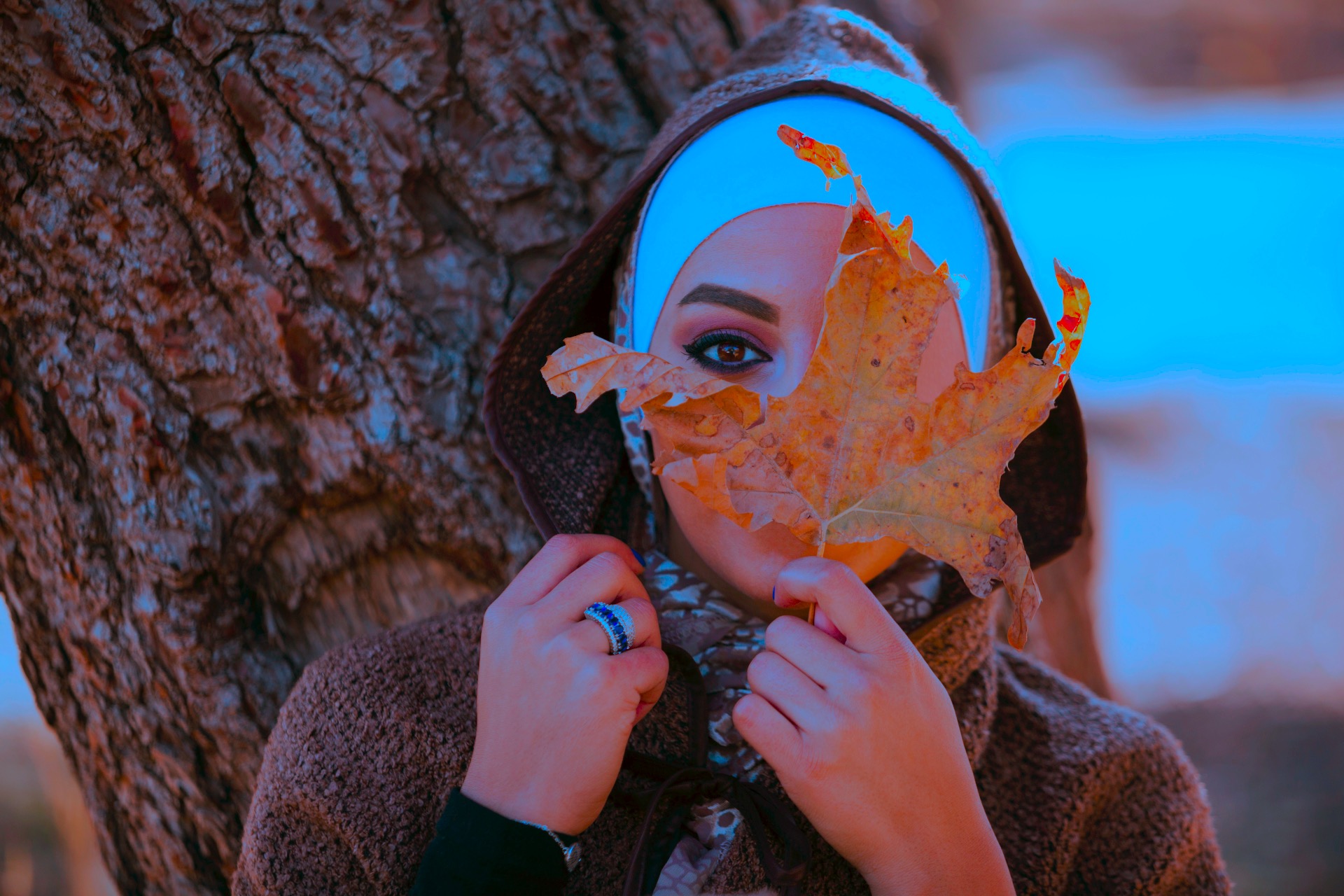}&  
     \includegraphics[width=0.22\textwidth]{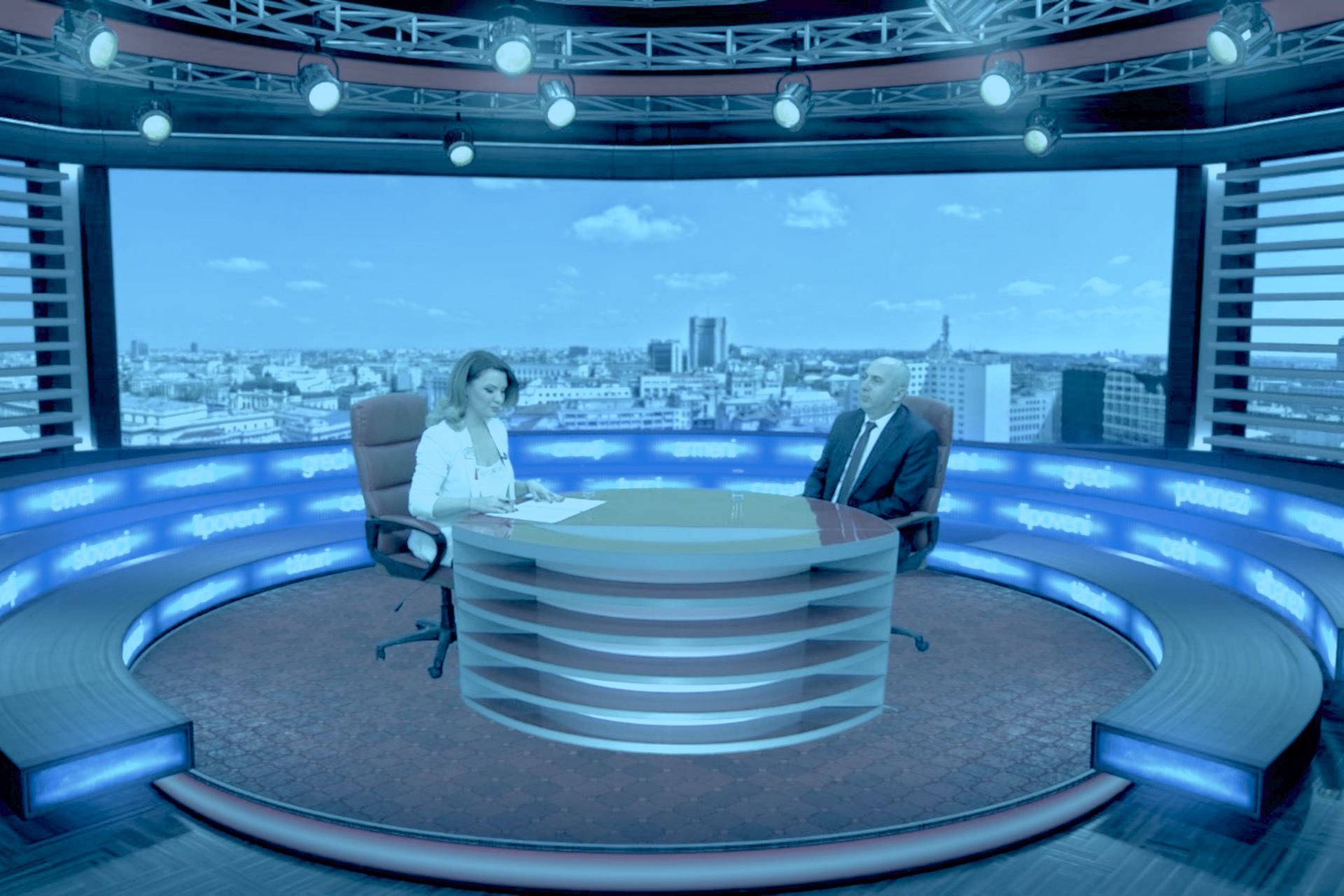}& 
     \includegraphics[width=0.22\textwidth]{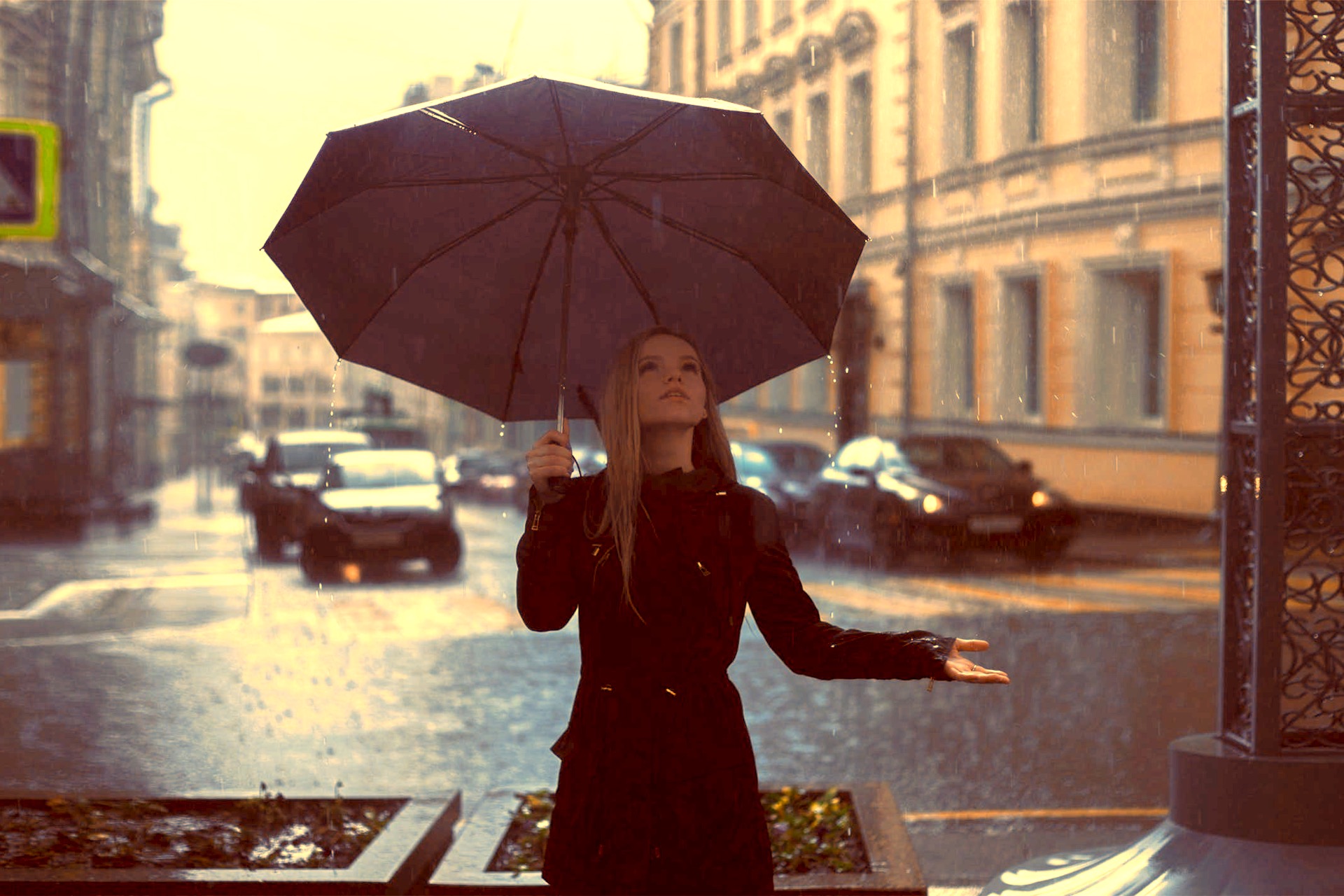} \\
    
    \includegraphics[width=0.22\textwidth]{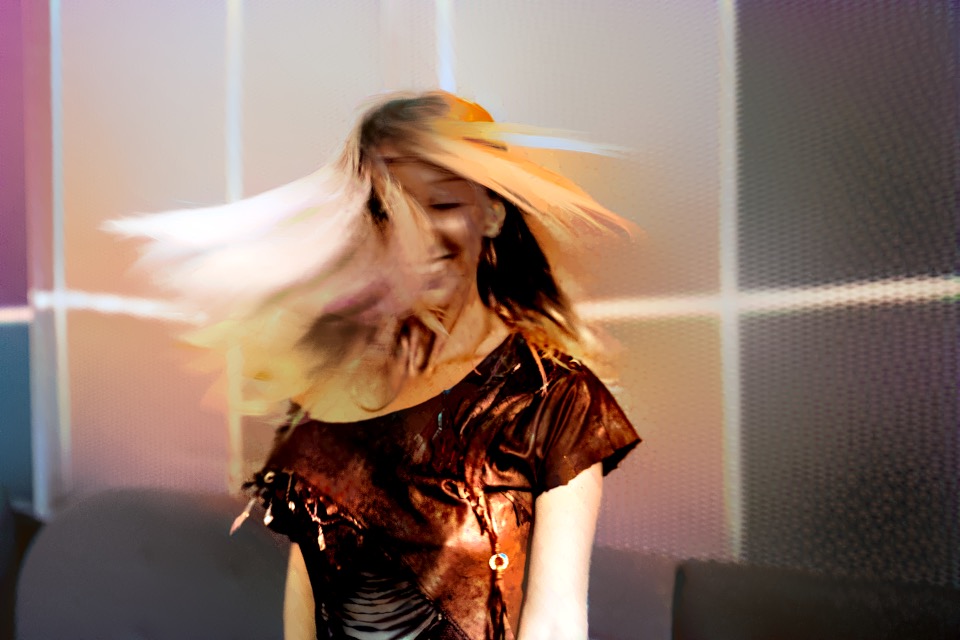}&
    \includegraphics[width=0.22\textwidth]{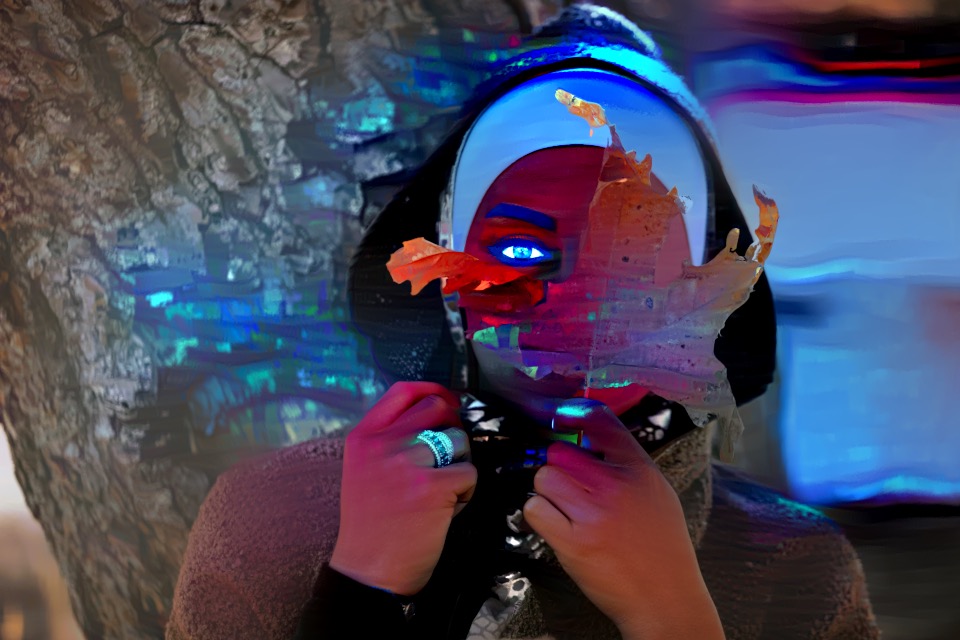}&  
     \includegraphics[width=0.22\textwidth]{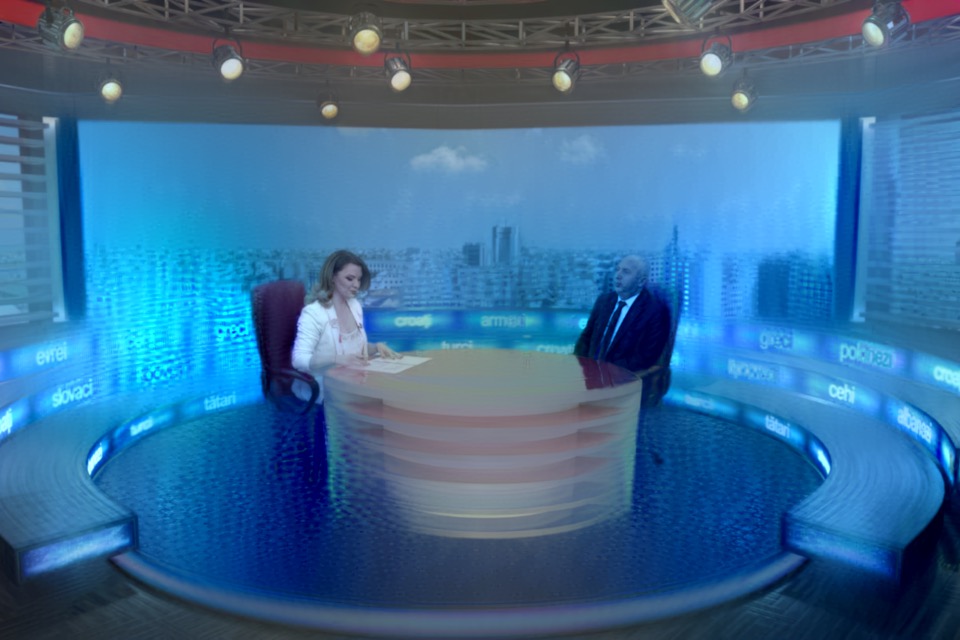}& 
     \includegraphics[width=0.22\textwidth]{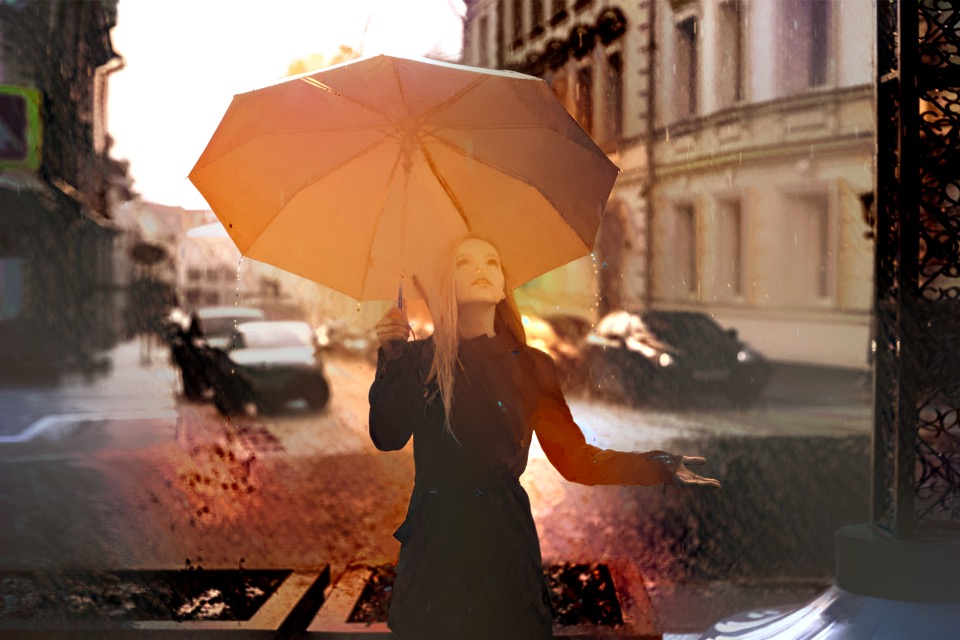} \\  

    \includegraphics[width=0.22\textwidth]{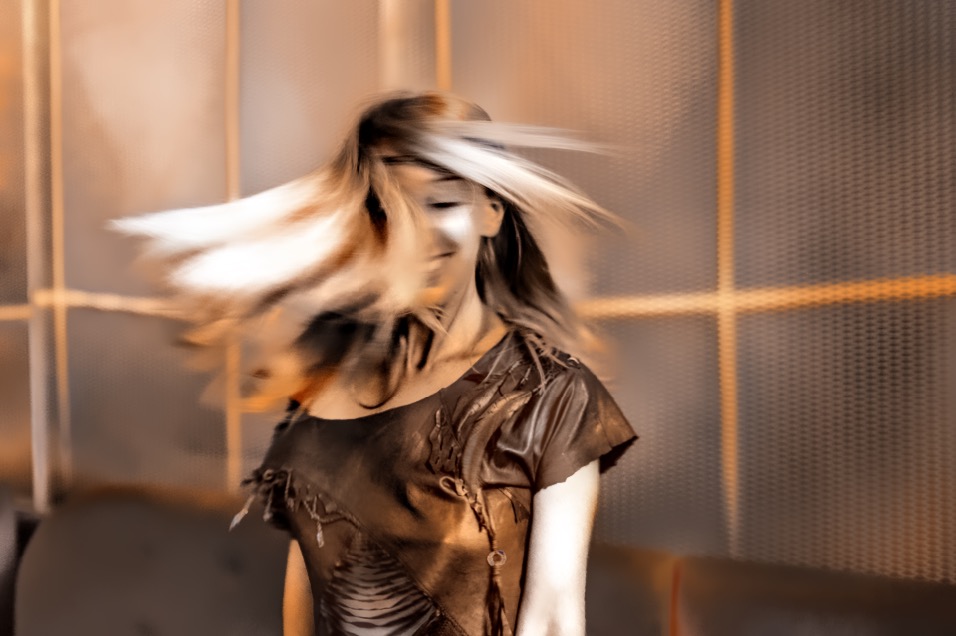}&
    \includegraphics[width=0.22\textwidth]{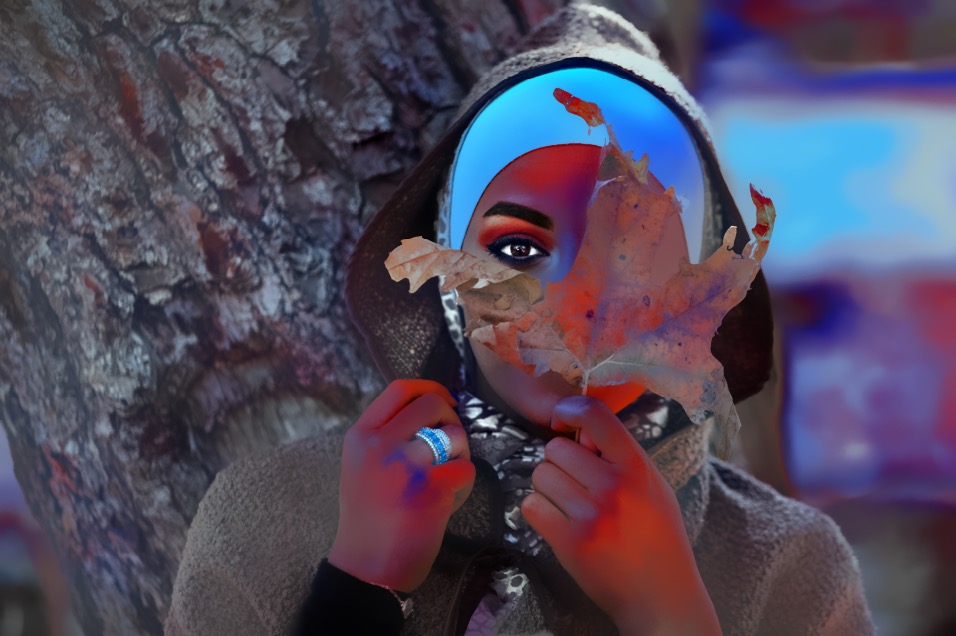}&  
     \includegraphics[width=0.22\textwidth]{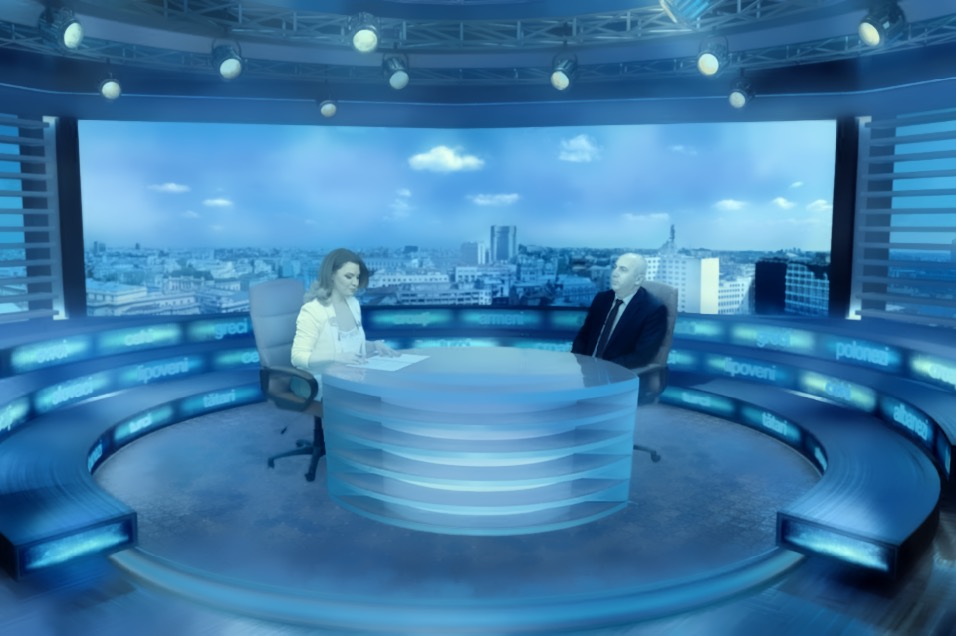}& 
     \includegraphics[width=0.22\textwidth]{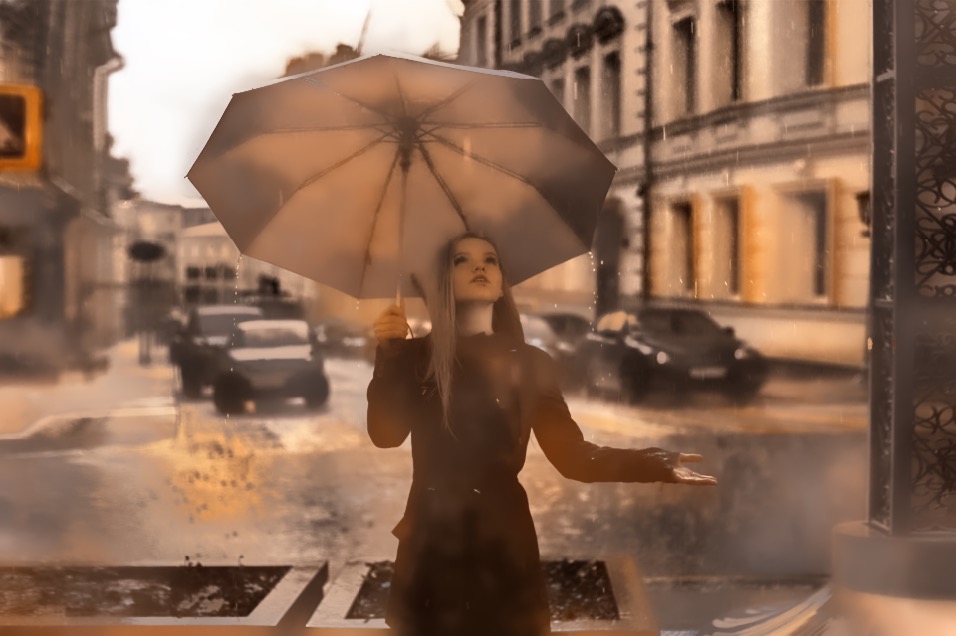} \\   
     
    \includegraphics[width=0.22\textwidth]{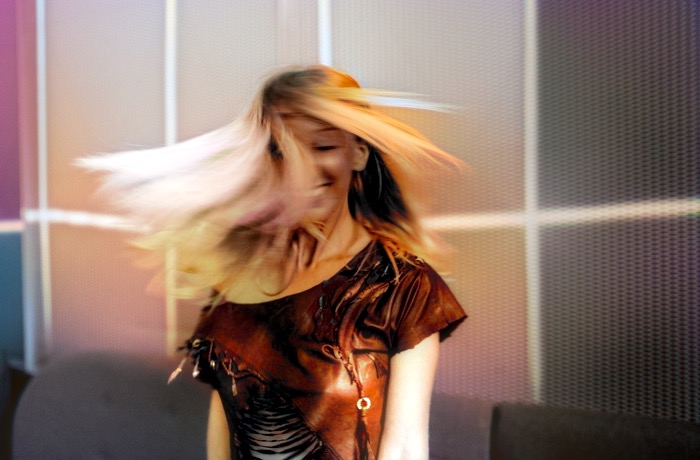}&
    \includegraphics[width=0.22\textwidth]{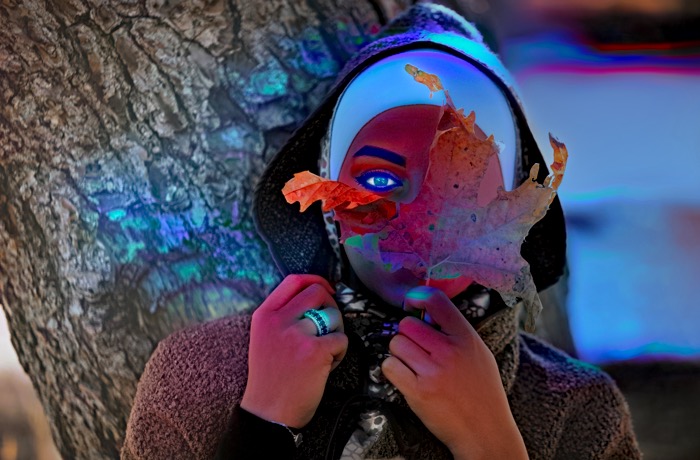}&  
     \includegraphics[width=0.22\textwidth]{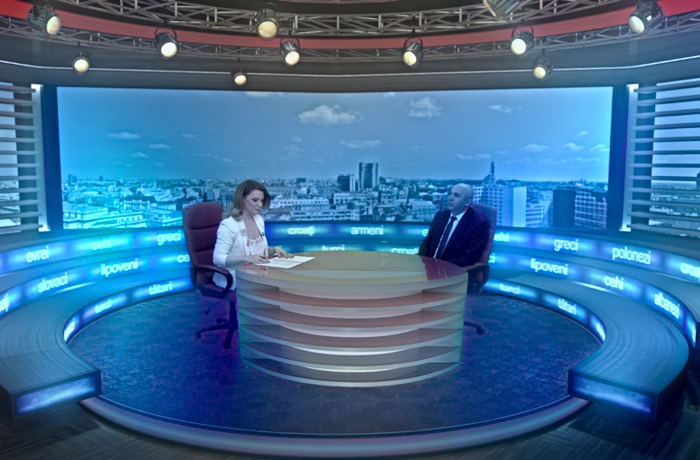}& 
     \includegraphics[width=0.22\textwidth]{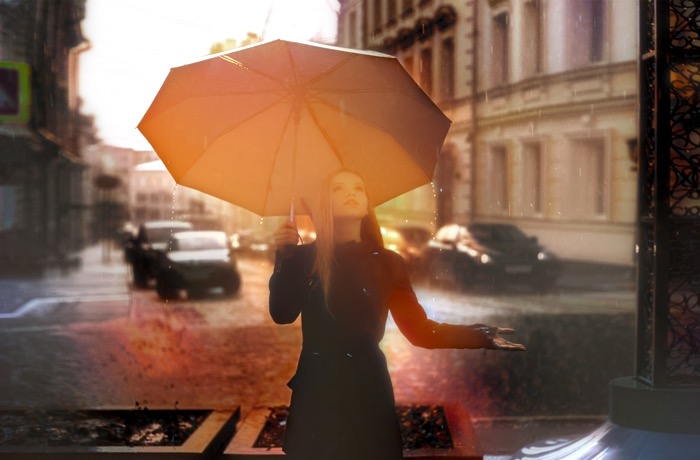} \\
     
    \includegraphics[width=0.22\textwidth]{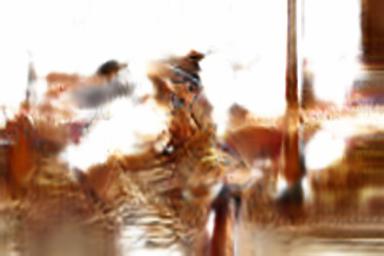}&
    \includegraphics[width=0.22\textwidth]{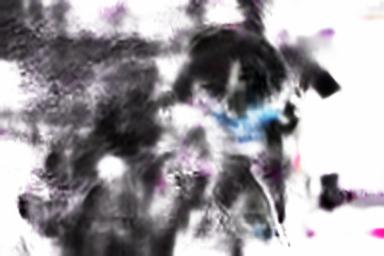}&  
     \includegraphics[width=0.22\textwidth]{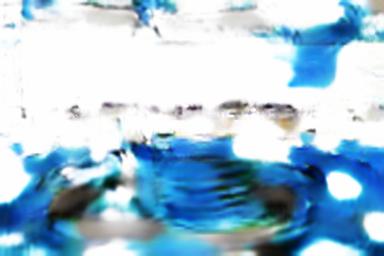}& 
     \includegraphics[width=0.22\textwidth]{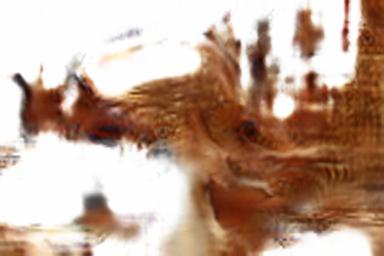} \\

  \end{tabular}
  \caption{From top to bottom, sources, targets, proposed result, Luan result \cite{luan17}, Li result \cite{li18}, Mechrez result \cite{mechrez17} (taking Luan results as input), Huang result \cite{huang18}.}
  \label{NST}
\end{figure*}

In Figure \ref{NST}, the results of relevant NST approaches (\cite{luan17}, \cite{li18}, \cite{mechrez17}, \cite{huang18}) are shown in comparison to the results of the proposed method. Emulating how the methods would be used by a professional in image editing or color correction, neither pre-processing nor additional training steps were taken to refine the results, and thus the methods are used ``out of the box'', as they were provided in their demo repositories. The results of Mechrez et al. are computed using the results of Luan et al. as input, as this was the primary method their work was tailored to refine. One can see that these show egregious spatial distortions and spurious colors which make them unacceptable for professional photorealistic applications, even after automatic refinement steps are taken. 

\begin{figure*}[t]
\centering
\begin{tabular}{cccc}

    \includegraphics[width=0.22\textwidth]{selected/relatedWork/sources-references/blue_1920_1280.png.jpeg}&
    \includegraphics[width=0.22\textwidth]{selected/relatedWork/sources-references/orange_1920_1280.png.jpeg}&  
     \includegraphics[width=0.22\textwidth]{selected/relatedWork/sources-references/studio_1920_1280.png.jpeg}& 
     \includegraphics[width=0.22\textwidth]{selected/relatedWork/sources-references/umbrella_1920_1280.png.jpeg} \\

    \includegraphics[width=0.22\textwidth]{selected/relatedWork/sources-references/studio_1920_1280.png.jpeg}&
    \includegraphics[width=0.22\textwidth]{selected/relatedWork/sources-references/purple_1920_1280.png.jpeg}&  
     \includegraphics[width=0.22\textwidth]{selected/relatedWork/sources-references/orange_1920_1280.png.jpeg}& 
     \includegraphics[width=0.22\textwidth]{selected/relatedWork/sources-references/blue_1920_1280.png.jpeg} \\  
     
    \includegraphics[width=0.22\textwidth]{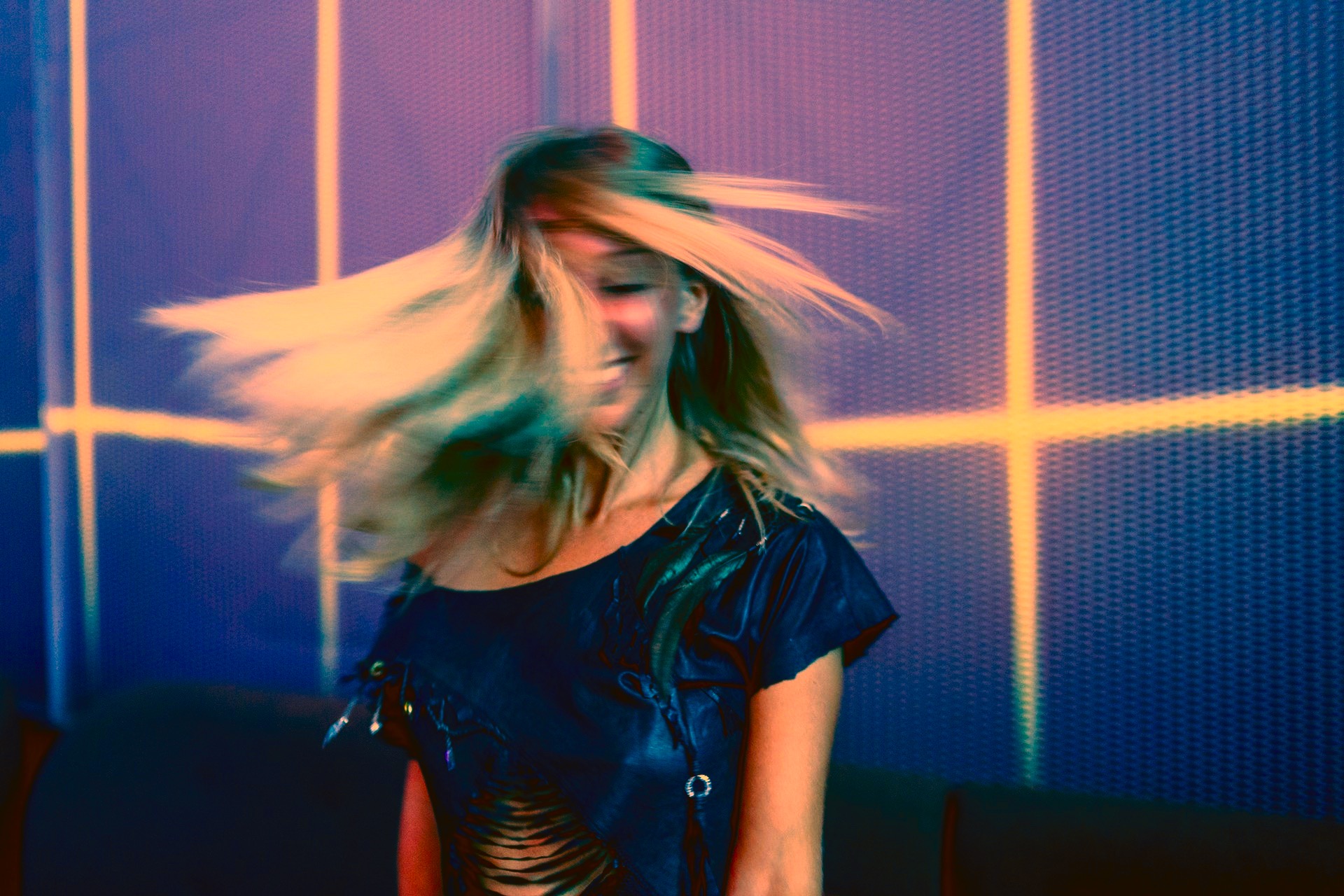}&
    \includegraphics[width=0.22\textwidth]{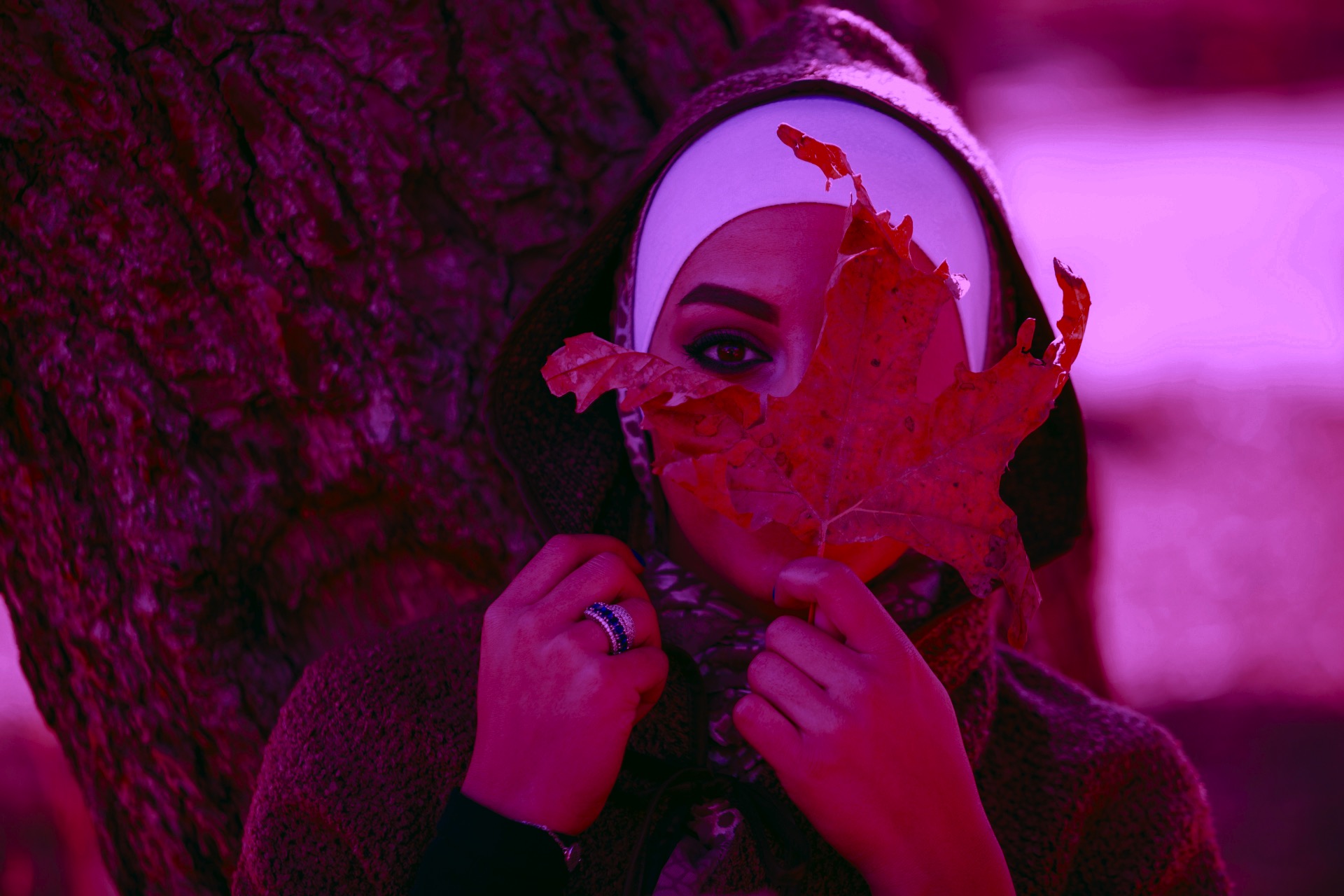}&  
     \includegraphics[width=0.22\textwidth]{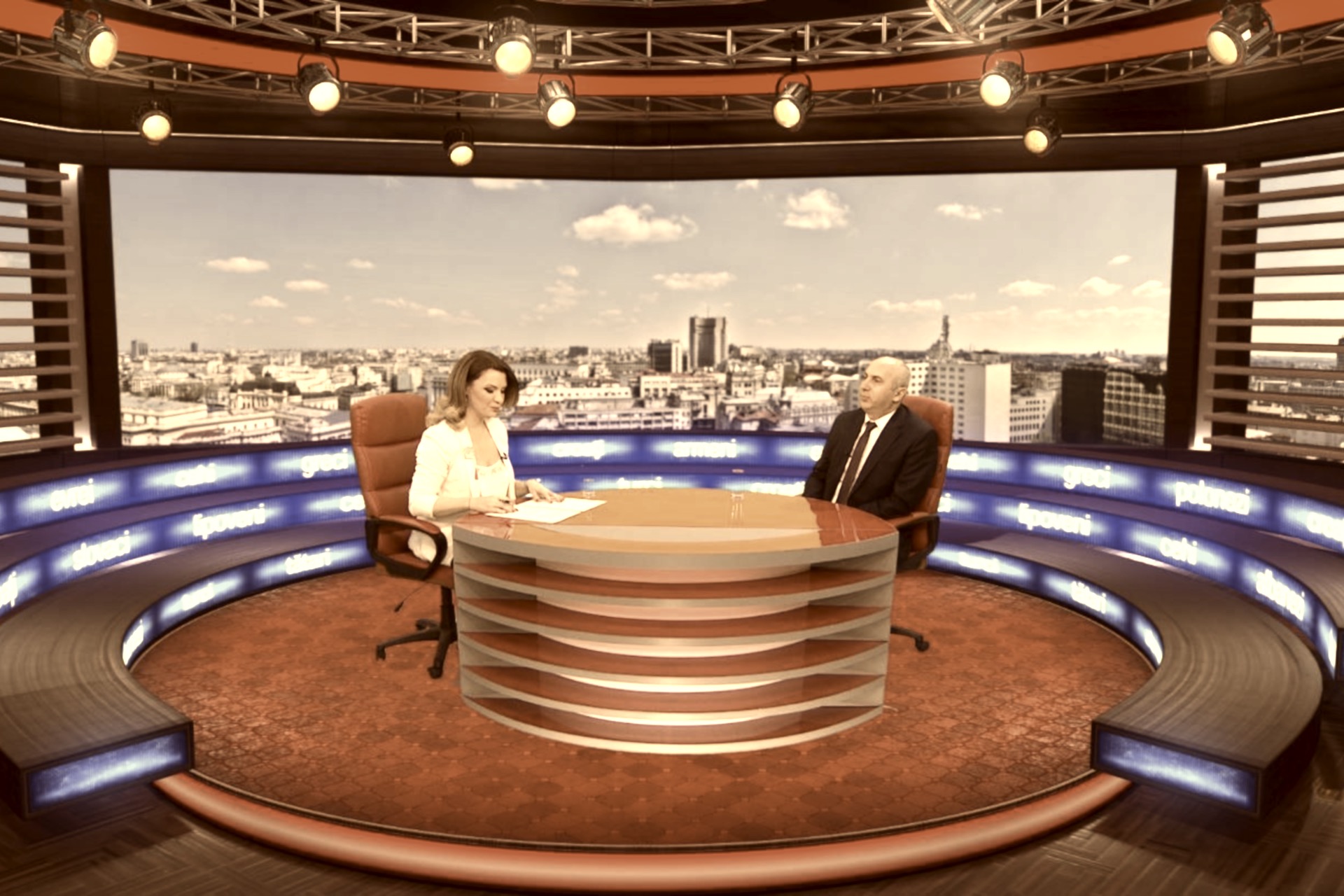}& 
     \includegraphics[width=0.22\textwidth]{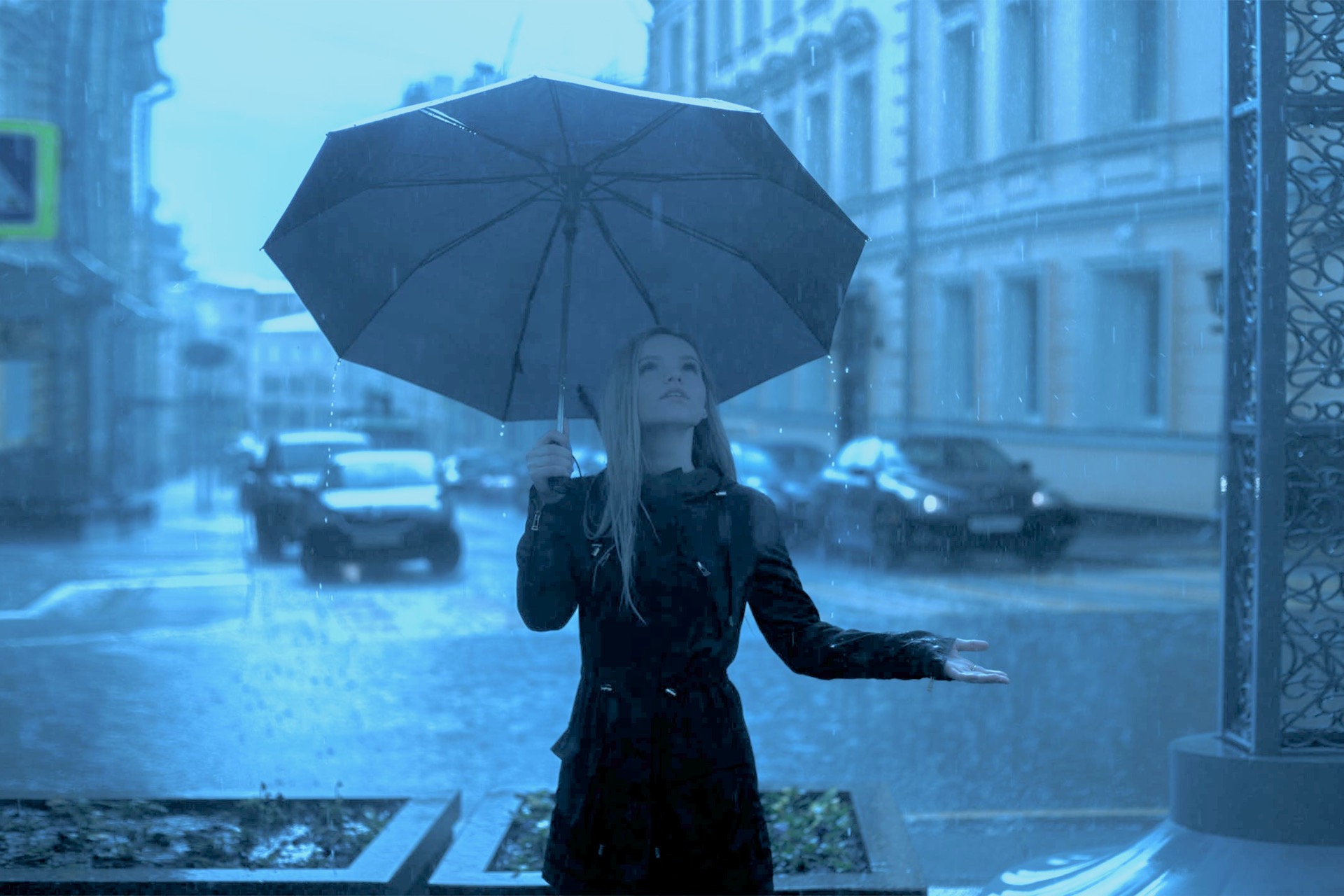} \\
    
    \includegraphics[width=0.22\textwidth]{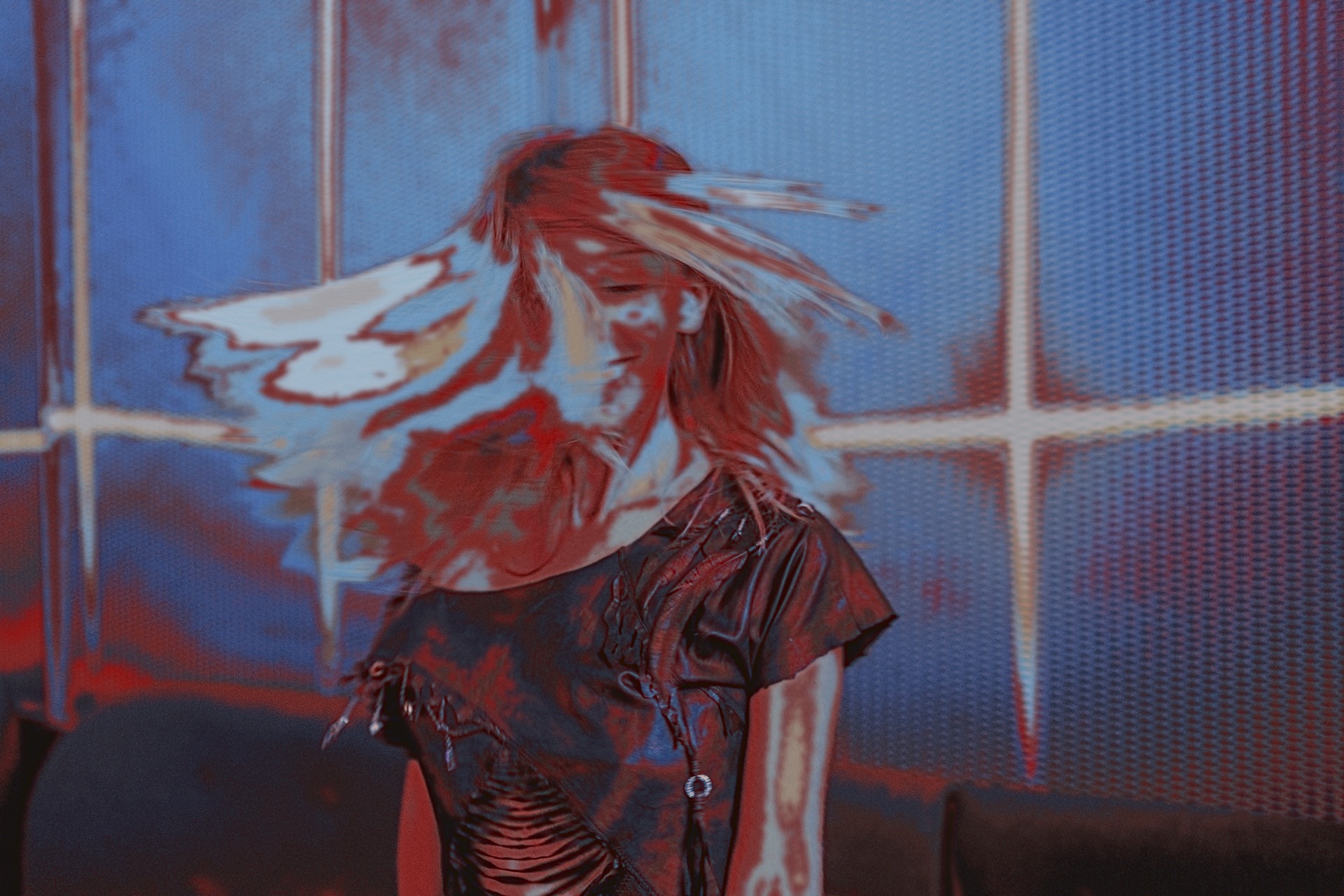}&
    \includegraphics[width=0.22\textwidth]{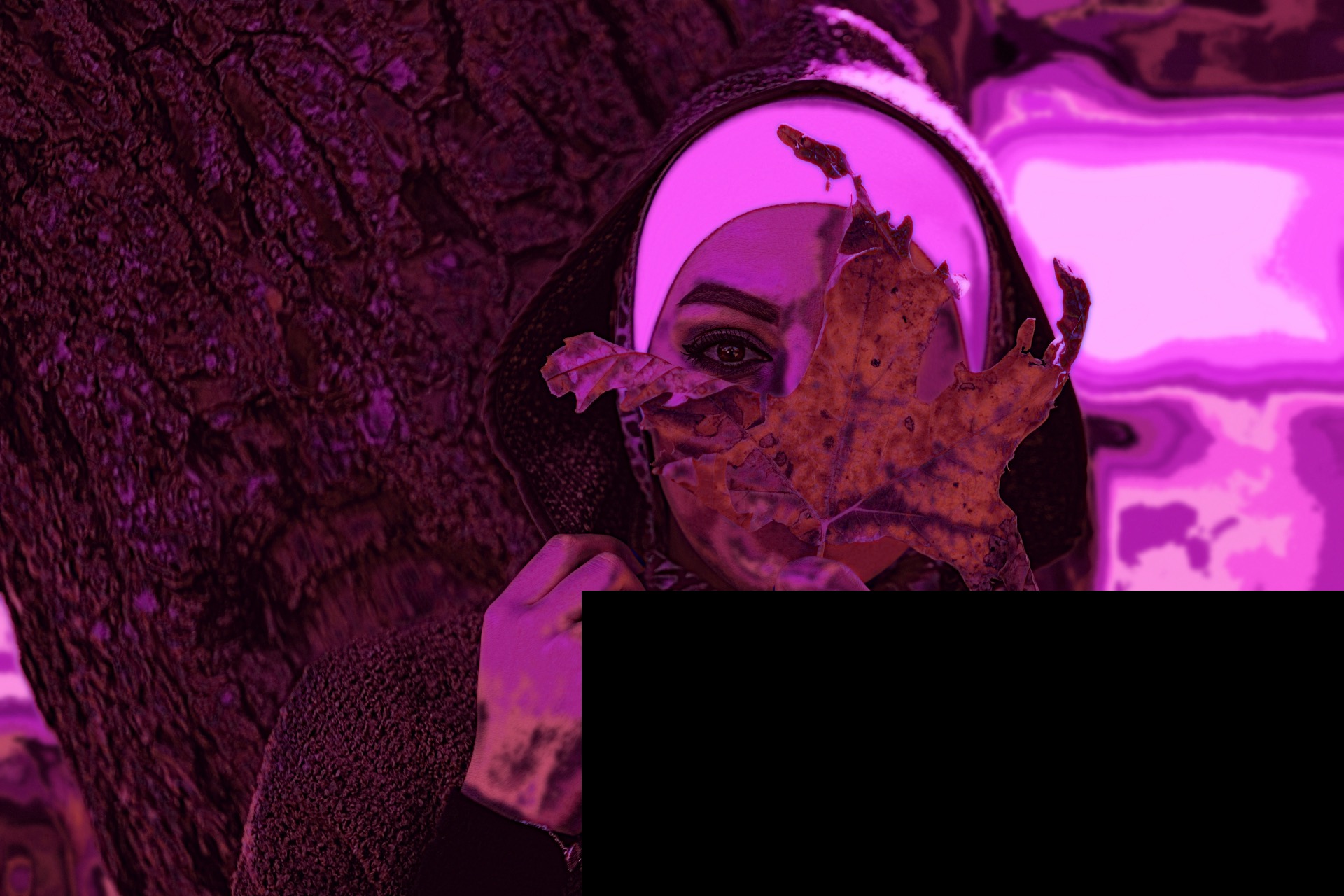}&  
     \includegraphics[width=0.22\textwidth]{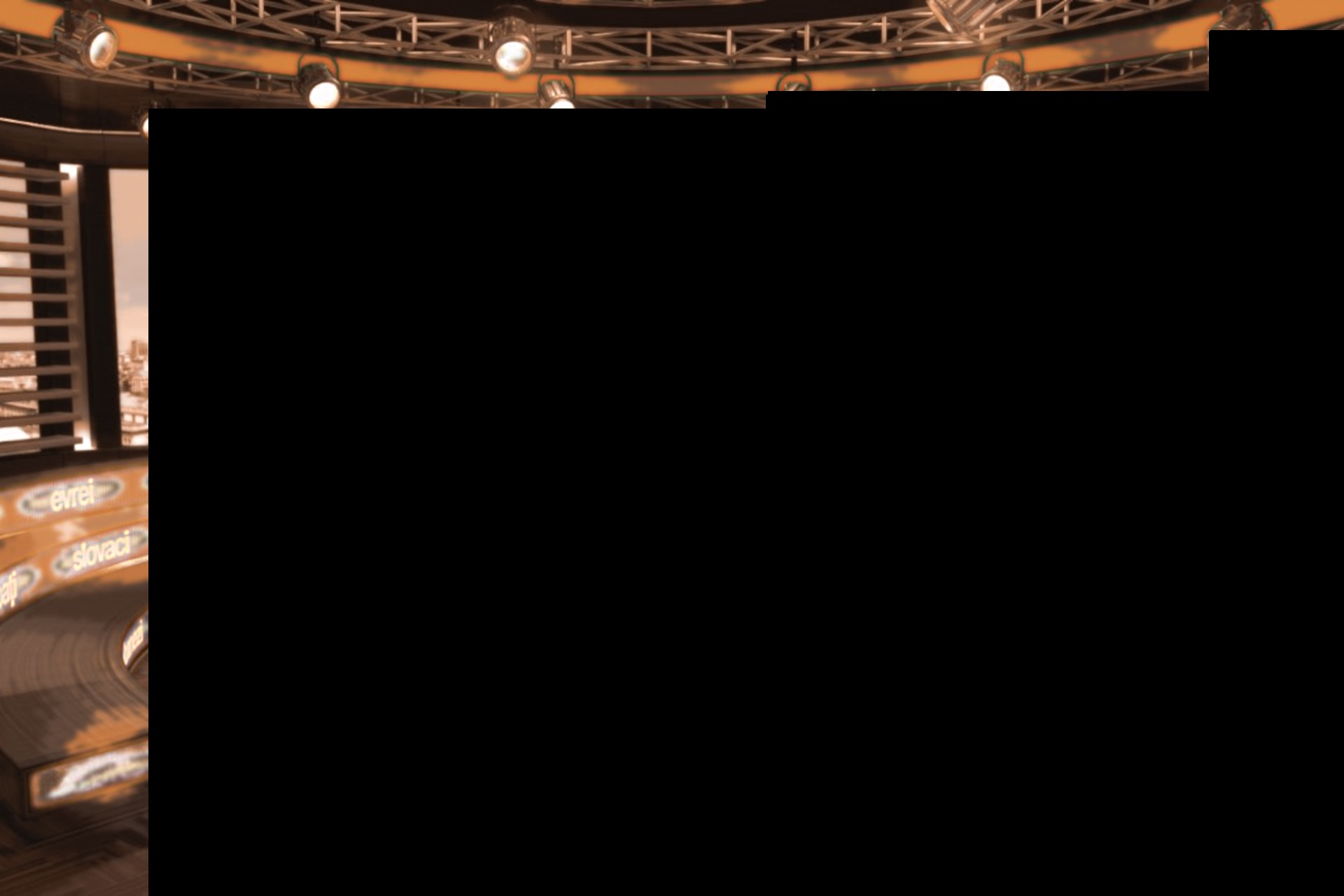}& 
     \includegraphics[width=0.22\textwidth]{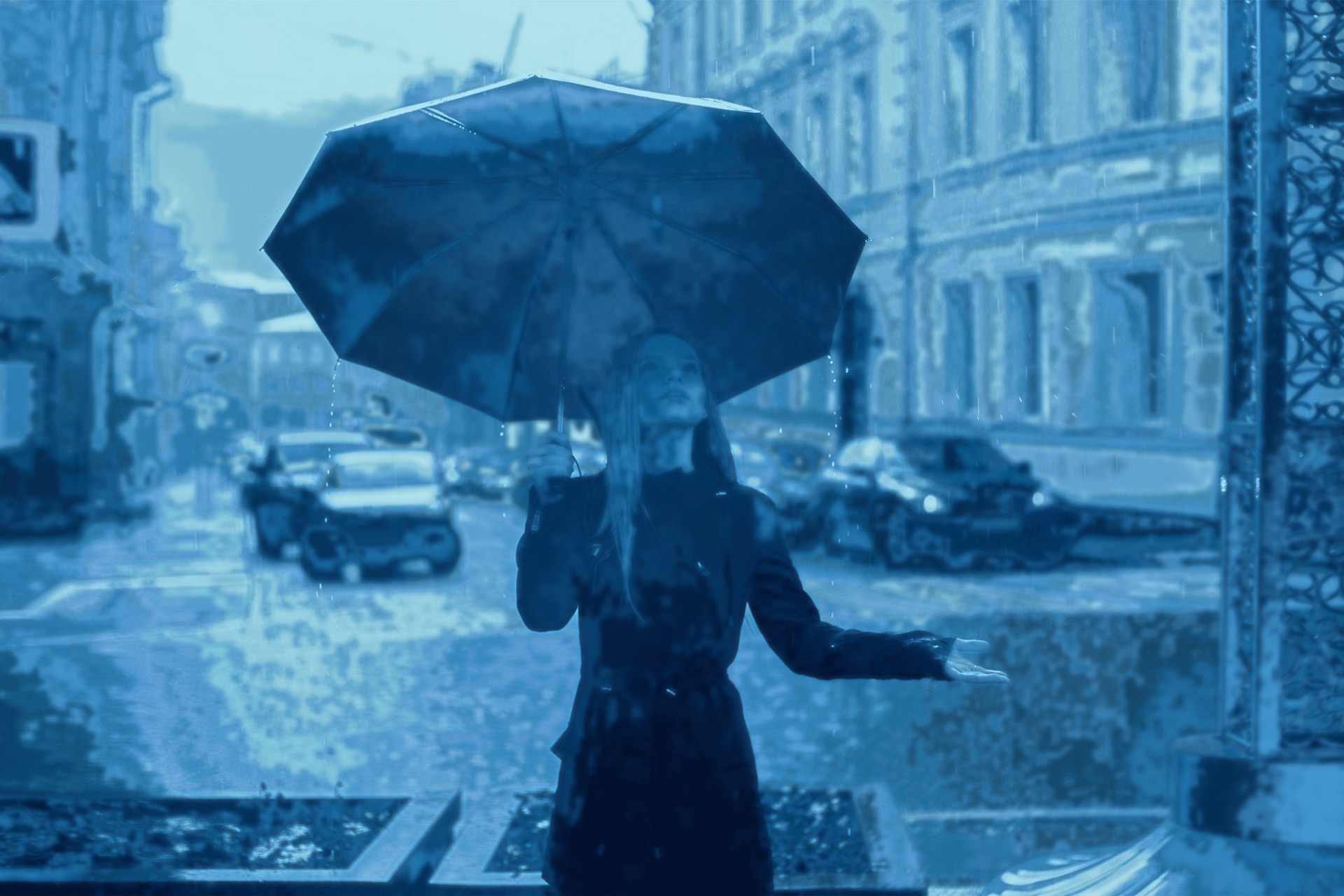} \\   

    \includegraphics[width=0.22\textwidth]{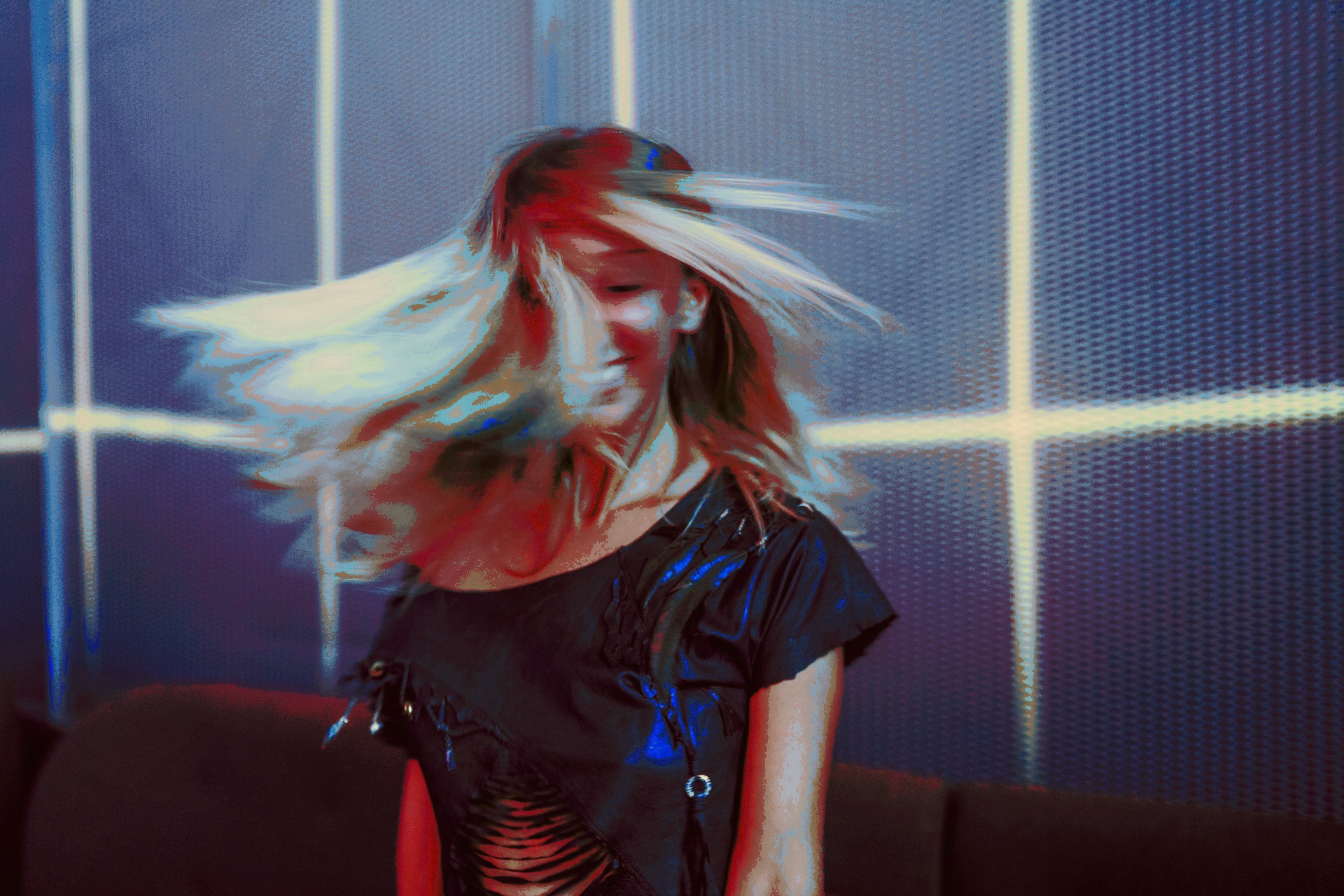}&
    \includegraphics[width=0.22\textwidth]{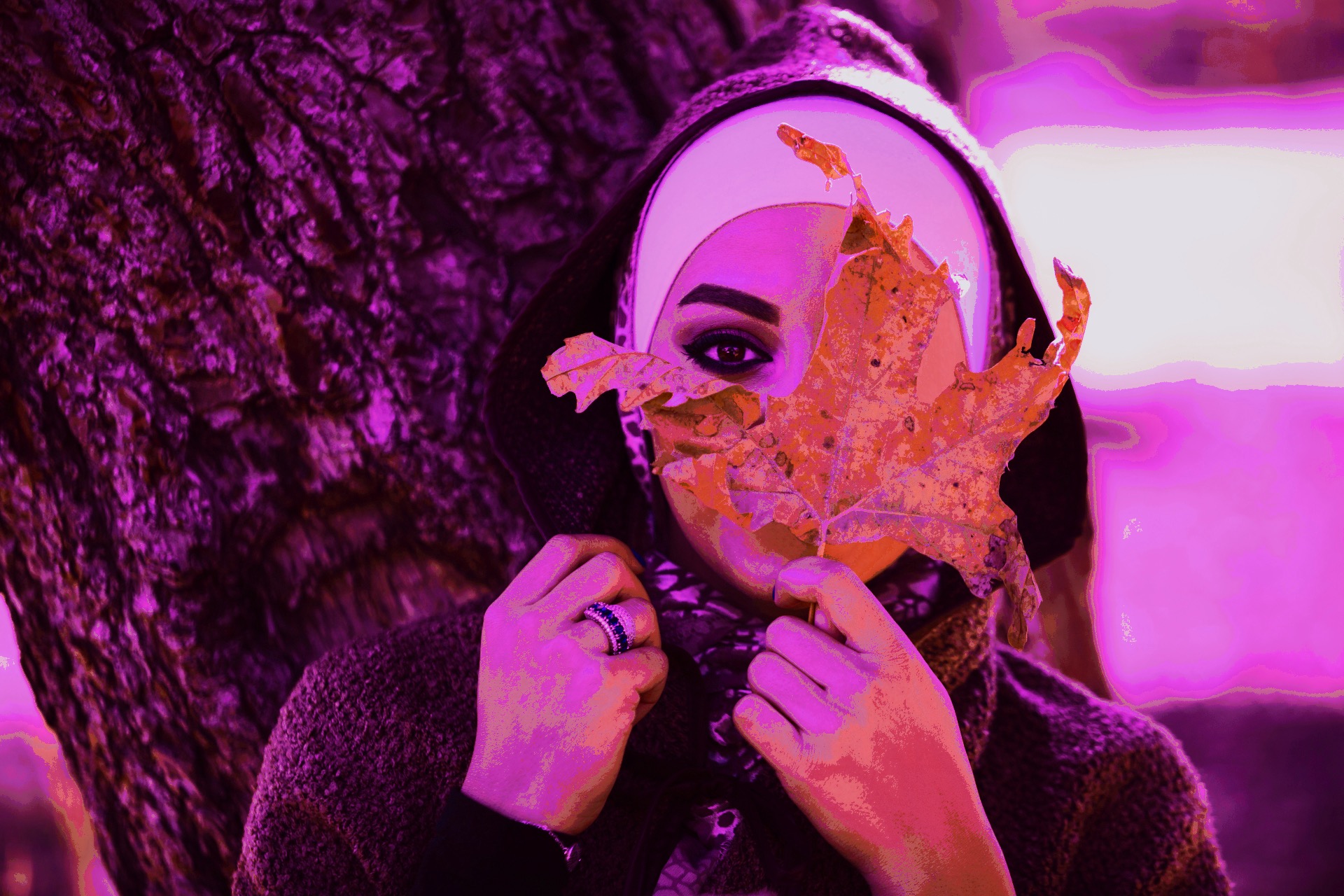}&  
     \includegraphics[width=0.22\textwidth]{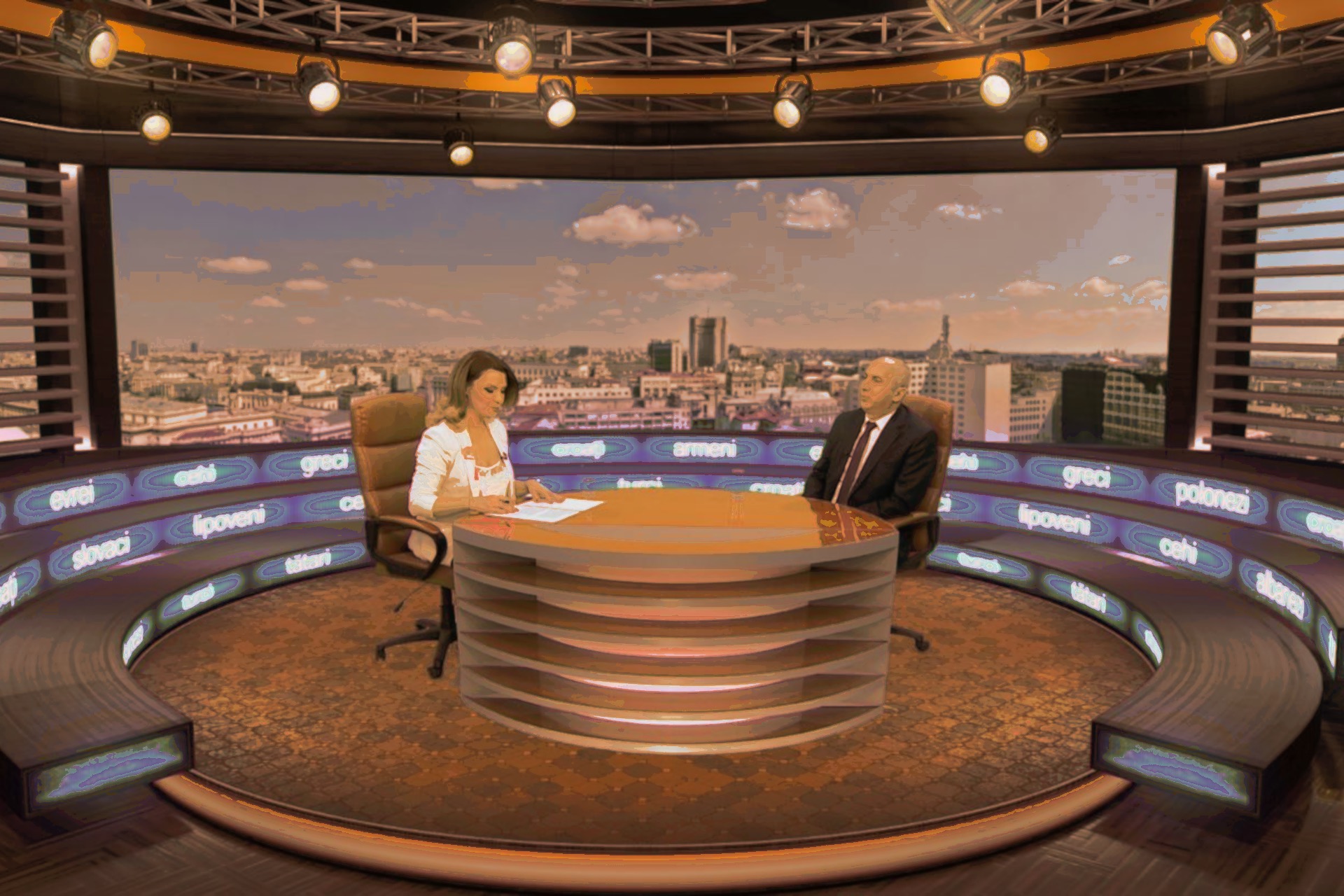}& 
     \includegraphics[width=0.22\textwidth]{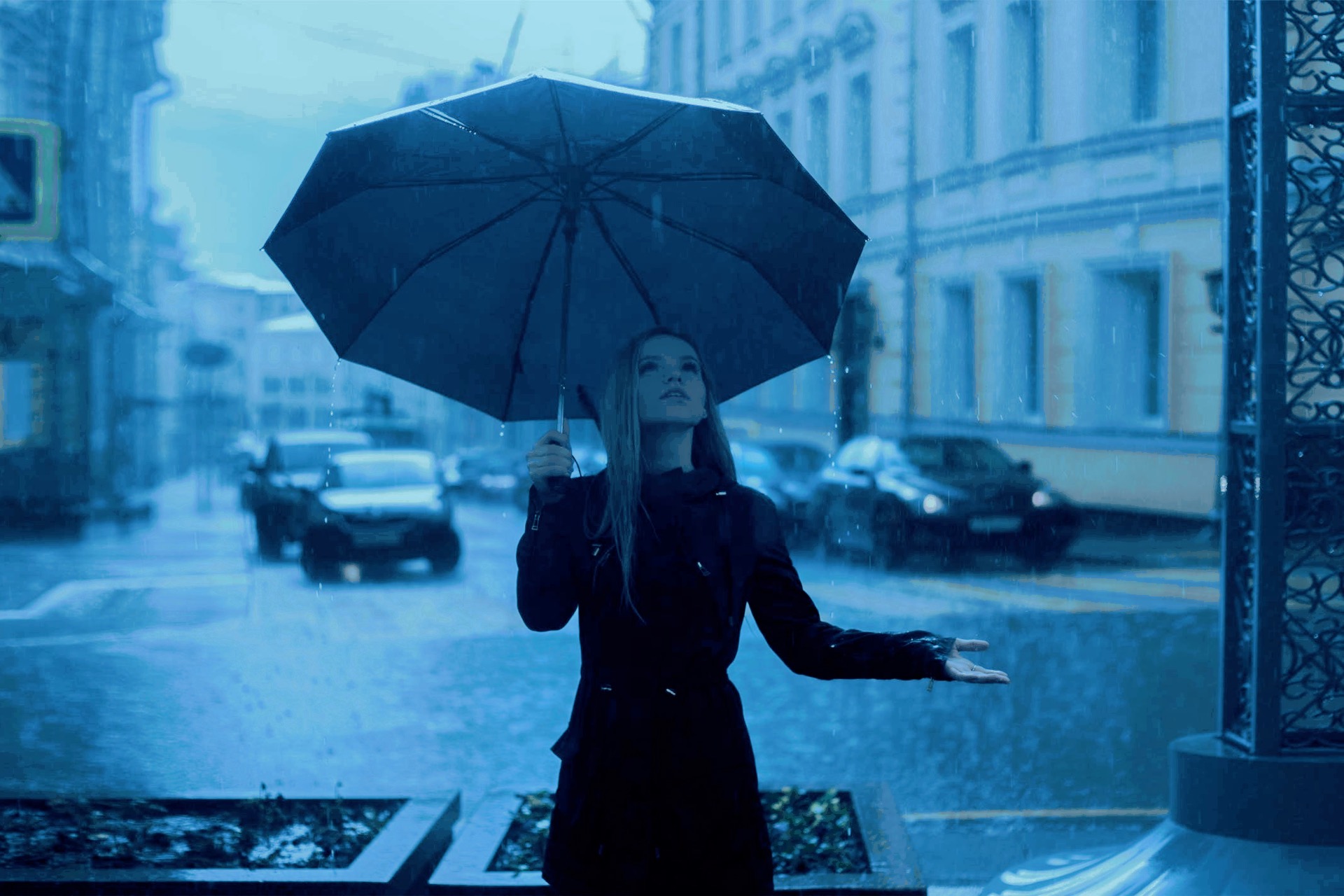} \\      
     
    \includegraphics[width=0.22\textwidth]{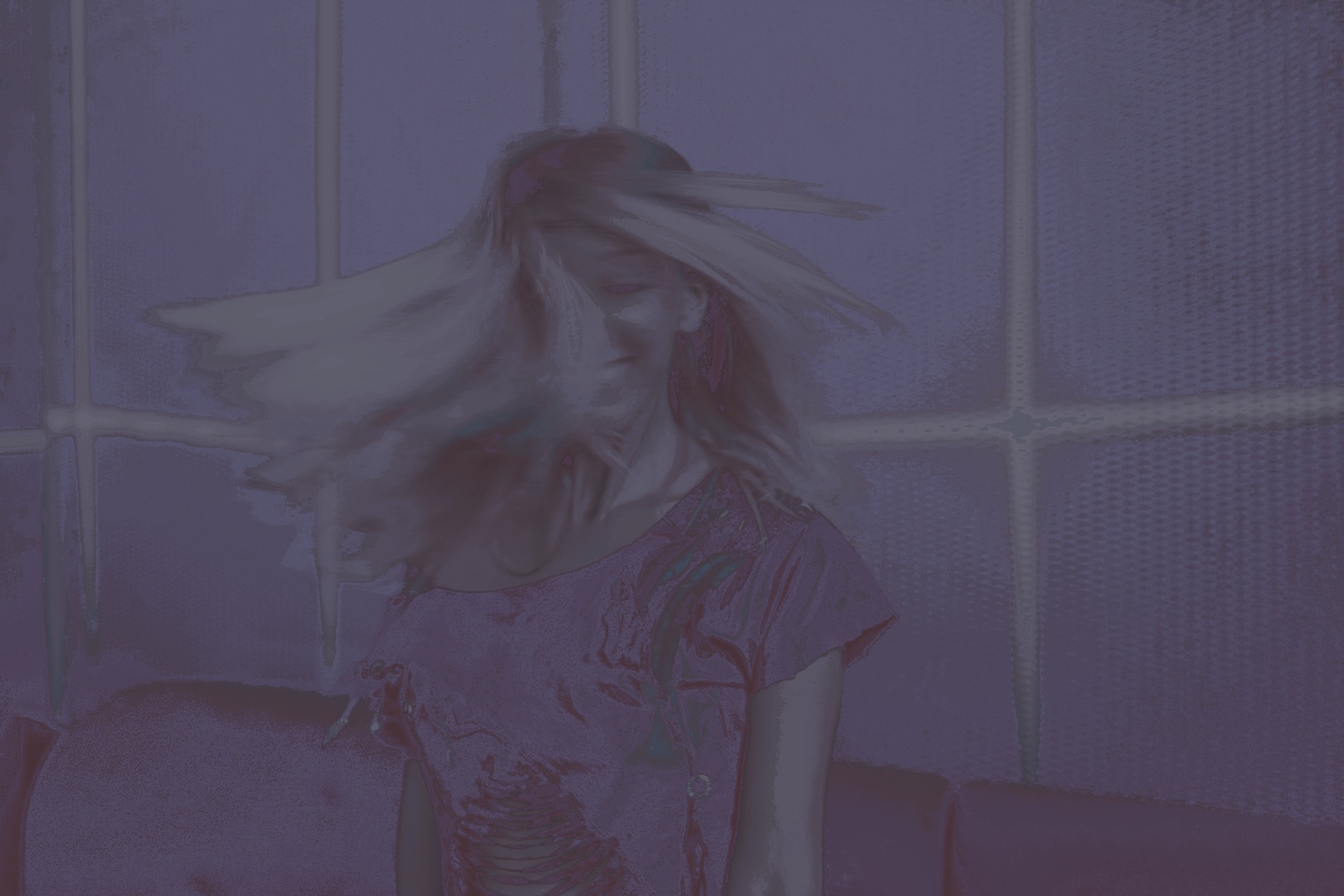}&
    \includegraphics[width=0.22\textwidth]{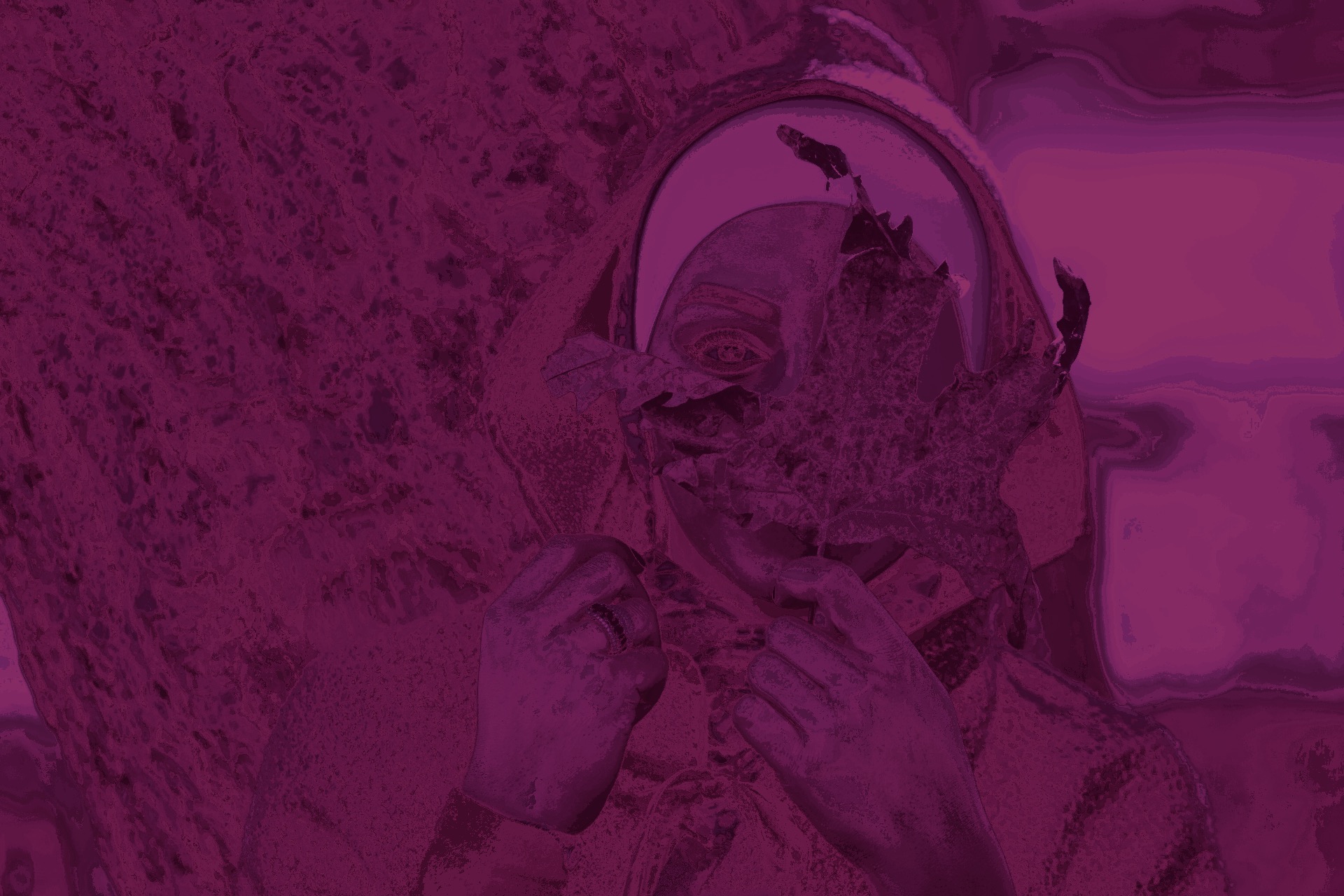}&  
     \includegraphics[width=0.22\textwidth]{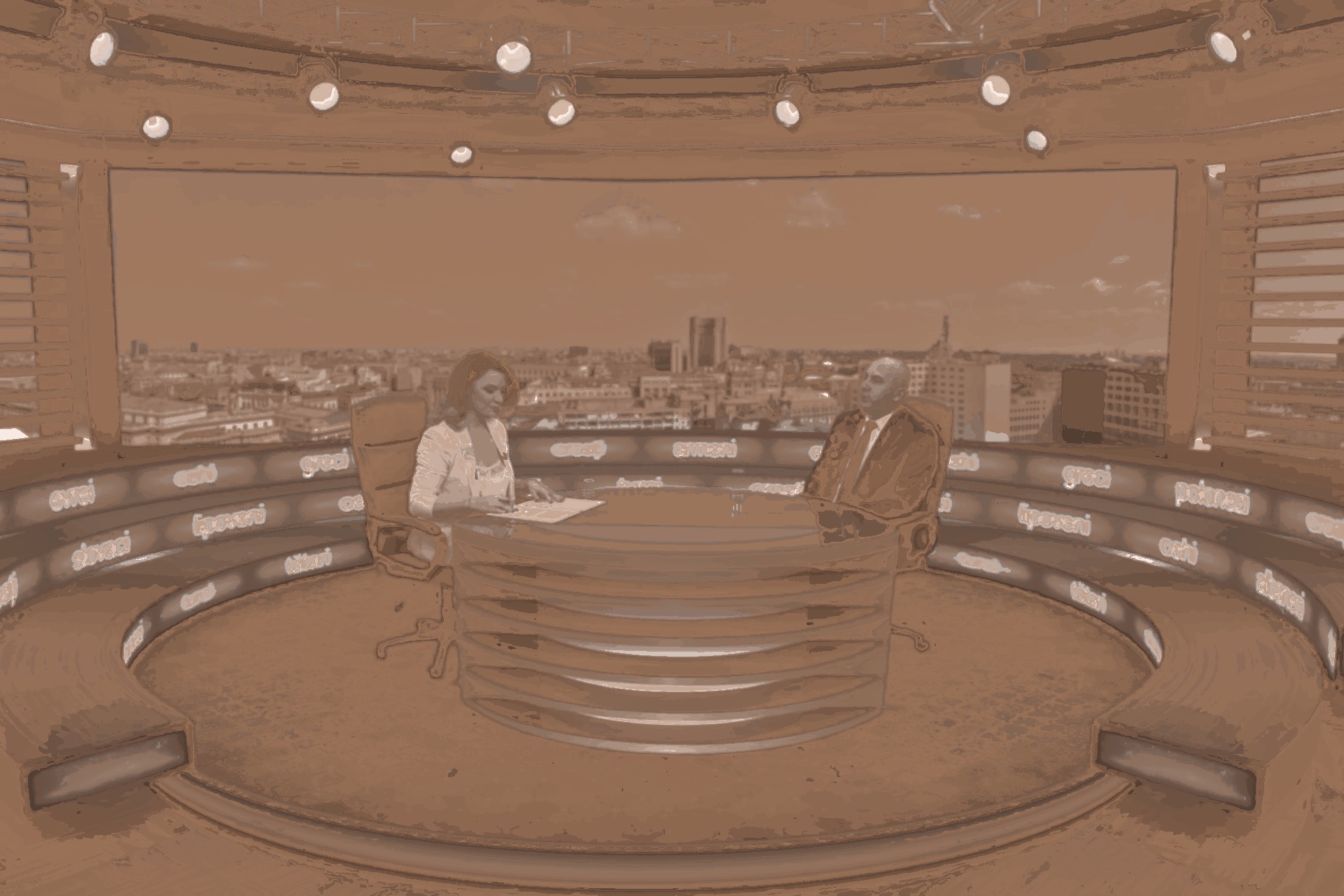}& 
     \includegraphics[width=0.22\textwidth]{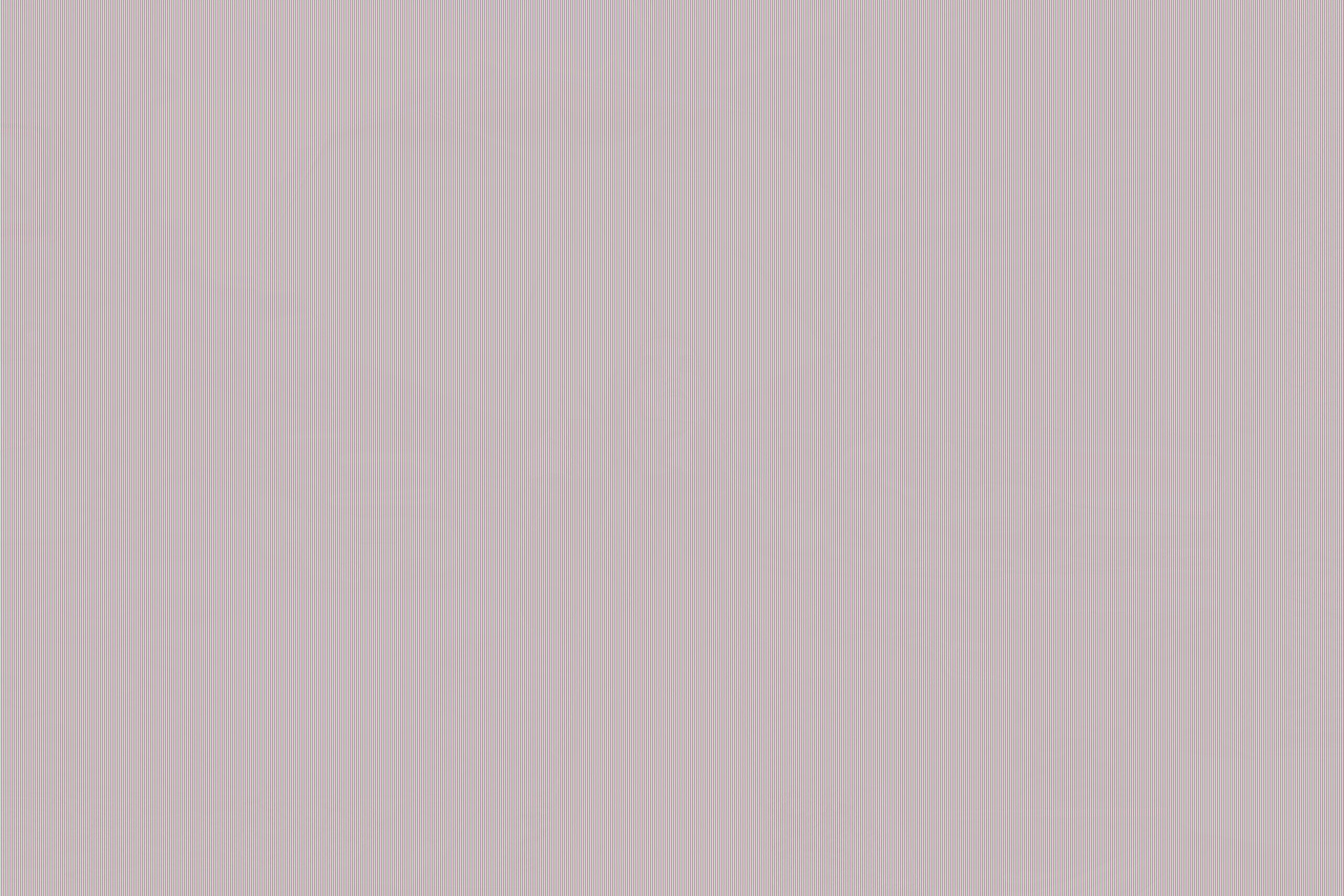} \\ 

  \end{tabular}
  \caption{From top to bottom, sources, targets, proposed result, Rabin result \cite{rabin14}, Ferradans result \cite{ferradans14}, Blondel result \cite{blondel17}.}
  \label{OT}
\end{figure*}

In Figure \ref{OT}, we compare the results of the SotA OT approaches (\cite{rabin14}, \cite{ferradans14}, \cite{blondel17}) to those of our method. One can see that when applied to arbitrary sources, they are brittle and frequently contain egregious artifacts which range from banding to entire image sections failing to render. In addition, these methods can be of high computational complexity, rendering them impractical for use in an image editing and color grading context.

\begin{figure*}[t]
\centering
\begin{tabular}{cccc}

    \includegraphics[width=0.22\textwidth]{selected/relatedWork/sources-references/blue_1920_1280.png.jpeg}&
    \includegraphics[width=0.22\textwidth]{selected/relatedWork/sources-references/orange_1920_1280.png.jpeg}&  
     \includegraphics[width=0.22\textwidth]{selected/relatedWork/sources-references/studio_1920_1280.png.jpeg}& 
     \includegraphics[width=0.22\textwidth]{selected/relatedWork/sources-references/umbrella_1920_1280.png.jpeg} \\

    \includegraphics[width=0.22\textwidth]{selected/relatedWork/sources-references/studio_1920_1280.png.jpeg}&
    \includegraphics[width=0.22\textwidth]{selected/relatedWork/sources-references/blue_1920_1280.png.jpeg}&  
     \includegraphics[width=0.22\textwidth]{selected/relatedWork/sources-references/orange_1920_1280.png.jpeg}& 
     \includegraphics[width=0.22\textwidth]{selected/relatedWork/sources-references/orange_1920_1280.png.jpeg} \\  
     
    \includegraphics[width=0.22\textwidth]{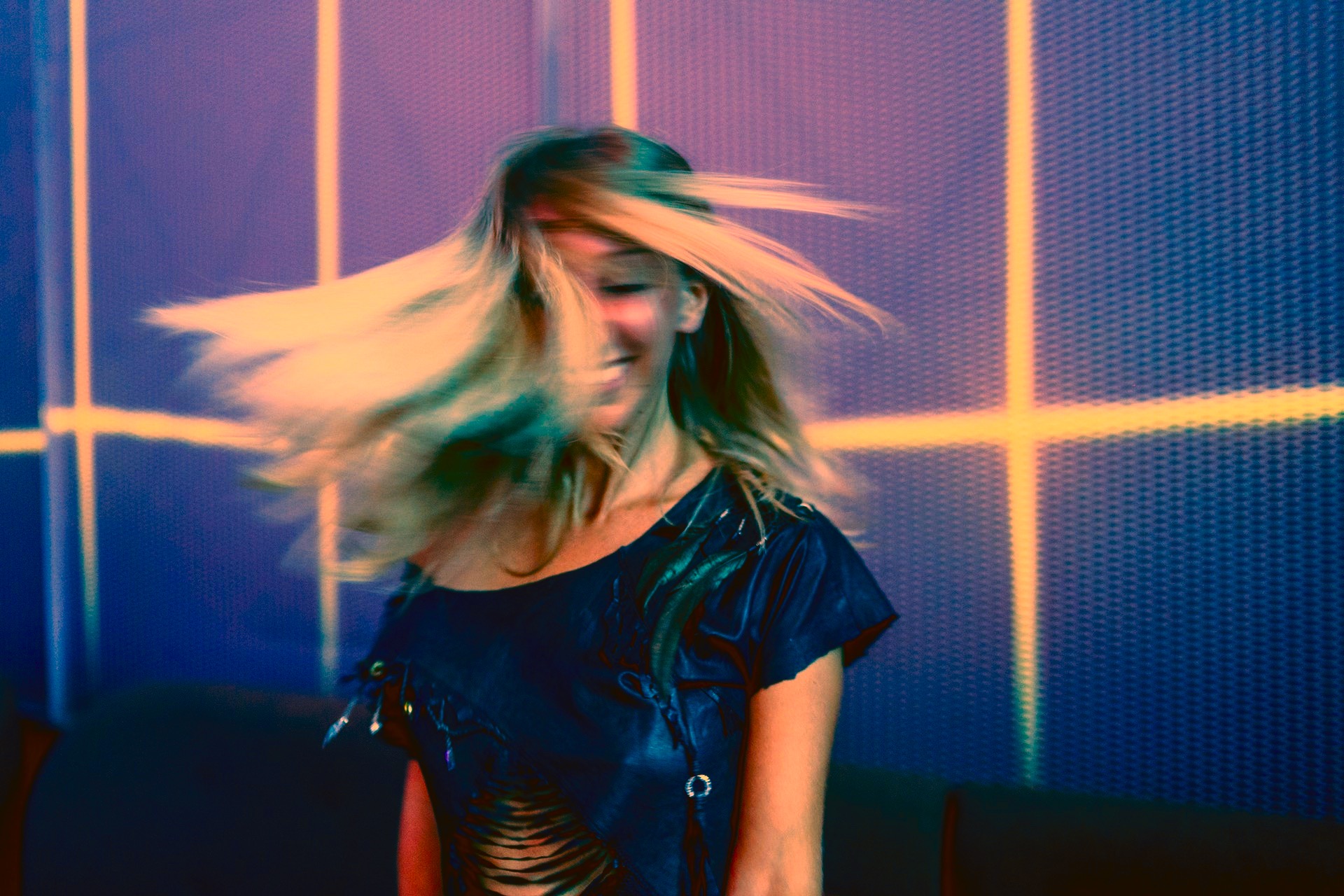}&
    \includegraphics[width=0.22\textwidth]{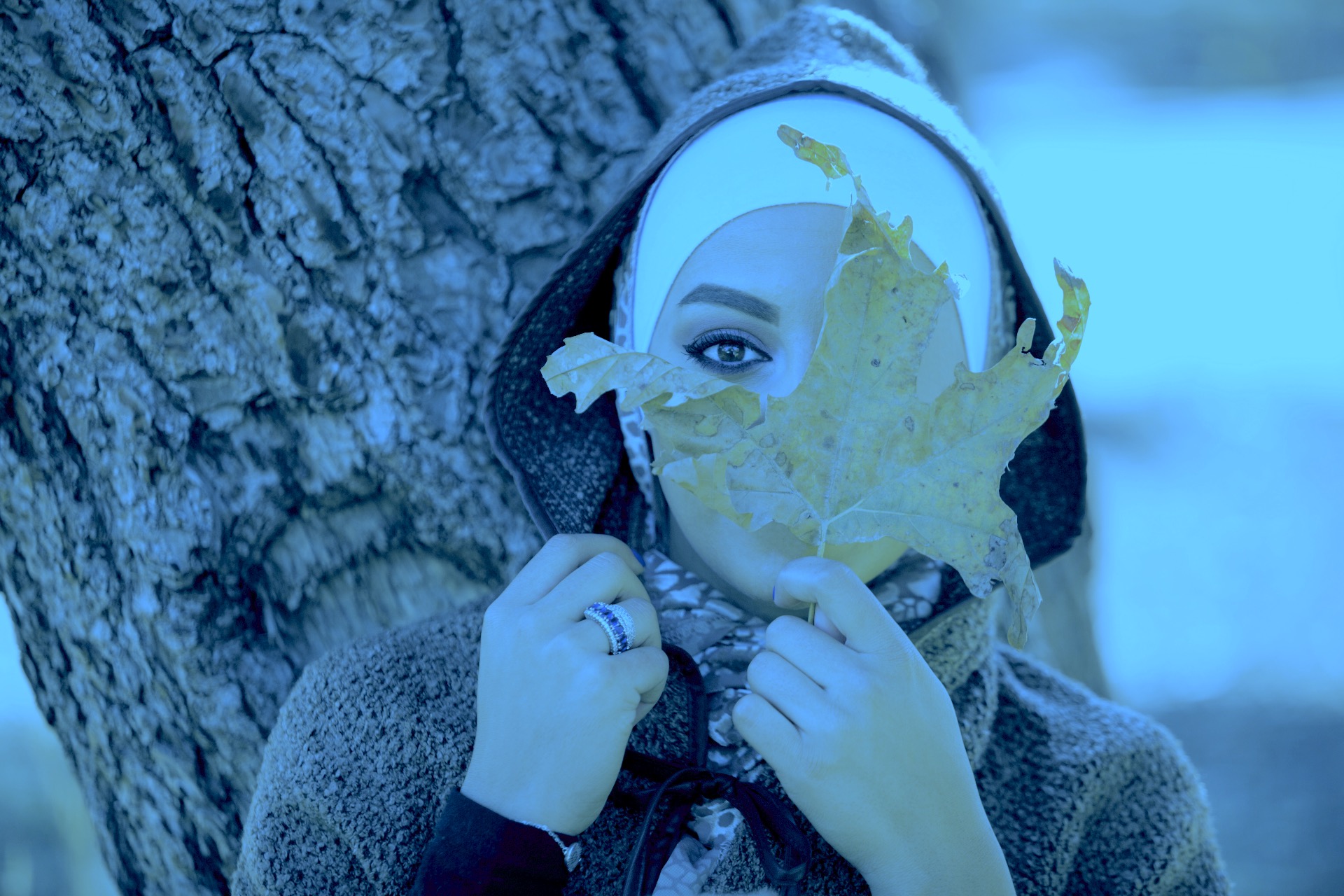}&  
     \includegraphics[width=0.22\textwidth]{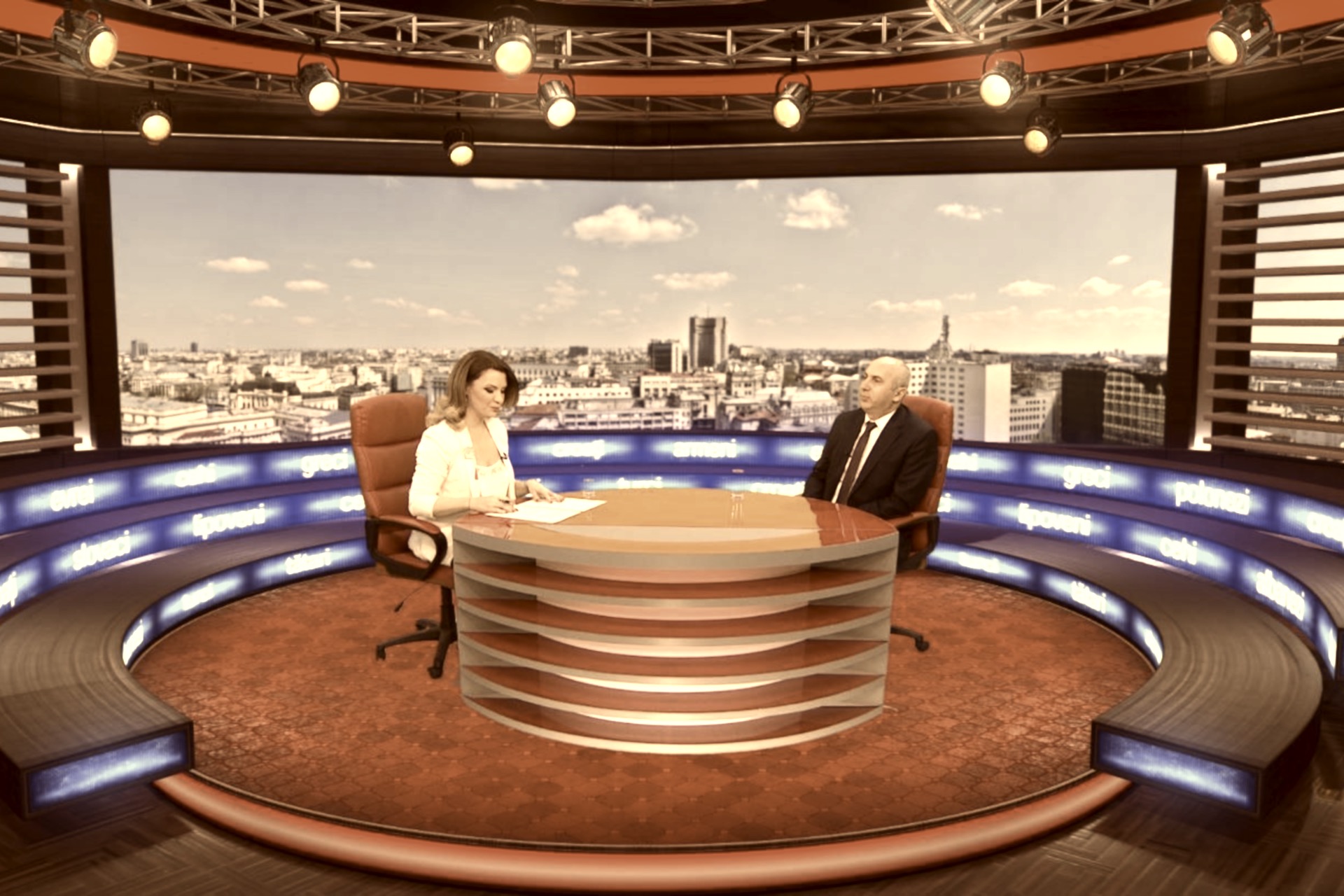}& 
     \includegraphics[width=0.22\textwidth]{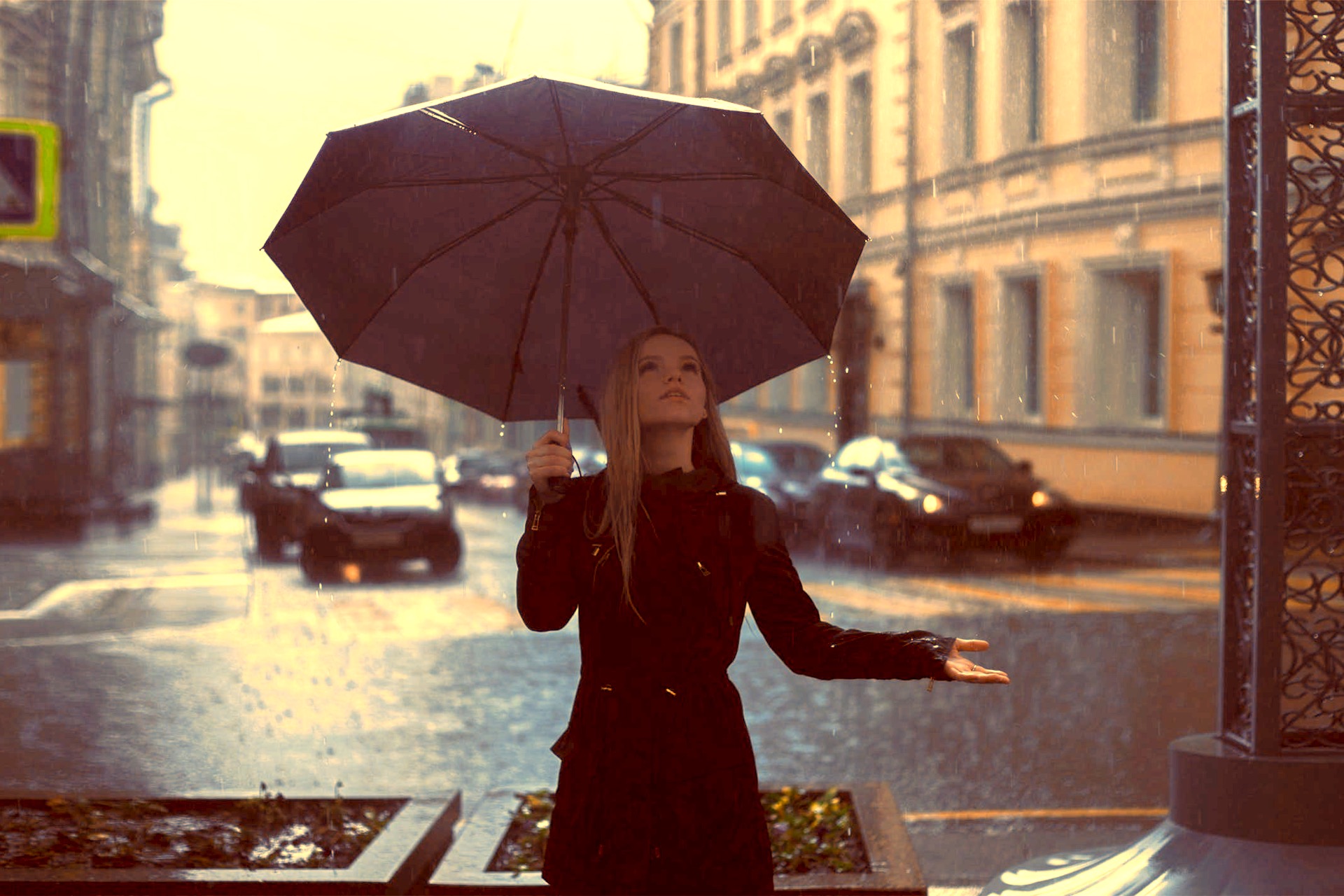} \\
    
    \includegraphics[width=0.22\textwidth]{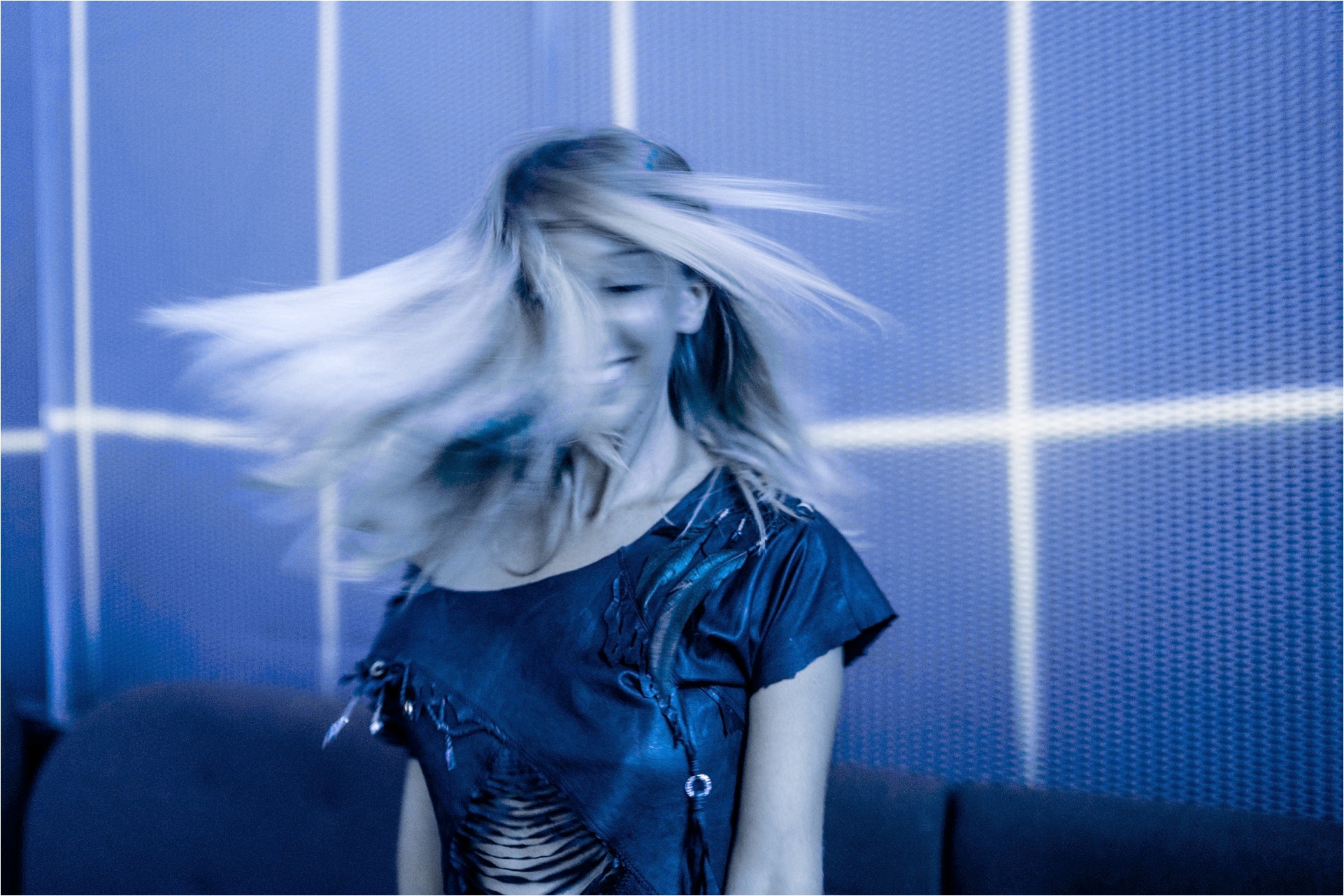}&
    \includegraphics[width=0.22\textwidth]{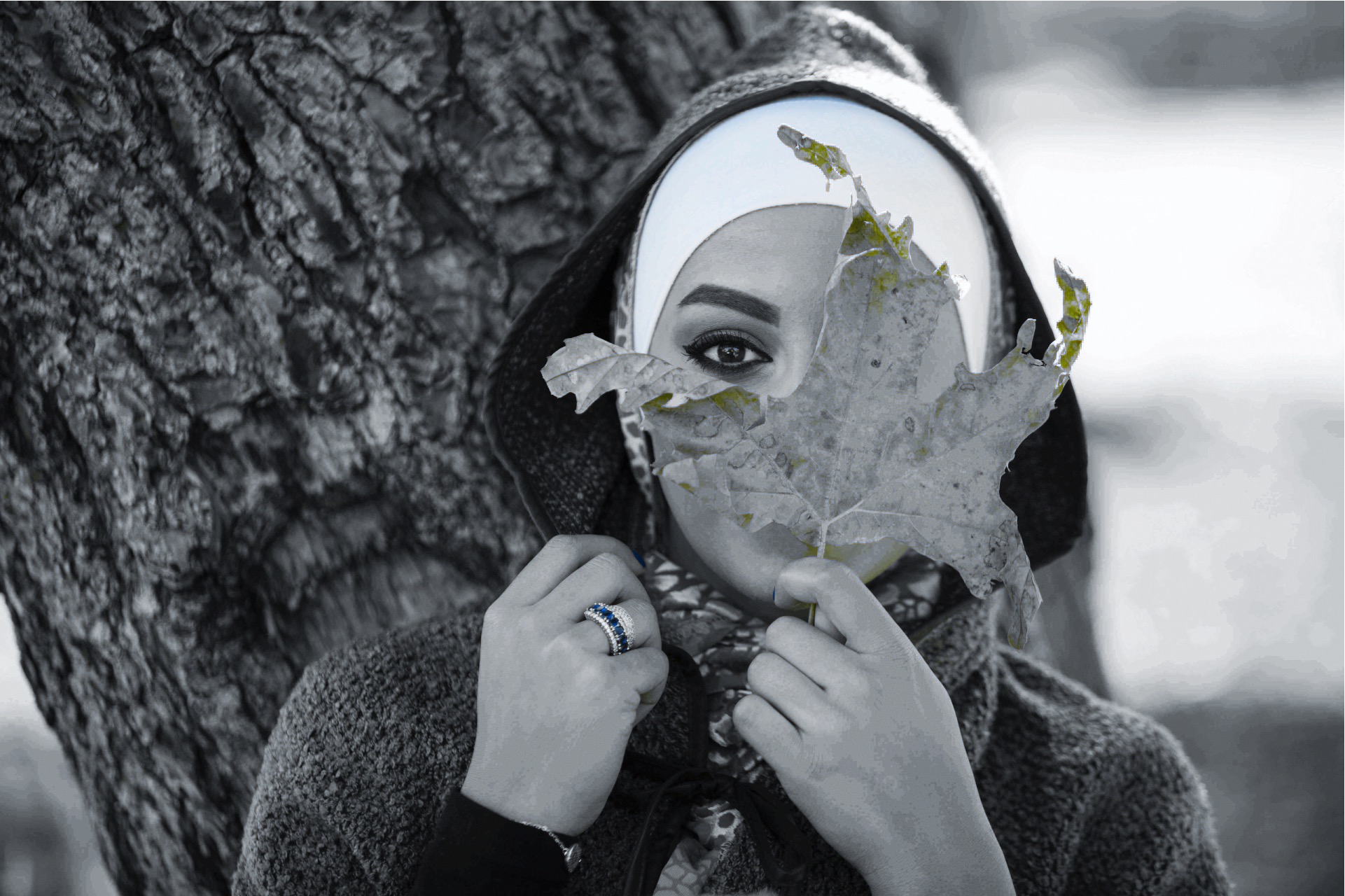}&  
    \includegraphics[width=0.22\textwidth]{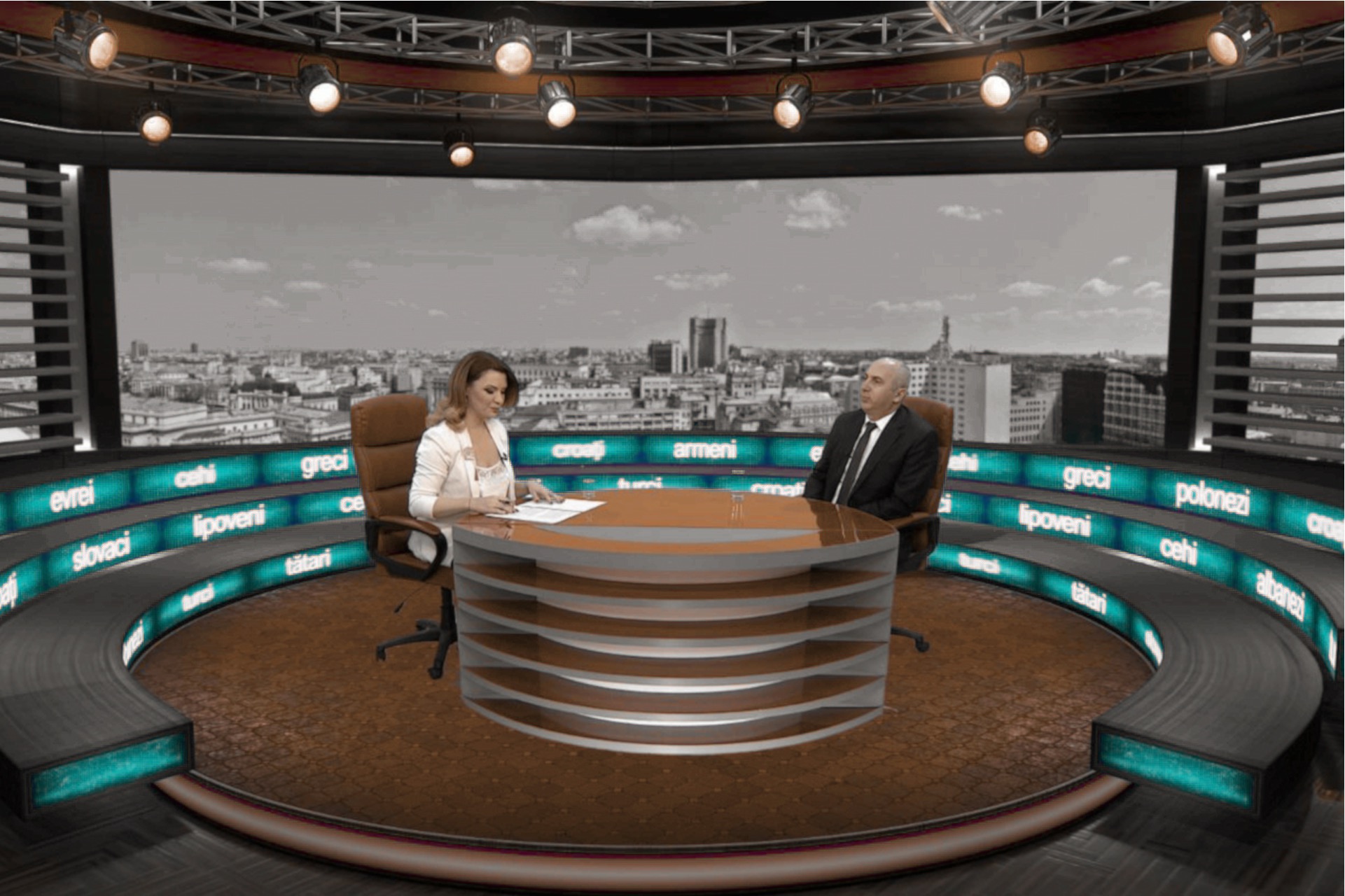}& 
    \includegraphics[width=0.22\textwidth]{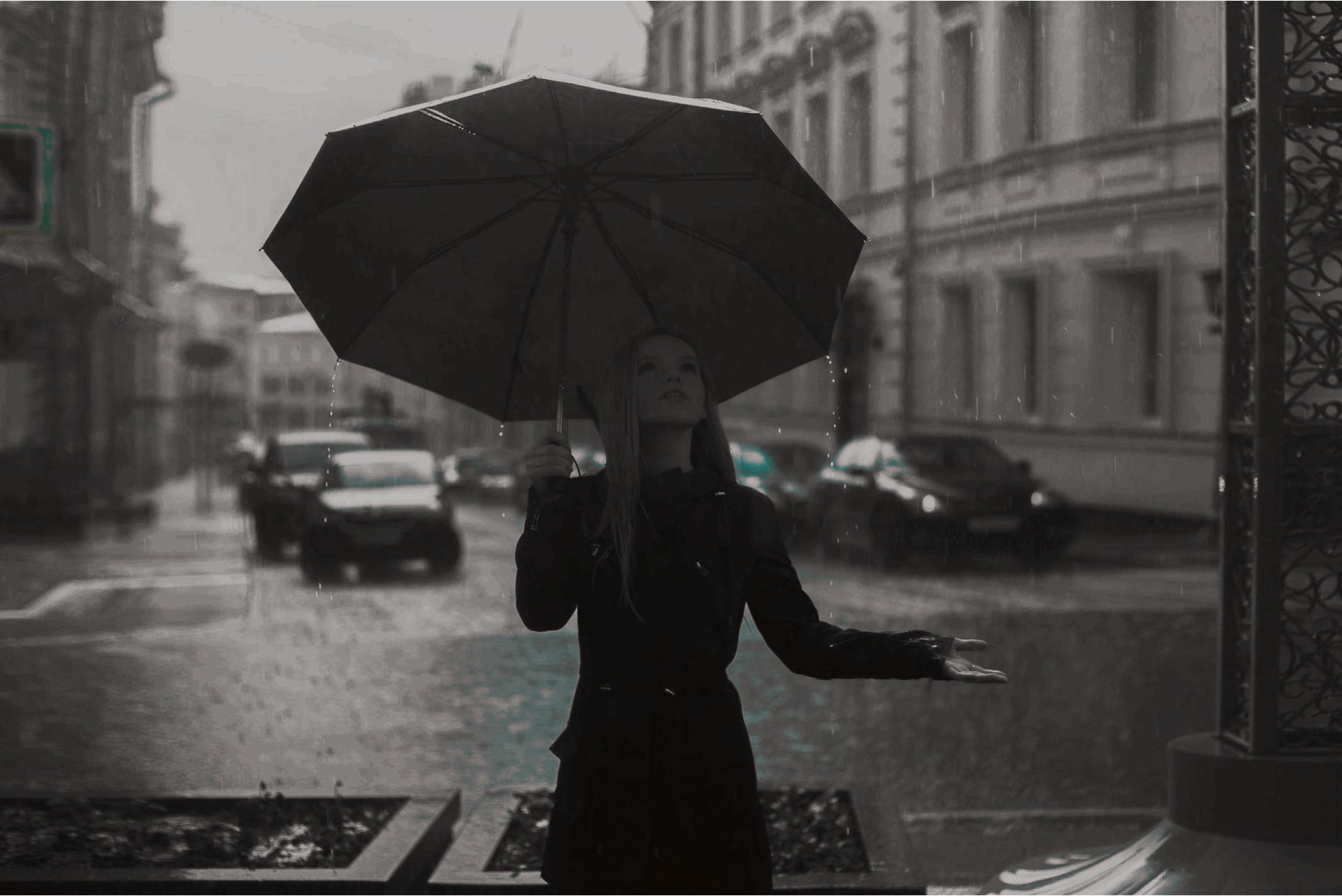} \\  

    \includegraphics[width=0.22\textwidth]{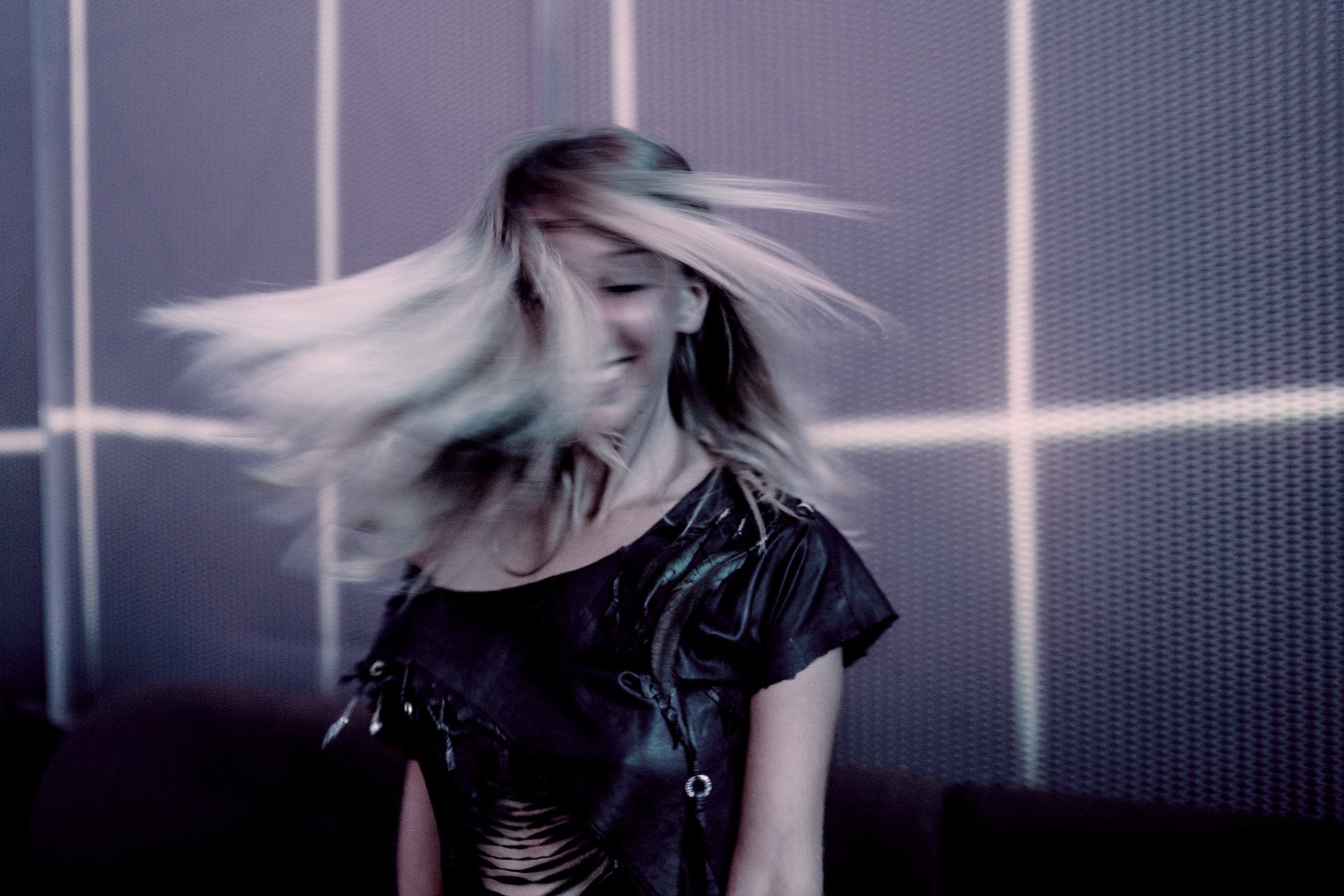}&
    \includegraphics[width=0.22\textwidth]{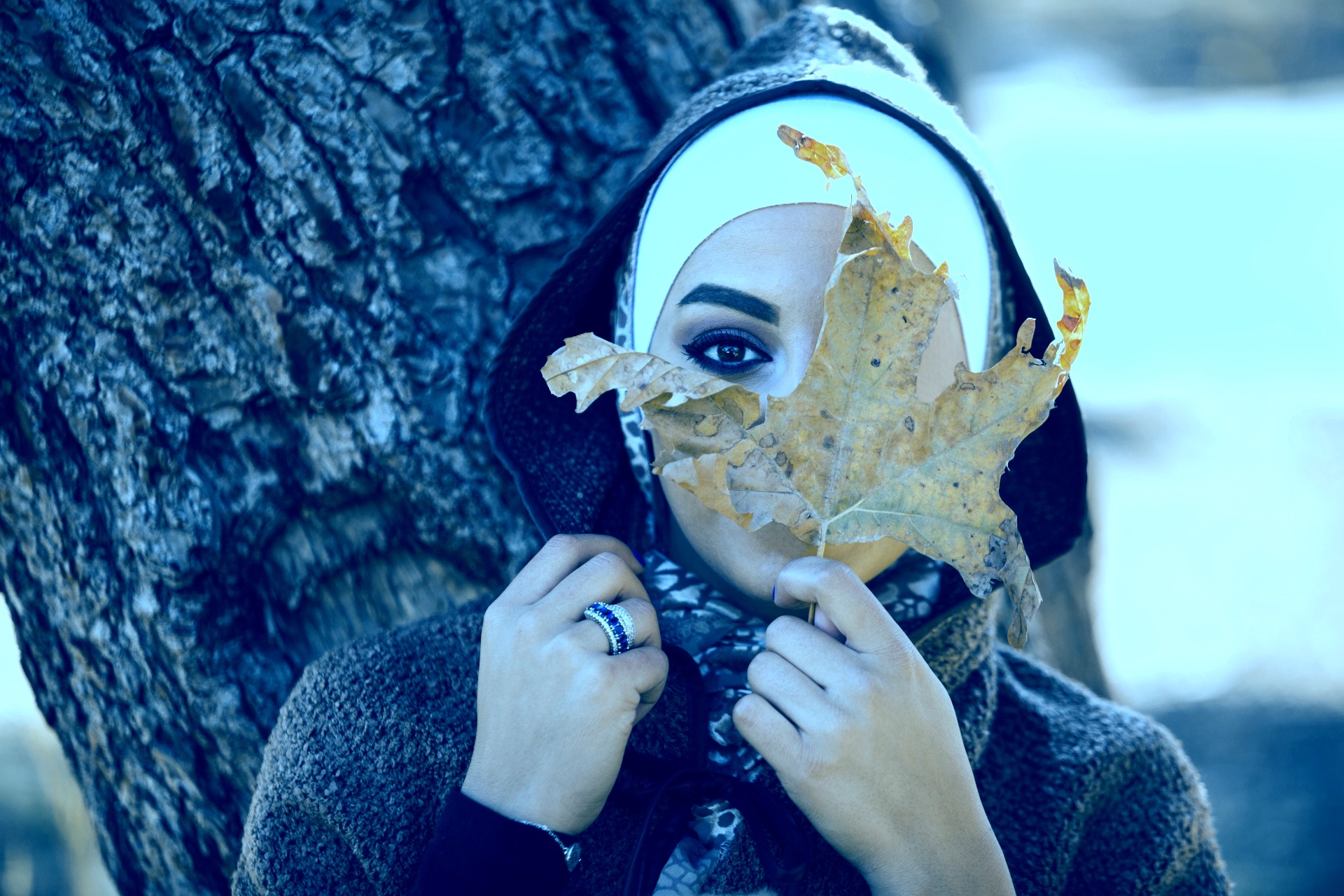}&  
     \includegraphics[width=0.22\textwidth]{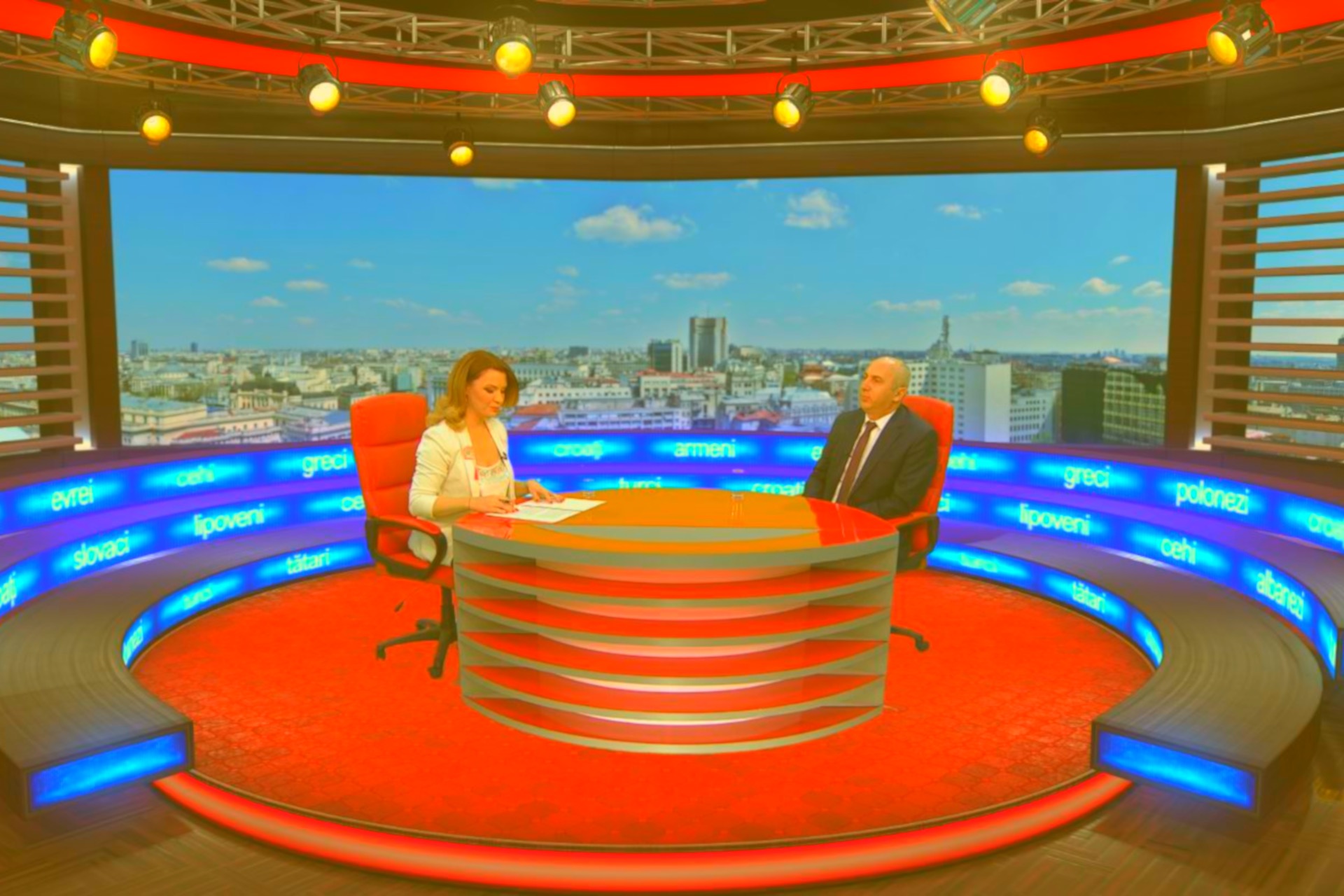}& 
     \includegraphics[width=0.22\textwidth]{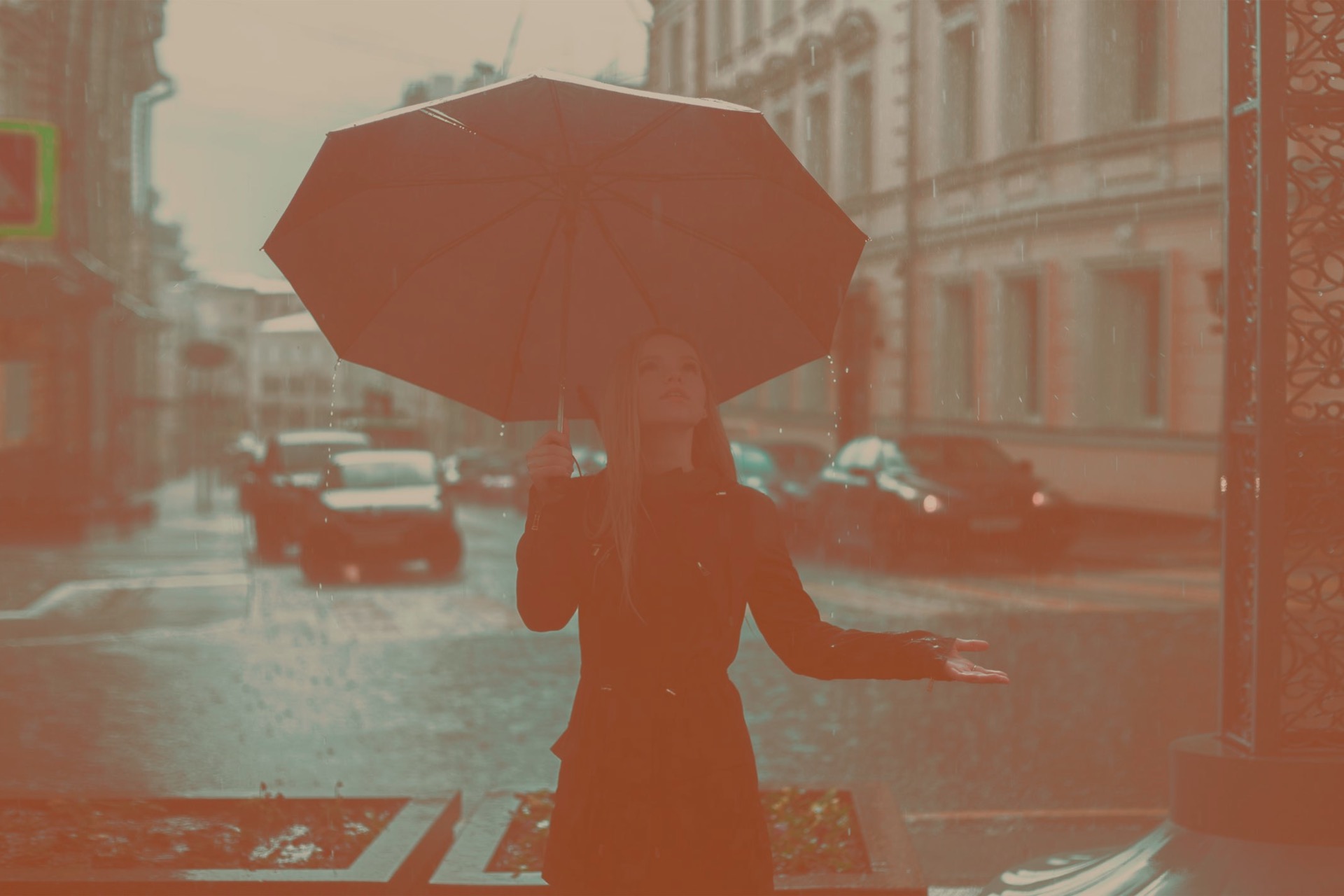} \\   
     
    \includegraphics[width=0.22\textwidth]{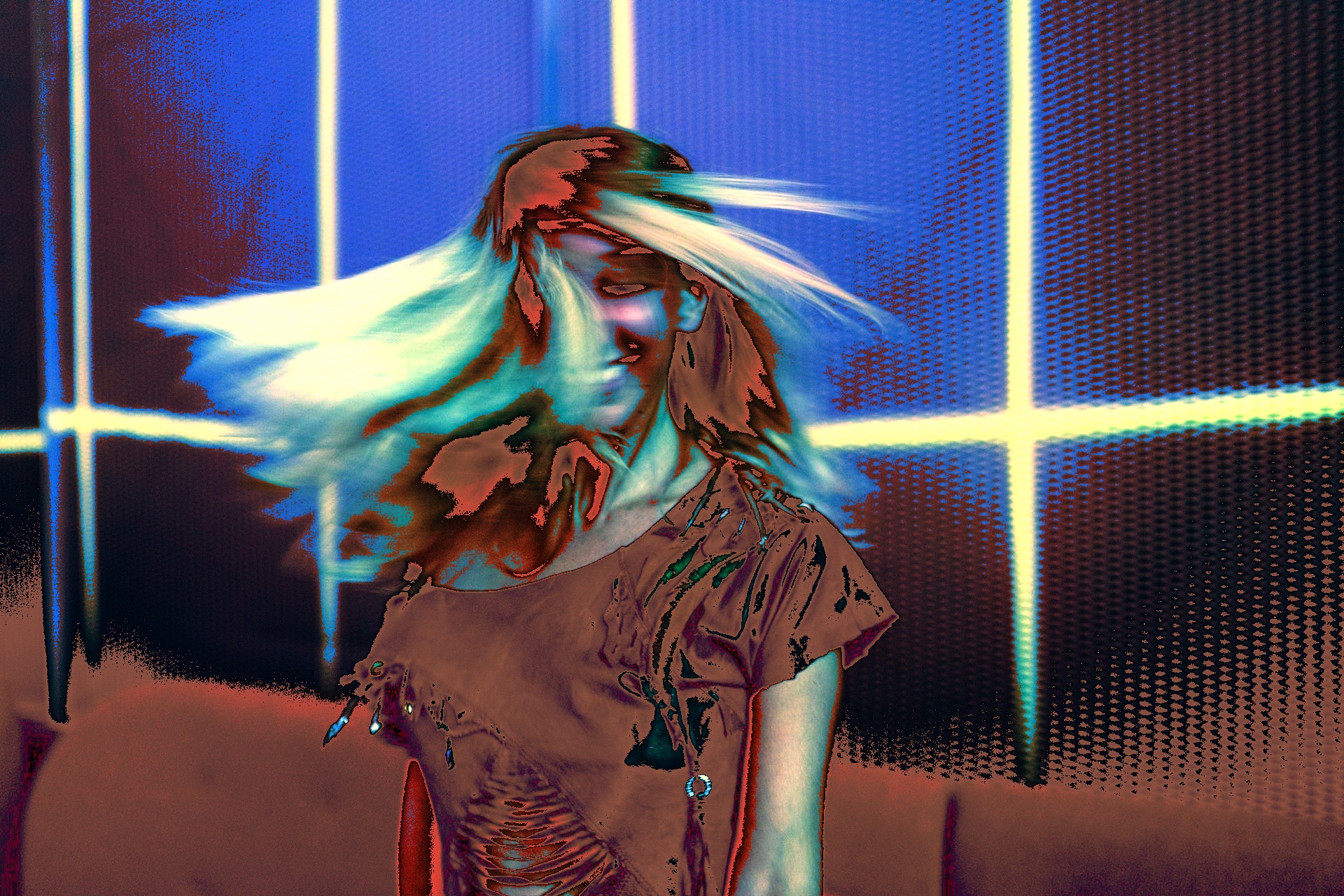}&
    \includegraphics[width=0.22\textwidth]{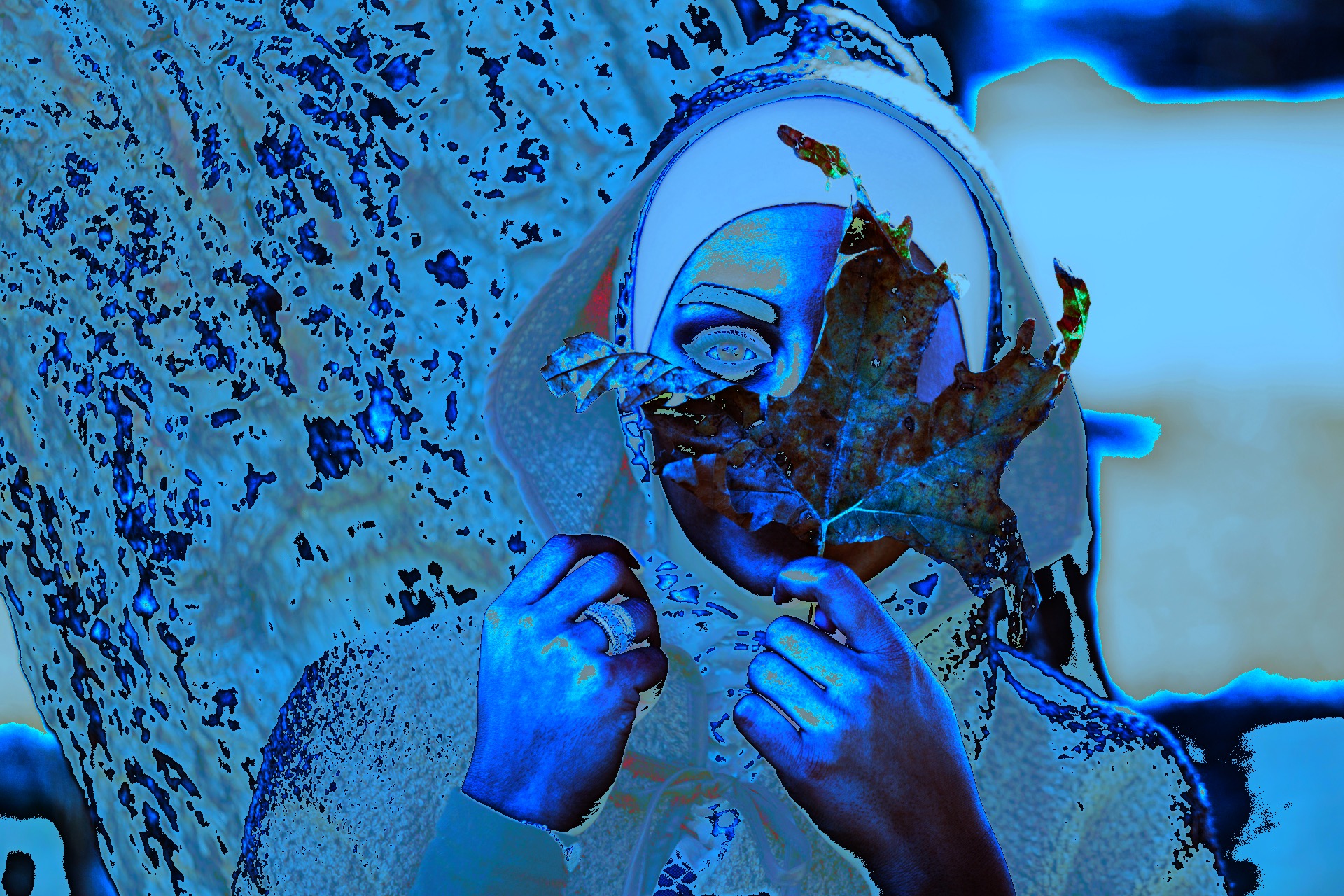}&  
     \includegraphics[width=0.22\textwidth]{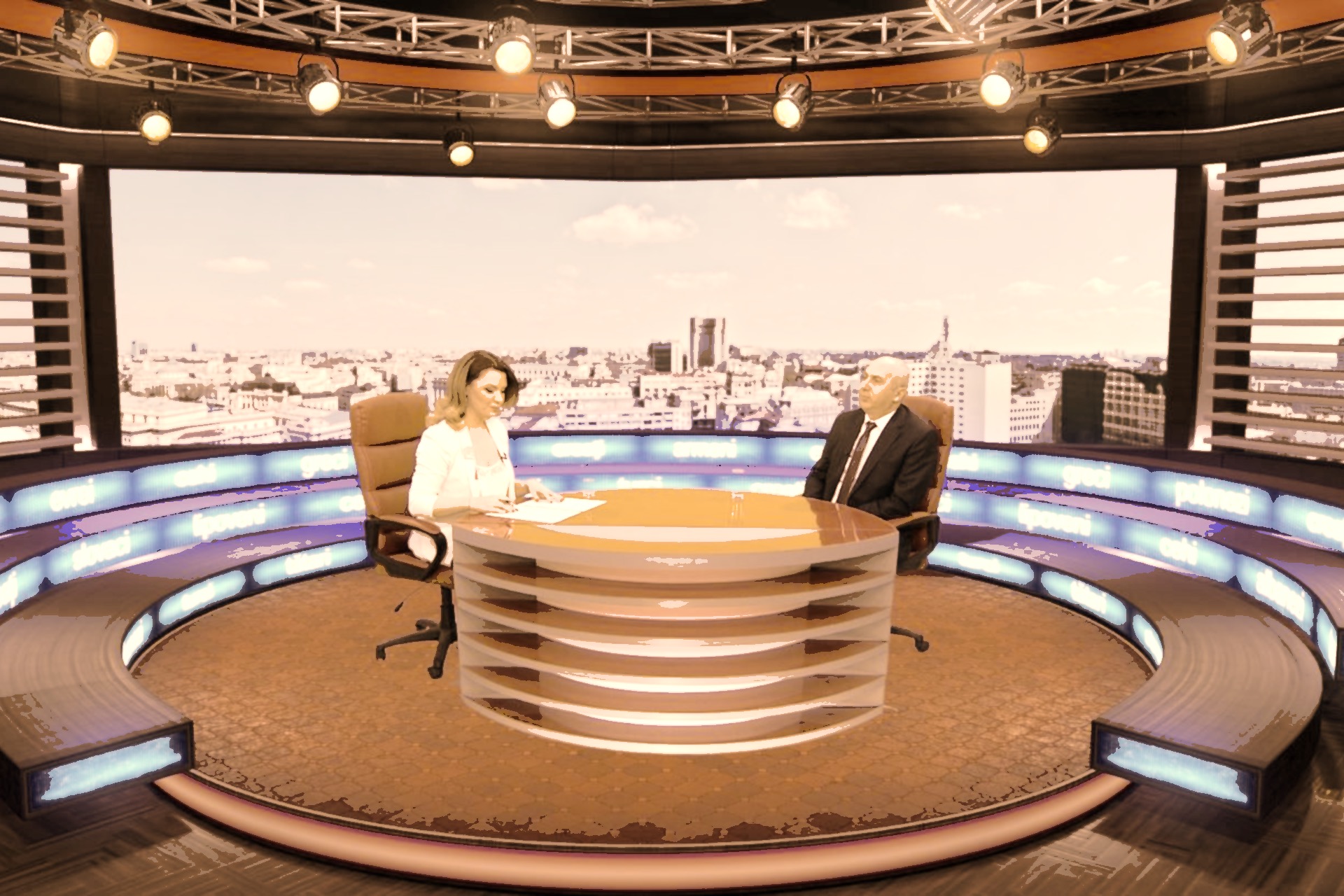}& 
     \includegraphics[width=0.22\textwidth]{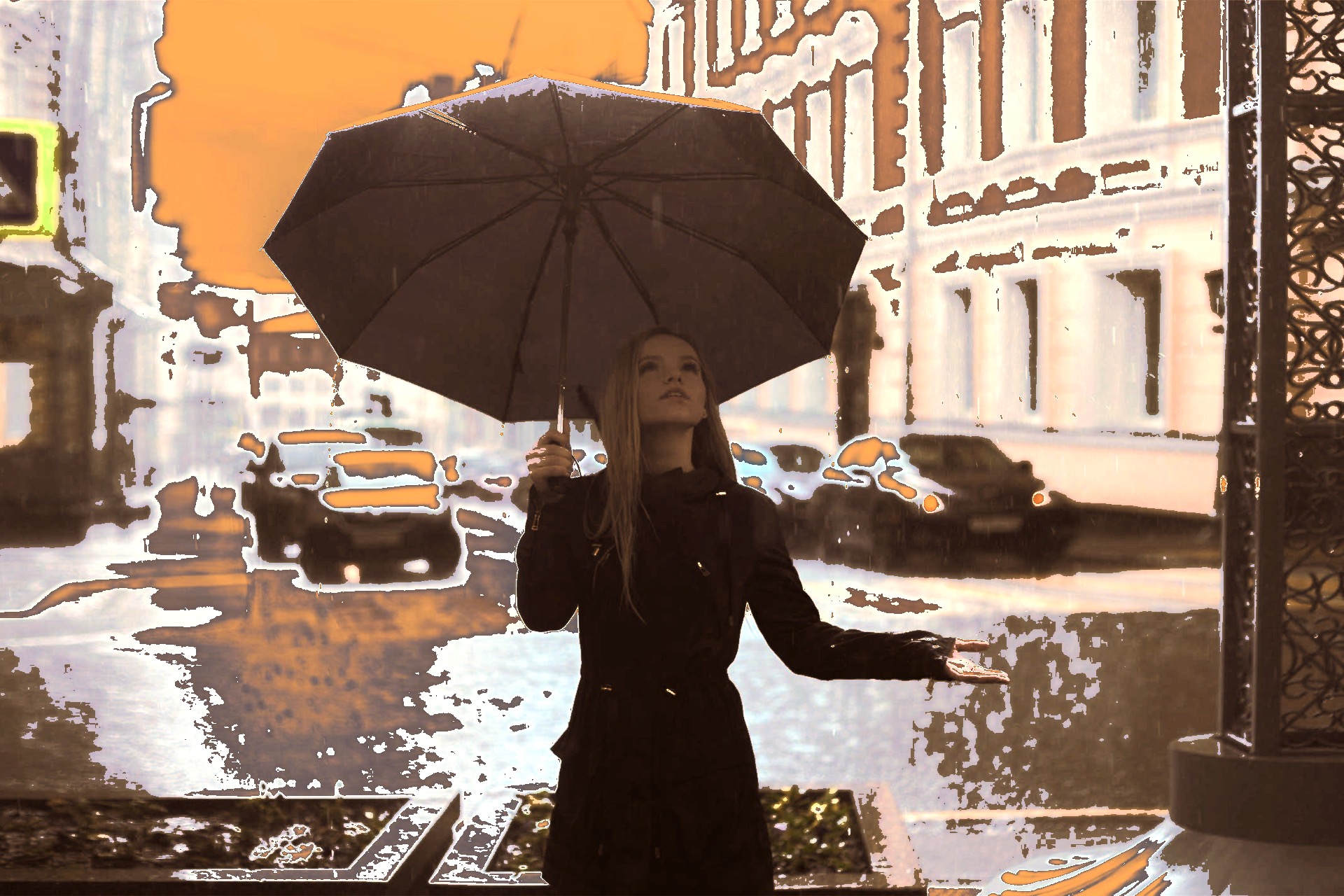} \\

  \end{tabular}
  \caption{From top to bottom, sources, targets, proposed result, Li result \cite{li19}, Wu result \cite{wu20}, Xie result \cite{xie20}.}
  \label{statTransfer}
\end{figure*}

The results of \cite{wu20} \cite{xie20} \cite{li19}, shown in Figure \ref{statTransfer}, demonstrate that these methods are also not robust outside of their test set and result in banding artifacts and poor color transfer, with the exception of the results of Wu et al. \cite{wu20}. It should be noted however, that using the saliency threshold parameter recommended by the authors, the entire image falls under the salient category and thus the method is simply performing the same transfer methodology as \cite{reinhard01}.

\begin{figure*}[t]
\centering
\begin{tabular}{cccc}

    \includegraphics[width=0.22\textwidth]{selected/relatedWork/sources-references/blue_1920_1280.png.jpeg}&
    \includegraphics[width=0.22\textwidth]{selected/relatedWork/sources-references/orange_1920_1280.png.jpeg}&  
     \includegraphics[width=0.22\textwidth]{selected/relatedWork/sources-references/blue_1920_1280.png.jpeg}& 
     \includegraphics[width=0.22\textwidth]{selected/relatedWork/sources-references/studio_1920_1280.png.jpeg} \\

    \includegraphics[width=0.22\textwidth]{selected/relatedWork/sources-references/purple_1920_1280.png.jpeg}&
    \includegraphics[width=0.22\textwidth]{selected/relatedWork/sources-references/umbrella_1920_1280.png.jpeg}&  
     \includegraphics[width=0.22\textwidth]{selected/relatedWork/sources-references/orange_1920_1280.png.jpeg}& 
     \includegraphics[width=0.22\textwidth]{selected/relatedWork/sources-references/purple_1920_1280.png.jpeg} \\  
     
    \includegraphics[width=0.22\textwidth]{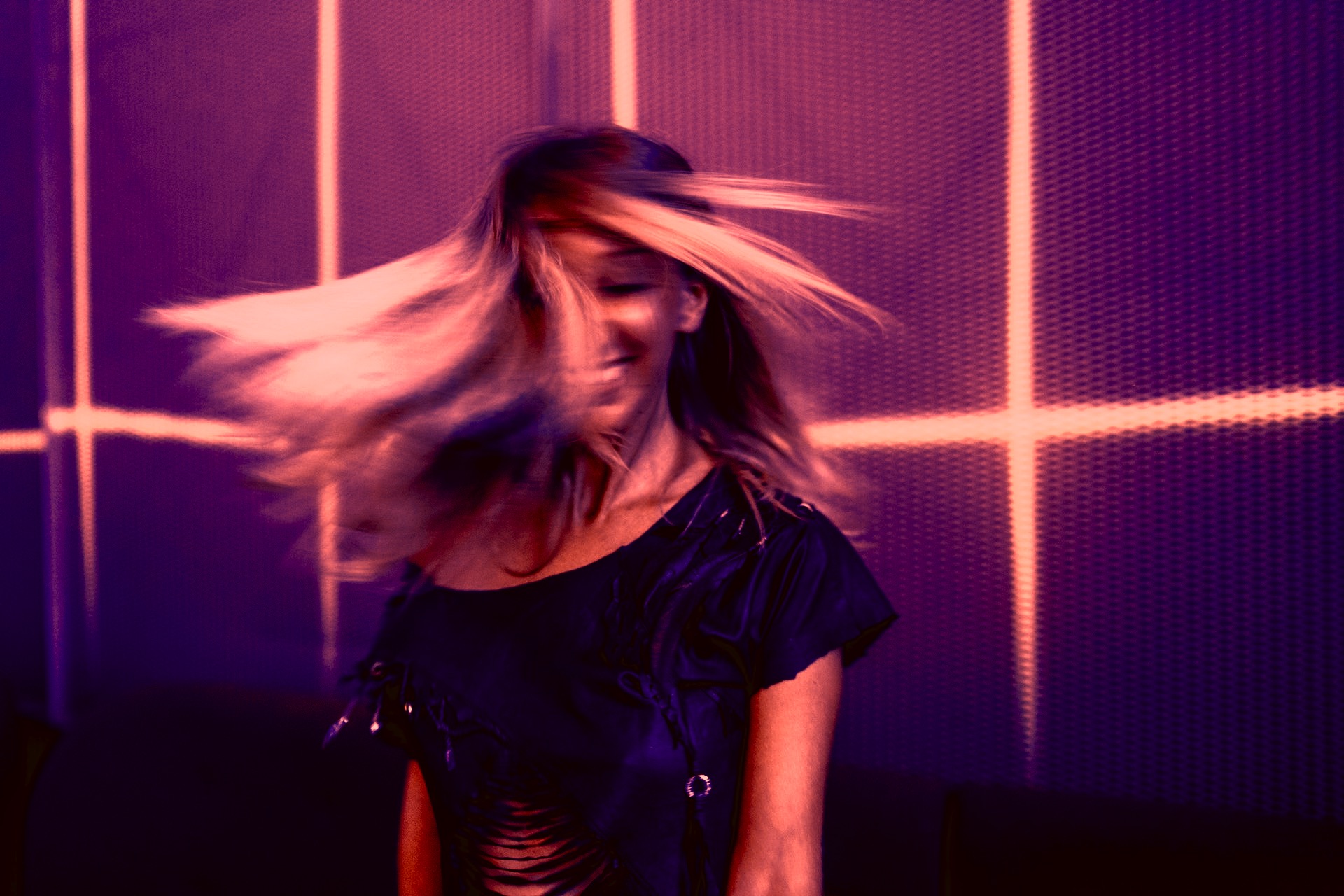}&
    \includegraphics[width=0.22\textwidth]{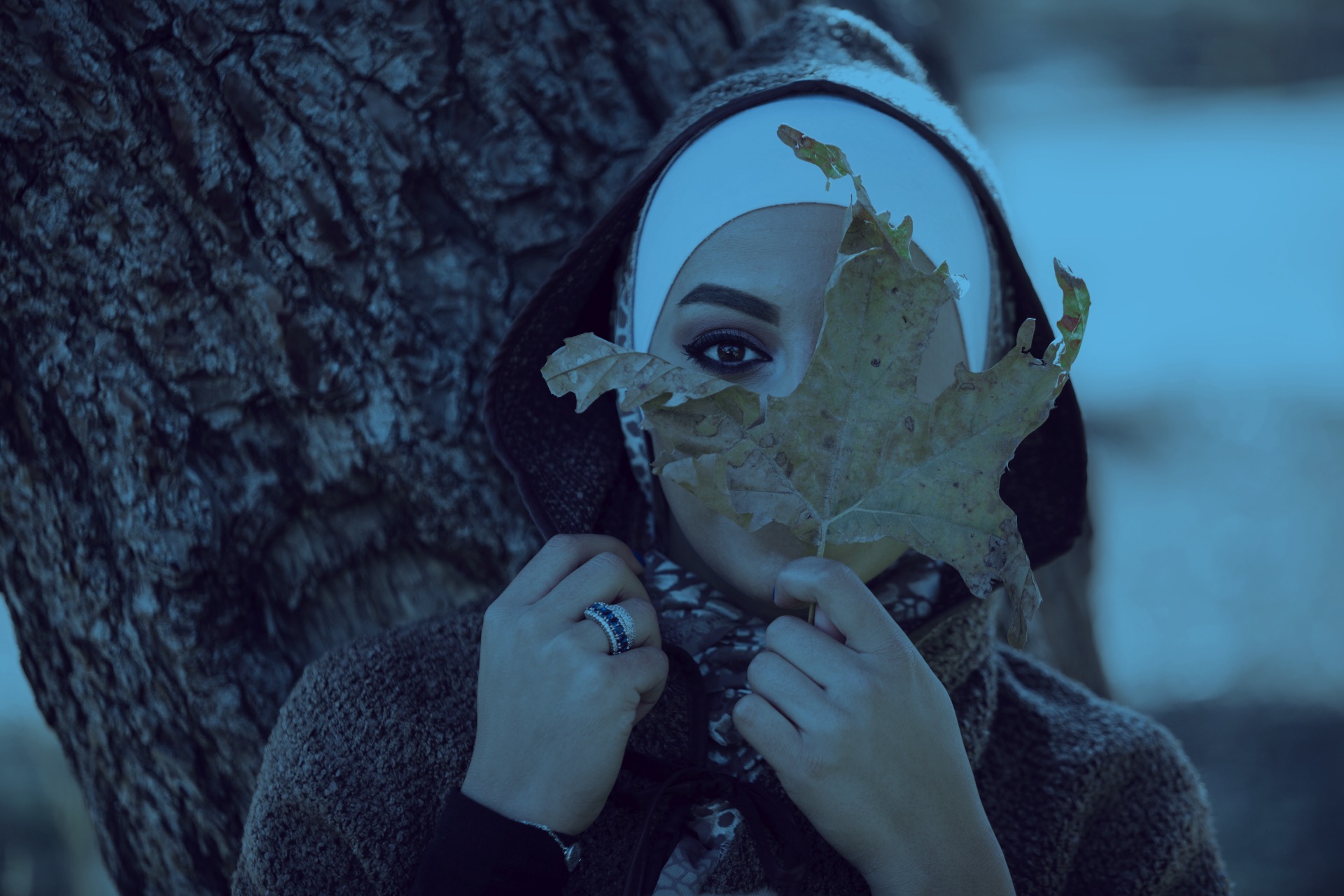}&  
     \includegraphics[width=0.22\textwidth]{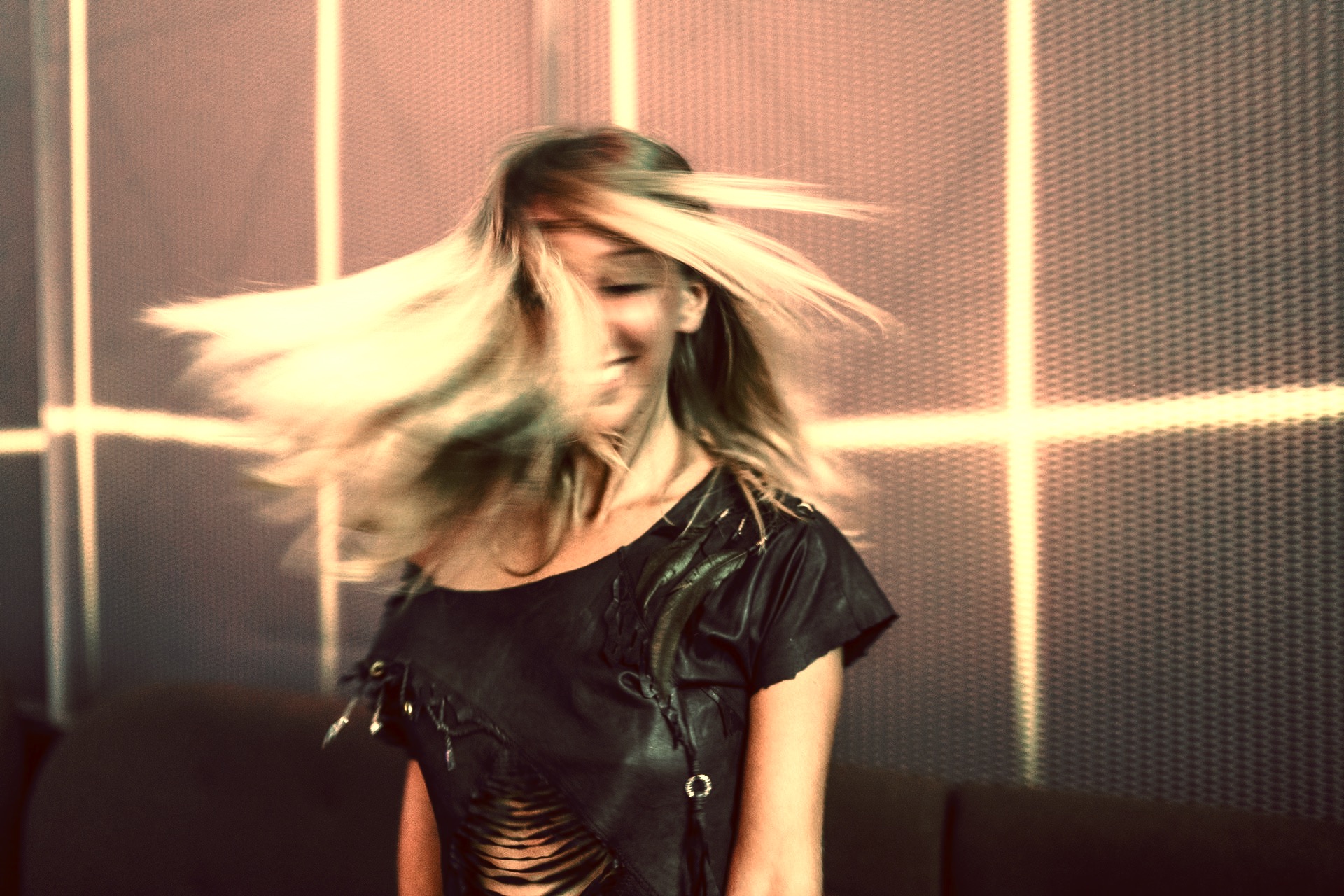}& 
     \includegraphics[width=0.22\textwidth]{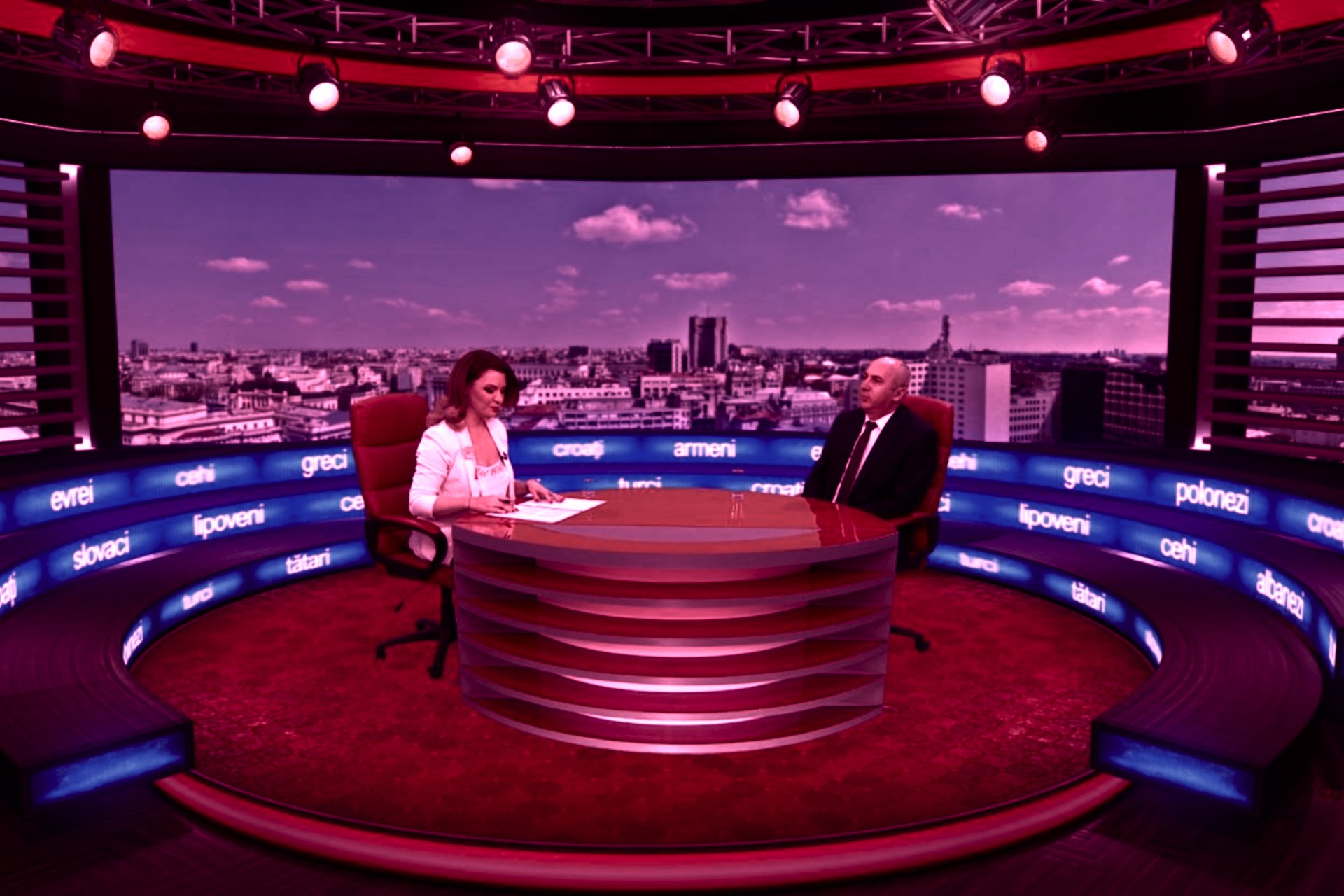} \\
    
    \includegraphics[width=0.22\textwidth]{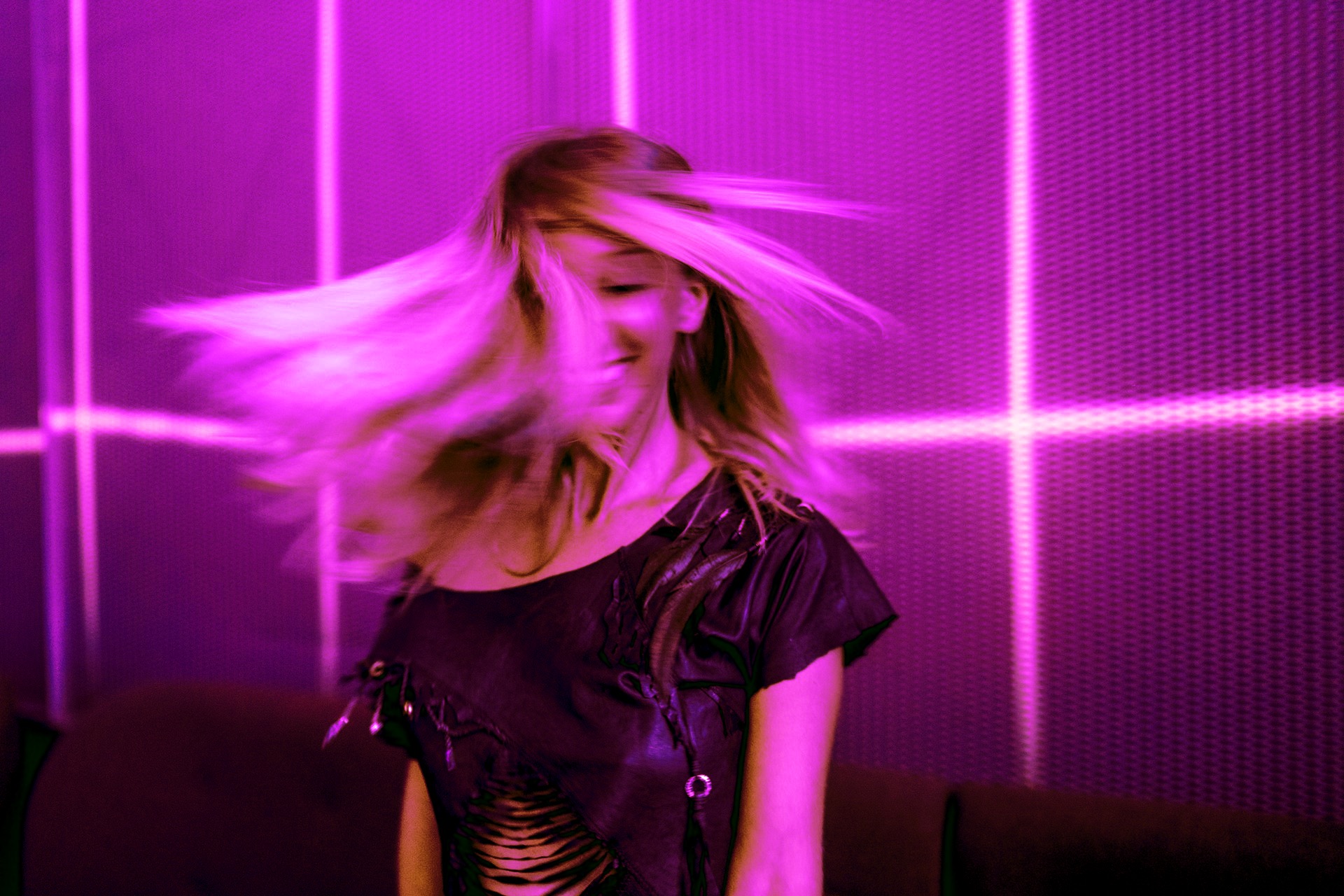}&
    \includegraphics[width=0.22\textwidth]{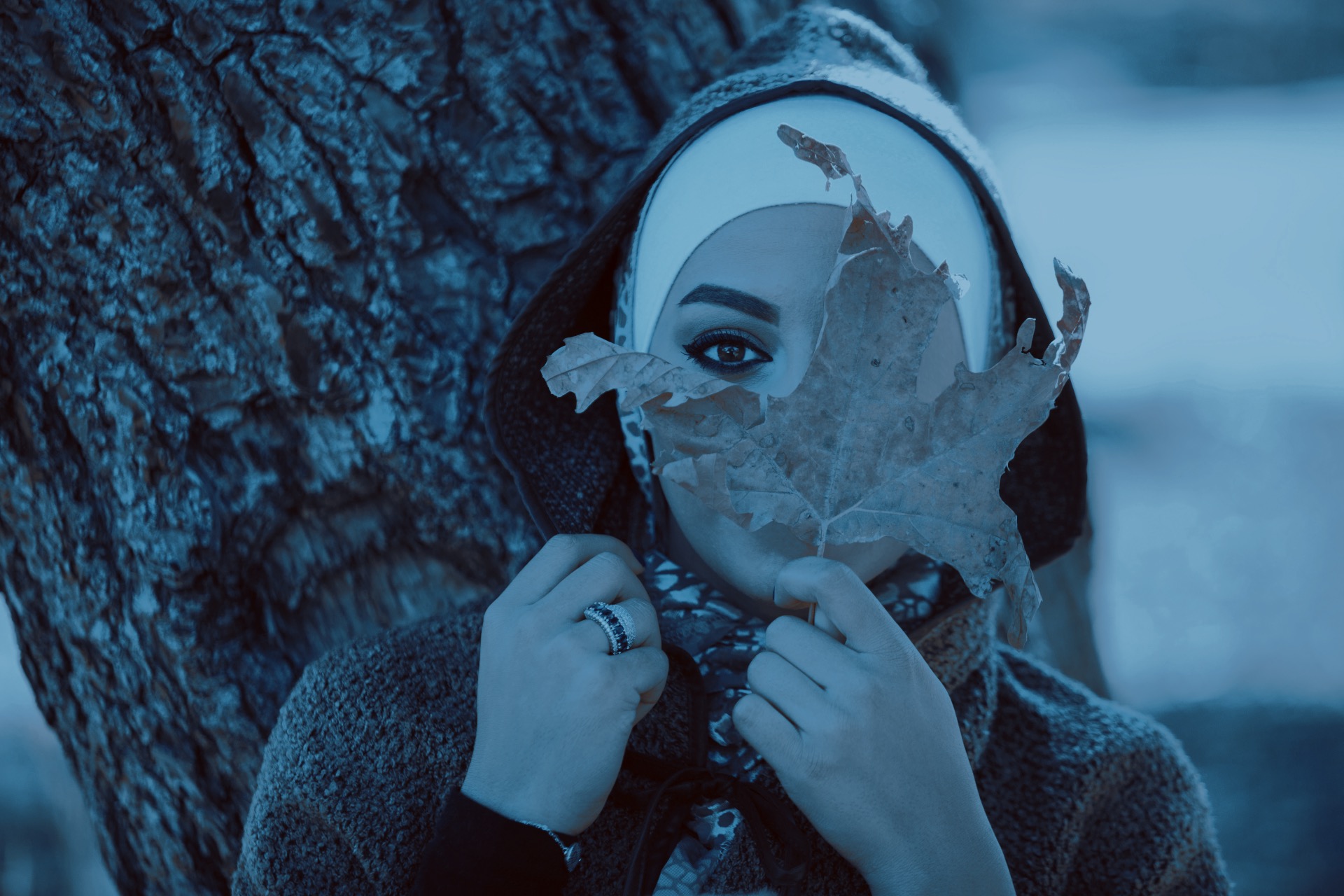}&  
     \includegraphics[width=0.22\textwidth]{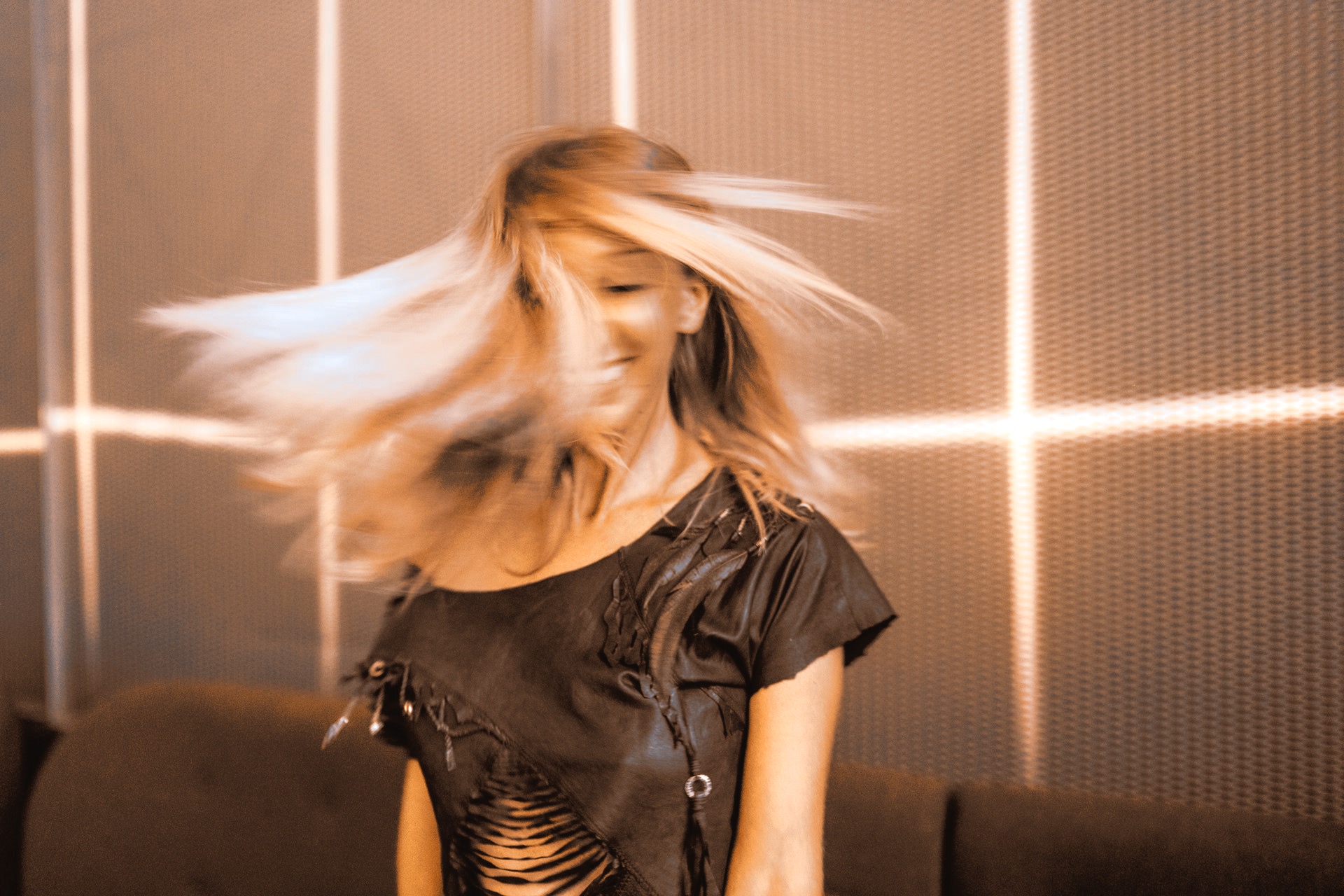}& 
     \includegraphics[width=0.22\textwidth]{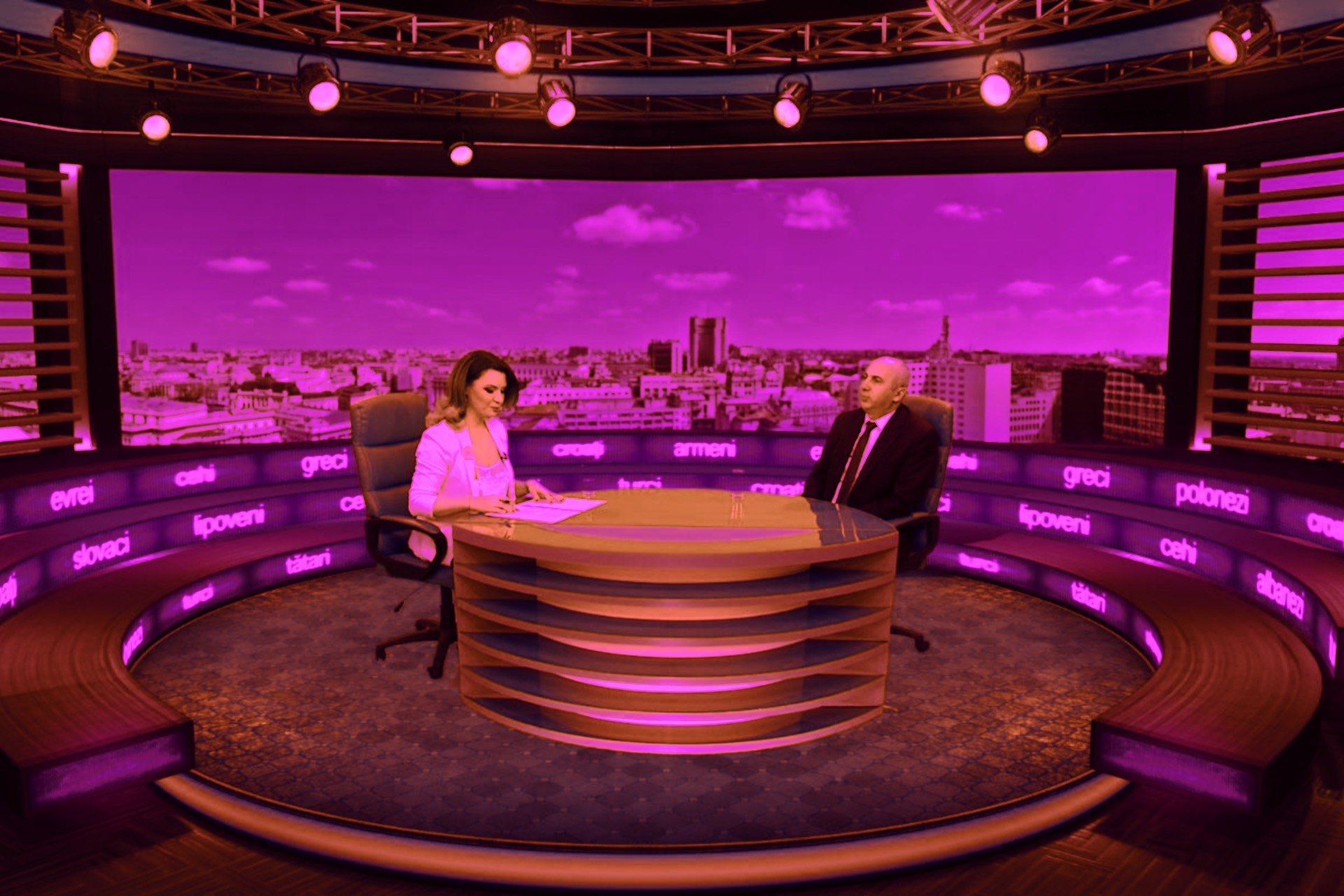} \\ 

    \includegraphics[width=0.22\textwidth]{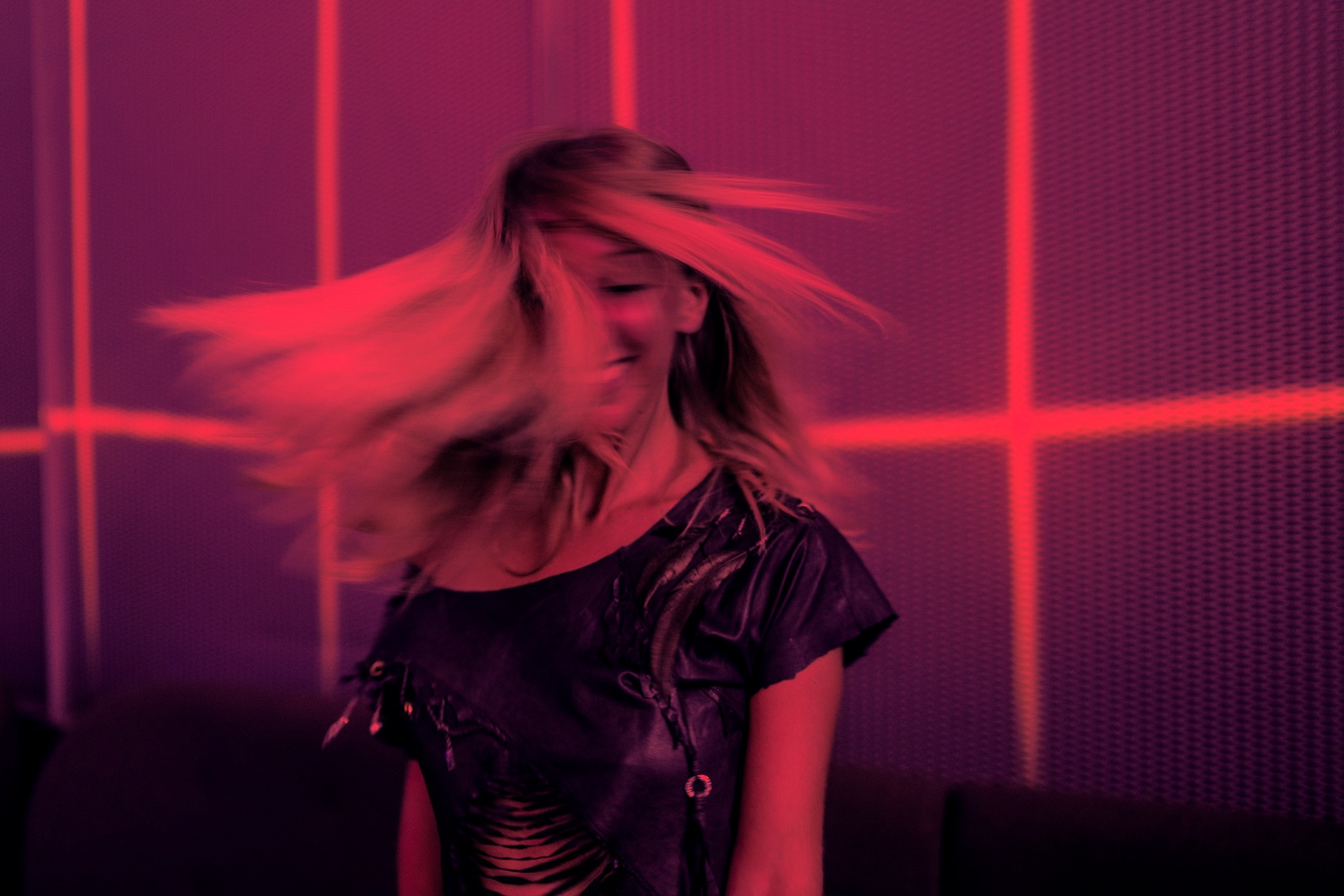}&
    \includegraphics[width=0.22\textwidth]{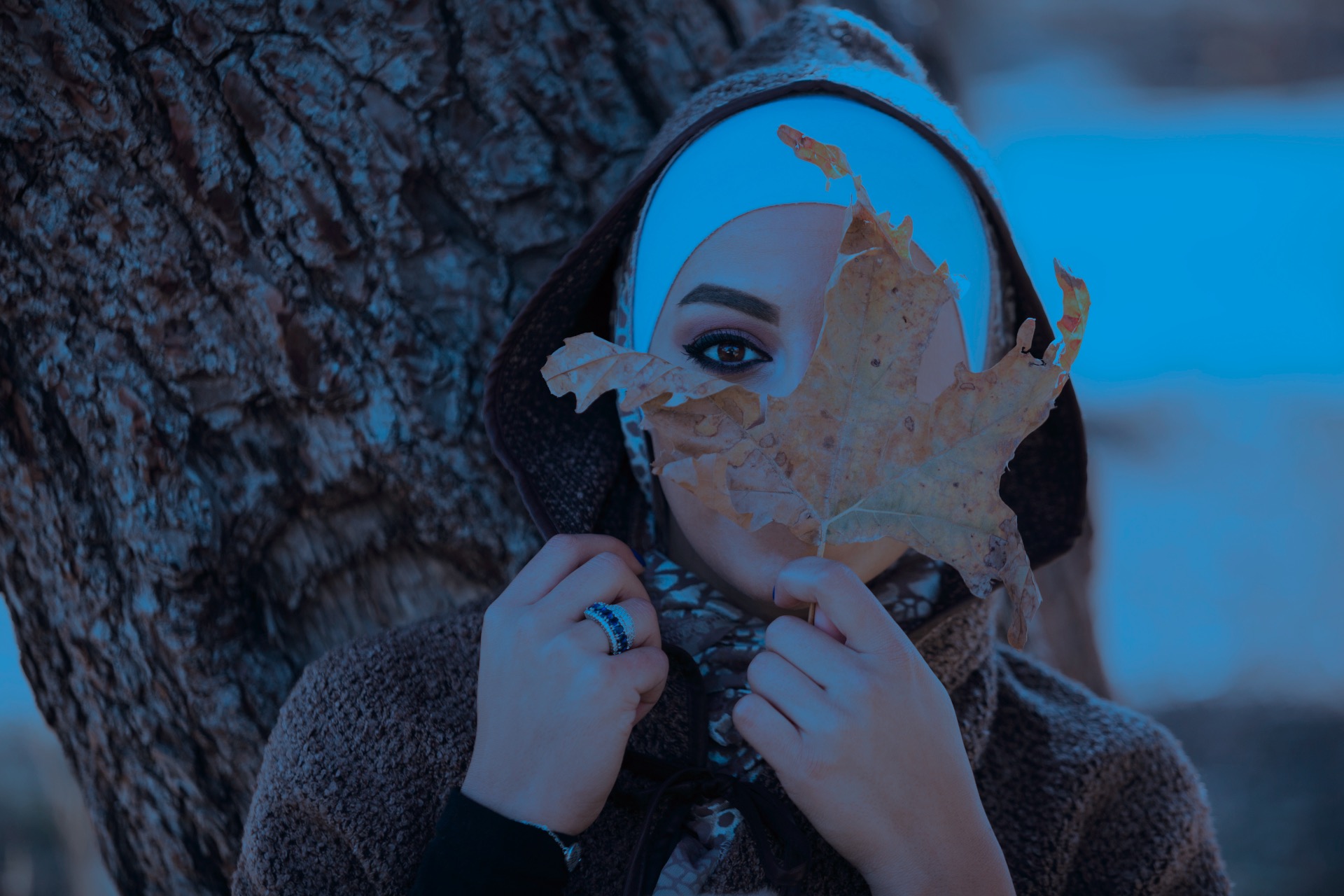}&  
     \includegraphics[width=0.22\textwidth]{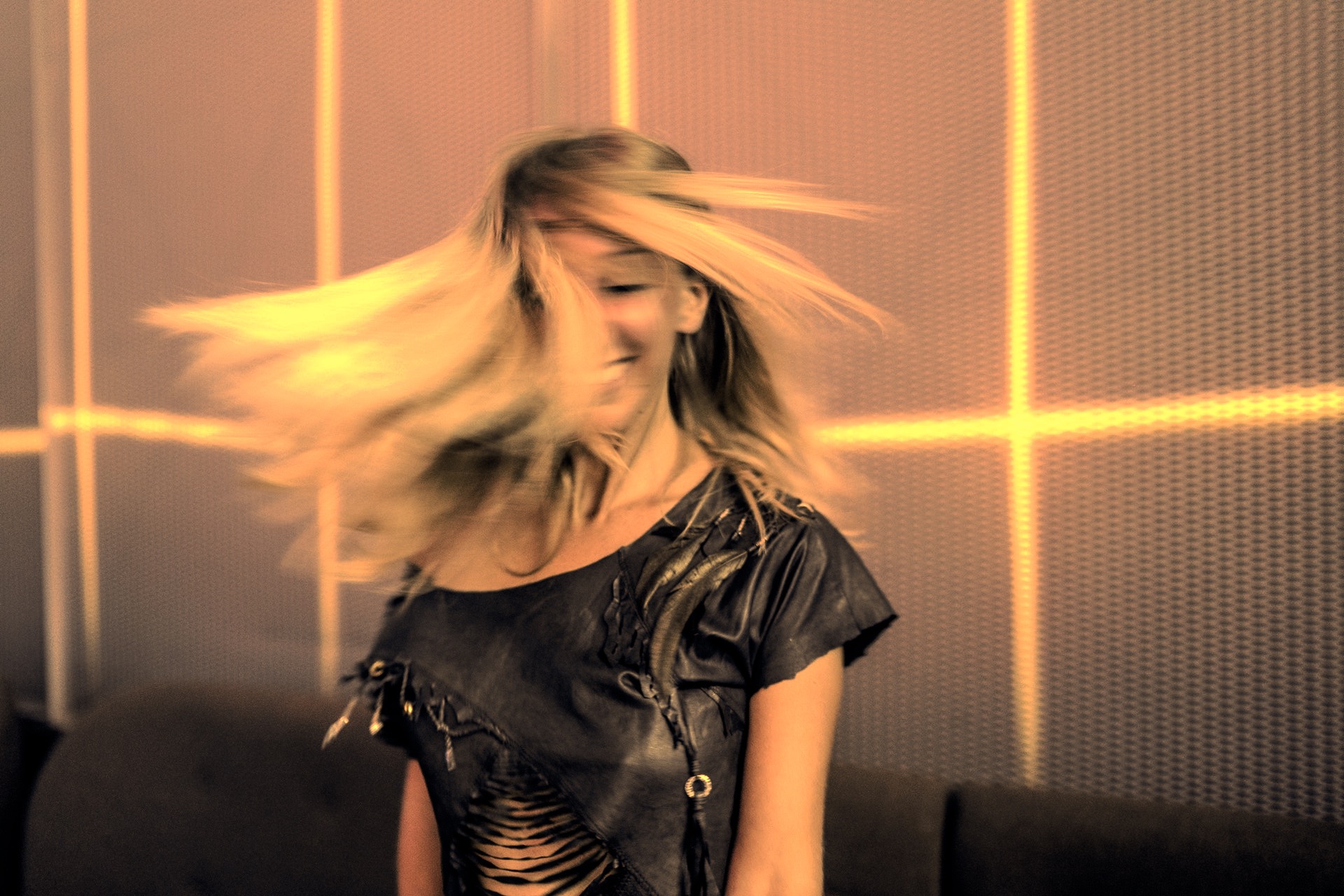}& 
     \includegraphics[width=0.22\textwidth]{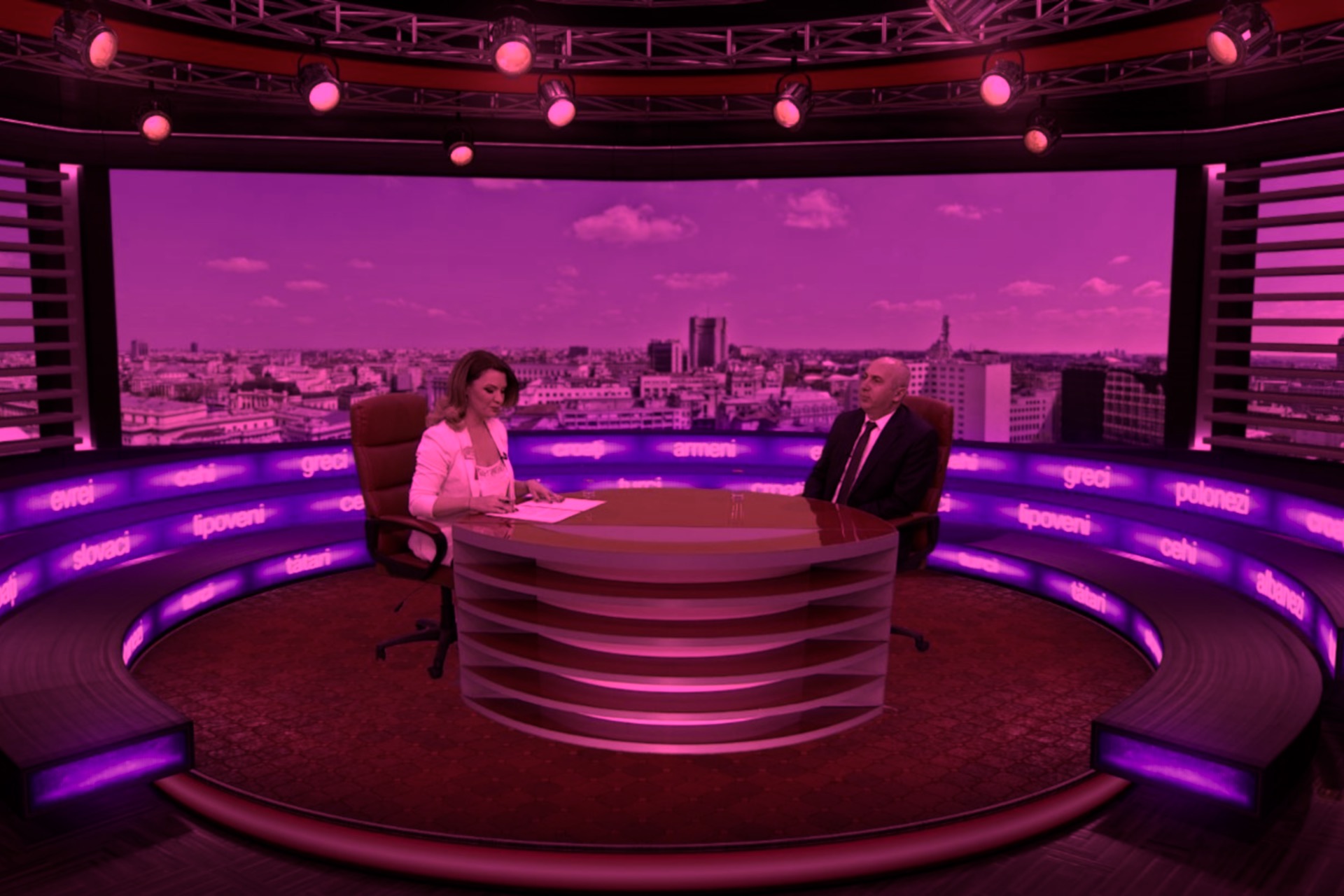} \\   
     
    \includegraphics[width=0.22\textwidth]{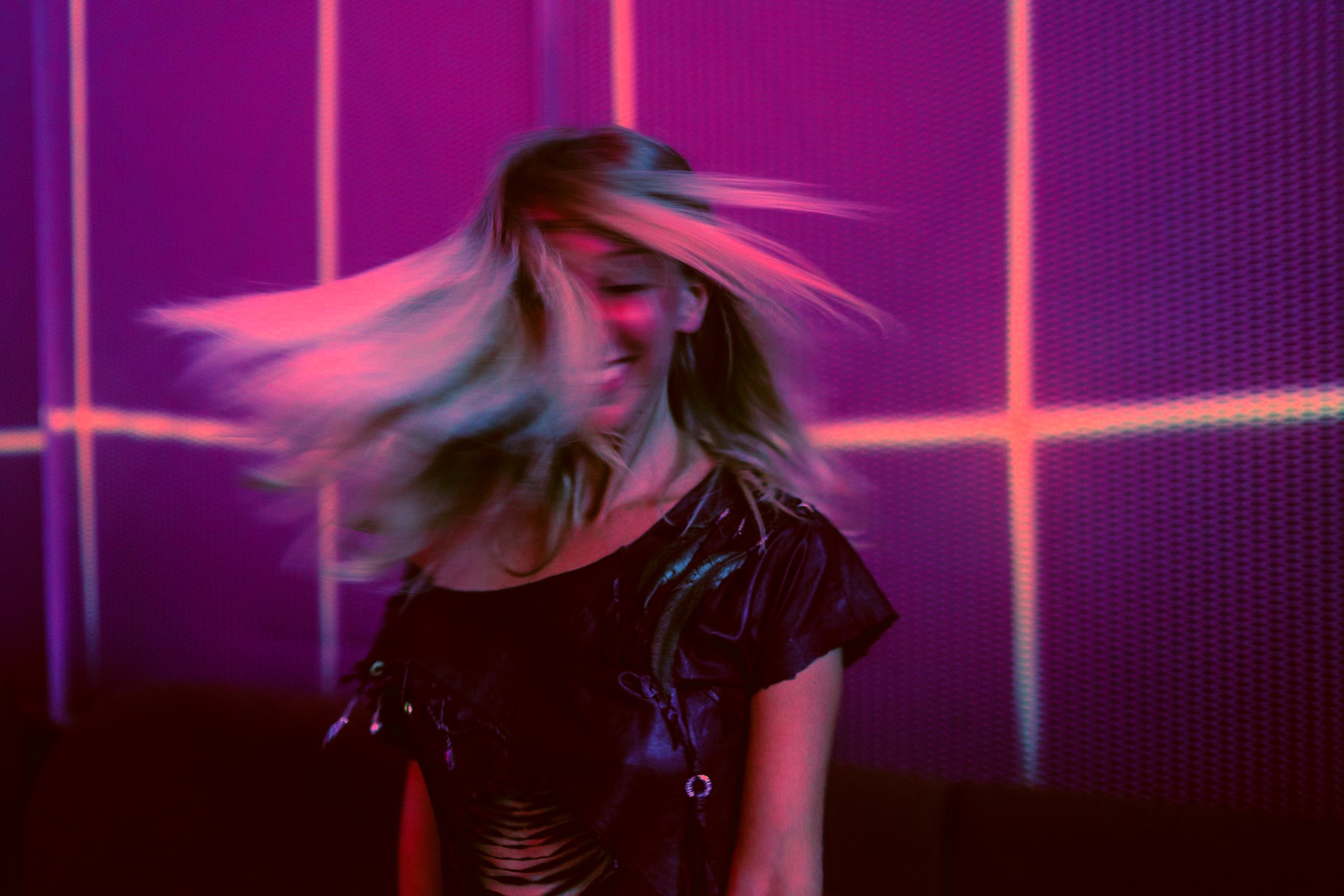}&
    \includegraphics[width=0.22\textwidth]{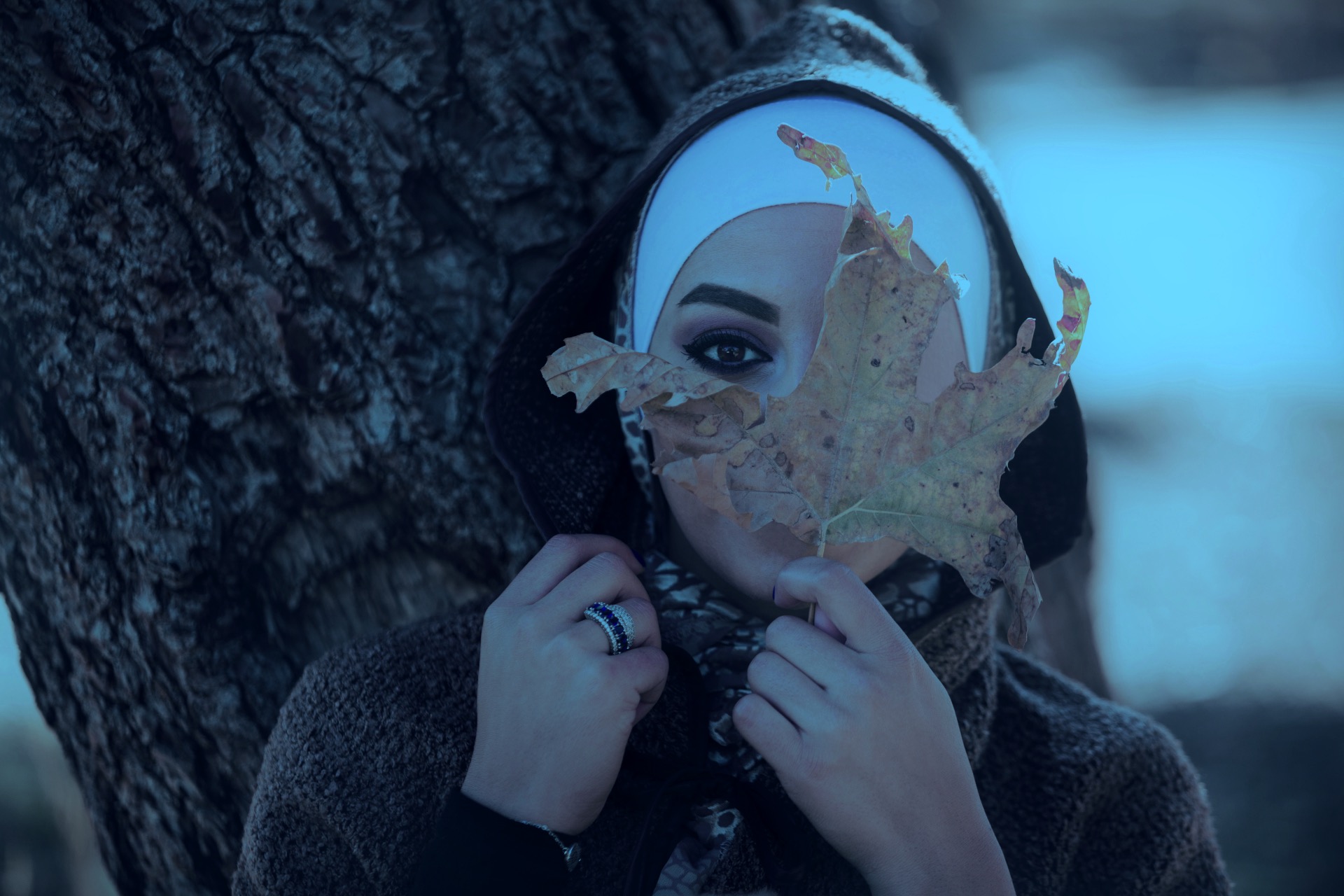}&  
     \includegraphics[width=0.22\textwidth]{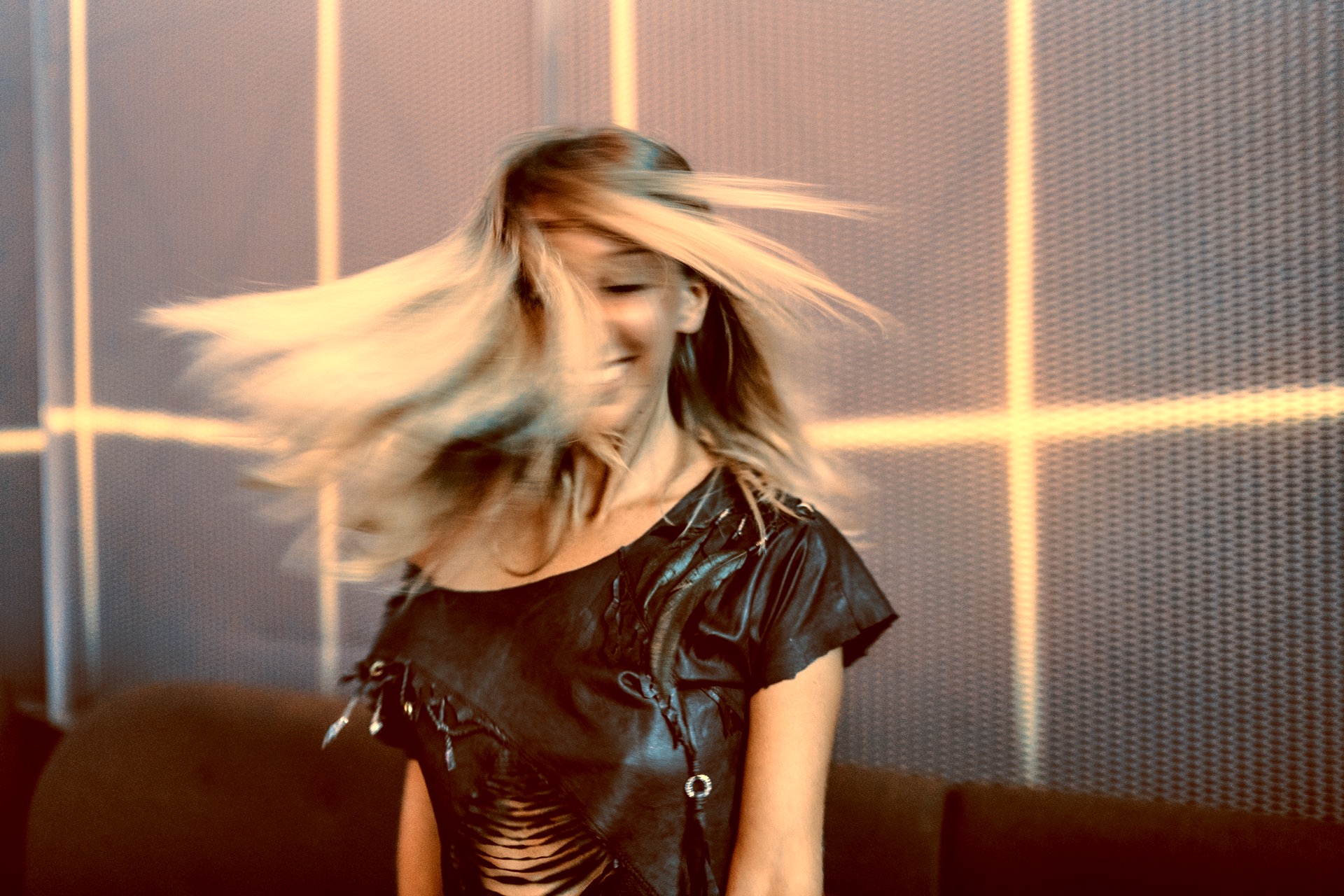}& 
     \includegraphics[width=0.22\textwidth]{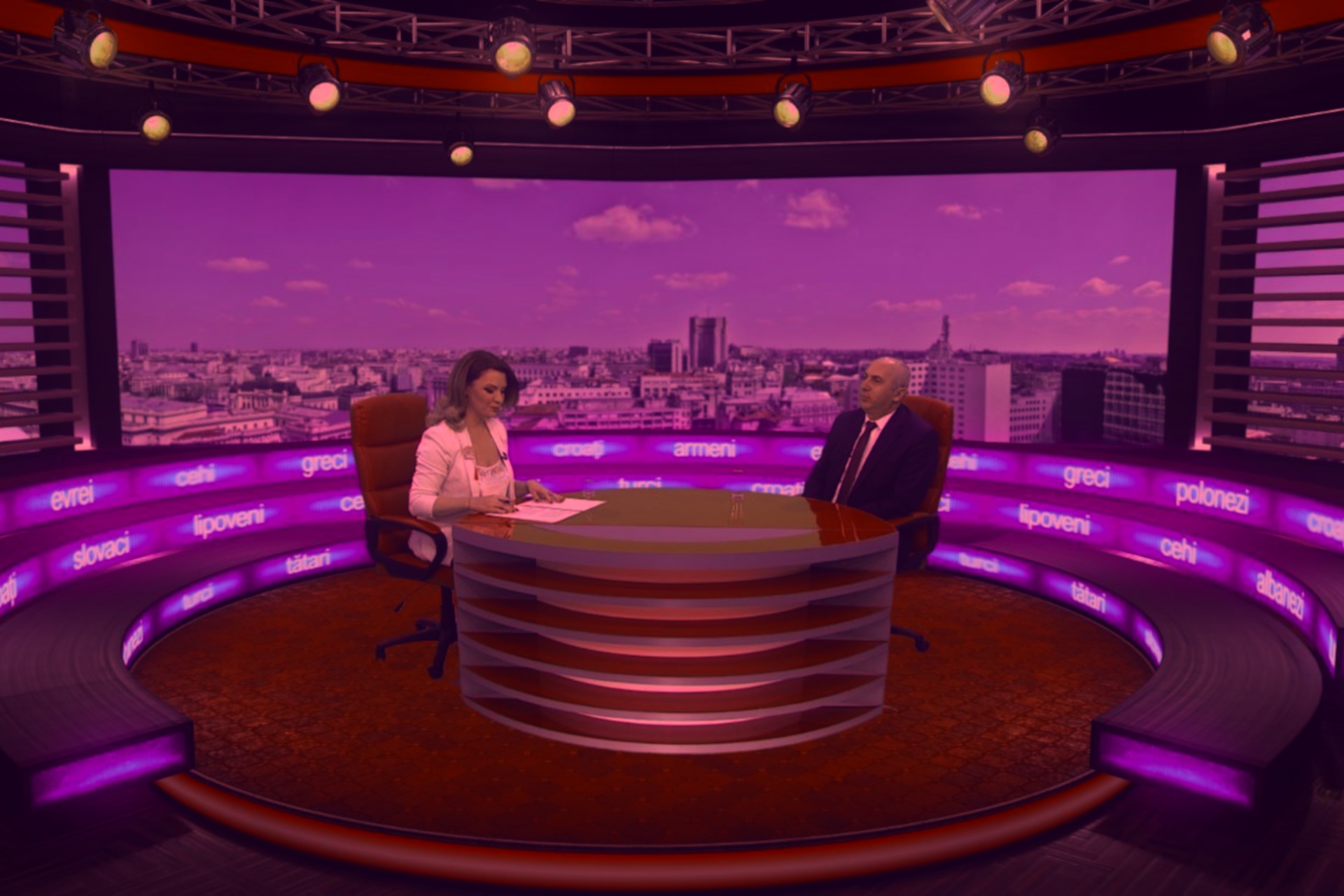} \\

  \end{tabular}
  \caption{From top to bottom, sources, targets, proposed result, Grogan result \cite{grogan17}, Photoshop result \cite{photoshop}, and Zabaleta \& Bertalmío result \cite{zabaleta21}.}
  \label{others}
\end{figure*}

Next, we compare the methods of \cite{grogan17}, \cite{photoshop}, \cite{zabaleta21} qualitatively in Figure \ref{others}, which we identified to be the most promising SotA methods for the photo-realistic application. One can see that while these methods still result in artifacts in some cases, they are for the most part correctable and limited to clipping or crushing, which can be recovered in pipelines where out of range values are allowed. 
\textcolor{black}{We specifically highlight the results of column 4, where our method is able to match the color of the clipped highlights in the overhead lighting and other white regions of the source and target image, while all other methods render them with the tint of the target illumination. This example provides an effective demonstration of the proposed method's ability to independently control distinct tone scale regions by separating clipped white areas from highlights where the illumination color is preserved.} 
In the section that follows we describe a psychophysical validation experiment in which observers compared their performance.


\subsection{Psychophysical experiment}
A total of 19 observers made subjective comparisons between the four most application-relevant methods (whose results are shown in Figure \ref{others}) in an online pair comparison experiment. Each method was used to make a style match between all combinations of the images in Figure \ref{sources}. Observers were shown the original source and target images and made a two-alternate-forced-choice (2AFC) judgment between the results of two competing methods based on their ability to translate the style of the target to the reference while maintaining a natural appearance, considering the original state of the source image. This was repeated for all image and method combinations, resulting in a total of 120 trials. A more detailed description of the experimental methods is given in Appendix \ref{subsec:psych_methods}.

The results of the experiment showed that, in comparison to the proposed method, the results of Photoshop \cite{photoshop} were preferred 32\% of the time, followed by those of Zabaleta and Bertalmío \cite{zabaleta21} which were preferred 41\% of the time, and the results of Grogan et al. \cite{grogan17} were preferred 58\% of the time. A full chart of the results is shown in Table \ref{tab:results} of the Appendix. Looking back to Figure \ref{others}, one can infer an explanation for the success of Grogan in the context of this experiment as the dominant color of the target image is closely reflected in all of their results. However, this is accomplished at the expense of maintaining the color discrimination boundaries between separate regions (e.g. foreground and background), resulting in the crushing of important memory colors (e.g. skin tones) to nearly monochromatic representations. In the context of professional color grading and image editing, users would consider this effect unacceptable because it prohibits further refinement and for this reason would prefer the non-destructive performance of the proposed method. In addition, the ranking reversal of \cite{grogan17} and \cite{zabaleta21} in this experiment versus the psychophysical validation of the latter brings to light the highly subjective nature of style transfer evaluation, where factors like the chosen image set and personal preferences of observers can heavily influence the results. For example, since many of the source images were already nearly washed out in highly chromatic light to begin with, these represented easy cases for the method of \cite{grogan17} to succeed and its shortcomings to be hidden.

\subsection{Applications}
\subsubsection{Broadcast Source Homogenization}

During the course of this work we adapted our method for a novel broadcast application where the image of remote participants captured on mobile phones and webcams can be composited onto the studio stream. In this application it is important to mimic the color properties of the studio lighting and capture equipment in order to best facilitate the illusion that the participant is present in the studio. Our proposed method was employed to make a color match between the user-submitted video and the broadcast stream, and then foreground extraction and super-resolution were applied to the user image to composite them onto an LED wall in the studio. This is a challenging color transfer problem, especially considering that studio broadcasters often have a virtual background or an LED wall that does not significantly impact the way the anchors are illuminated (as it is overpowered by the studio lights and is situated behind the anchors), but has a large influence on the statistics of the target image. So in order to account for this, we use a foreground extraction module to matte out the background of the user image and replace it with the studio background, such that the statistics of the background are more or less the same for the user and the anchor, and the color transfer can focus on the lighting characteristics which are relevant to their appearance. 

\begin{figure*}[t]
\centering
\begin{tabular}{cccc}

    \includegraphics[width=0.21\textwidth]{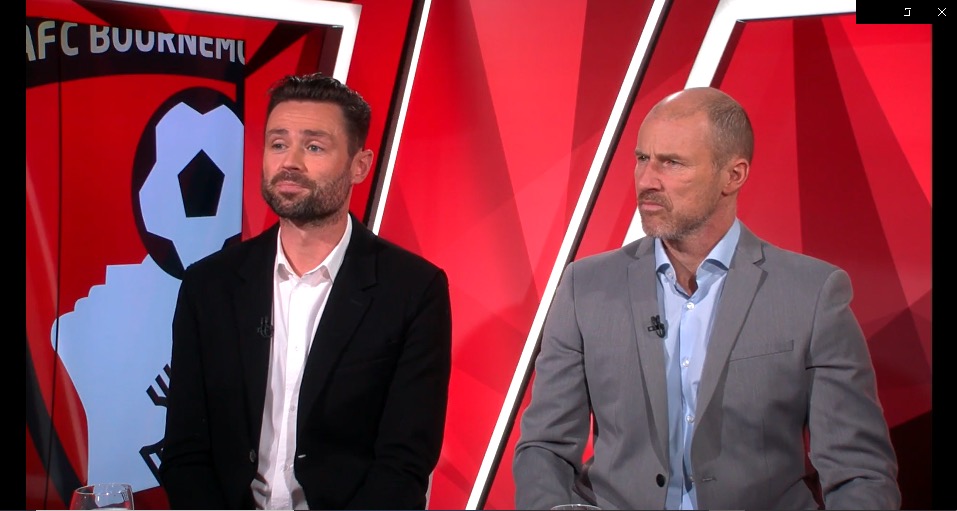}&
     \includegraphics[width=0.10\textwidth]{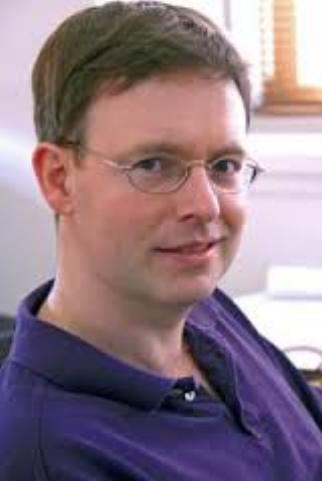}
    \includegraphics[width=0.10\textwidth]{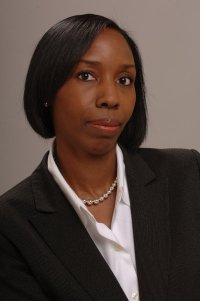}&    
     \includegraphics[width=0.21\textwidth]{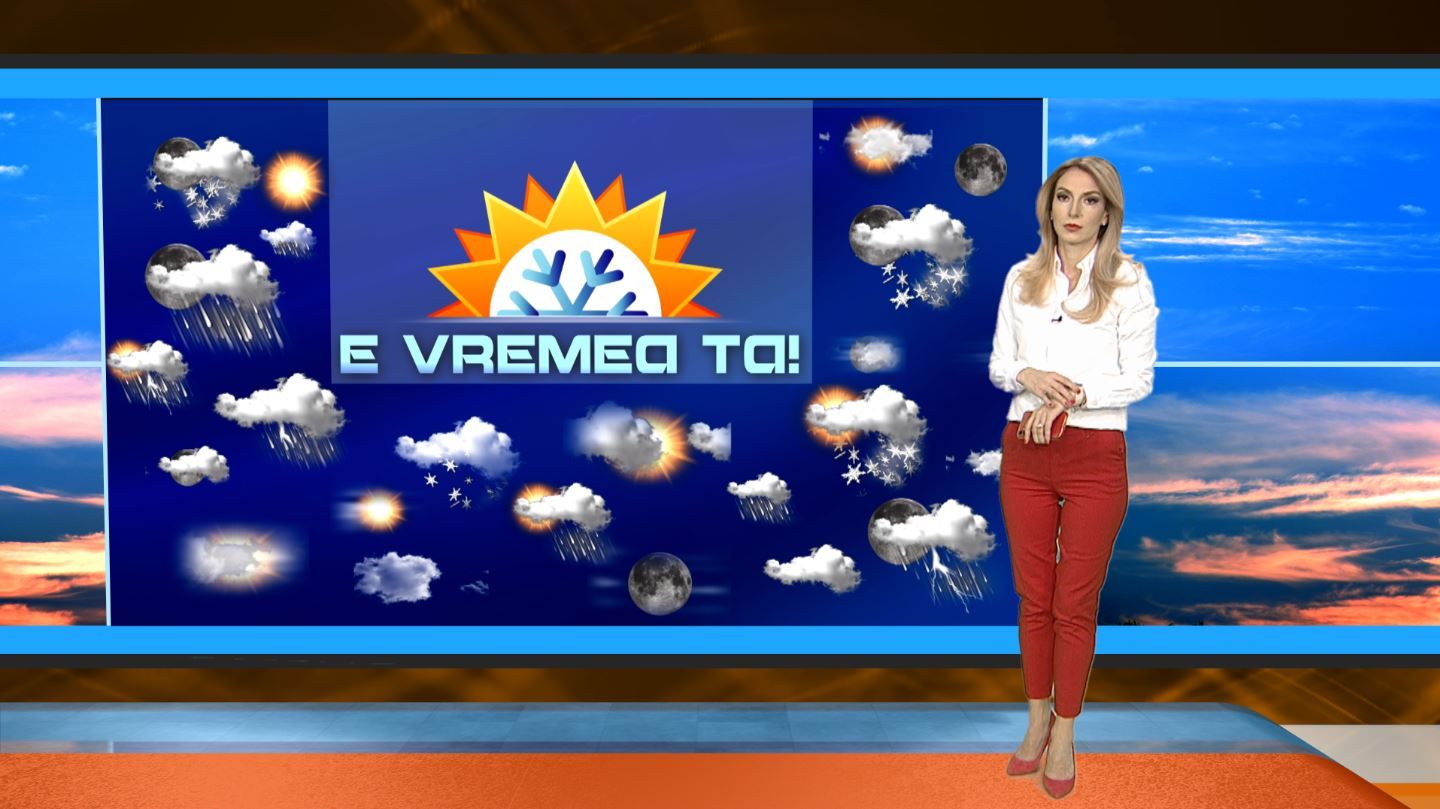} \\  
     
    \includegraphics[width=0.21\textwidth]{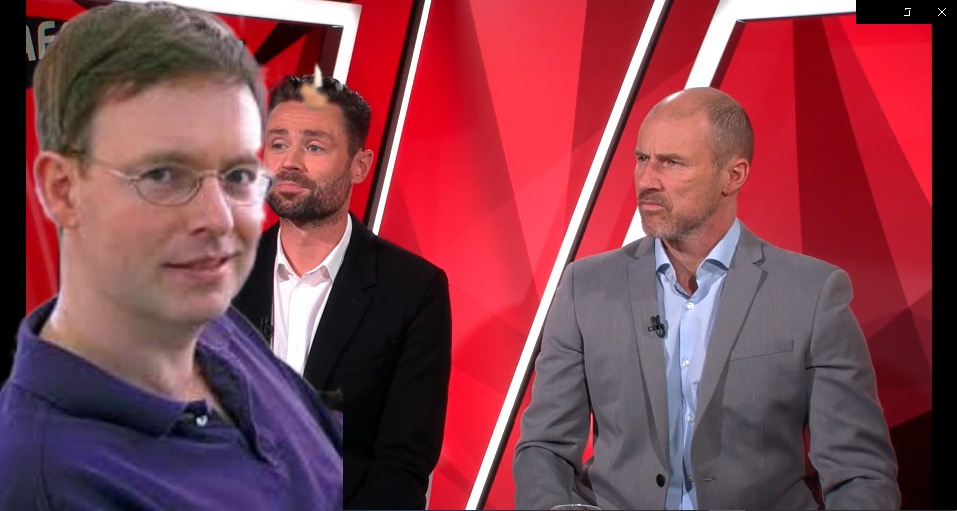}&
    \includegraphics[width=0.21\textwidth]{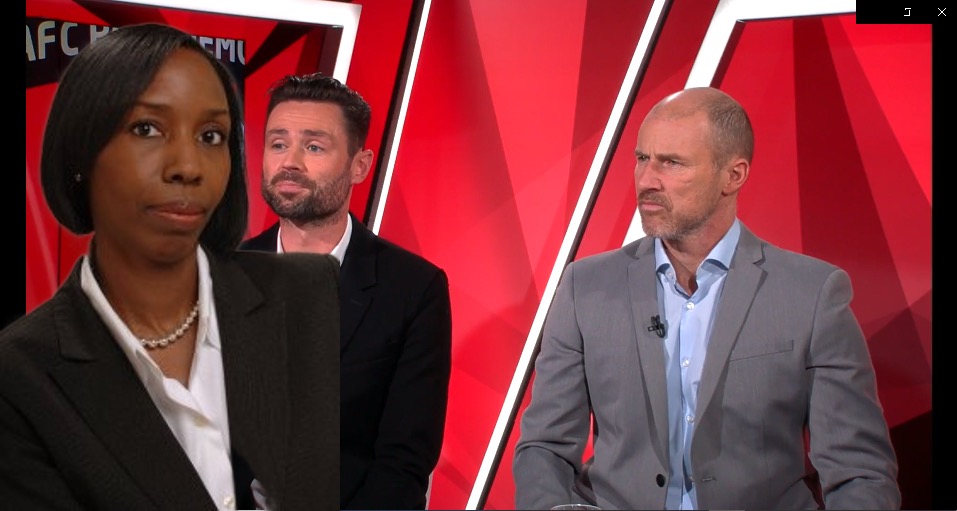}&  
     \includegraphics[width=0.21\textwidth]{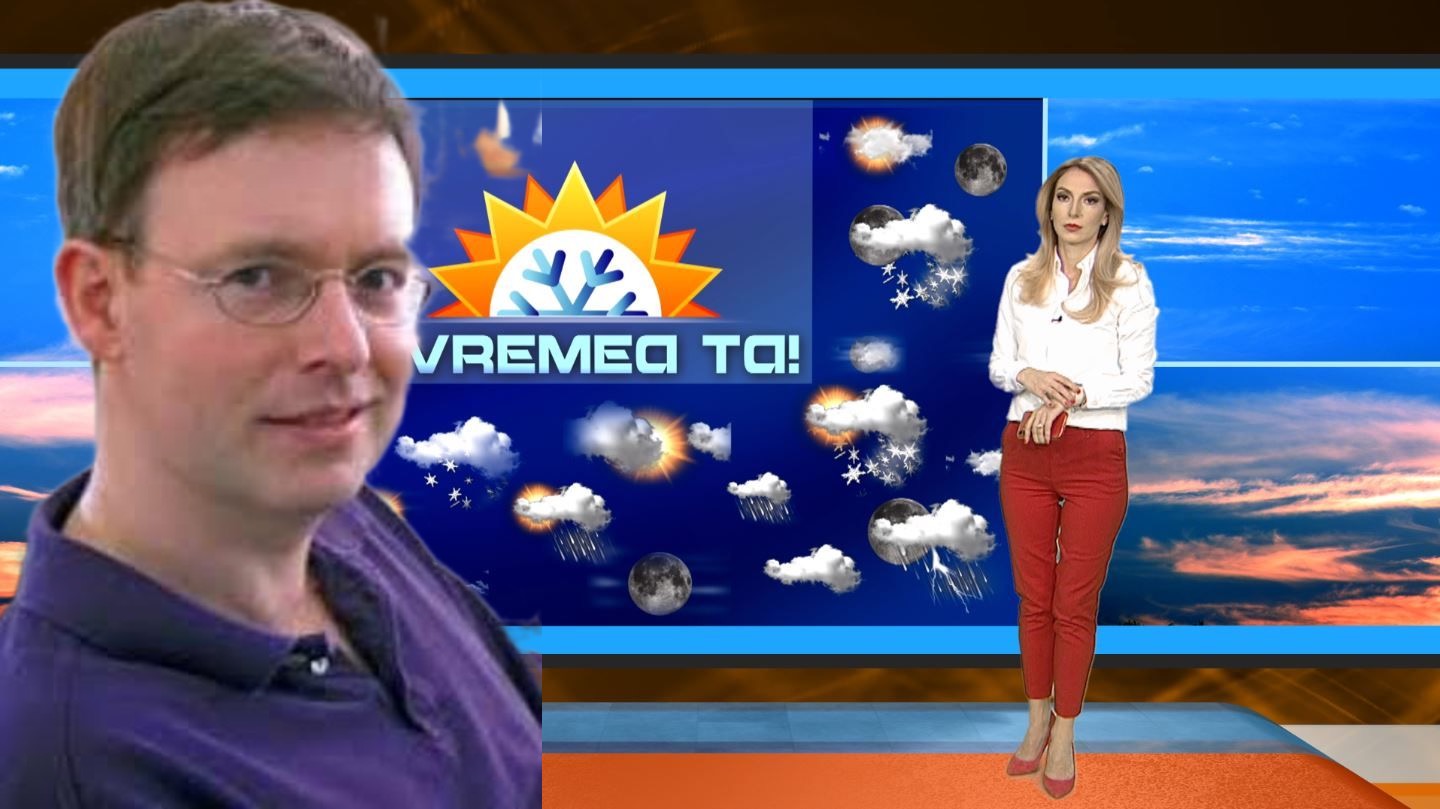}& 
     \includegraphics[width=0.21\textwidth]{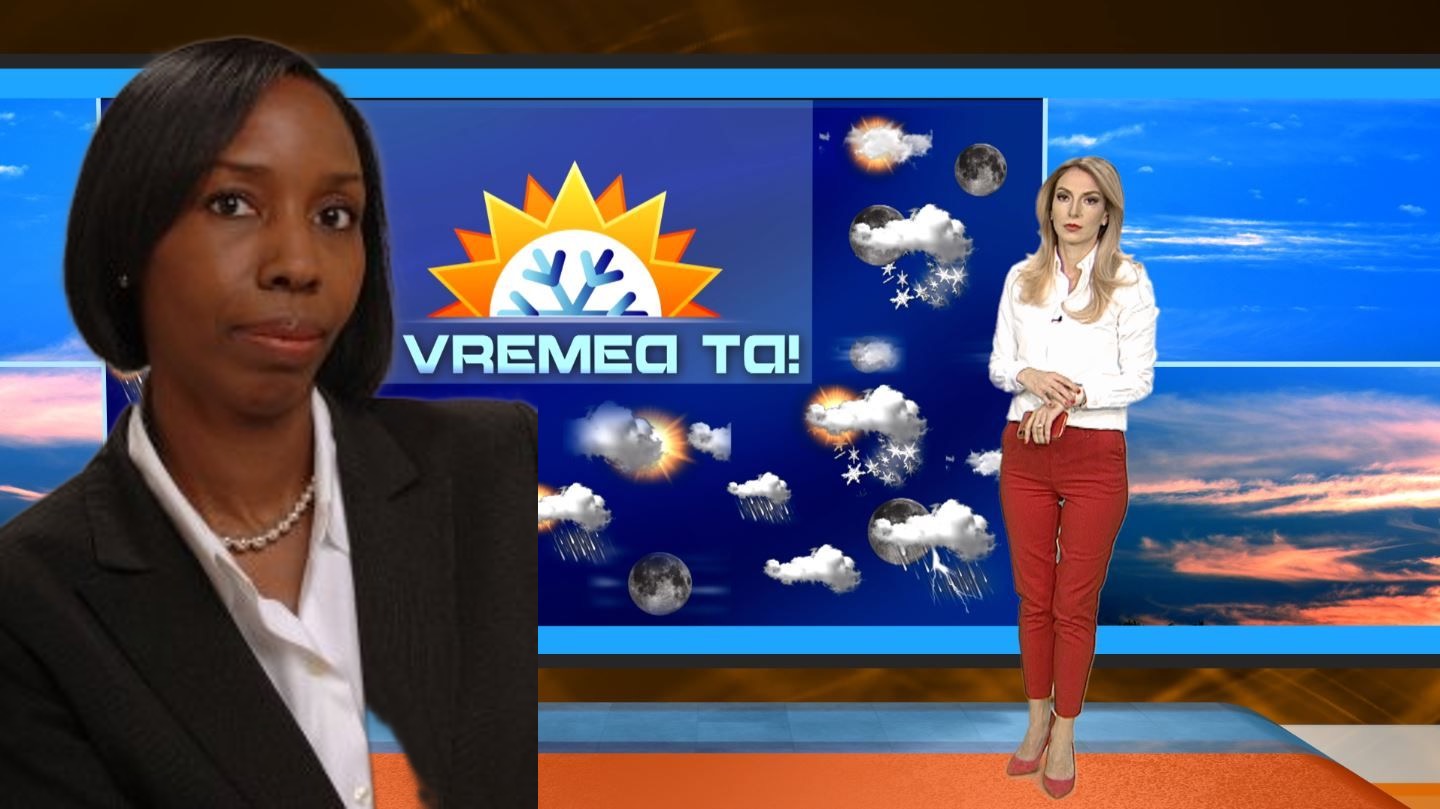} \\     
     
    \includegraphics[width=0.21\textwidth]{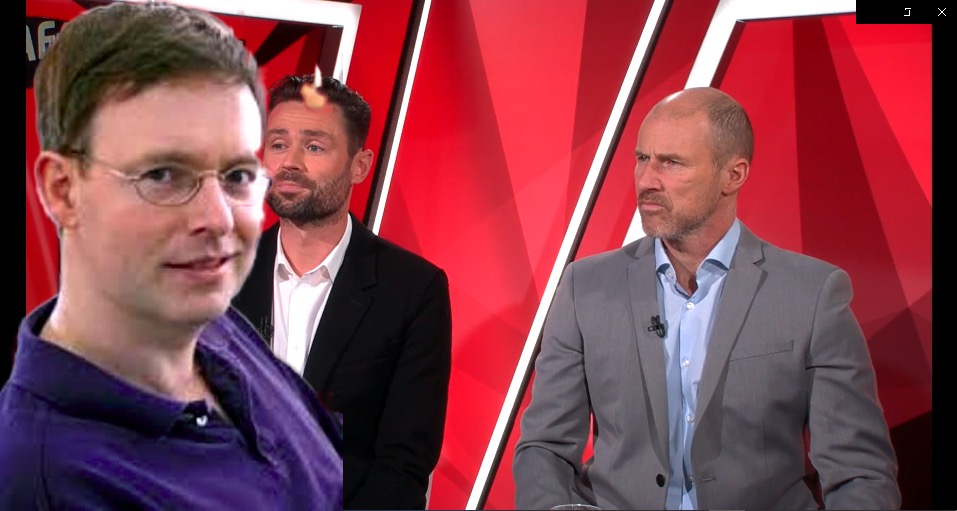}&
    \includegraphics[width=0.21\textwidth]{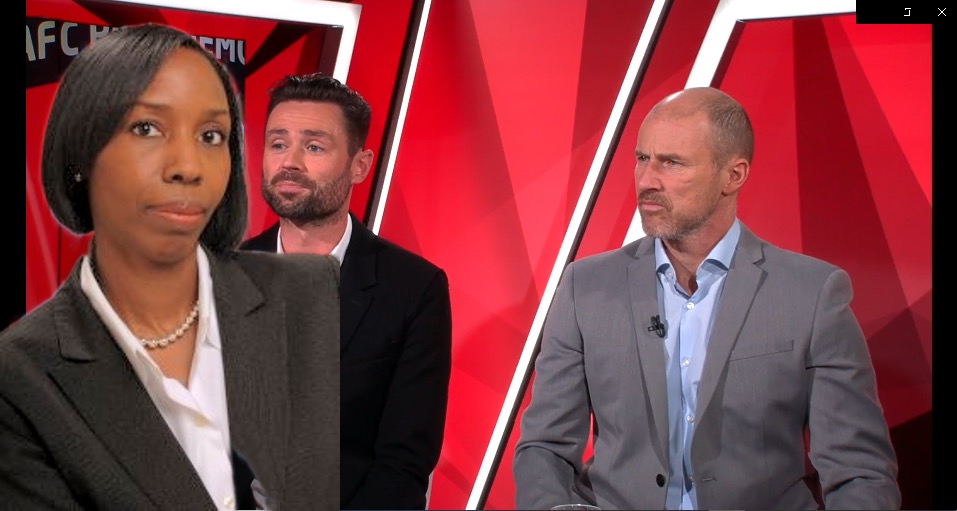}&  
     \includegraphics[width=0.21\textwidth]{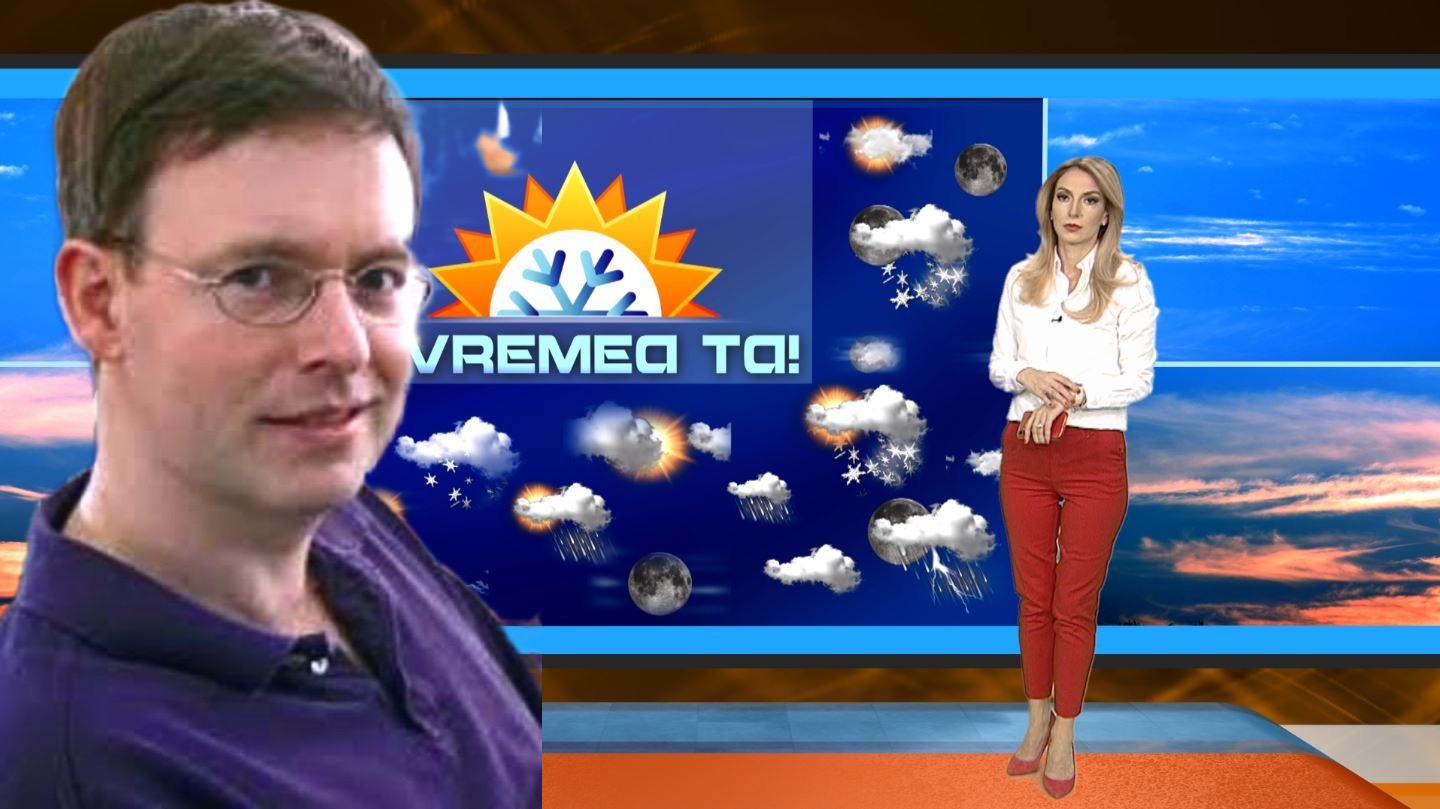}& 
     \includegraphics[width=0.21\textwidth]{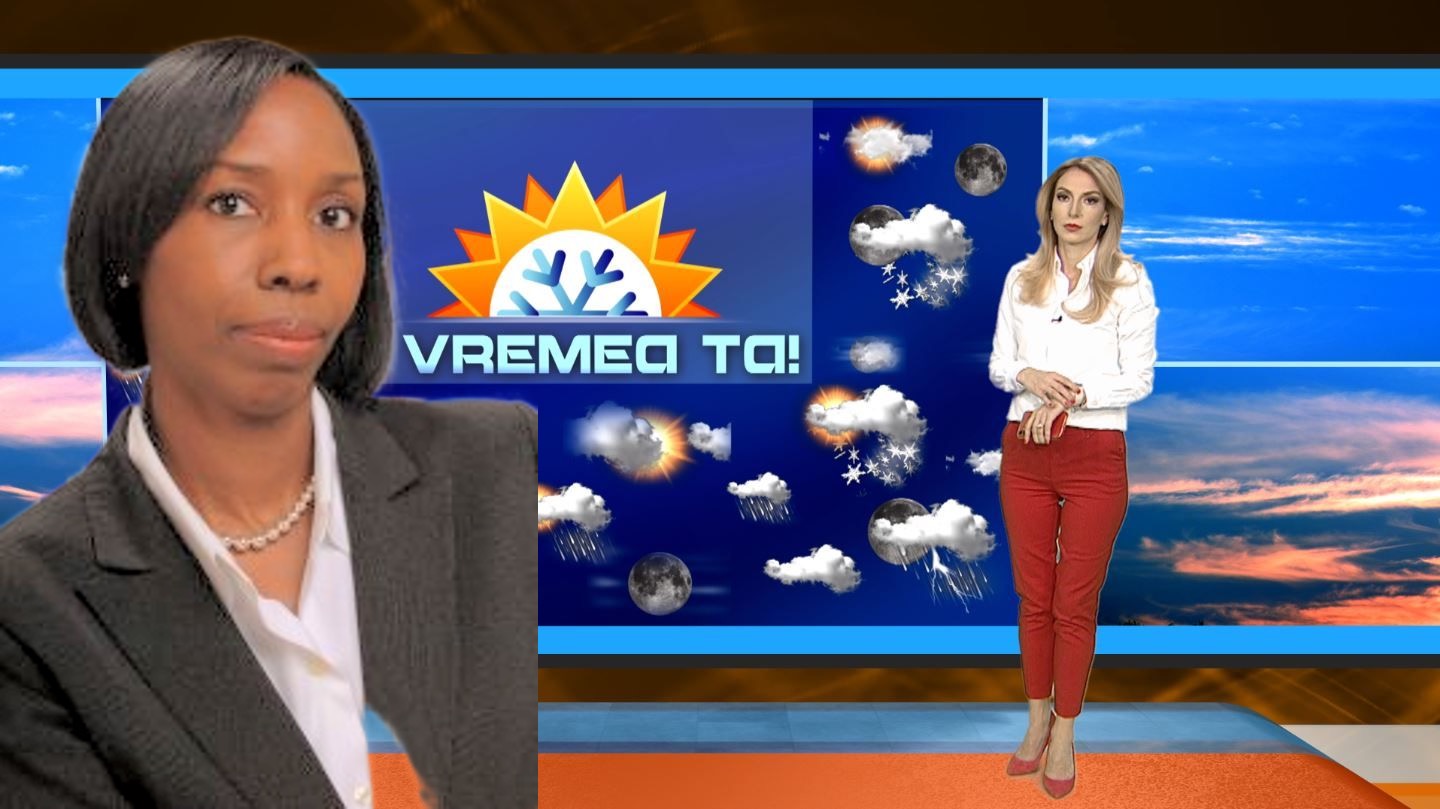} \\
    
    \includegraphics[width=0.21\textwidth]{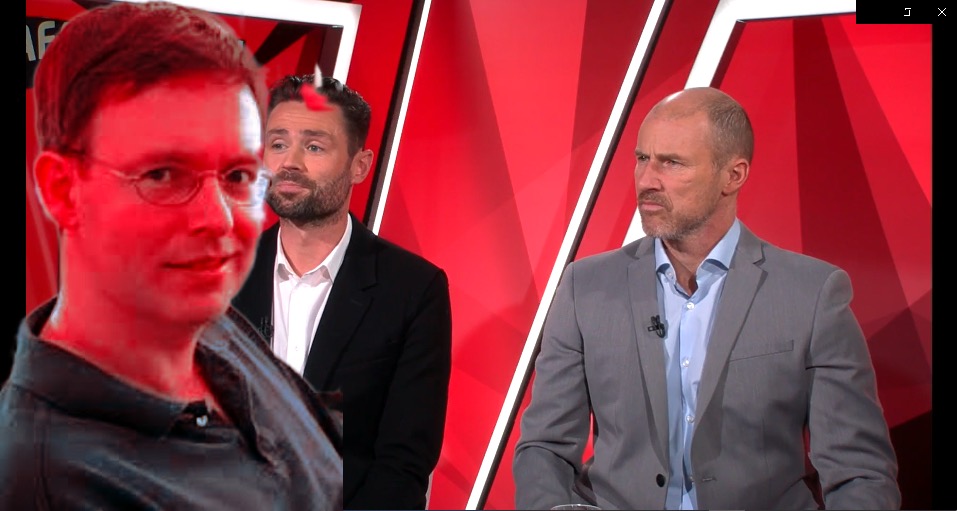}&
    \includegraphics[width=0.21\textwidth]{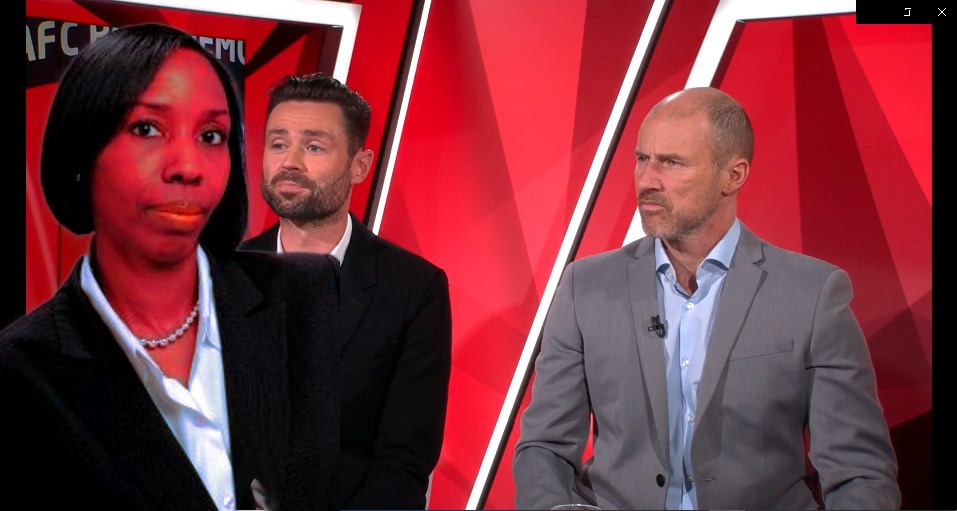}&  
     \includegraphics[width=0.21\textwidth]{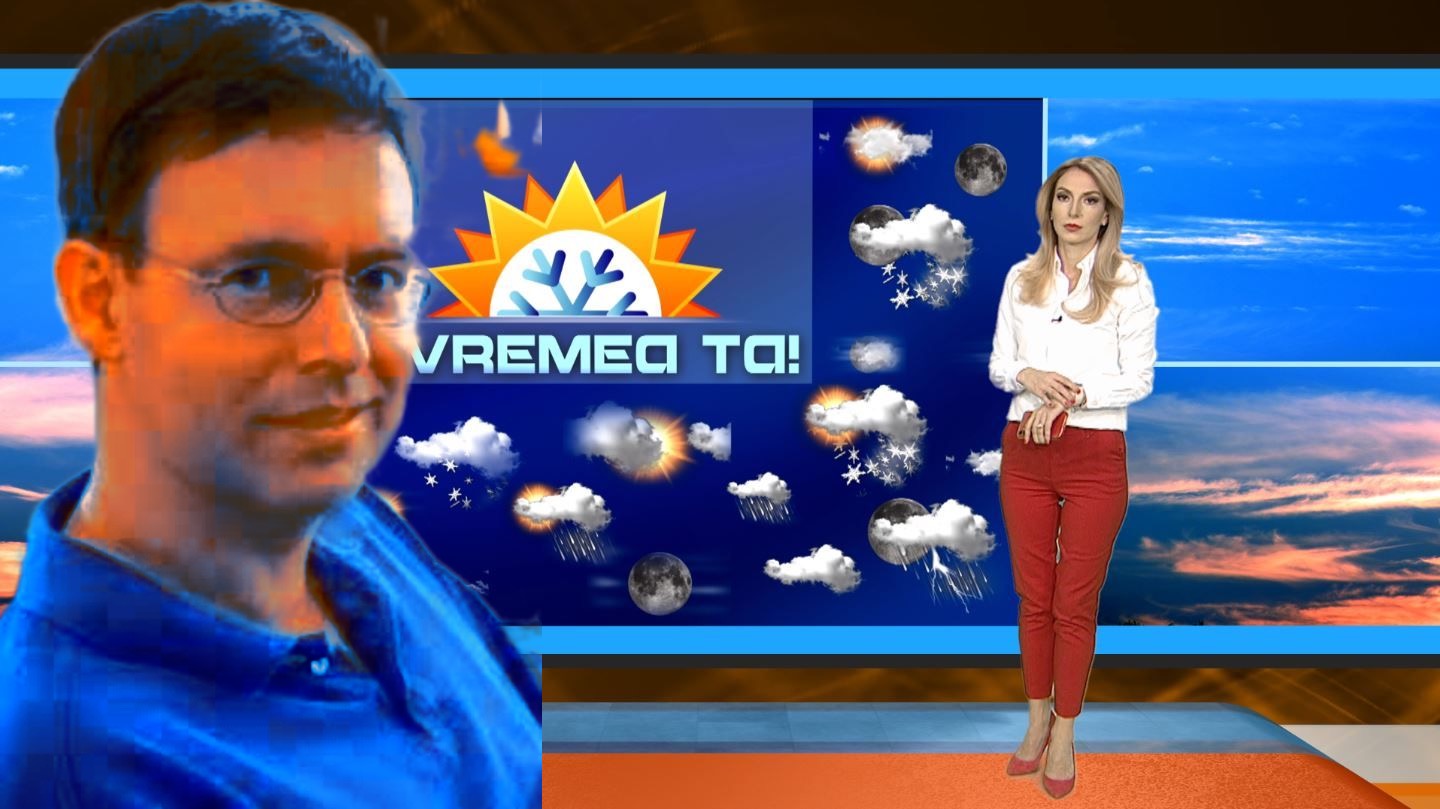}& 
     \includegraphics[width=0.21\textwidth]{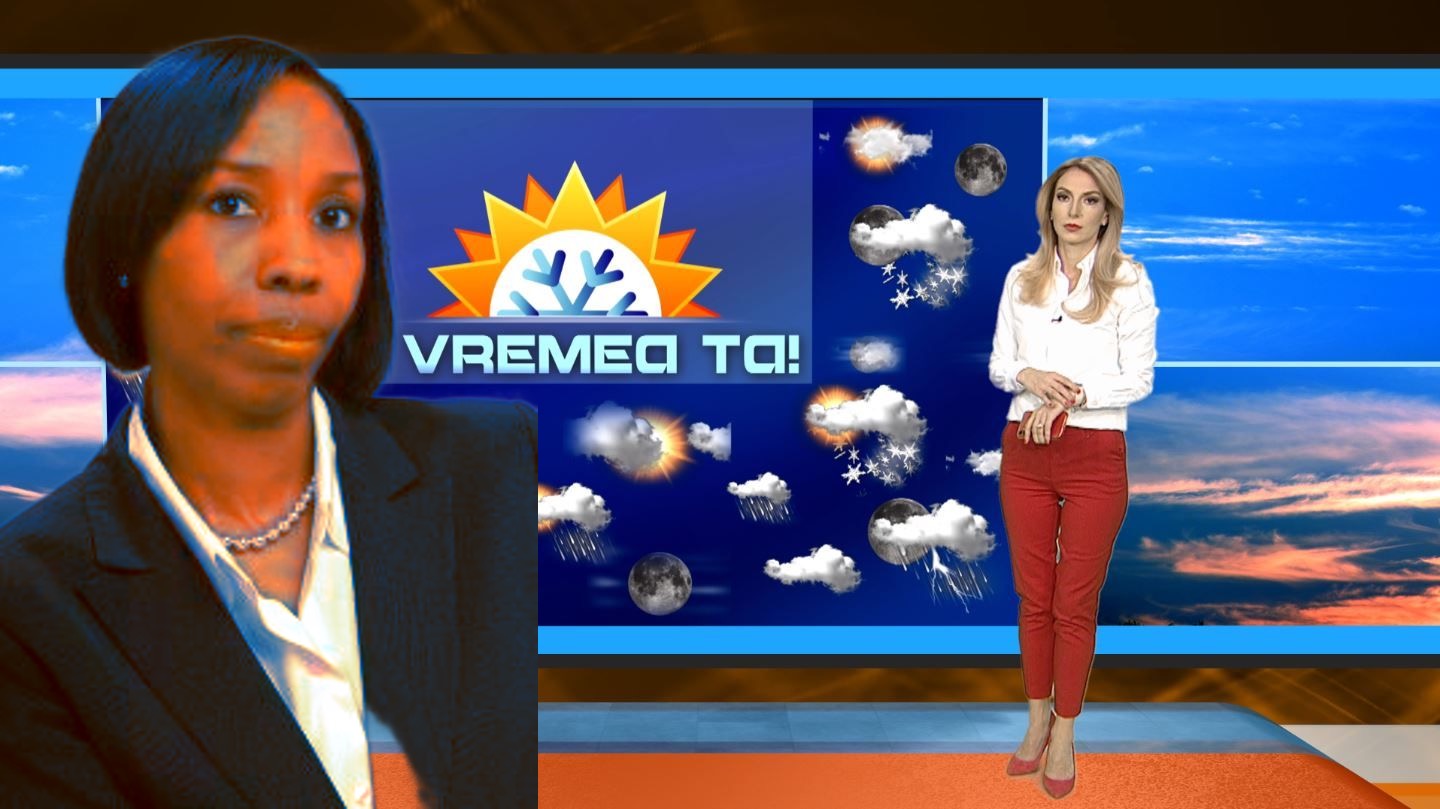} \\

    \includegraphics[width=0.21\textwidth]{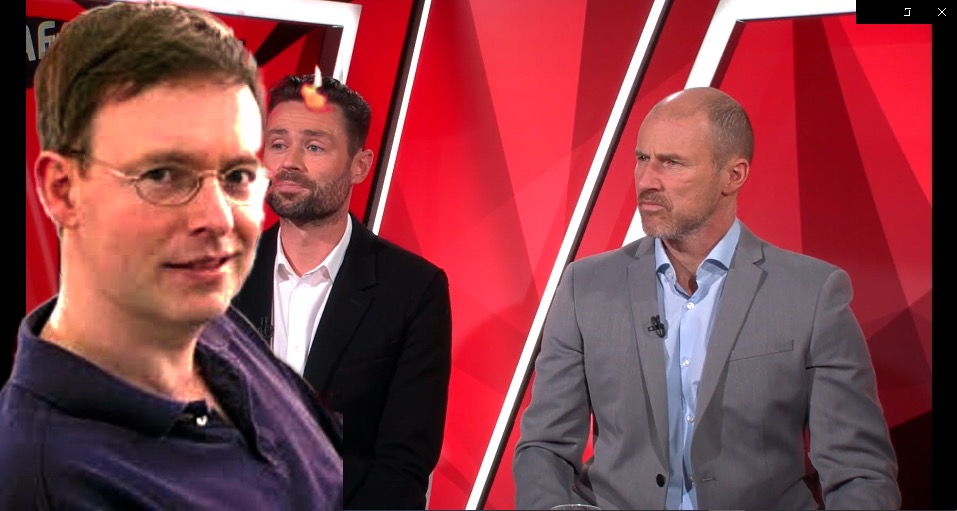}&
    \includegraphics[width=0.21\textwidth]{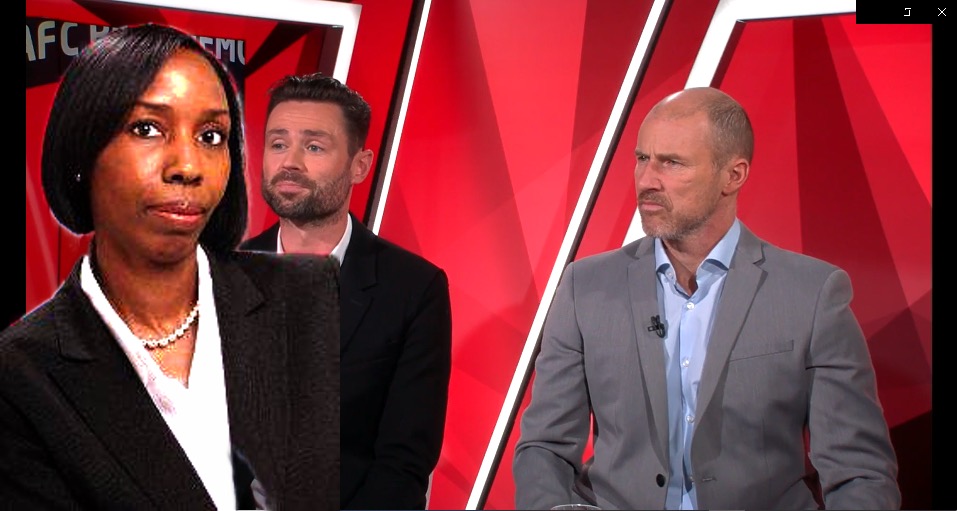}&  
     \includegraphics[width=0.21\textwidth]{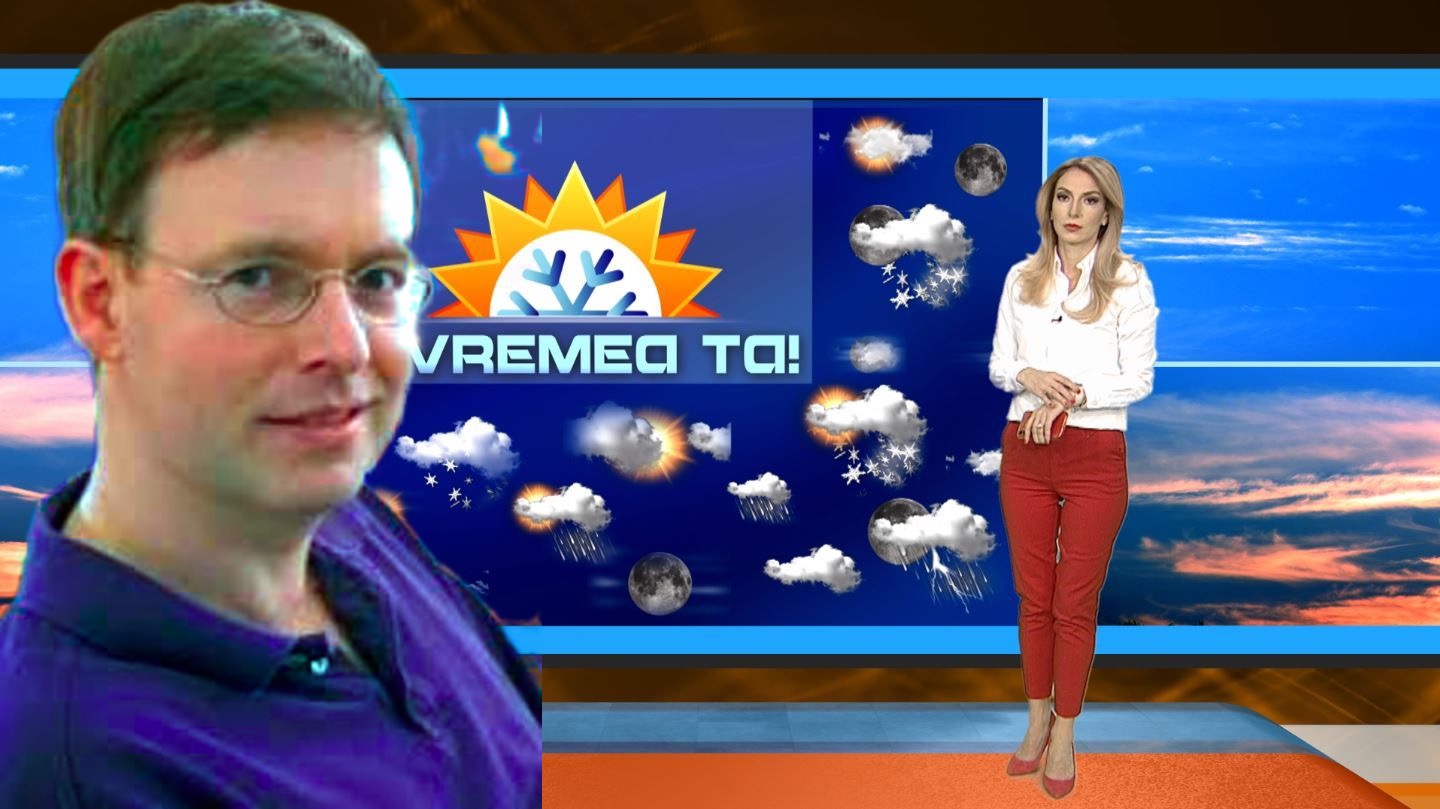}& 
     \includegraphics[width=0.21\textwidth]{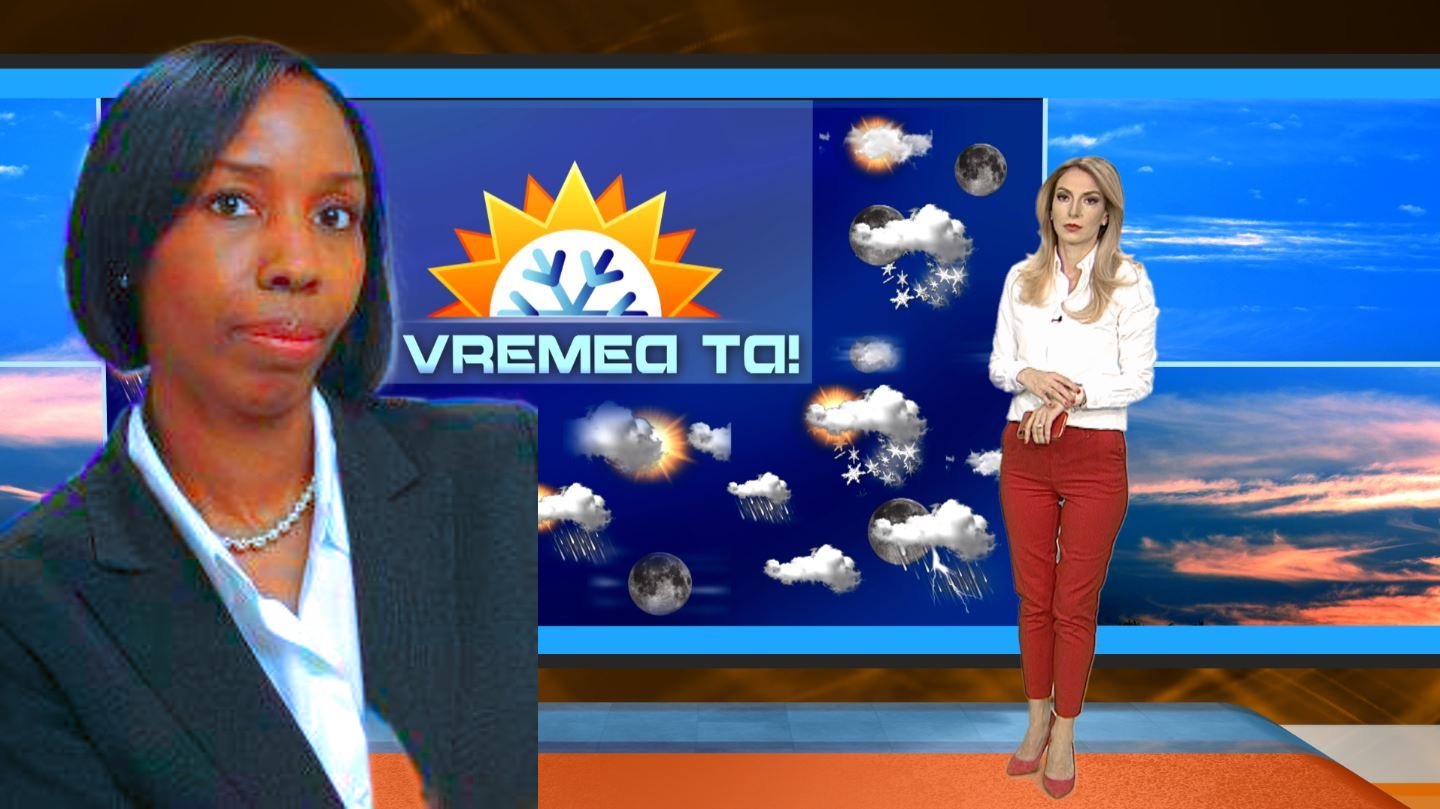} \\  
     
    \includegraphics[width=0.21\textwidth]{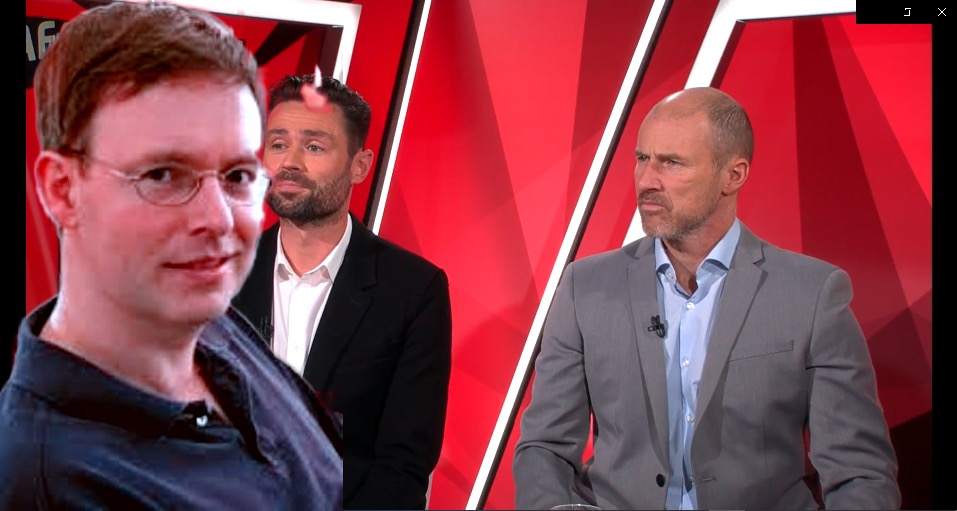}&
    \includegraphics[width=0.21\textwidth]{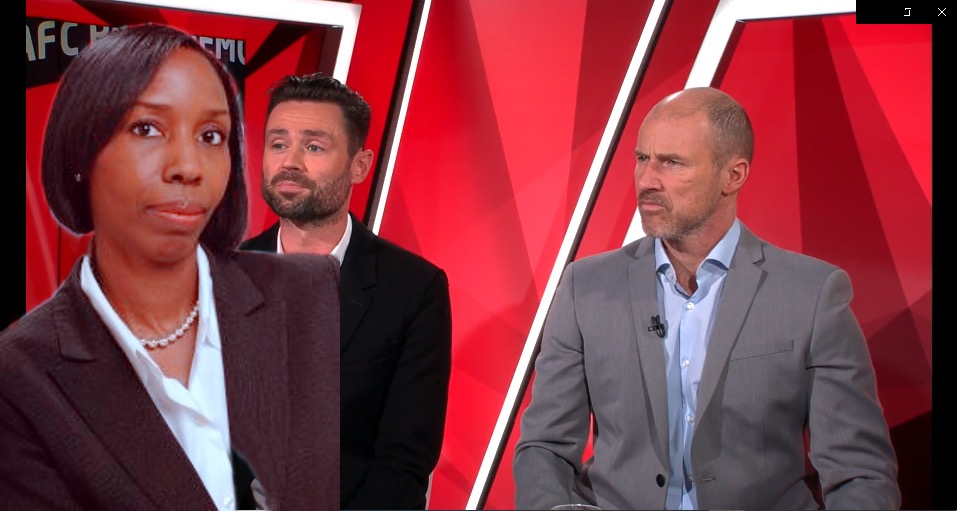}&  
     \includegraphics[width=0.21\textwidth]{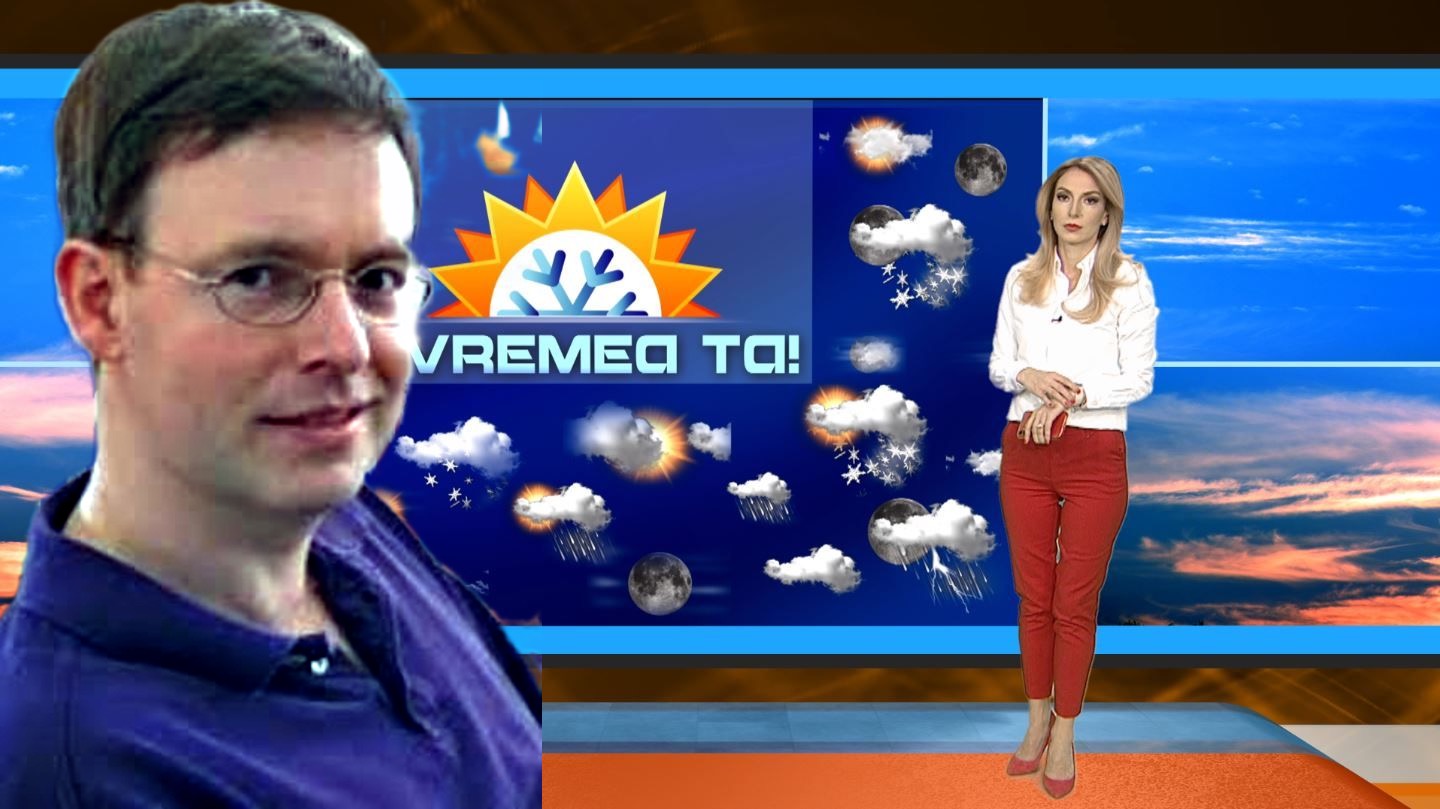}& 
     \includegraphics[width=0.21\textwidth]{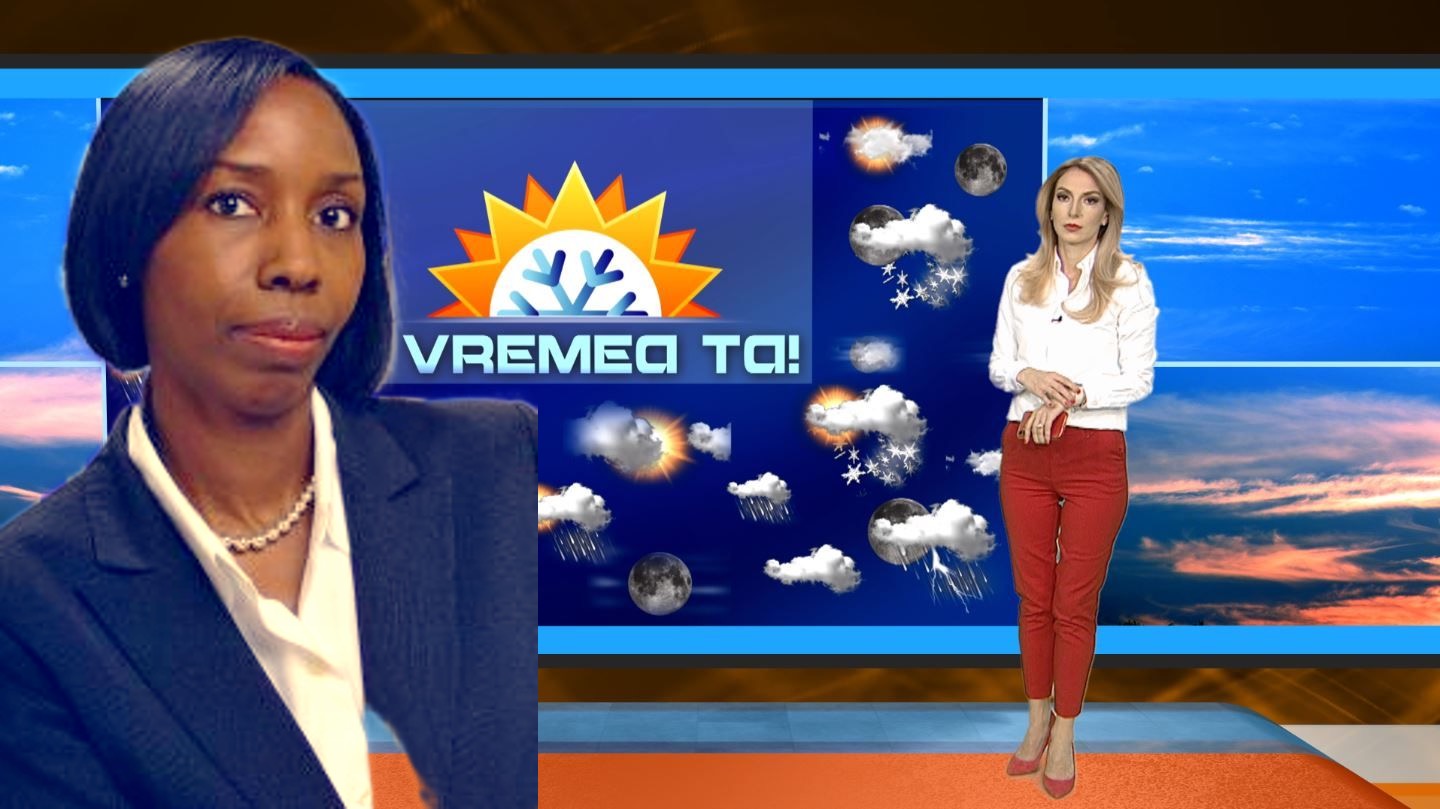} \\

  \end{tabular}
  \caption{Admire application example. From top to bottom, user source images, studio targets, control (no style transfer applied), proposed result, Grogan et al. result \cite{grogan17}, Photoshop result \cite{photoshop}, and Zabaleta \& Bertalmío result \cite{zabaleta21}. The user images are sourced from \cite{LFWTech}, the red studio image was provided by Premiere sports, and the blue studio image was provided by TVR Romania.}
  \label{admire}
\end{figure*}

For this application, our method is applied in the same settings as in the experiment. As a demonstration of the effectiveness of our method for this application, Figure \ref{admire} shows the results of replacing only the color transfer step of this process with the solutions of \cite{grogan17}, \cite{photoshop}, and \cite{zabaleta21}. It can be seen that while the methods of \cite{photoshop} and \cite{zabaleta21} perform acceptably, the proposed method maintains the greatest degree of integrity for the original user image. This is an important quality for this application as users may be alienated if their skin tone and the colors of their clothing are misrepresented. Additionally, one can see that the method of \cite{grogan17} is inappropriate for this application, as the user skin tones are rendered to perfectly match studio background colors.

\subsubsection{Diffusion}
\textcolor{black}{In photography and cinematography, 
an element of style often employed by artists 
is to use optical diffusion filters or particular types of lenses 
aiming to achieve a ``classy'' film look, with unique textures and soft images \cite{davie14}. 
But the use of optical diffusion filters requires careful handling of the camera when the filter is placed behind the optics, and good quality lenses that produce soft pictures are extremely expensive so there are often long waiting lists for their rentals.
This provides an interesting use case for the optical transfer capability of our method, as it can be used to emulate these optical effects at a fraction of the cost, and to allow artists to apply them in post. 
For instance, the artist may select an optical diffusion filter, take a photograph with it and another one without it, and from these two pictures our method will allow to estimate a kernel K that emulates the effect of the optical filter and which can then be applied to any other image or video, without the need to obtain access again to the optical filter.
}

\begin{figure*}[t]
\centering
\begin{tabular}{ccc}

    \textcolor{black}{Source} & \textcolor{black}{Target}  & Result\\

    \includegraphics[width=0.29\textwidth]{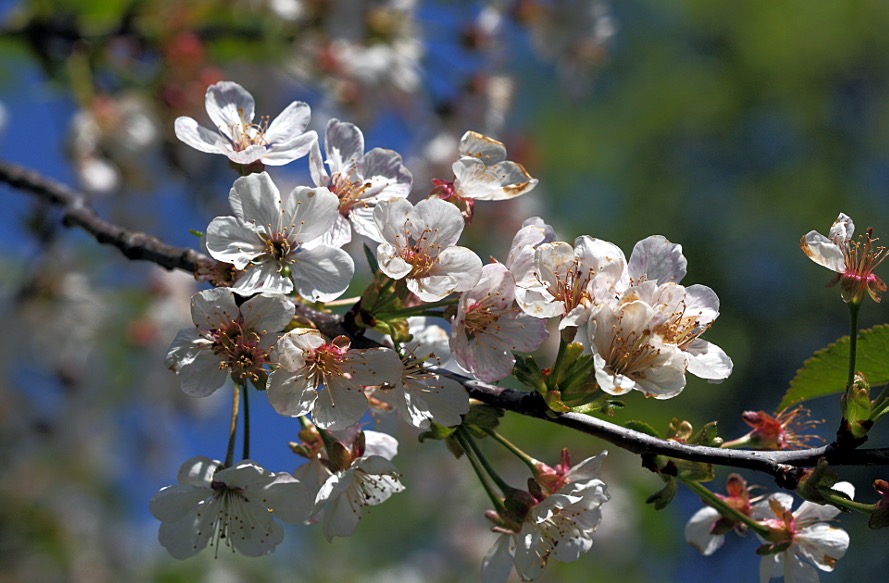}&      
    \includegraphics[width=0.29\textwidth]{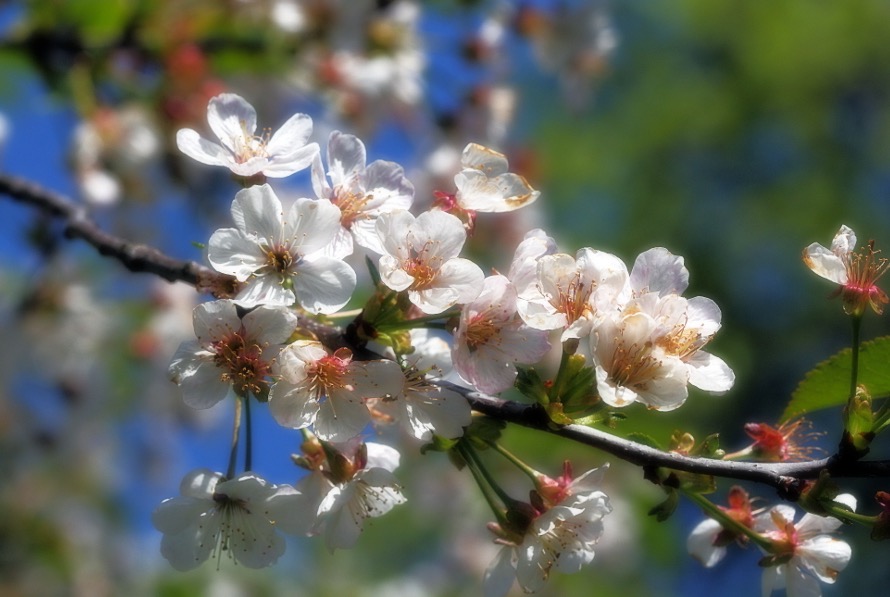}&
    \includegraphics[width=0.29\textwidth]{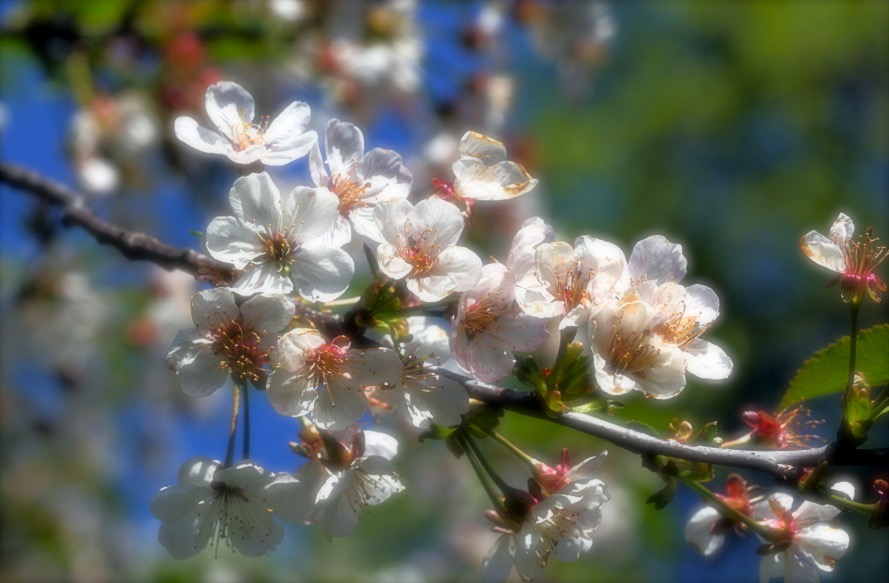} 

\end{tabular}
\begin{tabular}{cc}
    
    Source & \textcolor{black}{Result} \\

    \includegraphics[width=0.4\textwidth]{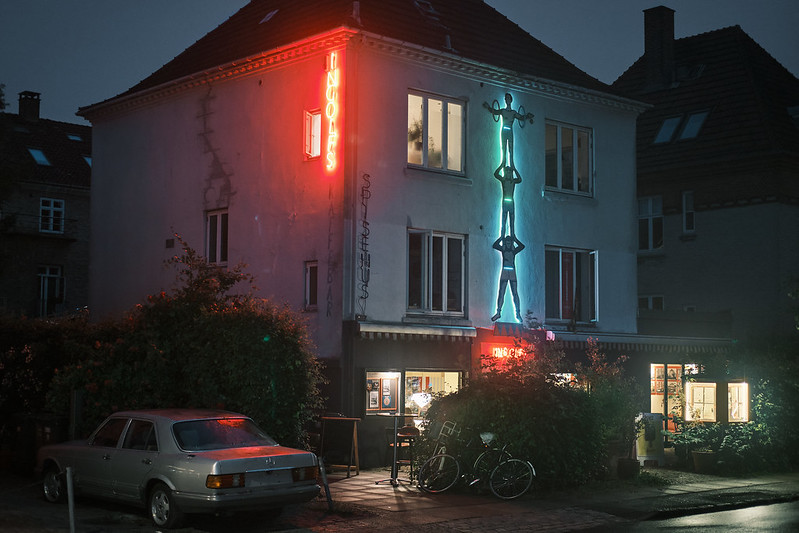}&
    \includegraphics[width=0.4\textwidth]{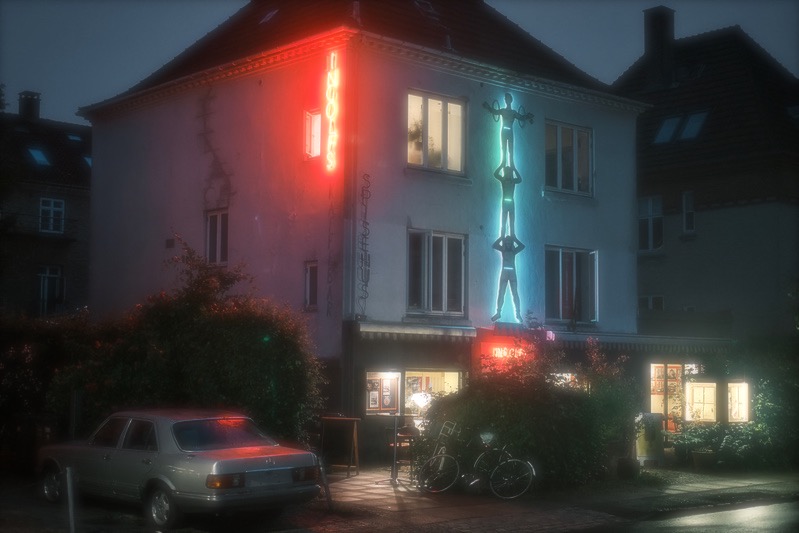} \\

    \includegraphics[width=0.4\textwidth]{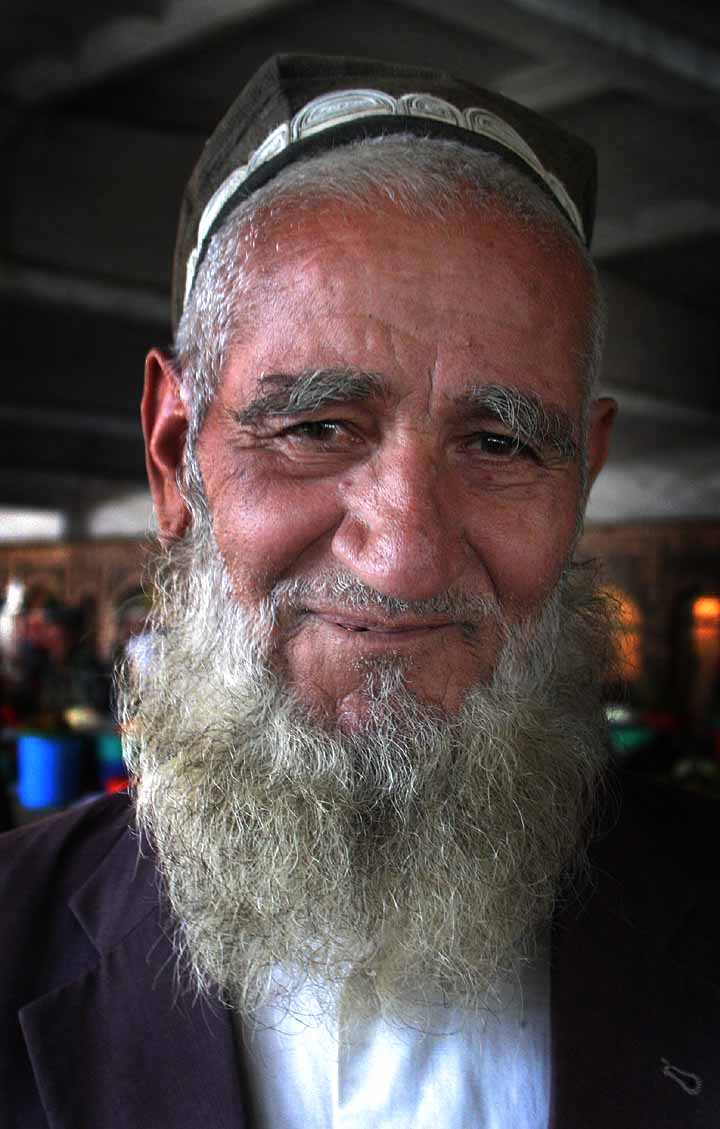}&
    \includegraphics[width=0.4\textwidth]{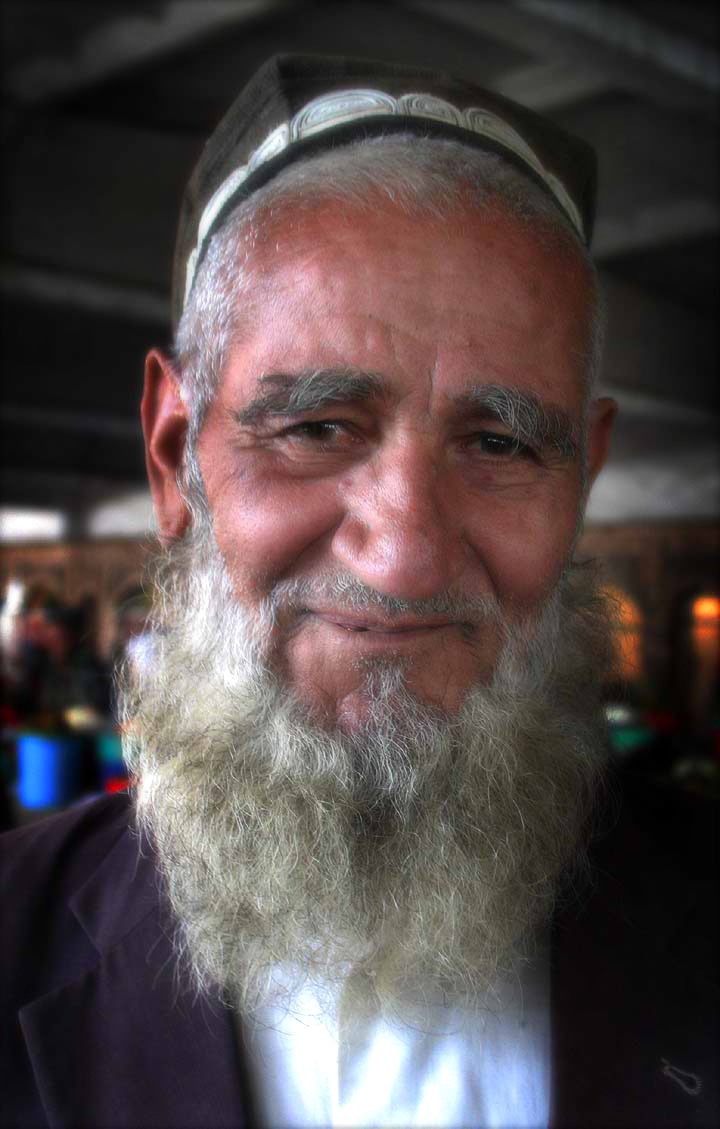} \\
  \end{tabular}
  \caption{Diffusion application examples, where an optical filter is simulated by the method. 
  Both color+luminance and diffusion transfers are made in the top row example. Flower diffusion target photo made available by Hedwig Storch, CC BY-SA 3.0. City source image uploaded to Flickr by Kristoffer Trolle (CC BY 2.0). Second and bottom row portraits sourced from \cite{LFWTech}.}
  \label{diffusion}
\end{figure*}

In Figure \ref{diffusion}, we show a series of examples where, using the method described in section \ref{subsec:optics}, a filter kernel is simulated from source and target images taken of the same subject with and without diffusion filters. One can see that, when used along with the color+luminance transformation, the appearance of the target image can be closely replicated, and that the extracted diffusion filter can be readily applied to any arbitrary image to emulate the effect. We note here though that diffusion effects which are dependent on the depth of field of the images would not be properly simulated by our method.

\begin{figure*}[t]
\centering
\begin{tabular}{cc}
    
     \textcolor{black}{Source} & \textcolor{black}{Target} \\
     \includegraphics[width=0.45\textwidth]{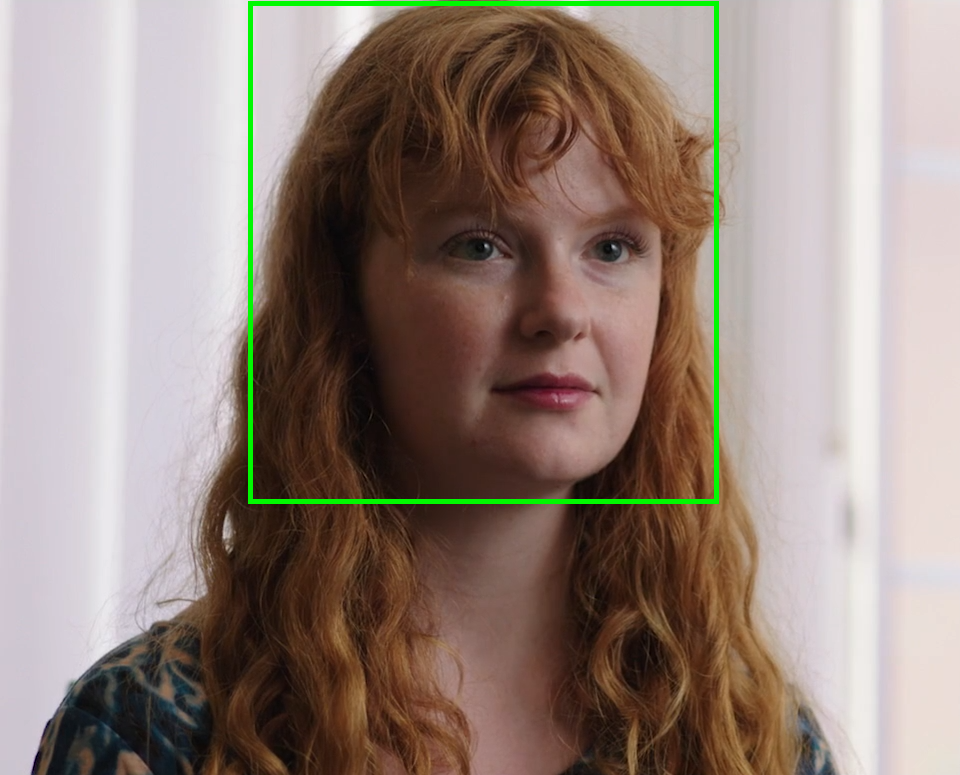}&
     \includegraphics[width=0.45\textwidth]{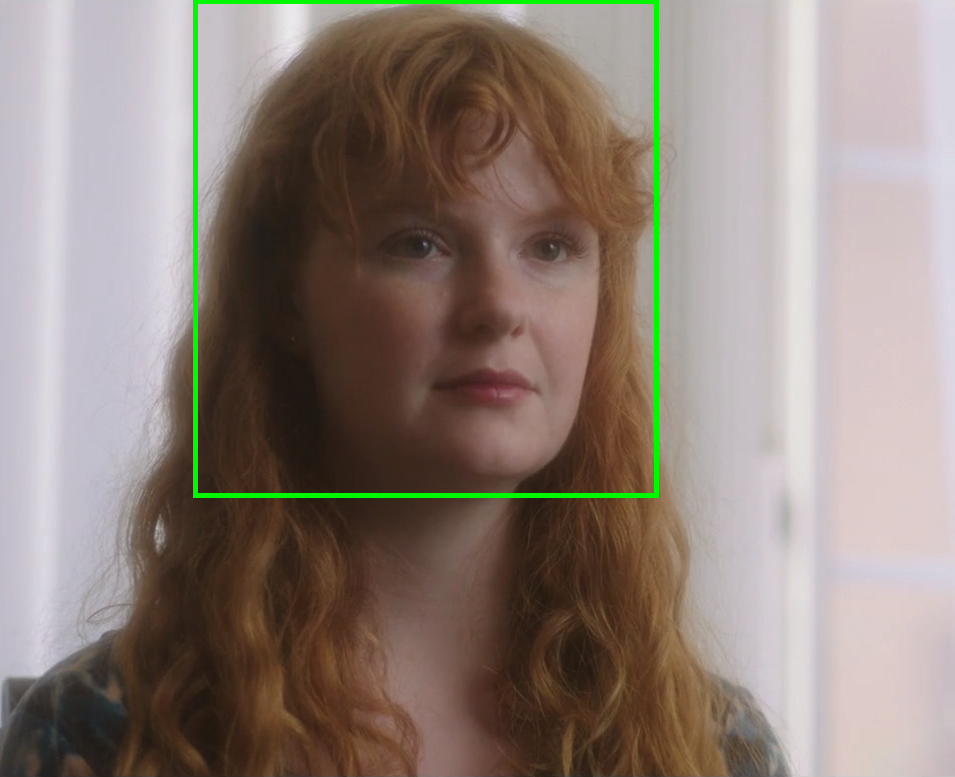} \\
     Result (no crop) & Result (crop) \\
     \includegraphics[width=0.45\textwidth]{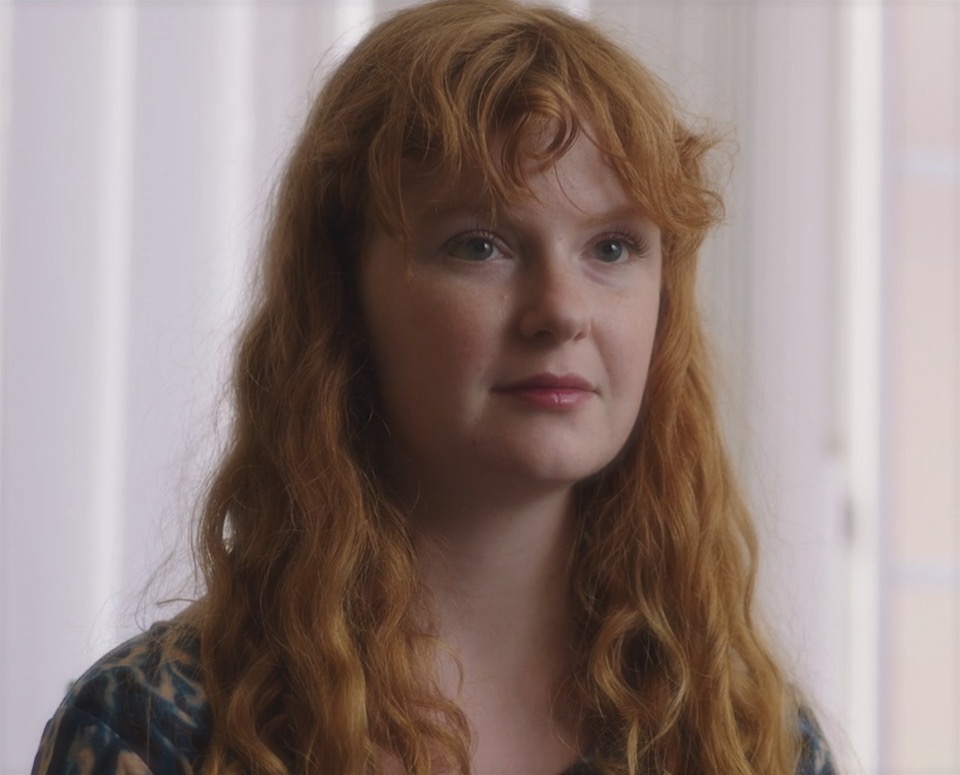}& 
    \includegraphics[width=0.45\textwidth]{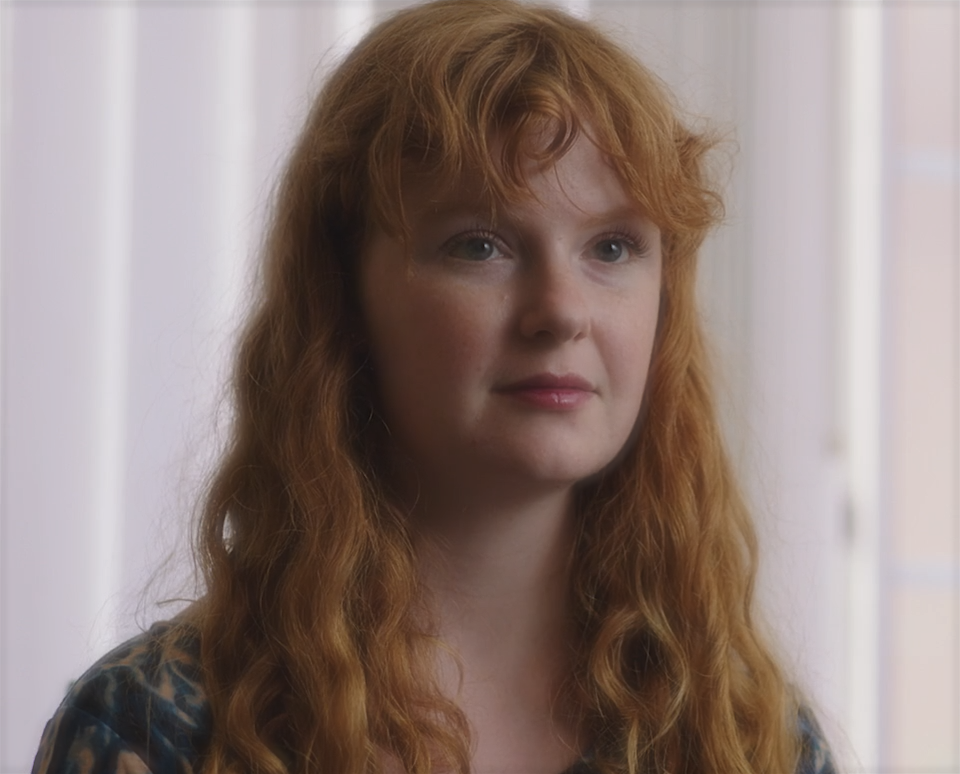}
    \\   
    
  \end{tabular}
  \caption{Example of diffusion simulation using cropping, replicating the effect of nylon stockings. The entire source and target images are used to derive the color transfer, while the diffusion simulation only takes into account the cropped window. 
  Images made available by Colton Davie \cite{davie14}.}
  \label{diffusionCrop}
\end{figure*}

In Figure \ref{diffusionCrop}, we show how cropping can also be useful in this application. Since the source and target images have strong depth of field effects, the effect of the filter is better derived from the flat, in-focus plane of the subject's face, rather than from the image as a whole. In this case, only the diffusion filter is derived from the cropped region, and the entire source and reference images are used to calculate the color+luminance transformation.

\subsubsection{Video}

It isn't trivial to extend an image-based ST method to video sequences, as a simple frame-by-frame application of the method would most probably produce noticeable artifacts, with visible flickering and temporal incoherence even if the changes in content and color between frames are not significant. Another aspect to consider is that it is computationally expensive to calculate the ST transform for each frame independently when it shouldn't change along the same sequence. In order to solve these issues, we have adopted the same approach used in \cite{zabaleta21} for the video extension of our method: we compute the transform for a single representative frame of the source video, \textcolor{black}{encode it as a 3D LUT,} and then apply the same transform to all the frames in the sequence. This guarantees temporal coherence and produces results that are free of flickering artifacts. 

\textcolor{black}{The process of applying 3D LUTs to images is implemented in all professional image/video editing suites and is of very low computational complexity, making this process amenable for real-time implementation in the current production ecosystem. Furthermore, this development allows for the transform to be computed with a lower resolution version of the source and target images and then quickly applied to the original, full resolution source image. The computation time for our full method (color+luminance and diffusion transfers) using a consumer level production workstation (2018 MacBook Pro with a 2.6 GHz intel core i7 processor and 16 GB 2400 MHz DDR4 memory, MATLAB 2021a) is 51 s for a 1920 x 1280 test image, 5.5 s at 50\% resolution, and 2.1 s at 25\%. While sub-sampling can cause a shift in sample statistics, we have found in informal tests that the resulting transformations derived at 25\% resolution have barely perceptible differences in comparison to the full resolution transform.} \textcolor{black}{In summary, by re-scaling and encoding the transform in a 3D LUT we can process HD video sequences in real time after a $\sim$2 s initial processing delay.}


In order to generate the LUT, a 33x33x33 3D table corresponding to an identity transform is passed through the implementation described above, and any operations which alter the image itself are identically applied to the LUT in sequence \textcolor{black}{(aside from the ortho-kurtosis transform which is accurately approximated with a spline function.)} This LUT is then saved in the Adobe {\it .cube} format \cite{cube} such that it can be carried over to any video editing or color correction platform and integrated into standard workflows. In the following link readers can view a series of demo videos where this process is applied: {\href{https://drive.google.com/drive/folders/15nKBljq0CCA9pm6i1fcISFSUvuYoBhac?usp=share_link}{LINK}}. One can see from the videos that the LUT approach allows for a video style transfer to be accomplished which is free from temporal artifacts. The diffusion transfer process can also be applied to video sequences by deriving a single PSF kernel and applying it to all frames in the sequence via convolution.

\subsubsection{Demo software}

Readers can access a \textcolor{black}{standalone} GUI implementation of the proposed method \textcolor{black}{and the MATLAB source code} through the following {\href{https://github.com/trevorcanham/dCoupST}{LINK}}. This implementation allows users to select the source and target images, toggle features to be transferred to specific color channels, perform cropping on the source and target input, encode the transform as a LUT, and batch process images for a particular transfer.

\subsection{Method limitations}
In order to supplement our discussion on usability, we offer a number of useful guidelines here for the best use of our method and highlight its limitations.

One artifact that can occur is the exacerbation of banding or other compression artifacts.
For this reason, we recommend the use of high bit depth, uncompressed images for best results.
Another is the clipping or crushing of detail in order to fit the target statistics.
To avoid this, we recommend working in a non-destructive space, where out-of-range values are preserved and clipping does not occur so that these details can be reintegrated back into the image at the user's discretion. 
\textcolor{black}{In addition,} the cropping function can be used as in Figure \ref{crop} to better align the source and target images to derive a compatible transfer.

Another limitation which we see highlighted in the comparison to \cite{grogan17} is that the color appearance of dominant colors in the target image is not always well preserved when integrating with the color details of the source image.
\textcolor{black}{We found our best results in this regard came from the use of the IPT color space. However, one might encounter scenarios where an alternative representation like RGB or HSL may be preferable.
In addition to modulating the color space, some applications may benefit from modulating the number of features transferred per color channel. For instance in our default configuration we recommend transferring the higher order moments only in the luminance channel, but it may be beneficial in some cases to also transfer these features on the chroma channels.}

As discussed previously, there exists an image compatibility factor in ST, where only a subset of possible source/target combinations will be acceptable. The range of this subset for a given source image depends on the degree to which it approaches its gamut boundaries, as well as its distance in feature space from a potential target. For example, common transfer mismatches include target images that are significantly darker or brighter than the source, causing details to be obscured via clipping and crushing. \textcolor{black}{Related to compatibility,} we only recommend the use of  spectral transfer \textcolor{black}{in the constrained fashion (via PSF estimation) described in section~\ref{subsec:optics}.} 
\textcolor{black}{When used outside of these constraints,} we found that this feature can vary greatly between images in a way that is difficult to compare visually and \textcolor{black}{relative to color transfers,} observers have a limited tolerance for the results - meaning only very subtle adjustments to the source image with respect to this feature are acceptable. 

\section{Conclusion}
\label{sec:conclusions}

We have proposed a photo-realistic style transfer method for image and video that is based on vision science principles and on a recent mathematical formulation for the deterministic decoupling of sample statistics. This method extends traditional approaches by increasing the number of features transferred, including higher order statistical moments which correspond to important cues for visual adaptation processes and serve as impactful and interpretable feature descriptors. In order to validate the method we make qualitative comparisons against state-of-the-art methods via a psychophysical experiment, and to aid in its adoption we offer insight into its usability for different applications.

We 
propose \textcolor{black}{as future work} a tool designed to give users the ability to adjust the statistical moments addressed here independently, without the need of a target image. Due to their decoupled and localized nature 
(with respect to the tone scale), we believe it would be useful for colorists to control them directly, particularly in the context of emerging formats where the tone scale range is extended. \textcolor{black}{The challenge in doing so, however, is in computing the feature adjustment such that users can receive real-time feedback from the controls.} Another interesting and challenging extension of this work would be to automatically characterize the optical PSF of a single image, without an additional reference. In more general terms, we want to characterize (and transfer) the {\em sharpness/smoothness} style of the images in a robust and visually meaningful way. Along the same line, it would be a valuable addition to characterize and reproduce the {\em texture} features, such as different types of noise and film grain. \textcolor{black}{We also propose the use of the method to gradually transition between displayed content which trigger varying states of adaptation in the viewer, as frequency and abrupt re-adaptation may be a cause of display fatigue.} Finally, by using a randomly distributed noise image as source, the method could be used to apply the statistical and power spectrum properties of a target image, which could then be used to pre-adapt observers before seeing the image itself (for example, in a psychophysical experiment) or to be used as a background for image presentation.

\section*{Acknowledgments}
The authors would like to acknowledge Liqin Cao for providing their method implementation, and the many observers who participated in the psychophysical experiments described here.

\appendix

\section{Psychophysical methods}
\label{subsec:psych_methods}
Our experimental image set (Figure \ref{sources}) represents a series of highly challenging cases for style transfer. First, all of the images are highly chromatic and stylized, and include scenes with artificial lighting like dance clubs and broadcast studios. Second, all images are compressed to 8-bit per color channel detail. These features make them susceptible to gamut clipping and banding and highlight the primary challenges of working with graded images ``in the wild.''

Also reflecting the qualities of an ``in the wild'' evaluation scenario, this experiment was conducted online using PsychoJS \cite{pierce19} and was distributed using the Pavlovia platform. While users were asked to run the experiment in a darkened viewing environment on a full-sized computer monitor (anything but a mobile phone or tablet), no other constraints were placed on the quality or calibration state of the viewing equipment. Observers were first presented with a series of instructions describing the task and then with a series of slides where the source and target images were shown at the top of the display and the results of two competing methods were displayed below. Observers were asked simply to select which of the two images they preferred within the context of the style matching application. They were presented with every unique style/target combination of a set of 5 images (20 unique pairs), which were compared between four methods (proposed, \cite{photoshop}, \cite{grogan17}, \cite{zabaleta21}), resulting in a total of 120 evaluations. This task took observers between 10 and 20 minutes to complete which was comfortably within the acceptable range considering experimental fatigue. In the end, a total of 19 observers participated in the experiment, including expert and non-expert observers from a range of occupations with varied affiliations, nationalities, gender and age groups. A full summary of the results is shown in Table \ref{tab:results}.

\begin{table}[!h]
\caption{Psychophysical experiment results, where the percentage which the method on the column label was preferred over the method on the row label, averaged over 20 image combinations and 19 observers.}
\label{tab:results}
\resizebox{1\textwidth}{!}{%
\begin{tabular}{c | c | c | c | c}
\hline
Methods &  Photoshop \cite{photoshop} & Grogan et al. \cite{grogan17} & Proposed &  Zabaleta and Bertalmío \cite{zabaleta21} \\\hline
 Photoshop \cite{photoshop} & N/A & 68\% & 62\% & 55\% \\\hline
Grogan et al. \cite{grogan17} & 32\% & N/A & 43\% & 38\%\\\hline
Proposed & 38\% & 58\% & N/A & 41\%\\\hline
 Zabaleta and Bertalmío \cite{zabaleta21} & 44\% & 63\% & 59\% & N/A \\\hline
\end{tabular}}
\end{table}

\section{The Nested Normalization (NeN) method for deterministic decoupling of features} 
\subsection{Method's foundations}
\label{subsec:method_foundations}
Let $\S$ be a set of $M$ given {\em features}, i.e., some differentiable real functions $f_j: D_\S \subset \R^N\rightarrow \R$. In our case, they are applied to discretized images, that we reorder as vectors $\xv\in \R^N$. For convenience, we will also use $\fv(\xv)\in\R^M$ to denote the ordered features, arranged in a vector in the feature domain in $\R^M$. Thus, $\fv:D_\S \subset\R^N\rightarrow\R^M$ is a lossy mapping (typically $M<<N$) representing the feature extraction applied to the images. In addition to being differentiable, we will make the following assumptions on the given features~\cite{Arxiv2022}:
\begin{enumerate}
    \item{$f_j(\xv) = \frac{1}{N}\sum_{n=1}^{N}{\left[\mv_j(\xv)\right]_n}$, where $\mv_j(\xv)$ is a shift-invariant or shift-equivariant mapping. This gives to these features the character of global, playing the role of sample statistics of the analyzed image.}
    \item{Features are not trivially coupled, i.e., no non-trivial function of them is a constant.}
    \item{Their gradients are vector fields jointly fulfilling the Frobenius condition, which allows defining integral submanifolds (that we term {\em invariance submanifolds}) whose tangent planes are the linear span of the features' gradients, at each point~\cite[Chapter 6]{Spivak}}.
    \item{All local extrema (maxima and minima) of the $f_j$'s are also global extrema.}
    \item{\textcolor{black}{The aggregated basins of attraction of saddle points of the features in $\S$ have a lower-than-$N$ dimension.}}
\end{enumerate}
Under the previous assumptions we define a normalization operation $\hat x_{\S}(\xv;\vv^{ref}):\R^N\rightarrow\R^N$, where $\vv^{ref}$ is an $M$-dimensional vector of 
algebraically compatible {\em reference values} for the features, such that: 
\begin{enumerate}
    \item{$\fv(\hat x_\S(\xv;\vv^{ref})) = \vv^{ref},\forall \xv \in D_\S$}
    \item{$\hat x_\S(\xv;\vv^{ref})$ is obtained by moving along {\em any} trajectory on the invariance submanifold ${\cal I_\S(\xv)}$ whose tangent plane is locally defined by the gradients $\nabla f_i(\xv)$, with the only two requirements of (i) taking the right sign direction for each gradient, and (ii) stopping where condition 1. holds.}
\end{enumerate}
For being able to apply such normalization we are forced to assume that neither $\xv$ is a critical point nor the {\em reference manifold} ${\cal R}_\S(\vv^{ref}) = \fv^{-1}(\vv^{ref})$ has critical points for the features in $\S$. 
\textcolor{black}{Because of assumption number 5. above, adding a tiny (infinitesimal, ideally) random perturbation $\epsilon$ to an arbitrary input vector $\xv_0$ (we set $\xv = \xv_0 + \epsilon$) takes it out from any possible basin of attraction of a saddle, with probability one.} Then, under previous conditions, first,  $\hat x_\S(\xv;\vv^{ref})$ exists and is unique in its connected domain; and, second, $\hat g(\xv) = g(\hat x_\S(\xv;\vv^{ref}))$ is decoupled from all features in $\S$, where $g(\xv)$ is any differentiable function non-trivially-coupled with the given features (see~\cite{Arxiv2022} for a proof).   
More formally, we can write:
\begin{eqnarray}
    \hat x_\S(\xv;\vv^{ref}) & = & {\cal I_\S(\xv)} \cap {\cal R}_\S(\vv^{ref}) \label{eq:normalization}  \\
    \hat g(\xv) & = & g(\hat x_\S(\xv;\vv^{ref})) 
    \label{eq:decoupling_definition}  \\
    \nabla \hat g(\xv) \cdot \nabla f_k(\xv) & = & 0,\,\, 
    k = 1,\dots,M, \forall \xv\in D'_\S, \nonumber 
\end{eqnarray}
where $D'_\S$ represents the common domain of the features in $\S$, $D_\S$, after the critical points and the basins of attraction of the saddle points have been removed from that domain (the latter through the ``perturbation trick" explained above), and the solution of Eq.~\eqref{eq:normalization} is obtained in the connected domain where $\xv$ belongs.    

The intuition behind this result comes from realizing that $\hat g(\zv) = \hat g(\xv)$ for all $\zv \in {\cal I}_\S(\xv)$ (because of Eqs.~\eqref{eq:normalization} and \eqref{eq:decoupling_definition}), a property that motivates naming ${\cal I}_\S(\xv)$ the {\em invariance} submanifold. Because of $\hat g$ being constant on the whole submanifold ${\cal I}_\S(\xv)$, its gradient $\nabla \hat g(\xv)$ must necessarily be orthogonal to that submanifold everywhere. Which, in turn, implies that $\nabla \hat g(\xv)$ is orthogonal to the tangent planes of ${\cal I}_\S(\xv)$, and, thus, to all gradients spanning those tangent planes, i.e., $\nabla \hat g(\xv)$ is orthogonal to $\{\nabla f_k(\xv)\}_{k=1}^{M}$ everywhere.

\subsection{Generic Algorithms for Analysis}
\label{subsec:gral_algs_for_analysis}
Once we know how to decouple from a given set $\S$ of features a new arbitrary feature, it is immediate to establish a 
hierarchical method for, given some features, take the first one as it is, then decouple the second from the first, then decouple the third from the two previously decoupled ones, and so on, resulting in a set of mutually decoupled features - see Algorithm~\ref{alg:hier_approach}. This procedure is what we termed in~\cite{Portilla:ICIP:2018} the Nested Normalization(s) (NeN) algorithm. In Algorithm~\ref{alg:nested_normalization_broad_path} we add more detail, 
by including the precise decoupling procedure explained in subsection~\ref{subsec:method_foundations}.
\begin{algorithm}
\begin{algorithmic}[1]
\REQUIRE Coupled features
$\{f_j\}$, 
Reference values
$\{v_j^{ref}\}$, $j = 1,\dots, M.$
\STATE {\bf Initialization:} $\g_1=f_1$
\FOR {$k=1$ to $M-1$}
    \STATE $\g_{k+1} \leftarrow \text{decouple}(f_{k+1},\{\g_i, i = 1\dots k\})$   \label{alg_step:multifeature_decouple}
\ENDFOR
\RETURN Decoupled features $\{\g_j, j = 1,\dots, M \}$
\end{algorithmic}
\caption{Nested Normalization: A hierarchical decoupling approach.}
\label{alg:hier_approach}
\end{algorithm}
\begin{algorithm}
\begin{algorithmic}[1]
\REQUIRE $\{f_j;v_j^{ref} \,\,j = 1,\dots, M \}$
\STATE {\bf Initialization:}  $\g_1(\xv) = f_1(\xv)$ 
\FOR {$k=1$ to $M-1$}
    \STATE $\fv_k(\xv) = [f_j(\xv)],\, \vv_k^{ref} = [v_j^{ref}],\, j = 1,\dots,k$
	\STATE $\RM_k = \fv_k^{-1}(\vv_k^{ref})$
    \STATE $\hat\S_k\, = \,\{\hat f_j,j=1\dots k\}\,\,\,\,$         [*\textcolor{gray}{{\em relaxed:}} $\S_k\, = \,\{ f_j,j=1\dots k\}$]
    \STATE Compute (ODEs) \,$\hat{\xv}_k(\xv; \vv_k^{ref}) = \I(\xv,\hat\S_k)\cap\RM_k\,\,\,\,$      [*\textcolor{gray}{{\em relaxed: }} $\I(\xv,\S_k)\cap\RM_k$] \label{alg_step:multifeature_decouple_}
    \STATE $\g_{k+1}(\xv) = f_{k+1}(\hat{\xv}_k(\xv))$   
    \label{decoupled_from_normalized}
\ENDFOR
\RETURN $\{\g_j, j = 1,\dots, M \}$
\end{algorithmic}
\caption{Nested Normalization, Analysis - broad path.}
\label{alg:nested_normalization_broad_path}
\end{algorithm}
%
At every step of this algorithm, the reference manifold ${\cal R}_k$ becomes a subset of the previous one, 
resulting in a nested structure: 
\[
\RM_{M-1} \subset \dots \subset \RM_1 \subset \RM_0 = D_\S \subset \R^N.
\]
This particular structure also makes that $\hat\fv_k^{-1}(\vv_k^{ref})= \fv_k^{-1}(\vv_k^{ref})=\RM_k$. 

The problem with the previous algorithm in its original form is 
the difficulty of dealing with the invariance submanifold ${\cal I}$ resulting from the previously computed decoupled features $\hat \S_k$.
In~\cite{Arxiv2022} we prove that, if the gradients of the original features fulfill the Frobenius condition, and the resulting features $\hat \S_k = \{\g_j, j = 1\dots k\}$ also fulfill the Frobenius condition, then the resulting set obtained by using $\S_k$ (the original features) instead of $\hat \S_k$ in Steps 5 and 6 of the algorithm 
will be the same as the result using the original version of the algorithm. 
We term this change the relaxation of the broad-path algorithm (see bracketed expressions in Alg.~\ref{alg:nested_normalization_broad_path}). 

An alternative way to implement the previous method in its relaxed version is the {\em  narrow-path} algorithm. It
 consists in setting a path made of a sequence of 1-D curves, each following a single gradient $\nabla f_k$ projected each time onto the orthogonal span of the previous gradients (thus not affecting the previously fixed reference values). That projection is equivalent, in this case, to projecting $\nabla f_k$ onto the reference manifold $\RM_{k-1}$,
 see Algorithm~\ref{alg:nested_normalization_narrow_path}.
\begin{algorithm}
\begin{algorithmic}[1]
\REQUIRE
$\xv \in  D'_\S$, \,
$\{f_j, v_j^{ref}, \,\,j = 1,\dots, M \}$
\STATE {\bf Initialization:} $\g_1 = f_1$; ${\cal R}_0 =  D'_\S$; $\hat \xv_0(\xv) = \xv$
\FOR {$k=1$ to $k=M-1$}
\label{step:project_gradient}
	\STATE $\yv_k(0) = \hat \xv_{k-1}(\xv)$
    \STATE $\gv_k(t) = P_{\RM_{k-1}}(\nabla f_k(\yv_k(t)))$ 
    \STATE Integrate $\gv_k(t)$ until $f_k(\yv_k(\hat t_k)) = v_k^{ref}$
	\STATE $\hat \xv_k(\xv)=\yv_k(\hat t_k)$ \label{step:reference_value_reached}
	\STATE $\g_{k+1}(\xv)=f_{k+1}({\hat \xv_k})$ \label{step:decoupled_from_normalized}
\ENDFOR
\RETURN $\{\g_j(\xv), j = 1,\dots, M \}$, $\hat x_{M-1}(\xv)$
\end{algorithmic}
\caption{Nested Normalization, Analysis - narrow path.}
\label{alg:nested_normalization_narrow_path}
\end{algorithm}
As explained in ~\cite{Arxiv2022}, when Frobenius conditions apply to the gradients of the original features, Algorithm~\ref{alg:nested_normalization_narrow_path} is mathematically equivalent to Algorithm~\ref{alg:nested_normalization_broad_path} in its relaxed version. However, the former has properties that make it appealing in some practical circumstances, like not requiring explicit calculation of the decoupling features or their gradients. The involved ODEs can be numerically integrated, only requiring the calculation of the original features' gradients. These are numerically projected each time onto the orthogonal span of the previous gradients and then integrated using efficient algorithms (e.g., Runge-Kutta with adaptive step). This provides very desirable flexibility, especially when lacking analytical solutions for the involved ODEs.
Finally, note that Algorithm~\ref{alg:nested_normalization_narrow_path} can be used even when Frobenius condition does not hold on the output "decoupled" features' gradients, 
obtaining higher-order "quasi-decoupled" features, that 
still provide perfect decoupling for images within the reference manifolds, and approximate decoupling outside of them. 
In this paper we combine the three explained algorithms (broad-path in its two versions, and narrow-path) to, strictly or approximately, decouple the involved features, as explained in sections 4.3 and 4.4. 

\subsection{De-normalizing for Transfer}
Step 3 in Alg.~\ref{alg:CFA} is the de-normalization procedure, see Algorithm~\ref{alg:NeN_reversed} , used for the feature transfer. 
It is very similar to the normalization itself, but in reversed order, and, instead of imposing a set of reference values to the decoupled features, it imposes a set of desired values (denoted with the super-index "des" in $\fv^{des}$), e.g., obtained analyzing a target image, in the style transfer case.
De-normalization Algorithm~\ref{alg:NeN_reversed} is presented here in its narrow-path version, but Algorithm~\ref{alg:nested_normalization_broad_path}, in its relaxed version, can be likewise reversed (omitted here for brevity).

\begin{algorithm}
\begin{algorithmic}[1]
\REQUIRE ${\hat \wv}_{M-1}\in \RM_{M-1}\subset D'_\S$, $\fv^{des}\in Rg(\hat\fv)$,
\STATE {\bf Initialization:} $\hat z_M = \hat w_{M-1}$
\FOR {$k=M$ to $1$}
	\STATE $\yv(0) = \hat z_{k}$ 
    \STATE Compute $\gv_k = P_{\RM_{k-1}}(\nabla f_k)$
	\STATE Follow $\gv_k$ until $f_k(\yv_k(t)) = v_k^{des}$
	\STATE $\hat z_{k-1}=\yv(t)$
\ENDFOR
\RETURN $\zv_f = \hat z_0$
\end{algorithmic}
\caption{De-normalization (narrow path).}
\label{alg:NeN_reversed}
\end{algorithm}

An important advantage of using decoupled vs. coupled features in a feature transfer frame is that, as proven in ~\cite{Arxiv2022}, the joint range of the decoupled features is also decoupled: $Rg(\hat\fv) = \prod_{j=1}^{M}{Rg(\g_j)}$,
which allows imposing arbitrary features' values (each within their fixed valid range), e.g., for feature-based creative processing or synthesis~\cite{MMSP2020}. 


\bibliographystyle{siamplain}
\bibliography{references}

\end{document}